\documentclass[aps,prb,nolongbibliography,notitlepage,twocolumn,superscriptaddress,nofootinbib]{revtex4-2}

\usepackage{amsmath}
\usepackage{amssymb}
\usepackage{bm}
\usepackage{multirow}

\usepackage[dvipsnames]{xcolor}
\usepackage{braket}
\usepackage{graphicx}
\usepackage{CJKutf8}
\usepackage{mathrsfs}
\usepackage{empheq}
\usepackage{etoolbox}
\usepackage{color}
\usepackage{epstopdf}
\usepackage{upgreek} 
\usepackage{mathtools}
\usepackage{soul}
\usepackage[colorlinks=true]{hyperref}
\AfterEndPreamble{\let\SMaddcontentsline\addcontentsline
  \renewcommand{\addcontentsline}[3]{}}   
\hypersetup{
    colorlinks=true,
    linkcolor=blue,
    citecolor=blue,
    urlcolor=blue
}

\def\balpha{{\bm{\alpha}}}
\def\bxi{{\bm \xi}}

\def\bk{{\bm k}}

\def\bv{{\bm v}}

\def\br{{\bm r}}

\def\bR{{\bm R}}

\def\ba{{\bm a}}

\def\bb{{\bm b}}
\def\bL{{\bm L}}
\def\bd{{\bm d}}          
\def\bc{{\bm c}}          
\def\bK{{\bm K}}          
\def\bkappa{{\bm \kappa}} 
\def\bbeta{{\bm \beta}}   
\def\bmeta{{\bm \eta}}    

\begin{document}

\title{Resonantly interacting $2e/3$ anyons: a controlled microscopic theory of anyon superconductivity}

\author{Qingchen Li}
\thanks{These authors contributed equally to this work.}
\affiliation{Department of Physics, Harvard University, Cambridge MA 02138, USA}

\author{Pavel A. Nosov}
\thanks{These authors contributed equally to this work.}
\affiliation{Department of Physics, Harvard University, Cambridge MA 02138, USA}

\author{Zhaoyu Han}
\affiliation{Department of Physics, Harvard University, Cambridge MA 02138, USA}

\author{Si-Yuan Chen}
\affiliation{Department of Physics, Harvard University, Cambridge MA 02138, USA}

\author{Taige Wang}
\affiliation{Department of Physics, Harvard University, Cambridge MA 02138, USA}
\affiliation{Materials Research Laboratory, Massachusetts Institute of Technology, Cambridge, MA 02139, USA}

\author{Eslam Khalaf}
\affiliation{Department of Physics, Harvard University, Cambridge MA 02138, USA}

\begin{abstract}
 We develop a microscopic theory of anyon superconductivity that connects electronic interactions in a lightly doped Laughlin state to the universal physics of a resonantly interacting anyon gas. Using geometric quantization of Laughlin quasiholes, we derive the effective many-body Hamiltonian of dilute charge-$2e/3$ anyons, including their valley structure, dispersion, and statistical and short-range interactions, together with an explicit map between anyon and electron observables. This reduces the dilute many-body problem to a few parameters fixed by few-body data computable, in principle, from few anyon numerics. In analogy with resonant Fermi and Bose gases, we organize the dilute anyon problem around a two-body binding resonance and its relevant and marginal perturbations. For a three-valley gas of charge-$2e/3$ anyons, we identify a two-body binding resonance whose long-distance theory is governed by an Aharonov--Casher anyon Hamiltonian where the magnetic field is the statistical Chern--Simons field whose ground state is exactly solvable. Using the ground state wavefunction, we demonstrate Cooper-pair off-diagonal long-range order directly in electron coordinates and determine the pairing symmetry and vortex-lattice structure. The leading perturbations at the fixed point are a relevant pair detuning and a marginal three-body contact in the valley-singlet channel, which determine the compressibility and sound velocity.
A repulsive three-body coupling is marginally irrelevant, allowing the fixed point to control a parametrically broad dilute regime near pair resonance, while attractive three-body interactions instead generate charge-$2e$ trimers and favor further clustering. More broadly, our results show that binding thresholds between competing anyon composites host universal resonant regimes that are not captured by the lowest-energy anyon species alone, providing a new setting for controlled theories of anyon matter.
\end{abstract}

\maketitle
\addtocontents{toc}{\protect\setcounter{tocdepth}{-1}}  %

\section{Introduction}
\label{sec:intro}

The discovery of fractional quantum anomalous Hall (FQAH) states in twisted MoTe$_2$ and rhombohedral graphene~\cite{Cai2023,Zeng2023,Park2023,Xu2023,Lu2024} and more general fractional Chern insulators (FCIs) stabilized in magnetic field~\cite{doi:10.1126/science.aan8458,xie2021fractional} has reignited interest in the physics of itinerant anyons. In a Landau level, continuous magnetic translation symmetry forbids anyons from dispersing, so that doped anyons are localized by arbitrarily weak disorder and the response remains that of the parent quantum Hall plateau. In a Chern band this symmetry is reduced to a discrete lattice subgroup, and anyons acquire a finite effective mass~\cite{Goncalves2026,Schleith2026,Yan2026,Iyer2026}. The fate of a dilute gas of doped anyons in a clean sample is then not necessarily dictated by disorder; instead, anyons can form genuine many-body phases of their own, including anyon metals and anyon superconductors~\cite{Laughlin1988PRL,Laughlin1988Science,Fetter1989,Chen1989,Lee1989,Wen1989,Hosotani1990,WenZee1990,Lee1991,Kitazawa1990,Fradkin1990,Girvin1990,Shi2025PRX,Shi2025PNAS,Shi2025SciPost,Shi2025NonAbelian,Shi2026Charge13,Nosov2026,Han2026OM,Fan2026,Wang2026TopSC,Wang2025Zaletel,Zhang2026Holon,Zhang2026Color,Nakajima2026,Mehta2026Coloring,Pichler2026,Kuhlenkamp2025,Divic2025,Kim2025,Han2026PRX,Wang2025Trilayer,Lotric2026,Seo2026,lotric2026fractional}.

This question has acquired direct experimental relevance with the observation of superconductivity in the vicinity of FQAH states in twisted MoTe$_2$~\cite{Xu2025SC} and rhombohedral graphene~\cite{Han2025Chiral}. In twisted MoTe$_2$, which serves as our main motivation, superconductivity emerges upon doping the $\nu=2/3$ FQAH state, separated by a narrow resistive region~\cite{Xu2025SC}. Two possible scenarios of the parent state of such a superconductor are: (i) it evolves out of a lightly doped gas of anyons whose density is measured relative to the $2/3$ state or (ii) it evolves out of the Fermi liquid state formed by doping the $\nu = 1$ state. The former has been the subject of several recent works aiming to understand the emergence of superconductivity in a dilute anyon gas \cite{Shi2025PRX, Shi2025PNAS, Nosov2026, Zhang2026Holon, Zhang2026Color, Mehta2026Coloring, Fan2026} whereas the latter corresponds to a more conventional Fermi surface instability, possibly enhanced by quantum geometry \cite{Xu2025Chiral,Guerci2025}. Direct measurements of the carrier density in this regime would serve to distinguish the two scenarios. While such direct measurement is currently lacking, the observed increase of the resistivity as the $\nu=2/3$ state is approached suggests that the carrier density is controlled by the proximity to $\nu=2/3$ \cite{Xu2025SC}, making a description in terms of dilute anyon doping more appropriate and an anyon mechanism for superconductivity a natural candidate. Several such mechanisms have been proposed, depending on whether the cheapest charged excitations are the elementary charge-$e/3$ anyons or charge-$2e/3$ composites~\cite{Shi2025PRX,Shi2025PNAS,Nosov2026,Pichler2026,Divic2025,Kuhlenkamp2025,Zhang2026Holon,Zhang2026Color,Mehta2026Coloring,Wang2026TopSC,Fan2026,Shi2026Charge13,Han2026OM,Kim2025, Senthil2026OM}. The former can lead to a variety of superconductors as well as to orthogonal metals~\cite{Shi2026Charge13,Han2026OM,Zhang2026Color,Mehta2026Coloring,Fan2026,Wang2026TopSC}, whereas the latter is generically expected to yield a superconductor~\cite{Shi2025PRX,Nosov2026,Pichler2026,Divic2025}. This state can be understood as an integer quantum Hall state of composite fermions and is closest in spirit to the anyon mechanism originally proposed by Laughlin~\cite{Kalmeyer1987,Laughlin1988PRL,Laughlin1988Science,Fetter1989}.

Earlier studies of anyon superconductivity typically started directly from a Hamiltonian of anyons with purely statistical interactions and no connection to an underlying electronic problem~\cite{Laughlin1988PRL,Fetter1989,Chen1989,Lee1989,Wen1989,Hosotani1990,WenZee1990,Lee1991,Kitazawa1990,Fradkin1990,Girvin1990}. In particular, such treatments did not account for the valley degree of freedom in the anyon band structure that arises from the fractionalized action of translation symmetry. More recent works have employed an effective field theory approach based on parton constructions~\cite{Shi2025PRX,Shi2025PNAS,Shi2025SciPost,Shi2025NonAbelian,Shi2026Charge13,Nosov2026,Han2026OM,Wang2026TopSC,Zhang2026Holon,Zhang2026Color,Mehta2026Coloring,Kim2025}. Such an approach is well suited to classify the possible phases and to extract their long-distance response, but it does not provide a direct connection to microscopics and cannot decide which phase is realized on energetic grounds. On the other hand, numerical studies using density-matrix renormalization group (DMRG) have found evidence for superconductivity in the vicinity of FQAH states~\cite{Wang2025Zaletel,Pichler2026}. However, these simulations address the full electronic problem, making it unclear whether the superconductors they find are related to anyons.

\begin{figure*}[t!]
    \centering
    \includegraphics[width=0.99\linewidth]{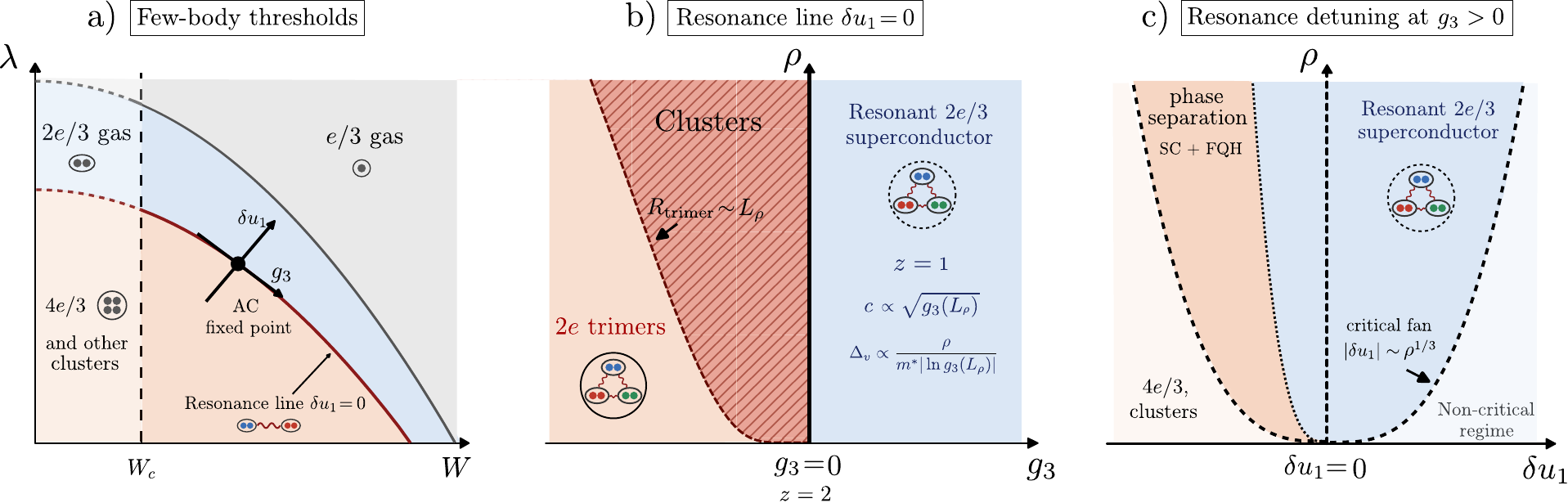}
    \caption{Schematic zero-temperature phase diagram near the resonant
Aharonov--Casher point (ACCS). (a) Few-body regimes as a function of the screening length $\lambda$ and the bandwidth $W$. Different regions indicate the preferred quasihole bound states, determined by their energy per added charge. The red curve marks the binding threshold of two charge-$2e/3$ preformed molecules. The relevant parameter corresponding to detuning away from the resonance is $\delta u_1$, while $g_3$ is a marginal three-body interaction along the resonance line. For illustration, we assume that $g_3$ changes sign along the resonance boundary, with the black dot corresponding to $\delta u_1=g_3=0$. $W_c$ marks the crossover to the regime where valleys are well-formed.
(b) Finite-density phase diagram at pair resonance, $\delta u_1=0$. At $g_3=0$ and finite anyon density $\rho$, the exact
ground state exhibits superconducting ODLRO, vanishing inverse compressibility, and a quadratic charge mode with $z=2$. Repulsive $g_3>0$ is marginally irrelevant and runs logarithmically to zero upon dilution, while generating finite compressibility and a linear long-wavelength sound mode with $z=1$. Here $c$ is the sound velocity and $\Delta_v$ denotes the gap for relative-valley excitations. Attractive $g_3$ produces charge-$2e$ bosonic trimers. The dashed curve $R_{\rm trimer}\sim L_\rho$, with interparticle spacing $L_\rho=\rho^{-1/2}$, marks the crossover between isolated and overlapping trimers. For $R_{\rm trimer}\gg L_\rho$, the uniform gas is unstable toward clustering or phase separation. For $R_{\rm trimer}\ll L_\rho$, there is a gas of charge-$2e$ trimers, which could form a BEC with residual Laughlin topological order if the trimer-trimer interaction is repulsive. (c) Phase diagram at finite pair detuning and fixed repulsive $g_3>0$. The outer dashed curves correspond to the crossover scale of the critical fan, $|\delta u_1|\sim(\rho l_{\rm pair}^2)^{1/3}$. Three-body repulsion stabilizes the superconductor slightly on the attractive side of the pair resonance until the pressure vanishes at the inner dotted line. To its left, the system separates into a superconducting liquid and the undoped parent Laughlin state. The coupling axes refer to values matched at the microscopic scale $l_{\rm pair}$, while $g_3(L_\rho)$ denotes the running coupling at the interparticle spacing.}
    \label{fig:phaseDiagram}
\end{figure*}

 Our approach connects these microscopic and long-distance field theory descriptions. Our main contributions are twofold. First, our formalism provides a direct route from the electronic Hamiltonian to the anyon degrees of freedom at the valley minima; it fixes the structure and symmetry data of the effective Hamiltonian describing dilute anyons close to the valley minima. The explicit parameters of the effective theory, which are inputs in our approach, are in principle computable by few-anyon numerics.
 The construction leads to an explicit mapping between anyon correlators and electron correlators, which are difficult to access in phenomenological effective gauge theories. Second, we identify anyon binding thresholds as a source of universal many-body physics. Near the boundaries between competing anyon composites, resonant scattering generates a critical dilute-gas regime that is not captured by specifying only which isolated excitation is cheapest. This is analogous to the organization of unitary Fermi and Bose gases around few-body resonances. Most strikingly, we show that this type of criticality can produce an exactly solvable fixed point for anyon superconductivity, with controlled corrections around it.

We carry out this program for charge-$2e/3$ anyons obtained by doping the $\nu = 1/3$ Laughlin state of an Aharonov–Casher band with a weak periodic potential and short-range interactions. We discuss how the anyon dispersion and their interactions arise from such microscopic description and derive the effective Hamiltonian describing a dilute gas of such anyons close to the minima of the dispersion. The parameters of such theory are expressible in terms of microscopic parameters and can be explicitly computed numerically, but here we focus on the general structure of such effective Hamiltonian and leave the detailed numerical extraction of such parameters to future works. In the dilute limit, the three-valley anyon gas is described by a many-body Aharonov–Casher Hamiltonian with Chern-Simons field describing the anyon mutual statistics, supplemented by short-range interactions. Tuning the intervalley $\ell = 1$ channel to its binding threshold yields a scale-invariant fixed point whose many-body ground state we find exactly. We argue that, because the intervalley $\ell = 1$ resonance has the widest critical window, this fixed point generically controls the dilute gas near the binding threshold of 2e/3 anyons, unless preempted by compact clustering. The logic parallels the treatment of resonant quantum gases, where the unitary Fermi gas is identified as a scale-invariant few-body fixed point around which the finite-density theory is organized~\cite{Nishida2007,Randeria2014,Zwerger2016}. We emphasize that our ground state is not a variational or parton ansatz. It is the exact ground state of the long-distance theory at the fixed point, and is therefore asymptotically exact in the dilute limit at resonance, with corrections controlled by the RG analysis.  Therefore it does not share the peculiarities associated with Gutzwiller projection of parton wavefunctions~\cite{sarma2026fieldtheorygutzwillerprojection}. 

The scale invariant fixed point has one relevant perturbation, the pair detuning, and one marginal perturbation, a valley-singlet three-body contact that is marginally irrelevant when repulsive and drives trimer $2e$ formation or clustering when attractive. Using the exact ground state wavefunction, we establish Cooper-pair off-diagonal long range order (ODLRO) in electron coordinates and determine the pairing angular momentum and the vortex lattice structure for each relative-angular-momentum channel. We also obtain the equation of state and collective modes from exact plasma correlations, and find that relative-valley fluctuations become gapless at the fixed point. The results are summarized in Sec.~\ref{sec:summary}.

\section{Summary of results}
\label{sec:summary}
In this section, we summarize our main results. Our approach builds on the formalism of anyon Hilbert space introduced in Ref.~\cite{Li2026Bound} and reviewed in Sec.~\ref{sec:setup}. We specialize to the case of $2e/3$ quasihole pairs, which we choose to be in the angular momentum $L = 2$ sector motivated by recent numerical results \cite{Li2026Bound}, but our approach is generalizable to any quasihole composites. The space of Laughlin quasiholes can be characterized by a set of coherent state wavefunctions $|\psi_\xi \rangle$, \eqref{eq:pair}, which depend holomorphically on the center-of-mass coordinates of the $2e/3$ anyons $\xi = (\xi_1,\dots,\xi_M)$ where $\xi_l = \xi_{l,x} + i \xi_{l,y}$. Projecting the Hamiltonian to that space can be achieved by promoting the anyon coordinates to dynamical variables and deriving a path integral written in terms of a simple Lagrangian, \eqref{eq:lagrangian}, which consists of two pieces (both related to overlaps that can be evaluated in Monte Carlo): a Berry phase term that incorporates both background and statistical magnetic field and a potential term that incorporates both a one-body periodic potential for the quasiholes and multi-body short-range interaction. The general form of the Lagrangian does not depend on the details of the projected Hamiltonian as long as the projection is justified, i.e. the many-body gap is large compared to the Hamiltonian matrix elements connecting the low subspace to higher energy excitations, and that it has the right symmetries, particularly breaking continuous magnetic translation symmetry down to a discrete subgroup. A concrete realization of this is an Aharonov-Casher band with weak periodic modulation and a short-range interaction \cite{AharonovCasher1979, morales-duran2024magic, TMDAharonocCasher, Yan2026}. We derive an effective Hamiltonian purely in terms of anyon coordinates by applying geometric quantization to the path integral as described in Ref.~\cite{Li2026Bound}. The anyon Hilbert space is identified with antiholomorphic functions $f(\bar \xi)$ whose inner product is defined via $\langle f|g\rangle=\int d\mu_{\bxi}\,\overline{f(\bar\xi)}\,g(\bar\xi)$ with the integration measure defined in terms of the K\"ahler potential $e^{\mathcal K(\bar \xi,\xi)} = \langle \psi_\xi|\psi_\xi \rangle$ through the reproducing kernel, \eqref{eq:RepKernel}. This Hilbert space formulation enables us to translate between anyon and electron wavefunctions \eqref{eq:ElectronWavefunction}. The Hamiltonian on the anyon Hilbert space is obtained by quantizing the potential  \eqref{eq:HamiltonianGQ}.\\

\subsection{Mapping between electrons and dilute dispersing anyons}
\begin{figure}
    \centering
    \includegraphics[width=\linewidth]{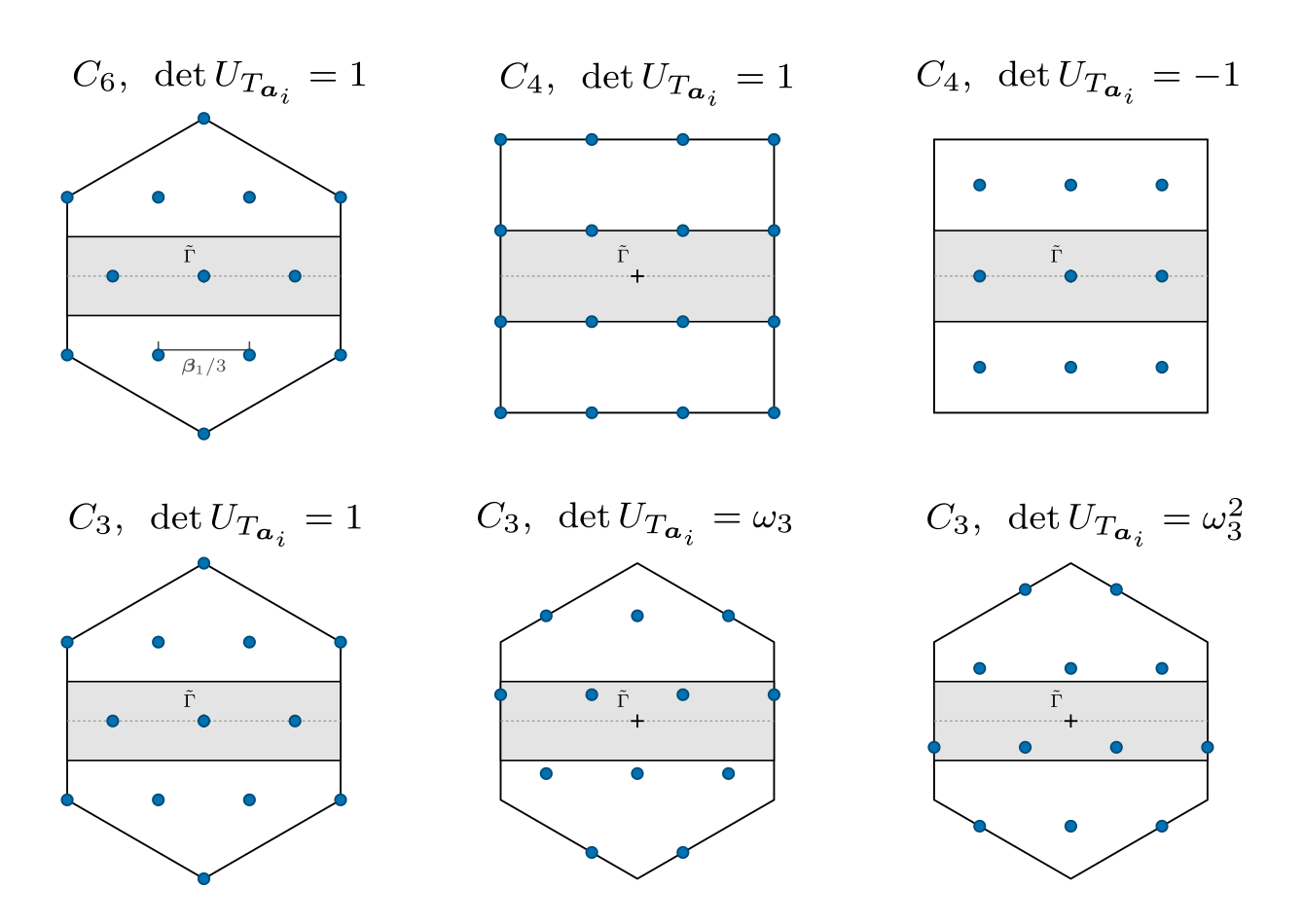}
    \caption{Possible $C_n$ symmetric valley placements for $n=3, 4, 6$. The valley configurations are symmetric relative to $\tilde \Gamma = -\frac{1}{6}\bbeta_1$ for $C_3$ and $C_6$ and $\tilde \Gamma = -\frac{1}{2}\bbeta_2$ for $C_4$.}
    \label{fig:ValleySym}
\end{figure}

In this work, we employ this formalism to develop a systematic approach to relate observables expressed in terms of the coordinates of dilute dispersing anyons, residing close to the minima of the anyon dispersion, to those expressed in electron coordinates. In the dilute limit, $l_{\rm pair}^2 \rho \ll 1$, the one-body term in the anyon Hamiltonian dominates: it generates a band dispersion for a single pair, with a bandwidth set by the periodic potential and the field modulation, whereas the statistical field and the short-range anyon-anyon interactions are both subleading. We analyze the dispersion of the $2e/3$ anyons assuming a generic symmetry-compatible one-body periodic potential for anyons. In contrast to the $e/3$ case considered in earlier works \cite{Yan2026, Iyer2026}, the band structure of the $2e/3$ anyons corresponds to two bands since the 3-fold enlarged unit cell contains 2 flux quanta. Each band is 3-fold degenerate in the reduced BZ leading generically to three band minima. We analyze the symmetry transformation properties of these bands and identify symmetry-compatible placement for the valleys for different possible rotation symmetries $C_{2,3,4,6}$. The results of this analysis are summarized in Fig.~\ref{fig:ValleySym} and Table \ref{tab:valleys}. In particular, we identify the symmetry irreps of the valley minima for different rotation symmetries and how they depend on the periodic potential felt by the anyons.

Since the band energy is the largest scale at dilute fillings, low-energy many-anyon states are built from Bloch states with momenta close to the valley minima. We project on this low-energy space by introducing a wavepacket expansion which expresses the anyon wavefunction $f(\bar \xi)$ in terms of wavepackets $\psi_{\bm \eta,\alpha}(\bar \xi)$ centered at coordinate $\eta$ and carrying valley label $\alpha$. The wavepackets are obtained by Fourier transforming the anyon Bloch states in the slow momenta measured relative to the valley momenta. Expanding $f$ in terms of wavepackets \eqref{eq:expansion} introduces the envelope wavefunction $w_\alpha(\bm \eta)$ which is a slowly varying function of a valley-resolved coordinate $\bm \eta = (\bm \eta_1,\dots,\bm \eta_M)$ and also depends on the valley labels $\alpha = (\alpha_1, \dots, \alpha_M)$, symmetric under their simultaneous exchange. Physically, $\bm \eta_a$ is the position of anyon $a$ coarse grained over the magnetic length. Importantly, the envelope states do not experience any background field, which is absorbed into the Bloch states and the band structure. The effective field experienced by the envelope function comes entirely from the statistical field, smoothly regularized at separations of order the pair size. The envelope thus describes anyons that feel only the statistical field of one another and no background field. This is where most earlier theories of anyon superconductivity usually begin \cite{Laughlin1988PRL, Laughlin1988Science, Fetter1989, Chen1989, Girvin1990, Fan2026}. Here, it is derived from the electronic problem, with the explicit map back to electron coordinates retained. The background field reappears when pairing is expressed in terms of electrons, and it forces the superconducting order parameter to carry a vortex lattice (Sec.~\ref{sec:vortex}).

\subsection{Resonant dilute gas analysis}
The effective Hamiltonian at the valley minima takes the form of a kinetic term minimally coupled to a smeared Chern-Simons statistical gauge field in addition to short-range interactions. We analyze this Hamiltonian using an approach similar to that employed for dilute Fermi and Bose gases, which we briefly review now. Consider, for definiteness, a 3D dilute Fermi gas with a short-range interaction. The analysis begins by considering the two-body problem for two low-energy spinful fermions in different angular momentum channels. In general, there are two possible solutions, $r^{\ell}$ and $r^{-\ell - 1}$, in any given angular momentum channel $\ell$. The normalization requirement at short distance selects one of them except for $\ell = 0$ ($s$-wave) where both solutions are admissible: the regular solution $r^0$ and the singular solution $r^{-1}$. This leads to a family of boundary conditions at coincidence labelled by the scattering length $a$ defined such that the short distance solution is $\sim 1 - a/r$. The scattering length introduces a new length scale that can be much larger than the interaction range and defines a `wide resonance'. The parameter $a$ describes the standard BCS-BEC crossover: as the interaction becomes more attractive $a<0$ decreases till it diverges at the threshold of binding, described by the scale-invariant unitary point, before becoming positive on the BEC side \cite{Randeria2014}.

We develop a similar analysis for the dilute $2e/3$ anyon gas. We find that the two-body problem in a bosonic gauge has two channels with two normalizable solutions at short distance: valley-antisymmetric $\ell = 1$ and valley-symmetric $\ell = 2$. These lead to four scale-invariant RG fixed points for the two-body problem labeled by $(\sigma_2, \sigma_1)$ with $\sigma_\ell = 1$ for the regular boundary condition $r^{\nu_\ell}$, with $\nu_\ell = |\ell - 4/3|$, and $\sigma_\ell = -1$ for the singular boundary condition $r^{-\nu_\ell}$. Earlier work discussing the anyon gas focused on the single valley problem \cite{Manuel1991,Bergman1994,Giovanni1995,Ghosh1996,Kim1997,Nishida2008,Turner2020,komargodski2026anyon}, where only the $\ell = 2$ channel can be studied for our statistics. In this work, we will focus instead on the $(+-)$ fixed point, which does not have an analog in the absence of valley degrees of freedom because $\ell = 1$ channel requires an antisymmetric valley
state. We will show that, on physical grounds, this fixed point is the most natural one close to a resonance where the $2e/3$ pairs form larger composites. Furthermore, we will show that this point is analytically tractable since it has only one marginal perturbation and its ground state is exactly solvable.

To identify the physical regime where such a fixed point is realized, we consider a generic phase diagram of the few-anyon problem in terms of the screening length $\lambda$ of the electron-electron interaction and the strength of the periodic potential $W$ which gives rise to the anyon dispersion. In principle, this phase diagram can be constructed explicitly for a given microscopic model using the anyon ED approach  \cite{Li2026Bound, Gattu2025}. This would allow us to use the solution of the few-body problem to extract the parameters that control the dilute anyon problem. Here, instead we will discuss the general features of such few-body problem, leaving the numerical investigation of this phase diagram to future works. 
As $\lambda$ is reduced from the unscreened Coulomb limit, the cheapest excitations evolve from $e/3$ to $2e/3$ pairs to more complicated composites \cite{Li2026Bound, Wang2026Molecules}. We expect a similar trend along the $W$-axis with the dispersion favoring lighter quasiparticles. The resulting schematic phase diagram is shown in Fig.~\ref{fig:phaseDiagram}a: it contains a region of free $2e/3$ dilute anyons separating the $e/3$ anyons and more complicated composites. Similar to dilute gases, we expect the transition between the free $2e/3$ gas and more complicated composites to be controlled by the wide resonances associated with the channels $\ell = 1$ or $\ell = 2$. For a finite detuning $\delta u_\ell$ away from the resonance, defined in Sec.~\ref{sec:twobody}, we can associate a length scale (analogous to the scattering length) $r_\ell\sim l_{\rm pair}|\delta u_\ell|^{-1/(2\nu_\ell)}$. Since $\nu_1=1/3<\nu_2=2/3$, for comparable detunings $|\delta u_1|\sim|\delta u_2|\ll1$, we have $r_1\gg r_2$, i.e., the $\ell=1$ resonance is parametrically wider. Alternatively, we can understand this at finite density which introduces a new length scale, the interparticle distance $L_\rho=\rho^{-1/2}$: we expect a `critical fan' in density which is widest for the $\ell = 1$ channel (see Fig.~\ref{fig:phaseDiagram}c and  Fig.~\ref{fig:fan}), suggesting this resonance controls the critical region irrespective of the details of the interaction that lead to binding. This motivates the study of the $(+-)$ fixed point.

Analyzing multi-particle interactions at the $(+-)$ fixed point, we find that, apart from the relevant perturbation that tunes to this point, there is a marginal valley-singlet three-body interaction $g_3$. For repulsive values, $g_3>0$ is marginally irrelevant flowing to zero logarithmically, whereas for attractive $g_3<0$ it leads to the formation of charge $2e$ trimers of size $R_{\rm trimer}$. For $g_3 < 0$ and $R_{\rm trimer} \gg L_\rho$, these trimers are overlapping, and the three-body contact gives negative compressibility. For short-range interactions, the general expectation is that the uniform resonant state is thermodynamically unstable toward clustering or phase separation. We caveat that this conclusion may not survive in the presence of long-range interaction or in a 2D gated sample where negative electronic compressibility is possible \cite{Eisenstein1992negative, eisenstein1994compressibility, ilani2000unexpected, zondiner2020cascade, pierce2021unconventional, Zeng2023}. For $R_{\rm trimer} \ll L_\rho$, charge-$2e$ trimers are compact and well-defined. If they condense, the resulting superconductor retains the Laughlin topological order and is distinct from the anyon superconductor, so the two must be separated by a phase transition. Near the $(+-)$ point this separation occurs through the phase-separated region; a direct transition through the $(+-)$ point itself would be non-generic, since that point requires tuning both $\delta u_1$ and $g_3$. The detailed many-body phase of such a dilute gas of trimers depends on the detailed interaction between them, which is left to future work. In contrast, the $\ell = 2$ resonance studied in recent works \cite{Turner2020} for single-valley anyons is considerably more complicated due to the presence of an infinite tower of relevant multi-particle perturbations.

\subsection{Hamiltonian for anyon SC with exactly solvable ground state}
The Hamiltonian at the $(+-)$ fixed point takes the form of an Aharonov-Casher Hamiltonian $H = \frac{1}{2m^*} \sum_a \Pi_a^\dagger \Pi_a$ with $[\Pi_a,\Pi_a^\dagger]=-2\,\delta\mathcal B_a$ where $\delta\mathcal B_a$ is the magnetic field felt by anyon $a$ due to the smeared statistical Chern-Simons field of the other particles. We call such a Hamiltonian Aharonov-Casher-Chern-Simons Hamiltonian (ACCS). Remarkably, the ground state of such a Hamiltonian can be obtained exactly since the Hamiltonian is a sum of positive semidefinite terms, so any zero mode of the Hamiltonian is an exact ground state. We show that, in the long distance theory, there is a unique zero mode and thus a unique ground state of this Hamiltonian \footnote{Accounting for regularization allows for other zero modes that are immediately pushed to high energy in the presence of repulsive short-range interaction in the $\ell = 0$ channel. See Sec.~\ref{sec:zeromode} for detailed discussion.}. Given the exact ground state, we identify the superconductor order parameter which corresponds to a three $2e/3$ anyon operator and show that it exhibits off-diagonal long-range order. We note an early precursor of that state appeared in Ref.~\cite{Girvin1990} where the spin playes the role of valley.

We emphasize two important distinctions from earlier discussions of similar wavefunctions \cite{Girvin1990, Choi1992, Seo2026}. First, the explicit mapping back to electron coordinates enables us to explicitly construct the pair order parameter in electronic language, determine its pairing symmetry and establish off-diagonal long-range order. For rotation symmetry $C_n$, the angular momentum of the order parameter is $6 + L_U \mod n$, where 6 comes from 3 $2e/3$ anyons each with $L=2$ and $L_U$ is fixed by the transformation properties of the valley wavefunctions which are determined by the properties of the quasihole periodic potential (Table \ref{tab:valleys}). Because the electrons experience a net effective magnetic flux of $2\pi$ per unit cell, the Cooper pair order parameter forms a vortex lattice \footnote{In ideal or Aharonov--Casher bands, the Bloch wavefunction decomposes into a momentum dependent part and a momentum-independent layer spinor \cite{ledwithFractionalChernInsulator2020a,wangExactLandauLevel2021b,morales-duran2024magic, TMDAharonocCasher}. The $2\pi$ flux is the effective field felt by the momentum dependent part of the wavefunction which is compensated by the winding of the spinor part. }. The pair wavefunction takes the form of a superposition of various relative pairing channels $\ell$ multiplied by a center-of-mass pair wavefunction with angular momentum $6 + L_U - \ell$. Each such center-of-mass piece corresponds to a specific symmetry-allowed vortex lattice (Table \ref{tab:vortices}) e.g. for $C_6$, the symmetry allowed vortex lattice is either a honeycomb lattice of single vortices or a triangular lattice of double vortices.

The second distinction is that, unlike previous studies of anyon superconductivity, we find valley excitations to be very soft and become gapless at the ACCS point. This contrasts with the argument of Ref.~\cite{Girvin1990} based on the single-mode approximation as well as mean field treatments that replace the Chern-Simons field with its average value; both assume valley fluctuations have a gap $\sim \frac{\rho}{m^*}$ where $m^*$ is the anyon mass. Here, we find that due to the special nature of the Chern-Simons field and the ACCS point, a valley excitation can be accompanied by smooth local density depletion to construct variational intervalley excitations whose energy can be arbitrarily small. This implies valley fluctuations can play an important role in the low-energy physics in the vicinity of the ACCS point.

\subsection{Response of the superconductor}
Similar to unitary dilute gases, scale invariance can be used to derive an effective field theory (EFT) close to the ACCS point which enables the calculation of various transport and thermodynamic response functions. However, in contrast to the unitary limit in dilute gases, where the ground state is not known and thus the coefficients of the EFT can only be determined from numerics \cite{Carlson2011} or comparison to experiments \cite{Navon2010,Mark2012}, here, we can compute such coefficients explicitly since we know the ground state wavefunction. Furthermore, we find that the local ground state correlation functions that determine the EFT coefficients can be computed exactly. These correlators take the form of a three-component plasma which can be mapped exactly to correlators of the integrable Toda theory \cite{Lukyanov1997,Fateev2002}.

We first note that the superconducting ground state at the ACCS point itself has a few unusual features that arise from the fact that the ground state energy density $e(\rho)$ vanishes identically at that point. These features include (i) infinite compressibility, $e''(\rho) = 0$, since the ground state has zero energy irrespective of particle number, (ii) quadratic dispersion of the sound mode, $z = 2$, which is a general feature for frustration-free Hamiltonians \cite{d4c4-5p2r}, (iii) gapless valley fluctuations and (iv) zero $T_c$ despite a finite zero-temperature stiffness. The latter is a consequence of the thermal occupation of the quadratic $z=2$ charge mode which diverges logarithmically at any $T > 0$ and destroys superconductivity. The inclusion of the marginally irrelevant repulsive 3-body interaction restores more conventional superconducting behavior: (i) the compressibility is finite $e''(\rho) > 0$, (ii) a linear sound mode with velocity $c^2=\frac{\rho e''(\rho)}{m^*}$ is recovered below a momentum crossover scale $k_* = 3\sqrt{m^* e''(\rho) \rho} \sim g_3(L_\rho) \frac{\rho}{c m^*}$ where $g_3(L_\rho)$ is the logarithmically flowing 3-body coupling evaluated at anyon interparticle distance $L_\rho$ \eqref{eq:main_g3_running}, (iii) valley fluctuations acquire a gap $\Delta_v \sim \frac{\rho}{m^* \ln[1/g_3(L_\rho)]}$, and (iv) finite critical temperature is restored. The standard analysis only accounting for density and phase fluctuations gives $T_c$ given by Eq.~\eqref{eq:thermo_tc}. However, we note that valley fluctuations can play a role close to $T_c$ since the valley gap is of the same order at $T_c$ and could modify the behavior close to the BKT transition.

The same three-body repulsion can also stabilize a superconducting liquid on the attractive side of the pair resonance,
$\delta u_1<0$. Its equilibrium density $\rho_*$ minimizes the energy per particle, $\frac{d}{d\rho}[e(\rho)/\rho]|_{\rho=\rho_*}=0$, which is the condition of zero pressure. When this minimum lies within the dilute resonant regime and competing bound phases are absent, the resulting phase diagram has two regions (see Fig.~\ref{fig:phaseDiagram}(c)). For $\rho>\rho_*$, the superconductor is homogeneous, but for $0<\rho<\rho_*$ the system phase separates into a superconducting liquid at density $\rho_*$ and the undoped parent Laughlin state. This coexistence of an undoped parent insulator and a superconducting liquid is reminiscent of the phase-separation scenario proposed for doped antiferromagnets in~\cite{Emery1990,DungHaiLee2003}.

\section{Main setup}
\label{sec:setup}

We begin by reviewing the projection onto the anyon Hilbert space introduced in Refs.~\cite{Schleith2026, Yan2026, Li2026Bound}. For definiteness, we consider doping quasiholes into the $\nu=1/q$ Laughlin state with $q=3$. The projection can be carried out in a controlled way for the following class of Hamiltonians. The single-particle Hamiltonian is taken to be of the form $H_{\rm AC} + U(\br)$ where $H_{\rm AC}$ is an Aharonov--Casher (AC) Hamiltonian in a nonuniform periodic magnetic field $B(\br)=2\partial_z\partial_{\bar z}K(z,\bar z)$, whose unit cell contains a $2\pi$ flux, and $U(\br)$ is a periodic potential with the same period. We set $\ell_B=1$ throughout; for a moir\'e band with one flux quantum per unit cell, $\ell_B$ is related to the moir\'e period $a_M$ through $2\pi\ell_B^2=\tfrac{\sqrt3}{2}a_M^2$. We will assume that we are at some partial filling of the LLL of $H_{\rm AC}$ and that the periodic potential $U(\br)$ is small compared to the cyclotron gap of the AC Hamiltonian, which we denote by $\Delta_{\rm AC}$. This justifies projecting onto the LLL of $H_{\rm AC}$. This model is a good description for the flat Chern bands of twisted MoTe$_2$ close to the magic angle in the adiabatic approximation~\cite{morales-duran2024magic, TMDAharonocCasher}. 

We consider an interaction of the form $V_1 + V$, where $V_1$ is the Haldane-Trugman-Kivelson pseudopotential \cite{Haldane1983,TrugmanKivelson1985} and $V$ is small compared to $V_1$, in a sense that will be made clear below. This form arises naturally when a short-range interaction is expanded in pseudopotentials. In the AC band, the zero modes of $V_1$ take the form of the Laughlin wavefunction and its multi-quasihole descendants, with the Gaussian factor replaced by $K$~\cite{ledwithFractionalChernInsulator2020a, wangExactLandauLevel2021b, ledwith_vortexability_2023}. These zero modes are separated from all other many-body states by a many-body gap $\Delta_{V_1}$. Assuming both $U$ and $V$ are small compared to the gap to the remote bands $\Delta_{\rm AC}$ and the many-body gap $\Delta_{V_1}$, we can project $U$ and $V$ onto the zero-mode manifold. 

We emphasize that the limit $\Delta_{V_1}, \Delta_{\rm AC} \to\infty$ is a theoretical device that renders the projection exact. The same construction applies whenever a manifold of anyon wavefunctions can be identified---for instance, variational quasihole or quasielectron states built from composite fermions---by projecting the interaction and the periodic potential onto it \cite{Gattu2025}. In that case, the approach is variational, but the structure of the projected theory is unchanged.

In this work, we are interested in the phase formed when the cheapest charged excitations upon doping the $\nu=1/3$ Laughlin state are charge-$2e/3$ composites. In the Landau-level context, this was shown to be the case for a range of screening lengths $\lambda$~\cite{Li2026Bound, Xu2025Clusters, Wang2026Molecules}. In a Chern band, both the periodic potential and the field modulation generate a dispersion, which is expected to disfavor larger bound states.\footnote{The dispersion generated by a periodic potential or field modulation is controlled by the Fourier components of the charge profile of the excitation at the reciprocal lattice vectors, which are increasingly suppressed for larger objects with more smeared charge. A cluster of $n$ quasiholes therefore gains less band energy than $n$ separated quasiholes, so that the periodic modulation shifts all binding thresholds toward stronger screening and favors smaller clusters. The $2e/3$ pair survives as long as its binding energy exceeds the band-energy gain of two separated quasiholes.} We will assume that the periodic modulations are small compared with the binding energy of the $2e/3$ composite, so that it remains a stable bound state, while larger bound states may be destabilized. As shown in Ref.~\cite{Li2026Bound}, the lowest energy $2e/3$ bound state lies in the relative angular momentum channel $L=2$, which we will assume to be the case. Thus, for a pair of quasiholes at positions $\xi_1=\xi_{\rm CM}+\xi_{\rm rel}$ and $\xi_2=\xi_{\rm CM}-\xi_{\rm rel}$, we can integrate out the relative coordinate by integrating the wavefunction against the $L=2$ state $\bar\xi_{\rm rel}^{\,2}$ with the appropriate measure. Since the two-quasihole factor $\prod_i(z_i-\xi_1)(z_i-\xi_2)=\prod_i[(z_i-\xi_{\rm CM})^2-\xi_{\rm rel}^2]$ is a polynomial in $\xi_{\rm rel}$, the integral extracts its $\xi_{\rm rel}^2$ coefficient, independently of the details of the measure up to normalization, and replaces this factor by $e_1\big((z-\xi_{\rm CM})^{-2}\big)\prod_{i=1}^{N}(z_i-\xi_{\rm CM})^2$,

where $e_1$ is the first elementary symmetric polynomial and $e_1((z-\xi)^{-2})\equiv\sum_i(z_i-\xi)^{-2}$, so that this factor equals $\sum_i\prod_{j\neq i}(z_j-\xi_{\rm CM})^2$. Thus, the manifold of states with $M$ $2e/3$ bound states is spanned by
\begin{multline}
\psi_{\xi}(\{z_i\})=\prod_{a=1}^{M}\Big[e_1\big((z-\xi_a)^{-2}\big)\prod_{i=1}^{N}(z_i-\xi_a)^{2}\Big] \\\prod_{i<j}(z_i-z_j)^{q}\;e^{-\frac12\sum_i K(z_i,\bar z_i)},
\label{eq:pair}
\end{multline}
labelled by the complex center-of-mass coordinates $\xi=(\xi_1,\dots,\xi_M)$ of the pairs, on which $\psi_{\xi}$ depends holomorphically and symmetrically. We write the wavefunctions on the plane for simplicity, since away from the boundaries the geometry does not matter for most of our discussion. The torus geometry becomes important when considering anyon Bloch states and valleys, as we will discuss later.

The projection is implemented by promoting the pair coordinates to dynamical variables. The states~\eqref{eq:pair} define a coherent-state path integral with integration measure $d\mu_{\bxi}$, defined by the resolution of unity $\int d\mu_{\bxi}\,|\psi_{\xi}\rangle\langle\psi_{\xi}|=1$ on the manifold~\eqref{eq:pair}, and the Euclidean Lagrangian
\begin{gather}
\mathcal L=i\sum_{a}\mathcal A_a(\bar\xi,\xi)\,\dot\xi_a+\mathcal U(\bar\xi,\xi),\qquad
\mathcal A_a=-i\partial_{\xi_a}\mathcal K, \nonumber\\
e^{\mathcal K(\bar\xi,\xi)}=\langle\psi_{\xi}|\psi_{\xi}\rangle,\qquad
\mathcal U(\bar\xi,\xi)=\frac{\langle\psi_{\xi}|\hat U+\hat V|\psi_{\xi}\rangle}{\langle\psi_{\xi}|\psi_{\xi}\rangle}.
\label{eq:lagrangian}
\end{gather}
The Berry connection $\mathcal A_a$ is derived from the K\"ahler potential $\mathcal K$, which simultaneously encodes the overlaps between the states~\eqref{eq:pair} and their Berry phases and defines a many-body quantum geometry on the space of anyons~\cite{Li2026Bound}. Equivalently, the anyon Hilbert space can be identified with the space of symmetric antiholomorphic functions $f(\bar\xi)$, with inner product $\langle f|g\rangle=\int d\mu_{\bxi}\,\overline{f(\bar\xi)}\,g(\bar\xi)$; the measure is fixed by the reproducing-kernel relation
\begin{align}
&\int d\mu_{\bxi}\;e^{\mathcal K(\bar{\omega},\xi)}f(\bar\xi)=f(\bar{\omega}),
\label{eq:RepKernel}\\
&[\hat H f](\bar{\omega})=\int d\mu_{\bxi}\;e^{\mathcal K(\bar{\omega},\xi)}\,\mathcal U(\bar{\omega},\xi)\,f(\bar\xi),
\label{eq:HamiltonianGQ}
\end{align}
where $\mathcal K$ and $\mathcal U$ are analytically continued to independent arguments, $e^{\mathcal K(\bar{\omega},\xi)}=\langle\psi_{\omega}|\psi_{\xi}\rangle$. The second relation defines the quantum Hamiltonian corresponding to the potential $\mathcal U$ through geometric quantization~\cite{Li2026Bound}. The electron wavefunction associated with $f$ is 
\begin{equation}
    |\Psi \rangle = \int d\mu_{\bxi}\,f(\bar\xi)\,|\psi_{\xi} \rangle
    \label{eq:ElectronWavefunction}
\end{equation}
 and conversely $f(\bar\xi)=\langle\psi_{\xi}|\psi\rangle$.

The projected potential can be organized into a sum of $k$-body terms, $\mathcal U=\sum_a\mathcal U_1(\bar\xi_a,\xi_a)+\sum_{a<b}\mathcal U_2(\bar\xi_a,\bar\xi_b;\xi_a,\xi_b)+\dots$, according to the number of pairs that lie within the range of the interaction. In a Landau level and in the absence of a periodic potential, continuous magnetic translation symmetry forces the one-body term to be a constant. Here it is a lattice-periodic function, receiving contributions from $\hat U$ directly and from $\hat V$ through the nonuniform quantum geometry of the AC band~\cite{Yan2026}. This term generates a dispersion for a single $2e/3$ anyon as we will discuss in Sec.~\ref{sec:anyonBS}. For short-range $\hat V$, the terms with $k\geq2$ are short-ranged: two pairs interact only when their charge profiles overlap, so $\mathcal U_2$ decays over a distance set by the size of the pair $l_{\rm pair} \sim \ell_B =1$ and the range of the interaction, and admits an expansion in pseudopotentials.

The K\"ahler potential encodes the magnetic field experienced by the anyons: the field felt by the $a$-th pair is $\mathcal B_a=2\partial_{\xi_a}\partial_{\bar\xi_a}\mathcal K(\bar\xi,\xi)$. When all pairs are separated by distances large compared to $l_{\rm pair}$, $\mathcal K$ can be evaluated with the plasma analogy
\begin{equation}
\mathcal K_{\rm plasma}(\bar\xi,\xi) = \sum_a\Big[\frac{s}{2}|\xi_a|^2+Q(\bar\xi_a,\xi_a)\Big]-p\sum_{a<b}\ln|\xi_a-\xi_b|^2,
\label{eq:plasma}
\end{equation}
where $Q$ is lattice periodic and
\begin{equation}
s=\frac23,\qquad p=\frac43 .
\label{eq:sp}
\end{equation}
For a more general composite object consisting of $n$ anyons each with charge $1/q$, the charge coefficient $s = n/q$ is linear in $n$ because the composite carries $n$ units of the elementary flux deficit, while the statistics coefficient $p = n^2/q$ is quadratic because each constituent of one composite braids with each constituent of the other. The field thus consists of a background part---a fraction $2/3$ of the field seen by the electrons, reflecting the charge $2e/3$, together with a periodic modulation inherited from the nonuniform quantum geometry---and of flux tubes of strength $-2\pi p$ bound to the other pairs, which encode the mutual statistics $\theta=\pi p=4\pi/3$ and reduce at long distances to the Chern--Simons description of anyons. Two remarks are in order. First, the plasma form~\eqref{eq:plasma} holds only for separations large compared to the size of the pair; at shorter distances the logarithm is smoothly regularized by the internal structure of the pair. Second, the modulation $Q$ is a one-body term: it modifies the single-anyon band problem, which we take as an input from Ref.~\cite{Yan2026}, but it leaves the mutual part untouched. For simplicity, we will drop that part which can generally be absorbed into the single-anyon band structure.

The coherent states~\eqref{eq:pair} are symmetric in the pair coordinates, so the anyon wavefunctions $f(\bar\xi)$ are symmetric, i.e.\ we work in a bosonic representation. An equivalent fermionic representation follows from writing any symmetric antiholomorphic function as $f(\bar\xi)=g(\bar\xi)/\Delta(\bar\xi)$, where $\Delta(\bar\xi)=\prod_{a<b}(\bar\xi_a-\bar\xi_b)$ and $g$ is antisymmetric; conversely, every antisymmetric antiholomorphic $g$ vanishes at coincident points and is divisible by $\Delta$, so the map is one-to-one. This modifies the K\"ahler potential via
\begin{equation}
K_F(\bar\xi,\xi)=K(\bar\xi,\xi)+\sum_{a<b}\ln|\xi_a-\xi_b|^2 .
\label{eq:KF}
\end{equation}
 In the plasma form~\eqref{eq:plasma}, the shift~\eqref{eq:KF} replaces $p\to p-1$: each pair now carries a statistical flux $-2\pi(p-1)$ which shifts the exchange phase by $\pi$. For $2e/3$ anyons, $p-1=1/3$, so they can equivalently be viewed as fermions carrying one third of a statistical flux quantum. We use the envelope version of this map, Eq.~\eqref{eq:Ugauge}, in Sec.~\ref{sec:zeromode}, where it makes the zero-mode structure manifest.

\section{Anyon Band structure and envelope expansion}

\subsection{Anyon band structure}
\label{sec:anyonBS}

In the dilute doping limit, the physics is mostly determined by the kinetic energy, i.e., the anyon dispersion, since both the interactions and the Chern--Simons magnetic field scale with the anyon density. It is then justified to expand our wavefunctions in terms of products of single-anyon Bloch states with momenta close to the band minima, $\Psi_{\bk}(\bar\xi)=\prod_{a=1}^{M}\psi_{\bk_a}(\bar\xi_a)$, where $\psi_{\bk}(\bar\xi)$ are the anyon Bloch states, which we now describe.

The anyon Bloch states can be defined by considering a torus spanned by $\bL_1=N_1\ba_1$ and $\bL_2=N_2\ba_2$ with $N_\Phi=N_1N_2$ flux quanta, so that the unit cell of the periodic potential encloses one flux quantum, $\ba_1\wedge\ba_2=2\pi$, with $\ba\wedge\bb=a_xb_y-a_yb_x$. We write the one-body potential as $\mathcal U_1(\bar\xi,\xi)=\sum_{\ba}V_{\ba}e^{i\ba\wedge\bxi}$, where $\ba=n_1\ba_1+n_2\ba_2$ runs over the lattice vectors. For an anyon of charge $s$, the single-particle Hilbert space is the lowest Landau level of a particle of charge $s$, with magnetic length $\ell^2=1/s$, and the Hamiltonian~\eqref{eq:HamiltonianGQ} corresponding to $\mathcal U_1$ is a sum of magnetic translations~\cite{SM},
\begin{equation}
    \hat h=\sum_{\ba}\tilde V_{\ba}\,\hat T_{\ba/s},\qquad
[\hat T_{\bd}f](\bar\xi)=e^{-s|d|^2/4}e^{-s d \bar\xi/2}f(\bar\xi+\bar d), 
\label{eq:hband}
\end{equation}
with $\hat T_{\bd}$ satisfying $\hat T_{\bd}\hat T_{\bd'}=e^{-is\bd\wedge\bd'}\hat T_{\bd'}\hat T_{\bd}$
and $\tilde V_{\ba}=e^{|a|^2/(4s)}V_{\ba}$.
The lattice translations $\hat T_{\ba_1}$ and $\hat T_{\ba_2}$ commute with $\hat h$ but not with each other, $\hat T_{\ba_1}\hat T_{\ba_2}=e^{-2\pi i s}\hat T_{\ba_2}\hat T_{\ba_1}$, since the anyon sees only $s$ flux quanta per unit cell. For both $s=1/3$ and $s=2/3$, the smallest representation of the magnetic translation group is three dimensional which implies that the band minima of $\hat h$ generically come in three valleys related by magnetic translations. 

For a single $e/3$ anyon, $s=1/q=1/3$, and the Bloch states are the simultaneous eigenstates of the commuting translations $\hat T_{\ba_1}$ and $\hat T_{q\ba_2}$,
\begin{equation}
\psi_{\bk}(\bar\xi)=\mathcal N\,\overline{\sigma\big(\xi-iqk\,\big|\,a_1,qa_2\big)}\,e^{\frac{i}{2}k\bar\xi}\,e^{-\frac{q}{4}|k|^2},
\label{eq:bloch}
\end{equation}
with $E_{\bk}=\sum_{\ba}\tilde V_{\ba}\,\eta_{q\ba}\,e^{iq\,\bk\cdot\ba}$. Here, $\sigma(z|\alpha_1,\alpha_2)$ is the modified Weierstrass sigma function of the magnetic unit cell $\balpha_1\times\balpha_2=\ba_1\times q\ba_2$~\cite{haldane2018modularinvariant,wangExactLandauLevel2021b,SM}, so that $\psi_{\bk}$ has a single zero per magnetic unit cell, and $\hat T_{\balpha}\psi_{\bk}=\eta_{\balpha}e^{i\bk\cdot\balpha}\psi_{\bk}$ for $\balpha\in\{\ba_1,q\ba_2\}$, where $\eta_{\balpha}=\pm1$ is the sign factor in the quasi-periodicity of the sigma function. Acting with $\hat T_{\ba_2}$ shifts the momentum by $\bbeta_1/3$, where $\bbeta_{1,2}$ are the reciprocal vectors of the magnetic unit cell, $\bbeta_i\cdot\balpha_j=2\pi\delta_{ij}$, so the three valleys are $\bK_\alpha=\bK_1-(\alpha-1)\bbeta_1/3$, $\alpha=1,2,3$. Expanding around a minimum, $E_{\bK_\alpha+\bkappa}=E_{\min}+|\bkappa|^2/2m^*+O(\kappa^3)$, with an isotropic mass $m^*$ enforced by the $C_n$ symmetry of the potential, $n=3,4,6$.

For $2e/3$ pairs, $s=2/3$ and the magnetic unit cell $\ba_1\times3\ba_2$ now encloses two flux quanta. As a result, we have two Bloch states for each momentum $\bk$ in the reduced Brillouin zone spanned by $\bbeta_1,\bbeta_2$, i.e. we have two bands. A convenient basis is provided by the single-flux Bloch functions of the smaller cell $\ba_1\times(\ba_2/s)$, shifted by half a lattice vector,
\begin{equation}
\psi_{\bk,\tau}(\bar\xi)=\mathcal N_{\bk,\tau}\,\overline{\sigma\big(\xi-ik/s-\tfrac{\tau}{2}a_1\,\big|\,a_1, a_2/s\big)}\,e^{\frac{i}{2}k\bar\xi}\,e^{\frac{\tau s}{4}a_1\bar\xi},
\label{eq:pairbasis}
\end{equation}
with $\tau=0,1$. These states are orthonormal with respect to the background Gaussian measure~\cite{SM}. The translations entering Eq.~\eqref{eq:hband} are by $\ba/s=\tfrac{3n_1}{2}\ba_1+\frac{n_2}{s}\ba_2$; for even $n_1$ these are lattice vectors of the small cell and act diagonally on $\tau$, whereas for odd $n_1$ they contain a half-translation along $\ba_1$ and exchange the two components. The projected Hamiltonian is therefore the $2\times2$ matrix $h(\bk)=\sum_{\ba}\tilde V_{\ba}\,e^{i\bk\cdot\ba/s}\,\Gamma_{\ba}(\bk)$ with
\begin{equation}
\Gamma_{\ba}(\bk)=\begin{cases}\eta_{\ba/s}\,\mathrm{diag}\big(1,\,e^{-\frac{i}{2}\ba_1\wedge\ba}\big), & n_1\ \text{even},\\[4pt]
e^{-\frac{i}{4} \ba_1\wedge\ba}\begin{pmatrix}0 & \eta^-_{\ba} e^{-\frac{i}{4}\ba_1\cdot\bk}\\ \eta^+_{\ba} e^{\frac{i}{4}\ba_1\cdot\bk} & 0\end{pmatrix}, & n_1\ \text{odd},\end{cases}
\label{eq:pairh}
\end{equation}
where $\eta^\pm_{\ba}\equiv\eta_{\ba/s\pm\frac12\ba_1}$ and the $\eta$'s are the sign factors of the small cell~\cite{SM}. Its eigenvalues $E_\pm(\bk)$ define the two pair bands, with eigenvectors $c_\pm(\bk)$ and Bloch states $\psi_{\bk,\pm}=\sum_\tau c_{\tau,\pm}(\bk)\psi_{\bk,\tau}$. We assume that the minima of the lower band are separated from the upper band by a gap $\Delta_+$ larger than all other energy scales in the problem, and retain only the lower band, denoting its Bloch states by $\psi_{\bk}(\bar\xi)\equiv\psi_{\bk,-}(\bar\xi)$. As for the $2e/3$ anyon, $\hat T_{\ba_2}$ shifts $\bk$ by $-2\bbeta_1/3\equiv\bbeta_1/3$, so the minima form three valleys $\bK_\alpha=\bK_1-(\alpha-1)\bbeta_1/3$, and $E_-(\bK_\alpha+\bkappa)=E_{\min}+|\bkappa|^2/2m^*+O(\kappa^3)$ with an isotropic mass.

\subsection{Symmetries of the Bloch states}
\label{sec:symmetries}

The three states at the band minima, $\{\psi_{\bK_\alpha}\}_{\alpha=1}^3$, span the three-dimensional irreducible representation of the magnetic translation group: $\hat T_{\ba_1}$ acts diagonally, with eigenvalues $-e^{i\bK_\alpha\cdot\ba_1}$, while $\hat T_{\ba_2}$ permutes the valleys cyclically up to phases~\cite{SM}.
We denote the corresponding $3\times3$ matrices by $U_{T_{\ba_i}}$. A $C_n$ rotation about a lattice site also maps the valley triplet to itself and is represented by a $3\times3$ unitary $U_{C_n}$, $\hat C_n\psi_{\bK_\alpha}=\sum_\beta(U_{C_n})_{\beta\alpha}\psi_{\bK_\beta}$. Rotation symmetry also restricts the positions of the minima: the valley triplet must be mapped by $C_n$ onto an equivalent representation of the magnetic translation group, which fixes the allowed configurations $\{\bK_\alpha\}$ up to reduced reciprocal lattice vectors~\cite{SM}. 

For simplicity, we consider a potential with some $n$-fold rotation symmetry but without mirror. The symmetry group is then a wallpaper group, $p6$, $p4$ or $p3$, and besides the lattice site it contains further rotation centers: for $p6$ the two corners of the hexagonal cell ($C_3$) and the three edge midpoints ($C_2$); for $p4$ the plaquette center ($C_4$) and the two edge midpoints ($C_2$); for $p3$ the two corners ($C_3$). The rotation by $2\pi/m$ about a center $\bc$ is not an independent operation: the rotation $C_m$ about the origin maps $\bc$ to $C_m\bc$, which in complex notation is $\omega_mc$ with $\omega_m=\omega_n^{n/m}$ and $\omega_n=e^{2\pi i/n}$, and $\bd_c\equiv\bc-C_m\bc$ is a lattice vector. The rotation about $\bc$ can therefore be written as $\hat C_m^{(c)}=\hat T_{\bc}\hat C_m\hat T_{\bc}^{-1} = e^{\frac{is}{2}\,\bc\wedge C_m\bc} \hat T_{\bd_c} \hat C_m$.
Thus the symmetry data of the valley triplet at all rotation centers are fixed by $U_{T_{\ba_i}}$ and $U_{C_n}$.

As an example, consider first $C_6$. There is a single symmetry-allowed configuration, $(\bK_1,\bK_2,\bK_3)=(\tfrac16,-\tfrac16,-\tfrac12)\bbeta_1$. The six-dimensional space spanned by $\{\psi_{\bK_\alpha,\tau}\}$ decomposes into two three-dimensional irreducible representations of the magnetic translation group. Their $C_6$ spectra about the site are $\{1,\omega_6,\omega_6^4\}$ and $\{1,\omega_6^2,\omega_6^3\}$. The lower-band triplet $\{\psi_{\bK_\alpha,-}\}$ carries one of the two and cannot switch under symmetry-preserving deformations of the potential without a band crossing. About the corners, $U_{C_3}^{(2b)}$ has the full spectrum $\{1,\omega_3,\omega_3^2\}$ in both representations, and about the edge midpoints $U_{C_2}^{(3c)}$ has spectrum $\{1,1,-1\}$. The determinants are $\det U_{C_6}=\omega_6^5$ at the site, $1$ at the corners and $-1$ at the edges, identical for the two representations. So for $C_6$ the two representations are distinguished only by the single-anyon spectrum at the $C_6$ center, and not at all by valley-singlet states.

The corresponding data for $C_4$ and $C_3$ are shown in Table~\ref{tab:valleys}. Note that, for $C_4$ and $C_3$ several configurations of the minima are allowed, and which one is realized, as well as which of the two representations the lower band carries, is a property of the band structure that has to be determined from the harmonics of the potential. In addition, the determinants coincide between the two representations for $C_6$ and $C_3$ at every center, but not for $C_4$; for $C_4$ the choice of representation therefore enters the rotation eigenvalues of the valley-singlet many-body states as we will see later.

The symmetry irreps is determined by the anyon periodic potential, which are inherited from the electron periodic potential. The explicit expressions for the anyon dispersion and Bloch wavefunctions are provided in SM \cite{SM}. As an example, we discuss the first/second harmonic approximation for the anyon periodic potential. The valley placement determines the irrep of the magnetic translation, and the irrep of the rotation further depends on which of the two bands is the lower-energy band. In our case, the sign of the coefficient $V$ determines which of the two irreps is realized for the fixed valley placement. As is shown in Table~\ref{tab:valleys}, we label the irrep corresponding to $V>0$ and $V<0$ as (1) and (2), respectively. For $C_6$, there is only one valley placement, and it can be obtained from the first harmonic potential, $V_\ba = V$. For $C_4$, the valley placement with $\det U_{T_{\bm a_i}}=1$ ($\det U_{T_{\bm a_i}}=-1$) can be obtained from the first (second) harmonic potential $V_\ba = V$. For $C_3$, we consider the first harmonic potential $V_\ba = V e^{-i \phi}$ for $\ba = -\ba_1, -\ba_2, (\ba_1 + \ba_2)$. The valley placement with $\det U_{T_{\ba_i}} = 1, \omega_3, \omega_3^2$ corresponding to $\phi = 0, -\pi/3$ and $\pi/3$.

\begin{table*}
\begin{ruledtabular}
\begin{tabular}{cccccccc}
$C_n$ & $(\bK_1,\bK_2,\bK_3)$ & $\det U_{T_{\ba_i}}$ & Wyckoff position & $\operatorname{Spec} U_{C_m}^{(1)}$ & $\det U_{C_m}^{(1)}$ & $\operatorname{Spec} U_{C_m}^{(2)}$ & $\det U_{C_m}^{(2)}$ \\
\hline
$C_6$ & $\{\tfrac16,-\tfrac16,-\tfrac12\}\bbeta_1$ & $1$
 & $1a$: $0$ ($C_6$) & $\{1,\omega_6,\omega_6^4\}$ & $\omega_6^5$ & $\{1,\omega_6^2,\omega_6^3\}$ & $\omega_6^5$\\
 & & & $2b$: $\tfrac{2\ba_1+\ba_2}{3},\tfrac{\ba_1+2\ba_2}{3}$ ($C_3$) & $\{1,\omega_3,\omega_3^2\}$ & $1$ & $\{1,\omega_3,\omega_3^2\}$ & $1$\\
 & & & $3c$: $\tfrac{\ba_1}{2},\tfrac{\ba_2}{2},\tfrac{\ba_1+\ba_2}{2}$ ($C_2$) & $\{1,1,-1\}$ & $-1$ & $\{1,1,-1\}$ & $-1$\\
\hline
$C_4$ & $\{\tfrac16,-\tfrac16,-\tfrac12\}\bbeta_1$ & $1$
 & $1a$: $0$ ($C_4$) & $\{1,\omega_4,\omega_4^2\}$ & $\omega_4^3$ & $\{1,\omega_4^2,\omega_4^3\}$ & $\omega_4$\\
 & & & $1b$: $\tfrac{\ba_1+\ba_2}{2}$ ($C_4$) & $\{1,\omega_4^2,\omega_4^3\}$ & $\omega_4$ & $\{1,\omega_4,\omega_4^2\}$ & $\omega_4^3$\\
 & & & $2c$: $\tfrac{\ba_1}{2},\tfrac{\ba_2}{2}$ ($C_2$) & $\{1,1,-1\}$ & $-1$ & $\{1,1,-1\}$ & $-1$\\
$C_4$ & $-\tfrac12\bbeta_2+\{0,-\tfrac13,\tfrac13\}\bbeta_1$ & $-1$
 & $1a$: $0$ ($C_4$) & $\{1,\omega_4,\omega_4^3\}$ & $1$ & $\{1,\omega_4,\omega_4^2\}$ & $\omega_4^3$\\
 & & & $1b$: $\tfrac{\ba_1+\ba_2}{2}$ ($C_4$) & $\{1,\omega_4,\omega_4^3\}$ & $1$ & $\{1,\omega_4,\omega_4^2\}$ & $\omega_4^3$\\
 & & & $2c$: $\tfrac{\ba_1}{2},\tfrac{\ba_2}{2}$ ($C_2$) & $\{1,1,-1\}$ & $-1$ & $\{1,-1,-1\}$ & $1$\\
\hline
$C_3$ & $\{\tfrac16,-\tfrac16,-\tfrac12\}\bbeta_1$ & $1$
 & $1a$: $0$ & $\{1,\omega_3,\omega_3\}$ & $\omega_3^2$ & $\{1,1,\omega_3^2\}$ & $\omega_3^2$\\
 & & & $1b$: $\tfrac{2\ba_1+\ba_2}{3}$ & $\{1,\omega_3,\omega_3^2\}$ & $1$ & $\{1,\omega_3,\omega_3^2\}$ & $1$\\
 & & & $1c$: $\tfrac{\ba_1+2\ba_2}{3}$ & $\{1,\omega_3,\omega_3^2\}$ & $1$ & $\{1,\omega_3,\omega_3^2\}$ & $1$\\
$C_3$ & $\tfrac13\bbeta_2+\{\tfrac5{18},-\tfrac1{18},-\tfrac7{18}\}\bbeta_1$ & $\omega_3$
 & $1a$: $0$ & $\{1,\omega_3,\omega_3^2\}$ & $1$ & $\{1,\omega_3,\omega_3^2\}$ & $1$\\
 & & & $1b$: $\tfrac{2\ba_1+\ba_2}{3}$ & $\{1,1,\omega_3^2\}$ & $\omega_3^2$ & $\{1,\omega_3,\omega_3\}$ & $\omega_3^2$\\
 & & & $1c$: $\tfrac{\ba_1+2\ba_2}{3}$ & $\{1,\omega_3,\omega_3^2\}$ & $1$ & $\{1,\omega_3,\omega_3^2\}$ & $1$\\
$C_3$ & $-\tfrac13\bbeta_2+\{\tfrac7{18},\tfrac1{18},-\tfrac5{18}\}\bbeta_1$ & $\omega_3^2$
 & $1a$: $0$ & $\{1,\omega_3,\omega_3^2\}$ & $1$ & $\{1,\omega_3,\omega_3^2\}$ & $1$\\
 & & & $1b$: $\tfrac{2\ba_1+\ba_2}{3}$ & $\{1,\omega_3,\omega_3^2\}$ & $1$ & $\{1,\omega_3,\omega_3^2\}$ & $1$\\
 & & & $1c$: $\tfrac{\ba_1+2\ba_2}{3}$ & $\{1,1,\omega_3^2\}$ & $\omega_3^2$ & $\{1,\omega_3,\omega_3\}$ & $\omega_3^2$\\
\end{tabular}
\end{ruledtabular}
\caption{Rotation spectra and determinants of the two inequivalent representations carried by the valley triplet, for the rotation $C_m$ about each special Wyckoff position (all site symmetries are $C_3$ for the $C_3$ lattice; the lattice vectors of the hexagonal lattices are at $120^\circ$, $\ba_2=C_3\ba_1$, as in the SM). $U^{(c)}_{C_m}=e^{\frac{is}{2}\bc\wedge C_m\bc}\,U_{T_{\bd_c}}U_{C_n}^{n/m}$ with $\bd_c=\bc-C_m\bc$. For the $C_3$ lattice the two representations are distinguished at the site for the first configuration and at one corner for the other two. The superscript labels (1) and (2) follow the SM~\cite{SM}, where (1) is the representation carried by the lower band of the model potentials used there; for a given potential, which representation the lower band carries has to be determined from its band structure. Some classes were previously obtained in Ref.~\cite{BurkovBalents2005Triangular,BalentsEtAl2005CompetingOrders} in a different context.}
\label{tab:valleys}
\end{table*}

\subsection{Envelope functions}
\label{sec:envelope}
We now define the envelope functions
\begin{equation}
\psi_{\bm\eta,\alpha}(\bar\xi)=\int_{|\bkappa|\lesssim\Lambda}\frac{d^2\bkappa}{(2\pi)^2}\,e^{-i\bkappa\cdot\bmeta}\,\psi_{\bK_\alpha+\bkappa}(\bar\xi),
\label{eq:envelope}
\end{equation}
where the momentum cutoff $\Lambda\ll1$ restricts the integral to the vicinity of the valley $\alpha$. The coordinate $\bm\eta$ can be understood as a smeared or coarse-grained anyon coordinate: $\bar\xi\,\psi_{\bm\eta,\alpha}(\bar\xi)\simeq\bar\eta\,\psi_{\bm\eta,\alpha}(\bar\xi)$ up to corrections that are subleading for functions varying slowly on the lattice scale~\cite{SM}. We can then write
\begin{equation}
f(\bar\xi)=\sum_{\alpha}\int\prod_{a=1}^{M}d^2\bm\eta_a\;w_{\alpha}(\bmeta)\prod_{a=1}^{M}\psi_{\bm\eta_a,\alpha_a}(\bar\xi_a),
\label{eq:expansion}
\end{equation}
where $\alpha=(\alpha_1,\dots,\alpha_M)$ labels the valley of each anyon, $\bmeta=(\bm\eta_1,\dots,\bm\eta_M)$, and $w_{\alpha}(\bmeta)$ is a slowly varying envelope wavefunction, symmetric under the simultaneous exchange $(\bm\eta_a,\alpha_a)\leftrightarrow(\bm\eta_b,\alpha_b)$.

The inner product of two states of the form~\eqref{eq:expansion} is given by 
\begin{equation}
\langle f|f'\rangle\simeq\sum_{\alpha}\int\prod_{a=1}^{M}d^2\bm\eta_a\;e^{-W(\bar\eta,\eta)}\,\overline{w_{\alpha}(\bmeta)}\,w'_{\alpha}(\bmeta),
\label{eq:weight}
\end{equation}
where $e^{-W(\bar\eta,\eta)}\simeq\prod_{a<b}|\eta_a-\eta_b|^{2p}$
for separations large compared with the size of the pair and is smoothly regularized at shorter distances, as discussed in Sec.~\ref{sec:setup}.

Note that $w_{\alpha}(\bmeta)$ is neither holomorphic nor antiholomorphic, and that its weight $e^{-W}$ contains no Gaussian piece. The envelope thus describes anyons that feel only the statistical field of each other, $\delta\mathcal B_a=2\partial_{\eta_a}\partial_{\bar\eta_a}W\simeq-2\pi p\sum_{b\neq a}\delta^{(2)}(\eta_a-\eta_b)$, but no background field, the latter having been absorbed into the band dispersion and the valley structure. This is the point at which earlier theories of anyon superconductivity usually start~\cite{Laughlin1988PRL, Laughlin1988Science, Fetter1989, Chen1989}: anyons with purely statistical interactions and no background magnetic field. Here it is derived from a microscopic Hamiltonian in which the anyons do experience a background field. This has important implications when the pairing function is expressed in terms of the electrons, which have to form a vortex lattice, as we will discuss in Sec.~\ref{sec:vortex}.

\section{Effective Hamiltonian}
\subsection{Effective Hamiltonian for the envelope}

We now derive the effective Hamiltonian for the envelope function $w_{\alpha}(\bmeta)$ in the dilute limit. The main corrections to the single-particle dispersion come from two sources: the short-range interaction $\mathcal U_2$ and the many-body part $\delta\mathcal K$ of the K\"ahler potential, which has both a long-distance and a short-distance part. It is therefore convenient to split $\delta\mathcal K=\delta\mathcal K_{\rm smooth}+\delta\mathcal K_{\rm short}$, where $\delta\mathcal K_{\rm smooth}$ is the plasma form~\eqref{eq:plasma} smoothly regularized such that it varies slowly on the lattice scale, and $\delta\mathcal K_{\rm short}$ is the remainder, supported within the size of the pair; correspondingly, the weight in Eq.~\eqref{eq:weight} is identified with $W=\delta\mathcal K_{\rm smooth}$ evaluated at the coarse-grained coordinates. We begin with the smooth part. The derivation of the effective Hamiltonian following the geometric quantization procedure is described in detail in the SM~\cite{SM} and leads, up to corrections of order $O(\Lambda^3)$, to
\begin{equation}
[\hat H_{\rm kin}w]_{\alpha}(\bmeta)=-\frac{2}{m^*}\sum_{a=1}^{M}\partial_{\eta_a}\big[\partial_{\bar\eta_a}-\partial_{\bar\eta_a}W(\bar\eta,\eta)\big]\,w_{\alpha}(\bmeta),
\label{eq:Heff}
\end{equation}
which is Hermitian with respect to the weight $e^{-W}$ of Eq.~\eqref{eq:weight}.

In addition to the kinetic term derived above, the effective Hamiltonian contains short-distance interactions generated both by the short-range two-body term $\mathcal U_2$ and by $\delta\mathcal K_{\rm short}$ in combination with the periodic potential (see SM~\cite{SM} for details). We note that $\mathcal U_2$ itself contains, besides the contribution of the interaction $\hat V$, a contribution of the periodic potential $\hat U$ that arises because the charge profile of two nearby anyons is not the sum of the individual profiles. The contribution of $\delta\mathcal K_{\rm short}$ arises because it carries Fourier components at momenta of the order of the reciprocal lattice vectors, which the periodic potential converts into two types of terms: intravalley umklapp terms and intervalley terms with momentum transfer $\bK_\alpha-\bK_\beta$, which scatter anyons between valleys. Both contributions are associated with large momentum transfer and involve moments of the band structure other than its curvature, i.e. they are parametrically independent of the effective mass. The resulting short-range interaction $\mathcal V_{\rm short}$, which acts on the coordinates $\bmeta$ and on the valley labels, can be expanded into two-body, three-body, and higher terms, each of which can in turn be expanded in pseudopotentials. We note that all the data that specify the problem --- the effective mass and the expansion coefficients of few-body interactions in terms of pseudopotentials --- can be obtained from the few-anyon ED described in Ref.~\cite{Li2026Bound}. 

It is more convenient to pass to a representation in which the Hamiltonian is manifestly Hermitian and the wavefunctions have a flat weight, by defining $\tilde w_{\alpha}=e^{-W/2}w_{\alpha}$, for which the Hamiltonian takes the form
\begin{gather}
\hat{\tilde H}_{\rm eff}=\frac{1}{2m^*}\sum_{a=1}^{M}\Pi_a^\dagger\Pi_a + \mathcal V_{\rm short},\quad
\Pi_a=2\partial_{\bar\eta_a}-\partial_{\bar\eta_a}W, \nonumber \\
[\Pi_a,\Pi_a^\dagger]=-2\,\delta\mathcal B_a,\qquad
\delta\mathcal B_a=2\partial_{\eta_a}\partial_{\bar\eta_a}W
\label{eq:ACCS}
\end{gather}
The first term is a many-body Aharonov--Casher Hamiltonian in the Chern--Simons magnetic field of the other anyons (ACCS). Such Hamiltonians were considered before in Refs.~\cite{Girvin1990, JackiwPi1990, Choi1992}. 

We note that there is some ambiguity in dividing the terms between the kinetic and the interaction part. To see this, note that we can alternatively express the kinetic part as $\frac{1}{2m^*}\bm\Pi^2$ or $\frac{1}{2m^*} \Pi \Pi^\dagger$ which differs from $\frac{1}{2m^*} \Pi^\dagger\Pi$ by a term proportional to $\frac{1}{2m^*}\sum_a\delta\mathcal B_a$, which for a (regularized) CS field is a short-range interaction that can be absorbed into $\mathcal V_{\rm short}$. Within our formalism, the split naturally arose from separating the many-body part of the K\"ahler potential into a smooth and a short distance part. However, it is important to emphasize that nothing so far guarantees that $\mathcal V_{\rm short}$ is small compared to the kinetic energy on the pair scale $\sim 1/(m^* l_{\rm pair}^2)$. Nevertheless, the factorized structure of the kinetic term in 
Eq.~\eqref{eq:ACCS} suggests a possibility of exact solvability. In particular, it means any state satisfying $\Pi_a \tilde w = 0$ is an exact zero-energy state if $\mathcal V_{\rm short}$ can be neglected; in the next section we construct such zero modes; then, in Sec.~\ref{sec:resonant}, we identify the conditions under which they survive the addition of $\mathcal V_{\rm short}$.

\subsection{Zero modes of the ACCS Hamiltonian}
\label{sec:zeromode}

To construct the zero modes of the operator $\Pi_a$, it is instructive to first rewrite the problem in terms of fermions as discussed at the end of Sec.~\ref{sec:setup}. In the flat-weight representation this is a singular gauge transformation given by
\begin{equation}
\tilde w_F = \frac{\Delta(\bar \eta)}{|\Delta(\eta)|}\, \tilde w \qquad \Delta(\eta) = \prod_{a<b}\eta_{ab}
\label{eq:Ugauge}
\end{equation}
where $\eta_{ab} := \eta_a - \eta_b$. $\tilde w_F$ is antisymmetric under the simultaneous exchange $(\bmeta_a,\alpha_a)\leftrightarrow(\bmeta_b,\alpha_b)$ whenever $\tilde w$ is symmetric. The Hamiltonian has the same form as Eq.~\eqref{eq:ACCS} with

\begin{gather}
\Pi^F_a=2\partial_{\bar\eta_a}-\partial_{\bar\eta_a}W_F, \qquad W_F=W+\sum_{a<b}\ln|\eta_{ab}|^2, \nonumber \\ e^{-W_F}\simeq {\prod_{a<b}|\eta_{ab}|^{2(p-1)}
=\prod_{a<b}|\eta_{ab}|^{2/3}}.
\label{eq:fermionicW}
\end{gather}

The $2e/3$ anyons are thus equivalently fermions carrying one third of a statistical flux quantum.

The zero modes of the fermionic problem, $\Pi^F_a\tilde w_F=0$ for all $a$, are
\begin{equation}
\tilde w_F(\bmeta)=e^{W_F/2}\,h_F(\eta)\simeq\prod_{a<b}|\eta_{ab}|^{-1/3}\,h_F(\eta),
\label{eq:zeromodesF}
\end{equation}
with $h_F$ holomorphic in all coordinates and antisymmetric within each valley. To select the ground state among the zero modes, we need a condition at infinity. We do not require $\tilde w_F$ to be normalizable, only that it does not grow in any coordinate. The reason is that the statistical field $\delta\mathcal B$ consists of flux tubes with no uniform background \footnote{Strictly speaking, there is a slow Gaussian part that is negligible at the level of approximation used here}, so that the factor $e^{W_F/2}$ is scale free and cannot confine; the same relaxation was made by Girvin \emph{et al.}~\cite{Girvin1990}, whose zero modes have precisely the form~\eqref{eq:zeromodesF} with no Gaussian factor. Physically, non-growth rather than normalizability is the right condition for the wavefunction of a superconductor, which is normalized only by the volume.

Let us now assume the total number of anyons is divisible by 3, $M = 3M_{\rm v}$. Since $h_F$ is polynomial in $\eta$ and antisymmetric within each valley, the requirement that $\prod_{a<b}|\eta_{ab}|^{-1/3}\,h_F(\eta)$ does not grow at infinity selects uniquely the state 
\begin{equation}
h_F(\eta)=\epsilon(\alpha)\prod_{\alpha=1}^{3}\Delta_\alpha\big(\eta\big),\qquad
\Delta_\alpha\big(\eta\big)=\prod_{\substack{a<b\\ \alpha_a=\alpha_b=\alpha}}\eta_{ab},
\label{eq:hFfactor}
\end{equation}
where each valley has $M_{\rm v}$ anyons. Here $\epsilon(\alpha)$ is the sign of the permutation that groups particle indices by valley and it ensures the correct fermionic exchange antisymmetry as is shown in SM~\cite{SM}. Sending one anyon to infinity while fixing all other anyon positions leads to a $|\eta_a|^{-2/3}$, which is not square integrable as anticipated. More instructively, we can also consider what happens when we take one anyon from each valley and send them to infinity together, which, as we will see, corresponds to a Cooper pair. In this case, we find that the wavefunction does not decay but instead approaches a constant $|\tilde w|\sim \textrm{constant}$. This is the behavior expected for a superconductor. This state corresponds to a filling fraction of 3 and corresponds to fully filling the LLL in each valley as anticipated in mean-field treatments using composite fermions. If $M=3M_{\rm v}+r$ with $r=1,2$, the same counting fixes the populations up to permutations of the valleys, $r$ valleys carrying $M_{\rm v}+1$ anyons and the others $M_{\rm v}$, but leaves a residual freedom, a polynomial $g$ of degree at most one in each coordinate of the valleys with $M_{\rm v}$ anyons, and the corresponding wavefunctions approach constants rather than decaying in some coordinates. This finite degeneracy is associated with the $r$ anyons in excess of a multiple of three, which cannot be accommodated in the paired structure below and behave as free particles at zero energy; these are unimportant in the thermodynamic limit. 

In terms of the bosons
\begin{equation}
\tilde w=e^{W/2}h\simeq\prod_{a<b}|\eta_{ab}|^{-4/3}h, \ h(\eta)= \epsilon(\alpha)\Delta(\eta) \prod_{\alpha=1}^3 \Delta_\alpha(\eta),
\label{eq:wboson}
\end{equation}
We see that intravalley pairs have relative angular momentum $\ell=2$ and behave as $|\eta_{ab}|^{2/3}$, while an intervalley pair has $\ell=1$ and behaves as $|\eta_{ab}|^{-1/3}$. This is a state where particles from different valleys have enhanced probabilities of being close to each other. The ground state wavefunction in electron coordinates can be obtained by first moving from the envelope coordinate $\eta$ to the coordinate $\xi$, then integrating out $\xi$ to get an electron wavefunction as explained in the supplemental material \cite{SM}.  In most cases of interest, where we want to compute the expectation value of some electron operator, it will be more convenient to re-express such operator in terms of the coarse-grained anyon coordinates $\eta$ and evaluate it using the wavefunction (\ref{eq:wboson}) rather than using the electron wavefunction directly. Nevertheless, we provide the explicit expression for the electron wavefunction in SM for completeness.

The zero mode expression (Eq.~\ref{eq:hFfactor} and \ref{eq:wboson}) is the only normalizable zero mode assuming the expression for the weight $W$ is exact at short distance. The actual $W$, however, is regularized at short distances since the statistical flux tube has the finite width $l_{\rm pair}$ of the pair (Sec.~\ref{sec:envelope}). Then $e^{W/2}$ is finite at coincidence and every symmetric holomorphic $h$ gives a normalizable zero mode, whether or not it vanishes when two anyons meet. The zero-mode manifold is thus much larger than the family generated by $h$ given in Eq.~\ref{eq:wboson}: it also contains all states in which two anyons are in the relative $\ell=0$, whose relative wavefunction $e^{W/2}\to |\eta_{ab}|^{-4/3}$ is normalizable at large distances and rendered normalizable at short distances by the finite size of the pair. In the fermionic language they are the states in which $h_F=h/\Delta(\eta)$ has simple poles at coincidence, which are admissible because the fermionic weight $e^{W_F/2}=e^{W/2}\prod|\eta_{ab}|$ vanishes there; the naive classification by holomorphic $h_F$ is therefore not a consequence of the ACCS Hamiltonian alone.

The additional zero modes are lifted by the short-range interaction in the $\ell=0$ channel. In the presence of a repulsive short-range interaction in the $\ell = 0$ channel ( $U_0 \hat P_0$ with $U_0 > 0$), we find that the state (\ref{eq:wboson}) remains an asymptotically exact zero mode, up to small correction in the dilute limit, since it does not contain any pair of particles in the $\ell = 0$ channel, while all other states are expected to acquire an energy of order $U_0$. \footnote{The existence of an energy gap, while difficult to rigorously prove, relies on similar reasoning regarding the Laughlin state, which is the unique state with no particles in the relative $\ell=1$ channel; any other state has an energy gap of order 1. Existence of Laughlin gap was recently proven rigorously \cite{Scheer2026Laughlin}. Gap in the current setup may be established using similar methods.} 

\section{Resonant $2e/3$ dilute anyon gas}
\label{sec:resonant}
In this section, we analyze the conditions under which the physics is controlled by the Aharonov-Casher Hamiltonian. Our strategy is borrowed from dilute quantum gases near a scattering resonance \cite{Braaten2006,Sachdev2007,Nishida2007}. At distances much larger than $l_{\rm pair}$, the statistical flux tubes become point-like, while their core structure and the short-range interactions enter through collision boundary conditions in the infrared (IR) theory. Away from particle coincidences, the IR Hamiltonian has the structure
\begin{gather}
 \hat H_{\rm IR}
 =\frac{1}{2m^*}\sum_{a=1}^{M}
 \bigl(\Pi_a^{(\rm IR)}\bigr)^\dagger
       \Pi_a^{(\rm IR)}+(\text{boundary conditions}),
 \nonumber\\
 \Pi_a^{(\rm IR)}
 =2\partial_{\bar\eta_a}
   +p\sum_{b\neq a}\frac{1}{\bar\eta_a-\bar\eta_b},
 \qquad p=\frac{4}{3} .
 \label{eq:thin_flux_hamiltonian}
\end{gather}
 The collision boundary conditions complete the definition of $\hat H_{\rm IR}$ and are obtained by matching few-body scattering to the full Hamiltonian in Eq.~\eqref{eq:ACCS}. The matched conditions are independent of how short-range terms were distributed between the kinetic energy and $\mathcal V_{\rm short}$.

 A generic boundary condition carries a length, the analog of a scattering length, to be compared with the interparticle distance $L_\rho$. When that length is much larger or much smaller than $L_\rho$, the gas is governed by a scale-invariant boundary condition (a zero-density fixed point) with corrections controlled by the ratio of the two lengths. The zero mode of Sec.~\ref{sec:zeromode} is the ground state at exactly one such fixed point, the one whose boundary conditions are those of the zero mode itself: singular $r^{-1/3}$ behavior for intervalley pairs in $\ell = 1$, regular $r^{2/3}$ in $\ell = 2$. For any other boundary condition, the Hamiltonian is still self-adjoint but is no longer of the form $\Pi^\dagger \Pi$ on that domain, and the zero mode is not an eigenstate. This fixed point is not reached by generic interactions: it requires intervalley pairs to sit at their binding threshold, the analog of tuning a Fermi gas to unitarity. We establish this in three steps. In Sec. \ref{sec:twobody}, we classify the scale-invariant pair boundary conditions and show that reaching the resonant one requires tuning a single coupling $\delta u_1$. In Sec. \ref{sec:ac_structure}, we analyze collisions of three or more anyons and find that one three-body coupling $g_3$ in the valley-singlet channel is marginal while all others are irrelevant. In Sec. \ref{sec:microscopic_route}, we explore how to achieve the ACCS fixed point by tuning microscopic parameters and argue that it is generally expected in some region in parameter space at finite density.

\subsection{Two-particle boundary conditions}
\label{sec:twobody}

Let us consider the two-body problem. In the flat-weight representation, two anyons with relative coordinate $\eta=\eta_1-\eta_2=re^{i\theta}$ and relative angular momentum $\ell$ have $\tilde w\propto e^{i\ell\theta}R_\ell(r)$. For distances $r \gg l_{\rm pair}$, the relative Hamiltonian takes the Aharonov--Bohm form
\begin{equation}
\hat{\tilde h}_{\rm rel}=\frac{1}{m^*}\Big[-\partial_r^2-\frac1r\partial_r+\frac{(\ell-p)^2}{r^2}\Big],\qquad r\gg\ l_{\rm pair} ,
\label{eq:hrel}
\end{equation}
Its two linearly independent zero-energy solutions behave as $r^{-\nu_\ell}$ and $r^{\nu_\ell}$ with $\nu_\ell=|\ell-p|$. For $\ell=0$ and $\ell\geq3$, $\nu_\ell>1$ and only the regular solution $r^{\nu_\ell}$ is normalizable at short distances, so that the boundary condition at coincidence is unique. On the other hand, for $\ell=1$ and $\ell=2$, $\nu_1=1/3$ and $\nu_2=2/3$, respectively, and both solutions are normalizable at short distances. Note that in the bosonic representation, the wavefunction is symmetric intravalley, which only allows for even $\ell$ and hence $\ell=2$, while we can have both $\ell=1$ (valley antisymmetric) and $\ell=2$ (valley symmetric) intervalley states.

For $\ell = 1,2$, the general solution is a superposition,
\begin{equation}
R_\ell(r)=A_\ell\,r^{-\nu_\ell}+B_\ell\,r^{\nu_\ell},\qquad \ell=1,2,
\label{eq:Rell}
\end{equation}
with the ratio $A_\ell/B_\ell$ fixed by the short-distance details, i.e. by the internal structure of the pairs and by $\mathcal V_{\rm short}$, and defines a length scale 
\begin{equation}
r_\ell=|A_\ell/B_\ell|^{1/2\nu_\ell},
\label{eq:rell}
\end{equation}
that breaks the scale invariance of the ACCS Hamiltonian. Thus, there are four scale-invariant points, characterized by the vanishing of one of the two coefficients $A_\ell$, $B_\ell$ in each of the two channels, which we label by $(\sigma_2,\sigma_1)$ with $\sigma_\ell=+$ if $R_\ell\sim r^{\nu_\ell}$ (regular) and $\sigma_\ell=-$ if $R_\ell\sim r^{-\nu_\ell}$ (singular).

These fixed points can also be understood in the RG sense via the scale dependence of the effective two-body coupling $u_\ell(L)\equiv\frac{L R_\ell'(L)}{R_\ell(L)}$. It specifies the logarithmic slope of the zero-energy exterior
wavefunction at a matching radius $L\gg l_{\rm pair}$ and is
independent of its normalization. Changing $L$ while keeping
the ratio $A_\ell/B_\ell$ fixed gives
\begin{equation}
 \frac{du_\ell}{d\ln L}=\nu_\ell^2-u_\ell^2,
 \qquad \ell=1,2.
 \label{eq:pair_beta}
\end{equation}
The singular boundary condition $B_\ell=0$ gives the unstable 
fixed point $u_\ell=-\nu_\ell$, while the regular condition
$A_\ell=0$ gives the stable fixed point $u_\ell=+\nu_\ell$.
Linearizing about these points gives RG eigenvalues
$+2\nu_\ell$ and $-2\nu_\ell$, respectively. For $\ell=1,2$, we define the dimensionless bare detuning from the singular (unstable) fixed point by
$\delta u_\ell\equiv l_{\rm pair}^{2\nu_\ell}B_\ell/A_\ell$.

The fixed point $(++)$ is the fully regular fixed point, which does not require tuning any parameters. The fixed point $(+-)$ corresponds to the Hamiltonian (\ref{eq:ACCS}) with $\mathcal V_{\rm short} = 0$. It is associated with the zero modes of $\Pi$, $\tilde w_\ell=e^{W/2}h_\ell(\eta_1,\eta_2) = |\eta_{12}|^{-4/3} \eta_{12}^\ell$ which behave as $r^{2/3}$ (regular) in the $\ell=2$ channel and $r^{-1/3}$ (singular) in the $\ell=1$ channel. Note that this is an unstable fixed point since, for any finite $r_1$, the solution in the $\ell=1$ channel is dominated by the regular state $r^{1/3}$ at sufficiently long distances, i.e. $B_1/A_1$ is a relevant perturbation of scaling dimension $2\nu_1=2/3$, and generic short-distance physics flows to $(++)$. Thus, realizing the $(+-)$ point requires tuning the two-body potential in the $\ell = 1$ channel to a resonance, i.e. we need to be at the point where charge-$2e/3$ objects are at the threshold of binding in the $\ell=1$ intervalley channel. The fixed points $(-+)$ and $(--)$ can be analyzed similarly.

\subsection{Perturbations at the resonant Aharonov-Casher fixed point}
\label{sec:ac_structure}

We now classify weak local contact perturbations of the scale-invariant ACCS theory, corresponding to a factorizable form of Eq.~\eqref{eq:thin_flux_hamiltonian} with the $(+-)$ pair conditions, and higher-body collision conditions implicitly selected by the factorizable ACCS quadratic form. To determine contact scaling, we place the system in a harmonic trap and use the state--operator correspondence for
Schr\"odinger-invariant scaling theories, which identifies the scaling dimension of local operators with the energy of the corresponding trapped state in units of the oscillator frequency
\cite{WernerCastin2006,Nishida2007,Doroud2016,boisvert2025revisiting}. 

For an $M$-particle cluster, we define the center of mass $Z$ and relative hyperradius $R$ by $Z=\frac1M\sum_{a=1}^M\eta_a$, $R^2
     =\frac1M\sum_{a<b}|\eta_a-\eta_b|^2$. The angular coordinates $\Omega_a=(\eta_a-Z)/R$ specify the positions on a configuration sphere. The relative integration measure is $R^{2M-3}dR\,d\Omega$. For fixed occupations $n_1+n_2+n_3=M$, the minimal-degree ACCS zero mode in the bosonic gauge has the form
\begin{equation}
 \begin{gathered}
 \tilde w_{}=R^{s_M}\Phi_M(\Omega),\\
 s_M=-\frac{M(M-1)}6
       +\frac12\sum_{\alpha=1}^3 n_\alpha(n_\alpha-1), 
 \end{gathered}
 \label{eq:main_ac_homogeneity}
\end{equation}
where $\Phi_M$ includes the valley indices. The radial scaling follows from the fact that each pair supplies the factor $\eta_{ab}/|\eta_{ab}|^{4/3}$ of radial degree $-1/3$, while an equal-valley pair requires an additional numerator zero to make the total wavefunction even, giving the regular degree $2/3$. 

Next, we add a harmonic potential of frequency $\omega$ and confining length scale $L=(m^*\omega)^{-1/2}\gg l_{\rm pair}$, such that the total Hamiltonian is $H_{\rm ACCS}
       +\frac{m^*\omega^2}{2}(R^2+M|Z|^2)$. The statistical potential has no radial or center-of-mass
component, and thus the relative motion is decoupled. Multiplying the homogeneous zero mode by the Gaussian factor gives the exact confined state
\begin{equation}
 \begin{gathered}
 \tilde w_\omega\propto
 e^{-(R^2+M|Z|^2)/(2L^2)}R^{s_M}\Phi_M(\Omega),\\
 E_M=\omega(M+s_M).
 \end{gathered}
 \label{eq:main_trap_state}
\end{equation}
The Gaussian factor preserves the collision boundary conditions and supplies an infrared cutoff.

Suppose we now include a local contact interaction that couples to the same $M$-particle channel. Its linear RG eigenvalue is consequently $y_M=4-2E_M/\omega$, which can be deduced from its first-order energy shift in the normalized trapped state at fixed interaction strength
\begin{equation}
 \begin{gathered}
 \frac{\delta E_M}{\omega}\propto
       \left(\frac{L}{l_{\rm pair}}\right)^{y_M},\\
 y_M
     =4-\frac{4M}{3}
       -\sum_{\alpha=1}^3
          \left(n_\alpha-\frac M3\right)^2.
 \end{gathered}
 \label{eq:main_contact_scaling}
\end{equation}
Thus positive, negative, and zero $y_M$ eigenvalues identify relevant, irrelevant, and marginal perturbations, respectively \cite{Turner2020}.

Among these possible couplings, the intervalley pair channel has $y_2=2/3$, reproducing the relevant two-body detuning identified above. The valley-singlet trimer coupling has $s_3=-1$ and $y_3=0$, and thus it is marginal. A fourth particle necessarily repeats a valley and requires an extra zero: the least suppressed quartet has occupations $(2,1,1)$ and $y_4=-2$. All higher-order contacts with $M\geq4$ are irrelevant at the ACCS point.

For comparison, the single-valley problem at the same exchange statistics can be tuned to its ACCS fixed point with the even $\ell=2$ channel at its singular $r^{-2/3}$ branch
\cite{Turner2020, komargodski2026anyon}. Assigning this branch to every pair formally gives $ s'_M=-\frac{M(M-1)}3$ and $ y'_M=4-2M+\frac{2M(M-1)}3$. These formal exponents become increasingly positive as $M$ increases. Moreover, already at $M=3$, the profile is $R^{-2}$ and its normalization diverges as $\int dR/R$. Crucially, our weaker intervalley singularity and repeated-valley zeros avoid these obstructions.

To establish completeness of the classification discussed above, we must also consider collision channels outside the zero-mode ACCS family. For such a channel, let $R^{s_M}\Phi_M$ denote its own leading collision profile, with $y_M=4-2(M+s_M)$.  If $\Pi_a R^{s_M}\Phi_M\ne0$ for at least one particle, its action lowers the power-law exponent by one. The resulting positive contribution to the energy then contains $ \sim \int dR\,R^{2M-5+2s_M}=\int dR\,R^{-1-y_M}$. However, for $y_M\geq0$, this integral diverges. Consequently, every admissible relevant or marginal channel must have a leading collision profile satisfying all the first-order AC equations. The pair-zero counting above then applies.

The fate of the marginal three-body coupling at long distances requires the nonlinear three-body analysis. With only the pair conditions imposed, the full angular calculation in the SM~\cite{SM} finds precisely one valley-singlet channel admitting two locally normalizable radial behaviors with
\begin{equation}
 F(R)\sim\frac{1}{R}\left(A+B\ln(R/l_{\rm pair})\right).
 \label{eq:main_triplet_boundary}
\end{equation}
The three-body contact interaction fixes the ratio $A/B$. The factorizable AC form selects $B=0$, whereas a nonzero logarithmic coefficient changes the radial three-body boundary condition while preserving the pair conditions.

Define the running contact as the logarithmic derivative
of $RF(R)$ at $R=L$, we find the following RG flow
\begin{equation}
 \begin{gathered}
 g_3(L)\equiv1+\frac{LF'(L)}{F(L)},\quad 
 \frac{dg_3}{d\ln L}=-g_3^2,\\
 g_3(L)=
 \frac{g_3^{(0)}}
      {1+g_3^{(0)}\ln(L/l_{\rm pair})}.
 \end{gathered}
 \label{eq:main_g3_running}
\end{equation}
Here $g_3^{(0)}=g_3(l_{\rm pair})$ is the microscopic
matched coupling. The absence of a linear term in the beta function is a consequence of marginality. The scale-invariant ACCS condition $B=0$ corresponds to the fixed point $g_3=0$. Repulsive coupling, $g_3^{(0)}>0$, is marginally irrelevant. On this side, reaching the ACCS point in the infrared
requires only tuning the relevant pair interaction, while the three-body coupling runs logarithmically to zero. Weak attraction, $g_3^{(0)}<0$, instead generates a length scale $ R_{\rm trimer}=l_{\rm pair}e^{-1/g_3^{(0)}}\gg l_{\rm pair}$ associated with the trimer bound energy  $E_{\rm trimer}\simeq-1/(m^*R_{\rm trimer}^2)$.

\subsection{Microscopic route to the resonant regime}
\label{sec:microscopic_route}

\begin{figure*}
    \centering
    \includegraphics[width=0.9\linewidth]{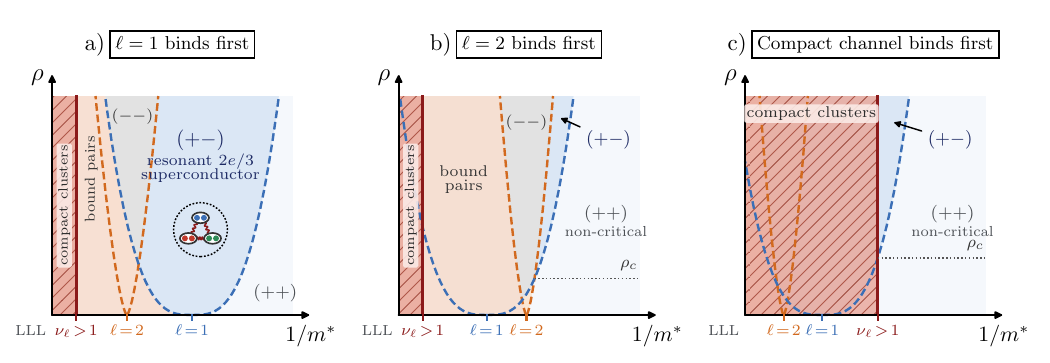}
    \caption{ Schematic illustration of the resonance windows as the anyon dispersion (characterized by the inverse mass $1/m^*$) is reduced from right to left, for three possible cases: (a) $\ell=1$ threshold channel is binding first, (b) $\ell=2$ threshold channel is binding first, and (c) a core-dominated channel with higher angular momentum $\nu_\ell>1$ is binding first. The vertical coordinate $\rho$ denotes the anyon density. Blue and orange curves denote the crossover at $r_\ell\sim L_\rho$ for $\ell=1$ and $\ell=2$. The corresponding dimensionless detuning widths scale as $(\rho l_{\rm pair}^2)^{1/3}$ and $(\rho l_{\rm pair}^2)^{2/3}$. Labels $(\sigma_2,\sigma_1)$ identify scale-invariant two-body boundary conditions reached at the interparticle scale, as defined in the text. The blue unhatched regime describes a resonant anyon superconductor. To the left of each fan, the corresponding pairs are bound with size $r_\ell\ll L_\rho$. The dark hatching marks preemption by compact composites with $\nu_\ell>1$, which have no fan.}
    \label{fig:fan}
\end{figure*}

The preceding discussion established that the resonant fixed point ${(+-)}$ 
requires a single tuning parameter $\delta u_1$, while the only other
coupling that survives in the dilute limit is the marginal three-body valley-singlet
contact term $g_3$. We now ask where in the space of microscopic parameters
this tuning is achieved, and which range of densities is controlled by the
fixed point.

Let us assume that we are at a generic point in parameter space where $2e/3$ anyons do not bind into larger composites. This point should generally be described by the regular fixed point $(++)$ with Hamiltonian $\frac{1}{2m^*} \Pi \Pi^\dagger + \sum_{\ell} U_\ell\,\hat P_\ell$, where $U_\ell$ denotes the short-range pseudopotential coefficient in the relative angular momentum channel $\ell$. At the regular fixed point, the leading boundary-condition perturbations in the $\ell=1,2$ channels are irrelevant, with RG eigenvalues $-2\nu_\ell$. To form a bound state in a channel $\ell$, $U_\ell$ needs to be negative and its absolute value has to exceed a binding threshold $|U_{\ell,c}| \sim 1/(m^* l_{\rm pair}^2)$. We define the microscopic detuning from the corresponding threshold as $\delta U_\ell\equiv U_\ell-U_{\ell,c}$. The dimensionless detuning $\delta u_\ell$ scales as $|\delta u_\ell|\sim m^*l_{\rm pair}^2|\delta U_\ell|$.

Now consider reducing the dispersion, by making the Berry curvature more uniform or switching off the periodic potential in the electronic problem. In the limit of zero dispersion, earlier work \cite{Li2026Bound, Wang2026Molecules} found a tendency for the formation of clusters of quasiholes if the screening length is sufficiently short. Thus, we expect some of the channels to approach their binding threshold as the dispersion is reduced. For channels with $\nu_\ell > 1$, the critical state corresponds to the solution $r^{-\nu_\ell}$ for distances larger than $l_{\rm pair}$, which yields a compact molecule whose size is $\sim l_{\rm pair}$. On the other hand, the bound state size for $\nu_\ell < 1$ ($\ell = 1, 2$) diverges upon approaching the critical point; for finite detuning, its size scales as $r_\ell\sim l_{\rm pair}|\delta u_\ell|^{-1/(2\nu_\ell)}\sim l_{\rm pair}(m^*l_{\rm pair}^2|\delta U_\ell|)^{-1/(2\nu_\ell)}$.

A finite, but dilute, density of $2e/3$ anyons introduces a new length scale: the interparticle distance scale $L_\rho = \rho^{-1/2} \gg l_{\rm pair}$. Once the interparticle distance is  smaller than the scale $r_\ell$, the physics is controlled by the critical point associated with a resonance in the $\ell$ channel. This defines a critical window given by $|\delta u_\ell|\sim(l_{\rm pair}/L_\rho)^{2\nu_\ell}=(\rho l_{\rm pair}^2)^{\nu_\ell}$. Since $\nu_{\ell = 1} = 1/3 < \nu_{\ell=2} = 2/3$, the $\ell = 1$ has a wider critical fan and is expected to control the physics close to the binding threshold of $2e/3$ objects at finite density, \emph{even if the main binding channel at zero density is not $\ell = 1$}. This is demonstrated in Fig.~\ref{fig:fan} which shows different scenarios for the binding threshold being achieved in different channels. In all cases, since the $\ell=1$ resonance is widest, it is expected to control the physics for some finite range of densities.  Moreover, as discussed in the preceding subsection, the $\ell=2$ resonance has a tendency towards clustering since higher-body contact interactions become relevant, and there is no  complete point-particle theory without further fine-tuning. In contrast, the inter-valley $\ell=1$ resonance, whose repeated-valley zeros make all higher-body operators irrelevant, is the only fixed point that comes with a controlled finite-density theory. 

Two comments are in order. First, the regularized factorized Hamiltonian with
$\mathcal V_{\rm short}=0$ sits exactly at the binding threshold of the
$\ell=0$ channel, even though this channel admits no additional boundary
condition in the point-particle theory. \color{black} This corresponds to a zero-energy solution: $\tilde w=e^{W/2} = |\eta_{12}|_{\rm reg}^{-4/3}$ which is an $\ell=0$ zero mode of $\Pi$, regular at coincidence and normalizable at large distances. Thus, any additional attraction in the $\ell=0$ channel binds it. We expect this to be unlikely since the $\ell = 0$ is typically repulsive as it requires the two pairs to sit on top of each other.  If the $\ell=0$ coupling is repulsive, then all such states acquire large energy shifts of the order of $U_0$ and disappear from the low energy theory, as expected from their non-normalizability in a point-particle description. At the same time, it does not affect the $\ell=1$ resonance in any essential way because it keeps the two-body channel unchanged and annihilates the entire three-body valley-singlet sector, whose spatial wavefunctions are odd under every pair exchange. It therefore does not itself generate the marginal singlet coupling $g_3$. Although it may not exactly preserve the full many-body ACCS wavefunction, its residual effect is  irrelevant in the RG sense: according to Sec.~\ref{sec:ac_structure}, its relative correction vanishes at least  as $\rho l_{\rm pair}^2$ at low density since it has a scaling dimension of $y_3=-2$ in a three-body sector beyond valley-singlets, and even more negative in higher-body sectors. This allows repulsive $\ell=0$ core physics to remove unwanted compact states without obstructing the ACCS dilute regime.\color{black}

Second, as shown in the preceding section, three-body interactions are marginal at the $(+-)$ fixed point and can also control its fate: an attractive three-body interaction is marginally relevant and leads to the formation of valley singlet charge $2e$ trimers with the scale $R_{\rm trimer}$, whereas a repulsive three-body interaction is marginally irrelevant. In principle, trimer binding could lead to a stable BEC. However, the actual fate of this regime strongly depends on trimer-trimer interactions and further clustering tendencies. In particular, there is a possibility of phase separation for $R_{\rm trimer}\gg L_{\rho}$, which we discuss in Sec.~\ref{sec:ac_thermo}, along with other finite-density consequences of two-body detuning and $g_3$ interactions.

Thus, writing the short-range two-body interaction defined in \eqref{eq:ACCS} as $\mathcal V^{(2)}_{\rm short}=\sum_{\ell}U_\ell\,\hat P_\ell$, the ground state is given by the $(+-)$ zero mode excluding the $\ell = 0$ channel provided that
\begin{gather}
    (m^* l_{\rm pair}^2|\delta U_1|)^3 \ll \rho l_{\rm pair}^2 \ll (m^* l_{\rm pair}^2|\delta U_2|)^{3/2}, \nonumber \\ 
    U_0 > 0, \quad \delta U_{\ell} > 0,\quad \ell\geq3
\end{gather}
As we will see in the next section, this will allow us to find the anyon superconductor ground state wavefunctions exactly.

\section{Properties of the anyon superconductor}
\label{sec:gs}

\subsection{Pair wavefunction and vortex lattice}
\label{sec:vortex}

We now turn to the superconducting order parameter. We first define the Cooper-pair order parameter
\begin{equation}
\Delta(\bR,\bv)=\langle\Psi^{(N-2,M+3)}|\hat c_{\bR+\bv}\hat c_{\bR-\bv}|\Psi^{(N,M)}\rangle,
\end{equation}
where the removal of two electrons is compensated by the creation of three $2e/3$ pairs. We decompose it into relative-angular-momentum channels, and the component in odd channel $\ell$ is
\begin{equation}
\Delta_\ell(\bR)=\langle\Psi^{(N-2,M+3)}|\hat\Delta_\ell(\bR)|\Psi^{(N,M)}\rangle,
\end{equation}
where $\hat\Delta_\ell(\bR)\propto\int d^2v\,\bar v^{\,\ell}\hat c_{\bR+\bv}\hat c_{\bR-\bv}$, with any odd $\ell$ being the relative angular momentum of the probing pair order parameter.
We now show that $\Delta_\ell(\bR)$ necessarily forms a vortex lattice and discuss the symmetry restrictions on such lattice.

The ground states are eigenstates of the magnetic translations, $\hat T_{\ba_i}|\Psi^{(N,M)}\rangle=e^{iM_{\rm v}\theta_i}|\Psi^{(N,M)}\rangle$ with $e^{i\theta_i}=\det U_{T_{\ba_i}}$ (Sec.~\ref{sec:anyonBS}), while a translation acts on the pair operator as $\hat T_{\bd}\hat\Delta(\bR,\bv)\hat T_{\bd}^{-1}=e^{i\bd\wedge\bR}\hat\Delta(\bR-\bd,\bv)$. Hence
 \begin{align}
    \Delta(\bR+\ba_i,\bv)&=e^{-i\theta_i}\,e^{i\ba_i\wedge\bR}\,\Delta(\bR,\bv),\\
    \Delta_\ell(\bR+\ba_i)&=e^{-i\theta_i}\,e^{i\ba_i\wedge\bR}\,\Delta_\ell(\bR).
    \label{eq:OPtranslation}
\end{align}
which is the quasi-periodicity of a charge-$2e$ field in a magnetic field with one flux quantum per unit cell for the electrons, i.e. two for the pairs. This corresponds to a phase winding of $4\pi$ around the unit cell meaning that the net vorticity, vortices minus antivortices, is two per unit cell.

This argument is purely topological and uses only the charge of the condensate and the flux through the cell. In a Chern band, the electrons obey ordinary periodic rather than magnetic boundary conditions. However, for ideal or Aharonov--Casher bands~\cite{ledwithFractionalChernInsulator2020a, wangExactLandauLevel2021b, morales-duran2024magic, TMDAharonocCasher}, relevant to twisted MoTe${}_2$, the Bloch wavefunctions factorize as $\psi_{\bm k,a}(\bm r)=\chi_a(\bm r)\,\phi_{\bm k}(\bm r)$, where $a$ labels the internal components (layers in twisted MoTe${}_2$), $\chi(\bm r)$ is a normalized $\bm k$-independent spinor Here, $\phi_{\bm k}$ obeys magnetic boundary conditions while the spinor carries the compensating winding, and their product obeys Bloch's theorem.  The quasi-periodicity and the vortex lattice are properties associated with the periodicity properties of $\phi_{\bm k}$. Our analysis therefore applies to superconductors formed in any Chern band admitting such decomposition. The vortex lattice is then not a response to an applied field but an intrinsic property of the state, with two vortices per moir\'e cell in $\Delta$. 

If the band is also ideal, i.e. equivalent to LLL in an inhomogeneous magnetic field, the dependence of the pair amplitude on the center-of-mass coordinate takes the form
\begin{equation}
\Delta_\ell(\bR)=f_\ell(R)\,e^{-\bR^2/2 + \phi(R)},
\label{eq:OPholo}
\end{equation}
where $\phi(\bR)$ is periodic and $f_\ell(R)$ is holomorphic. Holomorphicity implies that the zeros of $f_\ell$ all have positive winding, so antivortices are excluded and there are exactly two vortices per unit cell, and $f_\ell$ is determined by their positions,
\begin{equation}
f_\ell(R) \propto e^{c_\ell R}\,\sigma(R-R_1)\,\sigma(R-R_2),
\label{eq:OPsigma}
\end{equation}
with $\sigma$ the modified Weierstrass sigma function of the lattice $\ba_1\times\ba_2$~\cite{haldane2018modularinvariant, wangExactLandauLevel2021b}. The quasi-periodicity~\eqref{eq:OPtranslation} fixes $c_\ell$ and the sum $R_1+R_2$ in terms of the phases $\theta_i$. The relative position of the two vortices is fixed by rotation symmetry.

Under the rotation about a lattice site, we obtain 
\begin{align}
\Delta(C_n\bR,C_n\bv)=&\omega_n^{L_\text{tot}}\Delta(\bR,\bv) \\
\Delta_\ell(C_n\bR)=&\omega_n^{ L_{\Delta}(\ell)}\Delta_\ell(\bR)
\end{align}
with
\begin{align}
L_\text{tot}\equiv & 6+L_U \quad(\mathrm{mod}\ n)  \\
L_{\Delta}(\ell) \equiv &  L_\text{tot}- \ell
\label{eq:LDelta}
\end{align}
where $L_\text{tot}$ is the total angular momentum carried by each Cooper pair 
with $L_U$ the contribution of the valley triplet~\cite{SM}:
\begin{align}
    \det U_{C_n}\equiv \omega_n^{L_U}.
\end{align}
The mismatch between the internal angular momentum $\ell$ of the pair and $L_\text{tot}$ 
has to be carried by the center-of-mass motion of the pair, that is, by the vortex lattice: since $f_\ell$ is holomorphic, $L_{\Delta}$ is the multiplicity of the zero of $\Delta_\ell$ at the site modulo $n$. This fixes the positions of the vortices for each relative angular momentum channel $\ell$. We derive in SM that the same angular momentum prediction can arise from an effective field theory treatment. The vortex configuration corresponding to each $\ell$ is summarized in Table \ref{tab:vortices}. Following the discussion in Sec.~\ref{sec:symmetries} and table \ref{tab:valleys}, this result directly enables us to relate the angular momentum of the Cooper pair to the properties of the periodic potential generating the anyon dispersion.

Consider $C_6$, the two possible lattice vortices correspond to a double vortex at the site with $L_{\Delta} = 2$ (a triangular lattice of double vortices) or a single vortex at each corner with $L_{\Delta} = 0$ (a honeycomb lattice of single vortices). These correspond to the pairing channels $\ell\equiv3\ (\mathrm{mod}\ 6)$ and $\ell\equiv5\ (\mathrm{mod}\ 6)$, respectively. Note that there is no vortex configuration compatible with $\ell\equiv1\ (\mathrm{mod}\ 6)$, so such a component must vanish. Comparing to our wavefunction, we note that the lowest channel with a nonvanishing projection on the anyon-pair subspace is $\ell_{\min}=9$: the two electrons removed from the $\nu=1/3$ state carry relative angular momentum $3$, and the three pairs created carry internal angular momentum $-2$ each (Sec.~\ref{sec:odlro},~\cite{SM}). The smallest relative angular momentum component of the order parameter therefore has a double vortex at every lattice site, the next one, $\ell=11$, a honeycomb of single vortices, and $\ell=13$ vanishes. Since several relative-angular-momentum channels can contribute with comparable weight, the nodal structure of the full pair amplitude $\Delta(\bR,\bv)$ depends on the relative coordinate $\bv$, while the vortex configuration of each component $\Delta_\ell(\bR)$ is fixed by symmetry.

\begin{table}
\begin{ruledtabular}
\begin{tabular}{cccll}
$C_n$ & $\det U_{T_{\ba_i}}$ & $L_U$ & allowed $\ell$ & vortex configuration\\
\hline
$C_6$ & $1$ & $5$ & $\ell\equiv3\ (6)$ & double vortex at $1a$\\
 & & & $\ell\equiv5\ (6)$ & one vortex at each $2b$\\
\hline
$C_4$ & $1$ & $3$ (1) & $\ell\equiv1\ (4)$ & double vortex at $1b$\\
 & & & $\ell\equiv3\ (4)$ & double vortex at $1a$\\
 & & $1$ (2) & $\ell\equiv1\ (4)$ & double vortex at $1a$\\
 & & & $\ell\equiv3\ (4)$ & double vortex at $1b$\\
$C_4$ & $-1$ & $0$ (1) & $\ell\equiv1\ (4)$ & one vortex at $1a$ and one at $1b$\\
 & & $3$ (2) & $\ell\equiv1\ (4)$ & one vortex at each $2c$\\
\hline
$C_3$ & $1$ & $2$ & $\ell\equiv3\ (6)$ & double vortex at $1a$\\
 & & & $\ell\equiv5\ (6)$ & one vortex at $1b$ and one at $1c$\\
$C_3$ & $\omega_3$ & $0$ & $\ell\equiv3\ (6)$ & double vortex at $1c$\\
 & & & $\ell\equiv5\ (6)$ & one vortex at $1a$ and one at $1b$\\
$C_3$ & $\omega_3^2$ & $0$ & $\ell\equiv3\ (6)$ & double vortex at $1b$\\
 & & & $\ell\equiv5\ (6)$ & one vortex at $1a$ and one at $1c$\\
\end{tabular}
\end{ruledtabular}
\caption{Vortex configurations of the order parameter $\Delta_\ell(\bR)$ for the valley configurations of Table~\ref{tab:valleys}. Channels not listed have no compatible configuration and vanish identically. Wyckoff positions as in Table~\ref{tab:valleys}: $1a=0$; for the hexagonal lattices $1b,1c$ ($2b$) are the corners $(2\ba_1+\ba_2)/3$, $(\ba_1+2\ba_2)/3$; for the square lattice $1b=(\ba_1+\ba_2)/2$ and $2c=\ba_1/2,\ba_2/2$. For the first $C_4$ configuration the result depends on which representation, (1) or (2), the lower band carries.}
\label{tab:vortices}
\end{table}

\subsection{Off-diagonal long-range order}
\label{sec:odlro}

The order parameter of Sec.~\ref{sec:vortex} is a matrix element between states with different particle numbers. To establish superconductivity, we show that the pair correlation function of the ground state itself factorizes at long distances,
\begin{multline}
    C_\ell(\bR',\bR)=\frac{\langle\Psi^{(N,M)}|\hat\Delta_\ell^\dagger(\bR')\,\hat\Delta_\ell(\bR)|\Psi^{(N,M)}\rangle}{\langle\Psi^{(N,M)}|\Psi^{(N,M)}\rangle}\;\\\xrightarrow{|\bR-\bR'|\to\infty}\;\overline{\tilde\Delta_\ell(\bR')}\,\tilde\Delta_\ell(\bR),
\label{eq:odlro}
\end{multline}
i.e. the two-electron reduced density matrix has off-diagonal long-range order. We summarize the derivation of the SM~\cite{SM} for the lowest channel $\ell=9$; the higher channels follow the same steps.

The first step is the action of the pair operator on the anyon-pair manifold. Acting with $\hat c_{\bR+\bv}\hat c_{\bR-\bv}$ on the state~\eqref{eq:pair} and projecting back onto the manifold of $2e/3$ pairs, the two electrons removed at $\bR\pm\bv$ are replaced by three pairs at the same point, and for the lowest channel
\begin{equation}
\hat P_{\rm pair}\,\hat\Delta_9(\bR)\,|\psi_{\xi}^{(N,M)}\rangle\propto e^{-|R|^2/2}\prod_{a=1}^{M}(R-\xi_a)^4\;|\psi^{(N-2,M+3)}_{\xi,R,R,R}\rangle ,
\label{eq:projectedpair}
\end{equation}
where $\ell_{\min}=9$ arises because the two electrons carry relative angular momentum $3$ in the $\nu=1/3$ state while the three created pairs carry internal angular momentum $-2$ each. $\hat\Delta_9(\bR)\,|\psi_{\xi}^{(N,M)}\rangle$ containes many configurations other than those with only quasihole pairs with the relative angular momentum $L=2$, which has higher energy. We focus on the low-energy manifold and exclude them using $\hat P_{\rm pair}$ that selects the quasihole pairs with the relative angular momentum $L=2$. Therefore, what we calculate is the leading part of the correlation. We discuss in SM explicitly how this wavefunction corresponds to an insertion of a Cooper pair from an effective field theory perspective. The wavefunction for the state with a Cooper pair removed at position $R$ is obtained using Eq.~\eqref{eq:projectedpair} and \eqref{eq:ElectronWavefunction}
\begin{align}
    |\Gamma_9(\bR)\rangle = e^{-\frac{|R|^2}{2}}\int d\mu_{\xi}\, f(\bar{\bm\xi})\prod_{a=1}^M(R-\xi_a)^4|\psi_{\bm\xi,R,R,R}^{(N-2,M+3)}\rangle.
\end{align}
Then the projected electron two-particle reduced density matrix (2RDM) is
\begin{equation}
    C_9(\bR',\bR) \propto \langle \Gamma_9(\bR')|\Gamma_9(\bR) \rangle
\end{equation}
 Thus, $C_9(\bR',\bR)$ becomes an integral over two sets of pair coordinates  $\xi = (\xi_1,\dots,\xi_M)$ and $\xi' = (\xi'_1,\dots,\xi'_M)$ of the overlap of two $(M+3)$-pair coherent states $\mathcal{G}_{M+3}(\bar X';X) = e^{\mathcal K_{M+3}(\bar X';X)}$ (with some additional factors of $(R - \xi)^4$) where we defined $X = (\xi_1,\dots,\xi_M, R,R,R)$ and $X' = (\xi'_1,\dots,\xi'_M, R',R',R')$. This describes two states, one with three coordinates pinned at $R$ and one with three pinned at $R'$. In the plasma form of this off-diagonal overlap, each pinned coordinate is paired with a coordinate of the other set. The pairings fall into sectors according to the number $k\leq3$ of pinned coordinates paired with integrated ones. In the sectors with $k<3$, at least one coordinate pinned at $R$ is paired with one pinned at $R'$, which produces a Gaussian factor $e^{-|R-R'|^2/6}$ per such pairing leading to a decay in such sectors (the corresponding clusters for $k=1, 2$ reduce to 1-, or 2-RDM of anyons which are not valley neutral). In the fully exchanged sector $k=3$, the three pairs created at $R$ are identified with three pairs of the ground state and vice versa, and $\mathcal C_9$ reduces to the 3-RDM of the anyon ground state. Using the envelope approximation \ref{eq:envelope}, this can be expressed in terms of the 3-RDM of the envelope function which we denote by $\rho^{(3)}_{\beta'\beta}(\bm\zeta',\bm\zeta)$, where we used the shorthand notation $\beta = (\beta_1,\beta_2,\beta_3)$ and $\bm \zeta = (\bm \zeta_1, \bm \zeta_2, \bm \zeta_3)$. This leads to the expression
\begin{multline}
    C_9(R',R)\propto e^{-\frac{|R|^2+|R'|^2}{2}}\sum_{\beta,\beta'}\int\prod_{i=1}^{3}d^2\bm\zeta_i\,d^2\bm\zeta'_i\;\rho^{(3)}_{\beta'\beta}(\bm\zeta',\bm\zeta)\\\times\prod_{i=1}^{3}\overline{\psi_{\bm\zeta_i,\beta_i}(\bar R)}\,\psi_{\bm\zeta'_i,\beta'_i}(\bar R'),
    \label{eq:C3RDM}
\end{multline}

Finally, we use the plasma analogy ~\cite{Laughlin1983, Girvin1990} applied to the ground state of Sec.~\ref{sec:zeromode}, leading to the factorization of $\rho^{(3)}$
\begin{gather}
\mathcal Z_{\Psi} \propto \int\prod_{i=1}
^{M}d^2\bm\gamma_i \,e^{-H_{\rm pl}} \qquad
H_{\rm pl} = - 2\sum_{a<b} \bm{q}_{\alpha_a} \cdot \bm{q}_{\alpha_b} \ln |\gamma_{ab}|
\label{eq:plasma3}
\end{gather}
where $\bm{q}_{\alpha_a}$ are three vectors of length $\sqrt{2/3}$ at $120^\circ$ angle such that $\bm{q}_{\alpha_a} \cdot \bm{q}_{\alpha_b} = \delta_{\alpha_a \alpha_b}-1/3$ and $\sum_{\alpha_a} \bm{q}_{\alpha_a}=0$. This is a three-component plasma associated with the three valleys, and its interaction matrix has a null vector $(1,1,1)$, corresponding to one particle from each valley. $\rho^{(3)}$ is represented by a screening plasma with two impurity triplets at $\bm\zeta\simeq\bm R$ and $\bm\zeta'\simeq\bm R'$. Each triplet contains one particle from each valley and is neutral with respect to both the interacting plasma channels and the statistical gauge field. Since the two neutral triplets are asymptotically independent at large separation in the screening phase, $\rho^{(3)}$ takes the factorized form $\rho^{(3)}_{\beta'\beta}(\bm\zeta',\bm\zeta)\xrightarrow{|\bm \zeta-\bm \zeta'|\to\infty}\overline{\Phi_{\beta'}(\bm\zeta')}\Phi_{\beta}(\bm \zeta)$. Here $\Phi_{\beta}(\bm\zeta)$ encodes the free-energy cost of inserting an impurity triplet at $\bm\zeta$, and can be evaluated as a plasma expectation value (see SM~\cite{SM}). Combining Eq.~\eqref{eq:C3RDM} with the factorized form of $\rho^{(3)}$ gives the order parameter $\tilde\Delta_9(\bR)$ defined in Eq.~\eqref{eq:odlro} as
\begin{equation}
\tilde\Delta_9(\bR)\propto e^{-|R|^2/2}\sum_{\beta}\int\prod_{i=1}^3d^2\bm\zeta_i\;\Phi_{\beta}(\bm\zeta)\prod_{i=1}^3\overline{\psi_{\bm\zeta_i,\beta_i}(\bar R)}.
\label{eq:OPodlro}
\end{equation}
$\tilde\Delta_\ell(\bR)$ has the translation, rotation and holomorphic properties of Sec.~\ref{sec:vortex} and therefore the same vortex lattice~\cite{SM}.

The preceding calculation uses the exact ACCS wavefunction. At pair
resonance, $\delta u_1=0$, a finite repulsive three-body coupling $g_3$
changes the radial boundary condition of the unique critical
valley-singlet channel identified in Sec.~\ref{sec:ac_structure}.
While its angular wavefunction remains unchanged, its exterior radial profile is modified as $F(R)=[A+B\ln(R/l_{\rm pair})]/R$, with $B/A=g_3^{(0)}\equiv g_3(l_{\rm pair})$, and $R$ is the hyperradius. Consequently, fixing the outer amplitude $L_\rho F(L_\rho)$
at the interparticle scale $L_\rho$ fixes the microscopic amplitude through
\begin{equation}
 Z_3(\rho)
 =\frac{l_{\rm pair}F(l_{\rm pair})}
        {L_\rho F(L_\rho)}
 =\frac{1}{1+g_3^{(0)}\ln(L_\rho/l_{\rm pair})}.
 \label{eq:pair_source}
\end{equation}
The physical meaning of this renormalization is that the local electronic pair operator creates or removes a compact three-anyon singlet, and therefore
probes the collision amplitude at the microscopic scale. Each local pair insertion thus carries one factor $Z_3$, and the electronic ODLRO weight contains $Z_3^2$. For repulsive $g_3^{(0)}$, this reduces the overlap of the local electronic pair operator with the condensate by suppressing compact
triplet configurations.  As we will show below, the renormalization factor $Z_3^2$ is unrelated to the phase stiffness, which is constrained by the density conservation.

\subsection{Correlations in a three-component plasma from Toda theory}
\label{sec:toda_correlations}

Before proceeding with other physical properties of the ACCS
superconductor, we describe a method for calculating its plasma
correlations. The essential observation is that the logarithmic
pair factors in Eq.~\eqref{eq:plasma3} can be reproduced by
Gaussian-field correlations. For two real Gaussian fields
$\boldsymbol\phi=(\phi_1,\phi_2)$ with action
$S_0=(8\pi)^{-1}\int d^2x\,(\nabla\boldsymbol\phi)^2$,
the propagator is
$\langle\phi_i(\mathbf r)\phi_j(\mathbf0)\rangle_0
=-2\delta_{ij}\ln r$. Using the valley charge vectors defined
above, we obtain
\begin{equation}
 \begin{gathered}
 V_\alpha(\mathbf x)
 =:e^{i\mathbf q_\alpha\cdot\boldsymbol\phi(\mathbf x)}:,
 \\
 \left\langle\prod_a
 V_{\alpha_a}(\boldsymbol\gamma_a)\right\rangle_0
 =
 \prod_{a<b}|\gamma_{ab}|^{2\mathbf q_{\alpha_a}
                                    \cdot\mathbf q_{\alpha_b}}
 =e^{-H_{\rm pl}}.
 \end{gathered}
 \label{eq:toda_plasma_identity}
\end{equation}
Normal ordering removes contractions at the same point.
The Gaussian average is normalized to one, and integration
over the spatially constant fields imposes
$\sum_a\mathbf q_{\alpha_a}=0$, which requires equal valley
populations.

In the homogeneous screening phase, local correlations in the
fixed-density and grand canonical ensembles agree in the
thermodynamic limit. We can therefore sum the particle integrals
over valley-balanced particle numbers with an auxiliary fugacity
$\zeta$. Using Eq.~\eqref{eq:toda_plasma_identity}, the resulting
grand partition function becomes
\begin{equation}
 \begin{gathered}
 \Xi
 =\left\langle
 \exp\left[\zeta\int d^2x\sum_{\alpha=1}^3
 V_\alpha(\mathbf x)\right]\right\rangle_0
 =\frac{\int\mathcal D\boldsymbol\phi\,e^{-S_{\rm T}}}
        {\int\mathcal D\boldsymbol\phi\,e^{-S_0}},
 \\
 S_{\rm T}
 =\int d^2x\left[
 \frac{(\nabla\boldsymbol\phi)^2}{8\pi}
 -\zeta\sum_{\alpha=1}^3 V_\alpha(\mathbf x)
 \right].
 \end{gathered}
 \label{eq:toda_main_action}
\end{equation}
Expanding the exponential reproduces the plasma integrals and
their particle-number factorials. The three vectors
$\sqrt3\,\mathbf q_\alpha$ are the simple and affine roots of
$A_2$, so this is the imaginary $A_2$ affine Toda theory
\cite{Fateev2002}. The fugacity $\zeta$ is chosen to reproduce
the prescribed density $\rho$. This construction is analogous
to the field-theory representation of the two-component plasma
\cite{TwoComponentPlasma2000,Tellez_2005}, and we provide
further details in the SM~\cite{SM}.

We denote normalized averages with $e^{-S_{\rm T}}$ by
$\langle\cdot\rangle_{\rm T}$ and averages in the ACCS probability
weight by $\langle\cdot\rangle_{\rm pl}$. From Eq.~\eqref{eq:toda_main_action}, it is easy to see that inserting the microscopic
density $\rho_\alpha(\mathbf x)$  corresponds to inserting
$\zeta V_\alpha(\mathbf x)$ in the Toda representation \cite{SM}.
Consequently,
\begin{equation}
 \begin{aligned}
 \langle\rho_\alpha(\mathbf r)\rangle_{\rm pl}
 &=\frac{\rho}{3}
 =\zeta\langle V_\alpha\rangle_{\rm T},
 \\
 \langle\rho_\alpha(\mathbf r)\rho_\beta(\mathbf0)
       \rangle_{\rm pl}
 &=
 \frac{\rho}{3}\delta_{\alpha\beta}\delta^{(2)}(\mathbf r)
 +\zeta^2
 \langle V_\alpha(\mathbf r)V_\beta(\mathbf0)\rangle_{\rm T}.
 \end{aligned}
 \label{eq:toda_density_dictionary}
\end{equation}
The first line fixes $\zeta$ at the prescribed density. Similarly, inserting one particle from each valley gives $\langle\rho_1\rho_2\rho_3\rangle_{\rm pl}
=\zeta^3\langle V_1V_2V_3\rangle_{\rm T}$, with the same positions on both sides.

These identities allow us to calculate the short-distance
contact amplitudes. At separations much smaller than
$L_\rho$, Gaussian contractions give the leading
operator product
$V_\alpha(\mathbf r)V_\beta(\mathbf0)
\simeq r^{2\mathbf q_\alpha\cdot\mathbf q_\beta}
:e^{i(\mathbf q_\alpha+\mathbf q_\beta)
\cdot\boldsymbol\phi(\mathbf0)}:$.
The remaining one-point operator is averaged in the interacting Toda
theory. For two different valleys,
$\mathbf q_1+\mathbf q_2=-\mathbf q_3$, whereas a triplet containing one particle from each valley has zero total charge. Applying these products to Eq.~\eqref{eq:toda_density_dictionary} and its three-particle counterpart gives
\begin{equation}
 \begin{aligned}
 \langle\rho_1(\mathbf r)\rho_2(\mathbf0)\rangle_{\rm pl}
 &\simeq
 \frac{\zeta^2}{r^{2/3}}
 \left\langle:e^{-i\mathbf q_3\cdot\boldsymbol\phi}:
 \right\rangle_{\rm T},
 \\
 \langle\rho_1(\mathbf r)\rho_2(\mathbf r')
                  \rho_3(\mathbf0)\rangle_{\rm pl}
 &\simeq
 \frac{\zeta^3}
 {\bigl(|\mathbf r|\,|\mathbf r'|\,
                    |\mathbf r-\mathbf r'|\bigr)^{2/3}},
 \end{aligned}
 \label{eq:toda_main_contacts}
\end{equation}
where all pair separations involved in the collision are much
smaller than $L_\rho$. The three-point function contains no additional expectation values because its three vertex charges cancel.

The required one-point functions are known exactly
\cite{Fateev2002}. In our normalization,
\begin{equation}
 \begin{gathered}
 \langle V_3\rangle_{\rm T}
 =
 \frac{\sqrt{\pi}}{2^{2/3}}
 \left[\frac{\Gamma(1/3)}{\Gamma(2/3)}\right]^{3/2}
 \zeta^{1/2},
 \\
 \left\langle:e^{-i\mathbf q_3\cdot\boldsymbol\phi}:
 \right\rangle_{\rm T}
 =e^{\mathcal J}\langle V_3\rangle_{\rm T},
 \qquad
 \mathcal J\approx0.28021.
 \end{gathered}
 \label{eq:toda_main_one_point}
\end{equation}
The first expression determines $\zeta$ through
Eq.~\eqref{eq:toda_density_dictionary}.

The same representation determines the limiting behavior of the equal-time structure factor. For the total density $\rho(\mathbf r)=\sum_\alpha\rho_\alpha(\mathbf r)$, we define $S_{\rm ACCS}(k)=\frac1\rho\int d^2r\,e^{-i\mathbf k\cdot\mathbf r}
 [\langle\rho(\mathbf r)\rho(\mathbf0)\rangle_{{\rm pl}}-\rho^2]$. At small momentum, we find
\begin{equation}
 S_{\rm ACCS}(k)
 =\frac32-\frac{3k^2}{32\pi\rho}
 +O\!\left(\frac{k^4}{\rho^2}\right),
 \qquad k\ll 1/L_\rho,
 \label{eq:toda_main_structure_factor}
\end{equation}
 where we used $\int d^2r\,[\langle\rho(\mathbf r)\rho(\mathbf0)
\rangle_{{\rm pl}}-\rho^2]=3\rho/2$ and
$\int d^2r\,r^2 [\langle\rho(\mathbf r)\rho(\mathbf0)
\rangle_{{\rm pl}}-\rho^2]=3/(8\pi)$. The first relation follows by differentiating the mean-density identity
with respect to $\ln\zeta$ and using
Eq.~\eqref{eq:toda_main_one_point}. The second follows from
the exact second-moment sum rule for the screening plasma,
as detailed in the SM~\cite{SM}.

At large momentum, the leading correction is instead determined by the $r^{-2/3}$ collision singularity in Eq.~\eqref{eq:toda_main_contacts}. This gives
\begin{equation}
 \begin{aligned}
 S_{\rm ACCS}(k)
 &=1+
 2^{25/9}e^{\mathcal J}
 \left(\frac{\pi}{3}\right)^{2/3}
 \left[\frac{\Gamma(2/3)}{\Gamma(1/3)}\right]^2
 \left(\frac{\rho}{k^2}\right)^{2/3}
 +\cdots,
 \end{aligned}
 \label{eq:toda_structure_factor_tail}
\end{equation}
for $ 1/L_\rho \ll k\ll 1/l_{\rm pair}$.  The numerical coefficient in front of $(\rho/k^2)^{2/3}$ is approximately $2.39$.

\subsection{Collective response and thermodynamics}
\label{sec:ac_thermo}

The exact ACCS state constructed in Sec.~\ref{sec:zeromode} has
Cooper-pair off-diagonal long-range order and vanishing bulk energy
at every density. Its
compressibility is therefore infinite and the sound mode has a $z=2$ dynamical exponent. In order to restore finite compressibility and a linear sound mode we have to account for the perturbations discussed in Sec.~\ref{sec:ac_structure} which  generate density-dependent energy corrections. We derive both regimes from an effective density--phase action, working at zero temperature. Our approach will closely parallel superfluid effective field theory constructions developed for unitary gases \cite{SON2006197}.

Below the gap for relative-valley density fluctuations, the only soft variables are the total density and the charge-$2e$ phase. The prescribed mean density is $\rho=M/\mathcal A$, with
$\mathcal A$ the area. We write the local density as
$\rho(\mathbf x,t)=\rho+\delta \rho(\mathbf x,t)$, where fixed particle
number requires $\int d^2x\,\delta \rho=0$. The central quantity for our discussion will be the energy density, $e(\rho)$. Its derivative determines the anyon chemical potential, $\mu=e'(\rho)$. A smooth Cooper-pair phase multiplies the anyon wavefunction by $\exp[i\sum_a\varphi(\mathbf r_a)/3]$, giving the Berry term
$-\rho\partial_t\varphi/3$. The corresponding velocity is
$\mathbf v=\nabla\varphi/(3m^*)$, which contributes to energy as $m^* \rho\mathbf v^2/2$. The resulting real-time action is
\begin{equation}
 \begin{gathered}
\mathcal L=\rho(\mathbf x,t)X-e\bigl(\rho(\mathbf x,t)\bigr)
-\frac{(\nabla\rho)^2}{18m^*\rho(\mathbf x,t)}+\cdots,\\
 X=\mu-\frac{\partial_t\varphi}{3}
       -\frac{(\nabla\varphi)^2}{18m^*}.
 \end{gathered}
 \label{eq:thermo_density_gradient_action}
\end{equation}
The density-gradient term is retained because exactly at the ACCS point, $e( \rho)=0$ and $\mu=0$, and so density fluctuations cost only gradient energy, a feature generically expected in frustration-free models of superfluids~\cite{PhysRevB.111.174520}. The prefactor of the density-gradient term is fixed by matching with $S_{\rm ACCS}(k)$ in Eq.~\eqref{eq:toda_main_structure_factor} at $e(\rho )=0$. Thus, at the ACCS point, the inverse phase propagator has the $z=2$ structure
\begin{equation}
 G_{\varphi\varphi}^{-1}(\Omega,k)
 =m^*\rho \frac{\Omega^2}{k^2}
       +\frac{\rho k^2}{9m^*},\qquad
 \omega_k=\frac{k^2}{3m^*},
 \label{eq:thermo_phase_kernel}
\end{equation}
where $\Omega$ is Euclidean frequency.

Going beyond the ACCS point, we need to determine $e(\rho)$. The pair and triplet contacts generate the leading uniform energy
through their expectation values in the ACCS state:
\begin{equation}
 e(\rho)=\frac{\rho^2}{m^*}
 \left[C_2\delta u_1(L_\rho)+C_3g_3(L_\rho)\right]+\cdots,
 \label{eq:thermo_eos}
\end{equation}
where the dots denote subleading corrections. This expression involves two main ingredients. 
First, the constants $C_2,C_3>0$ measure short-distance correlations and can be computed as
$C_2=(4\pi/\rho^{5/3})\lim_{r\to0}
r^{2/3}\langle \rho_1(\mathbf r) \rho_2(0)\rangle_{\rm pl}$ and
$C_3=\lim_{\epsilon\to0}(\rho\epsilon)^{-2}
\int_{R<\epsilon}d^2r\,d^2r'\,
\langle \rho_1(\mathbf r)\rho_2(\mathbf r')\rho_3(0)\rangle_{\rm pl}$,
where $R^2=(r^2+r'^2+|\mathbf r-\mathbf r'|^2)/3$ defines the hyperradius. We evaluate these constants exactly in \cite{SM} using Eq.~\eqref{eq:toda_main_contacts} and additional results from the Toda theory. Second, we need the scale-dependent couplings $\delta u_1(L_\rho)$ and $g_3(L_\rho)$. Using the
matching conventions of Sec.~\ref{sec:twobody},
$\delta u_1(L_\rho)=\delta u_1(l_{\rm pair}^2 \rho)^{-1/3}$ and
$g_3(L_\rho)=\{[g_3^{(0)}]^{-1}+
\tfrac12\ln[1/(\rho l_{\rm pair}^2)]\}^{-1}$, where
$g_3^{(0)}=g_3(l_{\rm pair})$. The
pair contribution scales as $\rho^{5/3}$, whereas the triplet
contribution is quadratic with a logarithmic coefficient, giving $ m^*e''(\rho)=\frac{10}{9}C_2 \delta u_1(L_\rho)   +2C_3g_3(L_\rho)+\cdots$. We first discuss the case where both perturbations are repulsive, and then comment on attractive cases later.

For $e''>0$, the density-gradient term $(\nabla \rho)^2/\rho$ can be ignored at sufficiently long wavelengths, and integrating out the density fluctuations amounts to solving $e'(\rho)=X$. The resulting nonlinear phase-only theory is
\begin{equation}
 \begin{gathered}
 \mathcal L_\varphi=P(X),\qquad P(X)=X \rho(X)-e(\rho(X)),\\
 e'(\rho(X))=X.
 \end{gathered}
 \label{eq:thermo_pressure_action}
\end{equation}
The function $P$ can be associated with the thermodynamic pressure \cite{SON2006197}, and
$P'(\mu)=\rho$ and $P''(\mu)=1/e''(\rho)$. 
For repulsive pair detuning alone,
$e(\rho)=C_2\rho^{5/3}/(m^*r_1^{2/3})$, with
$r_1=l_{\rm pair}(\delta u_1)^{-3/2}$, hence $P\propto r_1X^{5/2}$. The exponent $5/2$ coincides with the three-dimensional unitary Fermi gas, while here it arises from the pair RG eigenvalue $2/3$ and retains the detuning length $r_1$. In contrast, exactly at pair resonance, $e(\rho)\simeq C_3 \rho^2g_3(L_{\rho})/m^*$ and
$P(X)\simeq m^*X^2/[4C_3g_3(L_{\rho(X)})]$.
This logarithmic quadratic pressure resembles a dilute two-dimensional repulsive Bose gas~\cite{Mora2009}, but its running interaction here is the three-anyon valley singlet.

The energy and the flow term determine the uniform
responses and sound velocity:
\begin{equation}
 \frac{\partial \rho}{\partial\mu}=\frac1{e''(\rho)},\quad 
 J_s=\frac{\rho}{9m^*},\quad
 c^2=\frac{\rho e''(\rho)}{m^*}.
 \label{eq:thermo_compressibility}
\end{equation}
Here $\partial \rho/\partial\mu$ is the thermodynamic compressibility and $J_s$ is the
coefficient of $(\nabla\varphi)^2/2$ in the energy. Pair-detuning-dominated regime gives $c^2=2\mu/(3m^*)$, while at pair resonance
$c^2\simeq\mu/m^*$ and
\begin{equation}
 \frac{\partial \rho}{\partial\mu}\simeq
 \frac{m^*}{2C_3}\left[
 \frac1{g_3^{(0)}}+\frac12\ln\frac1{ \rho l_{\rm pair}^2}\right].
 \label{eq:thermo_resonant_susceptibility}
\end{equation}
The susceptibility grows logarithmically upon dilution, while the sound velocity is nonzero for any finite density.

Retaining the density gradient allows to resolve the crossover to $z=2$ ACCS dynamics at shorter length scales. The linearized equations of motion are
$\partial_t\delta \rho=-\rho \nabla^2\varphi/(3m^*)$ and
$\partial_t\varphi=-3e''\delta \rho+
\nabla^2\delta \rho/(3m^*\rho)$, yielding
\begin{equation}
 \omega_k^2=c^2k^2+\frac{k^4}{9(m^*)^2},\qquad
 k_*=3\sqrt{m^*\rho e''(\rho)}.
 \label{eq:thermo_dispersion}
\end{equation}
The healing length $k_*^{-1}$ separates a proper sound regime with $z=1$ at $k\ll k_*$,
where Eq.~\eqref{eq:thermo_pressure_action} applies, from the quadratic $z=2$ mode at $k_*\ll k\ll\sqrt \rho$. The density susceptibility $G_{\rho\rho}(\Omega,k)$ and equal-time structure factor $S(k)$ are
\begin{equation}
G_{\rho\rho}(\Omega,k)=\frac{\rho k^2/m^*}{\Omega^2+\omega_k^2},\quad 
 S(k)
     =\frac32\frac{k}{\sqrt{k^2+k_*^2}}.
 \label{eq:thermo_density_response}
\end{equation}
The structure factor is linear in $k$ in the sound regime and
approaches $3/2$ above $k_*$. At ACCS it stays finite although the
static susceptibility diverges as $9m^*\rho/k^2$.

Relative-valley excitations require separate care near the ACCS point. Within a single-mode approximation, these excitations have energy $2\pi \rho/(3m^*)$, but their energy can be further lowered by depleting the surrounding total density. Exactly at the ACCS fixed point, such a polaronic cloud can have arbitrarily low energy (see \cite{SM} for an explicit variational state construction).  However, with small repulsive $g_3(L_\rho)$, the bulk energy is finite, and the valley gap becomes
\begin{equation}
 \Delta_v\sim\frac{8\pi\rho}
 {9m^*\ln[1/g_3(L_\rho)]}.
 \label{eq:thermo_relative_scale}
\end{equation}
 This gap scale is parametrically larger than the sound crossover energy $ck_*\sim g_3(L_\rho)\rho/m^*$, supporting the charge-only description of the zero-temperature sound regime. 

In order to describe the electromagnetic response we now couple our theory to an external vector potential $\mathbf A$. Minimal coupling replaces $\nabla\varphi$ by
$\nabla\varphi-2e\mathbf A$. In addition, there is also a magnetization contribution that can be extracted from the ACCS Hamiltonian. Specifically, we consider  $\Pi_a\mapsto\Pi_a-i(2e/3)(A_x+iA_y)(\mathbf r_a)$, with
a weak smooth field $\mathbf A=\hat{\mathbf z}\times\nabla\chi$ with periodic $\chi$ and zero added flux. This changes the exact ACCS state
to $\tilde{w}[\mathbf A]\propto
 e^{-(2e/3)\sum_a\chi(\mathbf r_a)}\tilde{w}$.
With $B=\partial_xA_y-\partial_yA_x$, its linear density response is
$\delta \rho(\mathbf k)/B(\mathbf k)=(4e/3)\rho S(k)/k^2$.
Matching this result from the field theory side at $e(\rho)=0$ and $k\rightarrow 0$ fixes the magnetization term to be $-2e\delta \rho B/(9m^*)$. After integrating out everything except $\mathbf A$, we find the following transverse response function 
\begin{equation}
 K_T(k)=\frac{4e^2\rho}{9m^*}
             \frac{k_*^2}{k^2+k_*^2}.
 \label{eq:thermo_magnetic_factorized}
\end{equation}
Positive curvature restores $K_T(0)=4e^2J_s$. Exactly at the ACCS point, density rearrangement cancels the
energy cost of a finite-momentum magnetic modulation, so $K_T(k\ne0)=0$ while the uniform-twist stiffness remains finite.

We also briefly comment on the cases where at least one of the perturbations is attractive. At pair resonance, a weak attractive $g_3$ instead gives $e''<0$ and an unstable long-wavelength density mode. The subsequent clustering leads to phase separation if a stable dense phase terminates the contraction. A stable fluid of local charge-$2e$ trimers could instead condense while retaining the parent's topological order. Its stability depends on interactions between trimers. The opposite case of pair attraction with repulsive $g_3$ produces a minimum of the energy per particle. Solving $\rho e'(\rho)-e(\rho)=0$ gives, to leading order,
\begin{equation}
 \rho_* =\frac1{r_1^2}
 \left[\frac{2C_2}{3C_3g_3(L_{\rho_{*}})}\right]^3,\quad
 e''(\rho_*)\simeq\frac{C_3g_3(L_{\rho_*})}{3m^*}>0.
 \label{eq:thermo_self_bound_liquid}
\end{equation}
where $r_1=l_{\rm pair}|\delta u_1|^{-3/2}$. When $\rho_*$ lies in the resonant dilute window and competing
microscopic binding thresholds are excluded, any mean density below $\rho_*$ leads to a phase separation of the resonant superconductor at density $\rho_*$ and the undoped FQH state.

So far, we have discussed zero-temperature properties. At finite temperature, valley-imbalance excitations, which are not included in our density--phase action, may also affect the thermal
transition. Although these modes are gapped, their thermal population contributes to the free energy and compressibility. Near the ACCS point, where $e''(\rho)$ is small, even a dilute
population could modify the equation of state. We leave this contribution for future work and consider below the transition predicted by the total-density and phase action, assuming that relative-valley fluctuations do not alter the nature of the finite temperature transition.

For weak repulsive interactions, thermal occupation of the nearly quadratic $z=2$ charge mode, $\omega_k\simeq k^2/(3m^*)$, produces a logarithmic reduction of the phase stiffness. These excitations contribute to the normal density, with the logarithm coming from the two-dimensional momentum integral between the crossover scale and the thermal momentum $k_T\sim\sqrt{m^*T}$. The perturbations $\delta u_1$ and $g_3$ cut off this infrared enhancement by restoring sound below a crossover scale $\sim k_*$. Combining the resulting reduction of the
zero-temperature stiffness $J_s=\rho/(9m^*)$ with the standard BKT condition for the long-distance stiffness,
$J_s^{\rm R}(T_c^-)=2T_c/\pi$, gives
\begin{equation}
 T_c\simeq\frac{8\pi\rho}{9m^*}
 \left[
 \ln\frac{\mathcal C}{m^*e''(\rho)}
 \right]^{-1},
 \label{eq:thermo_tc}
\end{equation}
to leading logarithmic accuracy for
$0<m^*e''(\rho)\ll1$. The dimensionless factor $\mathcal C$
depends on thermal fluctuations near the crossover and on
the microscopic structure of vortex cores.

For pair-detuning-dominated repulsion,
$m^*e''(\rho)\simeq(10C_2/9)( \rho r_1^2)^{-1/3}$,
where $r_1$ is the two-body scattering length defined above.
Thus $T_c\propto \rho/[m^*\ln(\rho  r_1^2)]$ within the
resonant window $r_1^{-2}\ll \rho \ll l_{\rm pair}^{-2}$.
At fixed nonzero detuning, further dilution eventually leaves
this window. Exactly at pair resonance, we have
$m^*e''(\rho)\simeq2C_3g_3(L_\rho)$, with
$g_3(L_\rho)^{-1}=g_3(l_{\rm pair})^{-1}
+\tfrac12\ln[1/(\rho l_{\rm pair}^2)]$.
When the density logarithm dominates the bare inverse coupling,
Eq.~\eqref{eq:thermo_tc} gives
\begin{equation}
 T_c\simeq
 \frac{8\pi \rho}
 {9m^*\ln\ln[1/(\rho l_{\rm pair}^2)]}.
 \label{eq:thermo_tc_triplet}
\end{equation}
The inner logarithm comes from the running three-body repulsion,
and the outer logarithm from thermal occupation of the nearly
quadratic charge mode. This is similar to the BKT temperature
of a dilute two-dimensional Bose gas with logarithmically
running repulsion.

\section{Discussion}

\subsection{Relation to BCS/BEC crossover and unitary limit}

It is instructive to compare our story to the BCS--BEC crossover of a dilute three-dimensional (3D) Fermi gas with a short-range $s$-wave interaction~\cite{Ho2004,Nishida2007,Giorgini2008,Randeria2014}. There, everything is organized by the dimensionless combination $k_Fa$, where $a$ is the scattering length. For weak attraction, $a$ is small and negative, and the ground state is a BCS superfluid with a coherence length $\xi\sim k_F^{-1}e^{\pi/(2k_F|a|)}\gg k_F^{-1}$. As the attraction gets stronger,  $|a|$ increases until it diverges at the two-body binding threshold corresponding to  $1/k_Fa=0$. This is the unitary limit where the coherence length is of the order of the interparticle distance. As the attraction is increased further, $a$ becomes positive and decreasing, and for $k_Fa\ll1$ the system is a BEC of dimers of size $a\ll k_F^{-1}$. The scattering length is a statement about boundary conditions: the $s$-wave solution outside the range of the potential is $1-a/r$, so $a$ is the ratio of the coefficients of the singular ($1/r$) and regular ($1$) branches. Thus, it plays a role similar to the ratio $|A_1/B_1|=r_1^{2/3}$ of our theory, with the anomalous dimension $2\nu_1=2/3$ replacing the dimension $1$ at three-dimensional unitarity. In this language the resonant fixed point we focus on, $r_1\to\infty$, is analogous to the unitary point of the $2e/3$ anyon gas, and the window of validity of the critical theory $L_\rho/r_1 \lesssim 1$ is the analog of $|1/k_Fa|\lesssim1$.

Nevertheless, we emphasize a few distinctions relative to the 3D gas. First, the ground state at our fixed point is known exactly, whereas that of the unitary Fermi gas is not. Second, the unitary Fermi gas at finite density has a finite energy per particle and a finite sound velocity, whereas the resonant anyon gas has zero energy at every density: its compressibility is infinite and its Goldstone mode quadratic. Third, the fate of the attractive side in our case is unclear. The two-body problem has a bound state, a $4e/3$ dimer, but the dimers carry valley charge and interact with each other and with the remaining anyons through the Chern--Simons field, so that the dimer gas is not obviously a BEC superconductor, and to first order in $\delta u_1$ the uniform state has negative compressibility, suggesting phase separation rather than a smooth crossover. Whether the same two-body interaction also produces larger composites is a separate question we leave to future works.

The three-body coupling defines a second axis through the fixed point, along which the analogy is instead with a two-dimensional (2D) Fermi gas with a contact interaction, since in both cases the coupling is marginal. In the 2D gas, a bound state forms for arbitrarily weak attraction with a size that decreases with increasing the magnitude of the attractive interaction. The repulsive coupling is marginally irrelevant so it flows to the free Fermi gas. Similarly, for the resonant anyon gas, the repulsive 3-body coupling is marginally irrelevant and flows back to the ACCS fixed point. The superconductor on this side resembles a BCS superconductor in that the few-body problem has no charge-$2e$ bound state. However, unlike in the weak-coupling BCS regime, the pair size is of the same order as the interparticle distance. Thus, the whole repulsive regime is similar to the 3D unitary gas, where the coherence length is also of the order of the interparticle distance and there is also no bound state. The contrast with the 2D gas is then that the repulsive side of the latter is a free fixed point, whereas ours is a superconductor, whose energy at the fixed point vanishes and is set entirely by the marginally irrelevant coupling.

An attractive three-body interaction, $g_3 < 0$, leads to a charge-$2e$ valley-singlet trimer in the few-body problem. The fate of a finite density of such $2e$ trimers depends on the details of their interaction and is beyond the scope of this work. We note that a possible BEC superconductor out of such trimers will be distinct from the $2e/3$ anyon superconductor since it retains the topological order of the parent Laughlin state and thus, we do not generally expect a direct transition between the two; a region of coexisting phases or phase separation is thus a more natural expectation for $g_3 < 0$.   

\subsection{Relevance to experiments on MoTe${}_2$}
Since twisted MoTe${}_2$ serves as the experimental motivation for this work, it is instructive to examine the possible connections between our analysis and the experimental phase diagram. The resonant ACCS fixed point we identified separates an anyon SC phase from a region with phase coexistence or phase separation characterized by negative compressibility. In theory, such transitions can be tuned through the two-body interaction in the $\ell = 1$ channel or by sitting at the two-body resonance and tuning the three-body interaction. Experimentally, it is unclear whether we can directly tune these parameters, but we can investigate possible experimental tuning knobs to exit the superconductor as possible candidates for this transition. For example, we can transition out of the SC phase by reducing the screening length or making the berry curvature more uniform at fixed filling. We note that the experimentally available tuning knob, namely the displacement field, likely moves us in the opposite direction, making the anyons lighter and breaking the $2e/3$ pairs rather than tuning to higher body resonances. This represents an alternative route to exist the $2e/3$ superconductor. A concrete prediction of our theory is that this transition will be associated with a compressibility peak with a characteristic density dependence. To see this, note that the compressibility close to the critical point is given by $$m^*e''(\rho) = \frac{10C_2}{9}  \delta u_1 (\rho l_{\rm pair}^2 )^{-1/3} + \frac{2 C_3 }{g_3^{-1} + \frac{1}{2} \ln[1/(\rho l_{\rm pair}^2)]}$$ Assuming the density dependence in the logarithm can be neglected, this takes the form $e''(\rho) = a_0 \rho^{-1/3} + a_1$ whose $\rho^{-1/3}$ scaling is a unique prediction to our theory.

One of our main observations is that the large compressibility of the SC close to the resonant point implies very soft valley fluctuations. This prediction deviates significantly from the mean-field picture, which predicts a gap controlled by the main energy scale in the problem $\rho/m^*$. The soft valley fluctuations are expected to strongly modify the finite temperature physics, leading to deviations from the standard BKT predictions. It will be interesting to study the effect of such fluctuations quantitatively in the future. In particular, measurements of the temperature dependence of the superfluid stiffness, similar to those done in trilayer graphene \cite{KimSuperfluid2025}, could provide important information about the role of such valley fluctuations. We note that current experiments in twisted MoTe${}_2$ ~\cite{Xu2025SC} exhibit a broad resistive transition, which could be relevant to the soft relative-valley fluctuations discussed here.

\section{Acknowledgements}
We acknowledge fruitful discussions with Ashvin Vishwanath, Dam T. Son, and Srinivas Raghu. This work is supported by NSF CAREER grant DMR award No. 2441781 and Alfred P. Sloan Research Fellowship to EK. TW acknowledges additional support from the Harvard Quantum Initiative Fellowship and DOE (USA) office of Basic Sciences Grant No. DE-FG02-03ER46076. Z.~H. is supported by the Gordon and Betty Moore Foundation EPiQS Award 8683. PN is supported by the Gordon and
Betty Moore Foundation through Grant GBMF14251 to the
American Chemical Society for the ACS-Moore Postdoctoral Bridge Award in Quantum Materials. This work uses the FASRC Cannon cluster supported by the FAS Division of Science Research Computing Group at Harvard University, the MIT Office of Research Computing and Data (ORCD). The authors acknowledge the use of AI tools Claude Opus 5.5, Fable 5.1 and ChatGPT 6.1 for searching the literature, assistance with derivations, and assistance with writing the manuscript.

\begingroup
\renewcommand{\addcontentsline}[3]{}
\bibliographystyle{apsrev4-2} 
\bibliography{refs}
\endgroup

\clearpage
\renewcommand{\bibliography}[1]{}
\renewcommand{\bibliographystyle}[1]{}

\onecolumngrid
\setcounter{secnumdepth}{2}
\setcounter{page}{1}
\setcounter{equation}{0}
\setcounter{figure}{0}
\setcounter{table}{0}
\setcounter{section}{0}

\renewcommand{\theequation}{S\arabic{equation}}
\renewcommand{\thefigure}{S\arabic{figure}}
\renewcommand{\thetable}{S\arabic{table}}
\renewcommand{\thesection}{S\arabic{section}}
\makeatletter
\renewcommand\p@section{} 
\makeatother
\begin{center}
{\large\bf Supplemental Material for}\\
\vspace{0.5em}
{\it ``Resonantly interacting $2e/3$ anyons: a controlled microscopic theory of anyon superconductivity''}
\end{center}

\let\addcontentsline\SMaddcontentsline   

\bigskip
\tableofcontents
\bigskip

\section*{Overview}

This Supplemental Material has two parts. Secs.~\ref{sec:review}--\ref{sec:pairs}
derive the low-energy anyon theory from the microscopic wavefunctions. Starting
from the Laughlin quasihole states in a periodic potential, geometric quantization
gives the Hamiltonian for wavepackets of $e/3$ quasiholes and of $2e/3$ quasihole
pairs near the three band minima (Secs.~\ref{single-particle-band-structure} and
\ref{sec:Hamiltonian}). It consists of an Aharonov--Casher--Chern--Simons kinetic
term, whose statistical weight is fixed by the K\"ahler potential, and a
short-range interaction. Within the degenerate zero-energy manifold of the ACCS
term, a repulsive short-range interaction selects the valley-singlet
state, which is the exact superconducting wavefunction. Sec.~\ref{sec:pairs} establishes the symmetry of
its Cooper-pair order parameter, its off-diagonal long-range order through the
plasma analogy, and its Chern--Simons field-theory description.

Secs.~\ref{sec:sm_fixed_points}--\ref{sec:seft} take only the infrared content of
this derivation as input: the point-anyon Hamiltonian, with the microscopic
structure of the short-range terms encoded in boundary conditions at particle
coincidence. Sec.~\ref{sec:sm_fixed_points} classifies the scale-invariant boundary
conditions, identifies the resonant fixed point whose zero mode is the exact
superconducting wavefunction, and shows that it has a single marginal coupling,
associated with three-body collisions. Sec.~\ref{sec:seft} derives the low-energy
superfluid effective action from this zero mode: its structure factor, the
equation of state with the exact contact amplitudes, the density--phase action and
electromagnetic response, and the gap of the relative-valley excitations.

\section{Review of geometric quantization}\label{sec:review}
In this section, we will present a self-contained derivation of the Hamiltonian operator from the path integral using geometric quantization (Berezin-Toeplitz quantization). This is valid since our path integral is defined on a K\"ahler manifold. A K\"ahler manifold is a $2M$-dimensional manifold with a complex structure, a symplectic structure, and a metric that are compatible with each other. In our problem, the K\"ahler potential is given by $\mathcal K(\bar \xi, \xi) = \ln \langle \xi|\xi \rangle$ where $\xi = (\xi_1,\dots,\xi_M)$ denotes the $M$ complex quasihole coordinates, $|\xi \rangle$ is a set of states spanning the Hilbert space of interest that depends holomorphically on $\xi$. The K\"ahler potential determines the effective magnetic field for the $l$-th quasihole via
\begin{equation}
    {\mathcal B}_l(\bar \xi,\xi) = 2\partial_{\bar \xi_l} \partial_{\xi_l} \mathcal K(\bar \xi, \xi).
\end{equation}
The Hilbert space can be identified with anti-holomorphic functions of $\xi$ with the appropriate boundary condition $f(\bar \xi) = \langle \xi|f \rangle$ and inner product that can be constructed from the knowledge of $\mathcal K$. This is done by introducing the measure $d\mu_{\bm \xi}$ defined by requiring that the reproducing kernel relation 
\begin{equation}
    \int d\mu_{\bm \xi} \, e^{\mathcal K(\bar \omega,\xi)} f(\bar \xi) = f(\bar \omega), 
    \label{RepKernel}
\end{equation}
is satisfied for any anti-holomorphic function $f(\bar \xi)$. Here, $\mathcal K(\bar \omega, \xi)$ is obtained from $\mathcal K(\bar \xi, \xi)$ by analytic continuation. This identity is equivalent to the existence of a resolution of unity of the form 
\begin{equation}
    \int d \mu_{\bm \xi} |\xi \rangle \langle \xi| = 1,
\end{equation}
which leads to (\ref{RepKernel}) upon the identification $f(\bar \xi) = \langle \xi|f \rangle$. We also have 
\begin{align}
    |f\rangle = \int d \mu_{\bm \xi} f(\bar \xi) |\xi\rangle.
\end{align}
The measure $d\mu_{\bm \xi}$ defines an inner product on the space of anti-holomorphic functions given by
\begin{equation}
    \langle f|g \rangle = \int d\mu_{\bm \xi} [f(\bar \xi)]^* g(\bar \xi).
\end{equation}

For a given quantum operator $\hat V$, the coherent state path integral associates a classical function known as the Q-symbol defined via
\begin{equation}
    \mathcal U(\bar \xi, \xi) = \frac{\langle \xi| \hat V|\xi \rangle}{\langle \xi|\xi \rangle} = e^{-\mathcal K(\bar \xi, \xi)} V(\bar \xi, \xi), \qquad V(\bar \xi, \xi) = \langle \xi| \hat V|\xi \rangle.
    \label{Q-symbol}
\end{equation}
One can also define a different classical function $\mathcal U_P$, known as the Glauber-Sudarshan P-symbol  via
\begin{equation}
    \hat V = \int d\mu_{\bm \xi}\; \mathcal U_P(\bar \xi, \xi) |\xi \rangle \langle \xi|.
\end{equation}
The two are related via
\begin{equation}
    \mathcal U(\bar \omega, \omega) = e^{-\mathcal K(\bar \omega, \omega)} \int d\mu_{\bm \xi} e^{\mathcal K(\bar \xi, \omega) + \mathcal K(\bar \omega, \xi)} \mathcal U_P(\bar \xi, \xi).
    \label{PtoQ}
\end{equation}

The Berezin-Toeplitz quantization provides a concrete recipe for constructing the Hilbert space operator from the P-symbol $\mathcal U_P(\bar \xi, \xi)$, or equivalently from the off-diagonal kernels $\mathcal U(\bar \omega, \xi)$ and $V(\bar \omega, \xi)$, via
\begin{equation}
    [\hat V f](\bar \omega) = \int d\mu_{\bm \xi} e^{\mathcal K(\bar \omega, \xi)} \mathcal U_P(\bar \xi, \xi) f(\bar \xi) = \int d\mu_{\bm \xi} \mathcal U(\bar \omega, \xi) e^{\mathcal K(\bar \omega, \xi)} f(\bar \xi) = \int d\mu_{\bm \xi} V(\bar \omega, \xi) f(\bar \xi),
    \label{QuantizationV}
\end{equation}
Here, $V(\bar \omega,\xi)=\langle \omega|\hat V|\xi\rangle$ and $\mathcal U(\bar \omega,\xi)=e^{-\mathcal K(\bar \omega,\xi)}V(\bar \omega,\xi)$ denote the off-diagonal analytic continuations of the corresponding diagonal quantities.

\subsection{Relation between the measure and the K\"ahler}
Consider a system of $M$ particles, each carrying charge $s$. They experience a uniform effective background magnetic field $\mathcal B = s = \frac{1}{\mu}$. The K\"ahler potential can then be written as
\begin{align}
    e^{\mathcal K(\bar\xi,\xi)} = \langle \xi|\xi\rangle,\qquad \mathcal K(\bar\xi,\xi) = \frac{\bar\xi \xi}{2\mu} + \delta\mathcal K(\bar\xi,\xi).
\end{align}
where $\bar\xi \xi = \sum_{i=1}^M \bar\xi_i \xi_i$. We also write the measure as 
\begin{align}
    d\mu_{\bm \xi} = \bigg(\prod_i\frac{d^2\bm\xi_i}{2\pi \mu}\bigg)\  \mathcal F(\bar\xi,\xi) e^{-\frac{\bar\xi \xi}{2\mu}}
\end{align}
$\mathcal F$ and $\delta \mathcal K$ relate each other through the reproducing relation. To see this, we define the Fourier transform 
\begin{align}
     \mathcal F(\bar{\xi},\xi) = \int d^2 \bm p\, F_{\bm p} e^{i\bm p\cdot \bm \xi},\qquad e^{\delta \mathcal K(\bar{\xi},\xi)} = \int d^2 \bm q\, K_{\bm q} e^{i \bm q\cdot \bm \xi}
     \label{F-K-Fourier-transform}
\end{align}
where $\bm p = (\bm p_1,\dots,\bm p_M)$ and $d^2\bm p = \prod_{i=1}^M d^2 \bm p_i$. The reproducing relation Eq.\eqref{RepKernel} now becomes 
\begin{align}
    f(\bar{\omega}) = \int d^2 \bm q\, e^{\frac{i}{2} q\bar{\omega}}f(\bar{\omega} + i\mu\bar{q}) \int d^2 \bm p\, F_{\bm p} K_{\bm q-\bm p} e^{-\frac{\mu}{2} p\bar{q}}
\end{align} 
which requires
\begin{align}
    \int d^2 \bm p\, F_{\bm p} K_{\bm q-\bm p} e^{-\frac{\mu}{2}p\bar{ q}} = \delta^{(2)}(\bm q)
\end{align}
Here $\delta^{(2)}(\bm q) = \prod_i \delta^{(2)}(\bm q_i)$. Here we use the shorthand $p\bar q=\sum_{i=1}^{M}p_i\bar q_i$.

In the long-distance limit, where $\bm p$ and $\bm q$ are small, we have $e^{-\frac{\mu}{2}p\bar q}\simeq 1$, so that $F$ is approximately the convolution inverse of $K$. Equivalently, in real space, we obtain
\begin{align}
    \mathcal F(\bar\xi,\xi) \simeq e^{-\delta\mathcal K(\bar\xi,\xi)},\qquad |\xi_i-\xi_j|\gg 1.
\end{align}

\section{Single-particle band structure} \label{single-particle-band-structure}

We consider the dispersion of a single particle in the presence of a periodic potential and an effective magnetic field $\mathcal B = s = \frac{1}{\mu}$. Here, a single particle refers to either a $1/3$ quasihole or a $2/3$ quasihole pair. The periodic potential has primitive lattice vectors $\bm a_1$ and $\bm a_2$, with a unit-cell area $\bm a_1\wedge\bm a_2=2\pi$.

The periodic potential takes the form
\begin{align}
    \mathcal U_1(\bar\xi,\xi)=\sum_{\bm a}V_{\bm a}e^{i\bm a\wedge\bm\xi}.
    \label{periodic-potential}
\end{align}
where $\bm a$ denotes the lattice vectors, $\bm a=n_1\bm a_1 + n_2\bm a_2$, and $\bm a\wedge \bm b  = \frac{\bar a b - a\bar b}{2i} = a_x b_y - a_y b_x$. We also define $\wedge \bm a = (a_y,-a_x)$. We consider periodic potentials with $C_n$ rotational symmetry, where $n=3,4,6$. The corresponding Hamiltonian acting on an anti-holomorphic wavefunction $f(\bar\xi)$ is obtained from Berezin--Toeplitz quantization Eq.~\eqref{QuantizationV} as
\begin{align}
    [\hat H f](\bar\omega)=\int d\mu_{\bm\xi}\,e^{\mathcal K(\bar\omega,\xi)}\mathcal U_1(\bar\omega,\xi)f(\bar\xi).
    \label{BT-Hamiltonian}
\end{align}
The K\"ahler potential and the measure corresponding to the magnetic field $\mathcal B$ are 
\begin{align}
    \mathcal K=\frac{\bar\xi\xi}{2\mu},\qquad d\mu_{\bm \xi} = \frac{d^2\bm\xi}{2\pi \mu}e^{-\frac{\bar\xi\xi}{2\mu}}
\end{align}

We define  the magnetic translation operator by
\begin{align}
    [\hat T_{\bm d}f](\bar\xi)=e^{-\frac{|d|^2}{4\mu}}e^{-\frac{d\bar\xi}{2\mu}}f(\bar\xi+\bar d),
    \label{magnetic-translation-definition}
\end{align}
which satisfies
\begin{align}
    \hat T_{\bm d}\hat T_{\bm d'}=e^{-\frac{i}{\mu}\bm d\wedge\bm d'}\hat T_{\bm d'}\hat T_{\bm d}=e^{-\frac{i}{2\mu}\bm d\wedge\bm d'}\hat T_{\bm d+\bm d'}.
\end{align}
The corresponding single-particle Hamiltonian in Eq.~\eqref{BT-Hamiltonian} can be written as
\begin{align}
    \hat h=\sum_{\bm a}\tilde V_{\bm a}\hat T_{\mu\bm a},\qquad \tilde V_{\bm a}=e^{\frac{\mu|a|^2}{4}}V_{\bm a},\qquad [\hat h f](\bar\xi)=\sum_{\bm a}V_{\bm a}e^{-\frac{a\bar\xi}{2}}f(\bar\xi+\mu\bar a)
\end{align}
We also find that the Hamiltonian commutes with the primitive lattice translations, $[\hat h,\hat T_{\bm a_1}]=[\hat h,\hat T_{\bm a_2}]=0$.

\subsection{1/3 quasihole}

For a $1/3$ quasihole, the effective magnetic field is $\mathcal B=s=\frac{1}{3}$, corresponding to $\mu=q=3$.

The normalized eigenfunctions of this Hamiltonian are given by
\begin{align}
    \psi_{\bm k}(\bar\xi)=\mathcal N\,\overline{\sigma\big(\xi-iqk\,\big|\,a_1,qa_2\big)}\,e^{\frac{i}{2}k\bar\xi}e^{-\frac{q}{4}|k|^2},
\end{align}
where $\mathcal N$ is the normalization constant. Here the modified Weierstrass sigma function $\sigma$ is defined as (here $2\pi\mathcal B_{\sigma}=\bm\alpha_1\wedge\bm\alpha_2$)
\begin{align*}
    \sigma(z|\alpha_1,\alpha_2)=\frac{\alpha_1}{\theta_1'(0)}\theta_1\bigg(\frac{z}{\alpha_1}\bigg|\frac{\alpha_2}{\alpha_1}\bigg)e^{\frac{\bar\alpha_1}{4\alpha_1\mathcal B_{\sigma}}z^2},\qquad \theta_1(z|\omega)=\sum_{n=-\infty}^{\infty}e^{i\pi\omega(n+\frac{1}{2})^2+2\pi i(n+\frac{1}{2})(z+\frac{1}{2})}
\end{align*}
which satisfies
\begin{align*}
    \sigma(z+\alpha|\alpha_1,\alpha_2)=\eta_{\bm\alpha}e^{\frac{\bar\alpha}{2\mathcal B_{\sigma}}(z+\frac{\alpha}{2})}\sigma(z|\alpha_1,\alpha_2)
\end{align*}
where $\eta_{\bm\alpha}=1$ if $\bm\alpha/2$ is a lattice vector and $\eta_{\bm\alpha}=-1$ otherwise,
\begin{align*}
    \eta_{\bm\alpha}=(-1)^{m+n+mn},\qquad \bm\alpha=m\bm\alpha_1+n\bm\alpha_2.
\end{align*}

For the wavefunction $\psi_{\bm k}$, we have the translation relation
\begin{align}
    \psi_{\bm k}(\bar\xi+q\bar a)=\eta_{q\bm a}e^{\frac{q|a|^2}{4}}e^{iq\bm k\cdot\bm a}e^{\frac{1}{2}a\bar\xi}\psi_{\bm k}(\bar\xi).
\end{align}
The magnetic translation operator acts on $\psi_{\bm k}$ as
\begin{align}
    \hat T_{\bm d}\psi_{\bm k}(\bar\xi)=e^{\frac{i}{2}\bm k\cdot\bm d}\psi_{\bm k-\frac{\wedge\bm d}{q}}(\bar\xi).
\end{align}

The dispersion corresponding to $\psi_{\bm k}$ is
\begin{align}
    E_{\bm k}=\sum_{\bm a}\tilde V_{\bm a}\eta_{q\bm a}e^{iq\bm a\cdot\bm k}.
\end{align}
Define the magnetic reciprocal lattice vectors by $\bm\beta_i\cdot\bm\alpha_j=2\pi\delta_{ij}$, with $\bm\alpha_1=\bm a_1$ and $\bm\alpha_2=q\bm a_2$. Since $\psi_{\bm k}$ is not an eigenstate of $\hat T_{\bm a_2}$, acting with $\hat T_{\bm a_2}$ generates three degenerate states whose momenta differ by $\bm\beta_1/3$. The three band minima are related by
\begin{align}
    \bm K_\alpha=\bm K_1-\frac{\alpha-1}{3}\bm\beta_1,\qquad \alpha=1,2,3.
\end{align}
Expanding the dispersion around each band minimum $\bm K_\alpha$ as $\bm k=\bm K_\alpha+\bm\kappa$, the $C_n$ symmetry with $n=3,4,6$ ensures that the effective mass is isotropic. Therefore, we obtain $E_{\bm K_\alpha+\bm\kappa}=E_{\min}+\frac{|\bm\kappa|^2}{2m^*}+O(\kappa^3)$, with
\begin{align}
    \frac{1}{m^*}=-\frac{q^2}{2}\sum_{\bm a}|a|^2\tilde V_{\bm a}\eta_{q\bm a}e^{iq\bm a\cdot\bm K_\alpha}.
\end{align}

\subsection{2/3 quasihole pair}

For a $2/3$ quasihole pair, the effective magnetic field is $\mathcal B=s=\frac{2}{3}$, corresponding to $\mu=\frac{1}{s}=\frac{q}{2}=\frac{3}{2}$. We define the magnetic unit cell by $\bm\alpha_1=\bm a_1$ and $\bm\alpha_2=3\bm a_2$, and the magnetic reciprocal lattice vectors by $\bm\beta_i\cdot\bm\alpha_j=2\pi\delta_{ij}$. They are given by $\bm\beta_1=\wedge\bm a_2$ and $\bm\beta_2=-\frac{1}{3}\wedge\bm a_1$. This magnetic unit cell encloses two magnetic flux quanta. Therefore, there are two bands in the reduced Brillouin zone spanned by $\bm\beta_1$ and $\bm\beta_2$.

We define two normalized basis states $\psi_{\bm k,\tau}(\bar\xi)$,
\begin{align}
    \psi_{\bm k,\tau}(\bar\xi)=\mathcal N_0e^{-\frac{\mu}{4}\left|k-i\frac{\tau a_1}{2\mu}\right|^2}\overline{\sigma\left(\xi-i\mu k-\frac{\tau}{2}a_1\,\big|\,a_1,\mu a_2\right)}e^{\frac{i}{2}\bar\xi k}e^{\frac{\tau a_1\bar\xi}{4\mu}},\qquad \tau=0,1.
\end{align}
where $\mathcal N_0$ is the normalization constant. The states $\psi_{\bm k,\tau}$ are mutually orthonormal with respect to the Bargmann measure $\frac{d^2\bm\xi}{2\pi\mu}e^{-\frac{|\xi|^2}{2\mu}}$. We now consider the action of $\hat T_{\mu\bm a}$ on the two-component basis. Writing $\bm a=n_1\bm a_1+n_2\bm a_2$, we have $\mu a=\frac{3n_1}{2}a_1+n_2\mu a_2$. For even $n_1$, the translation preserves the component index $\tau$, while for odd $n_1$, it exchanges the two components. Therefore, we divide the lattice vectors $\bm a=n_1\bm a_1+n_2\bm a_2$ into two sets,
\begin{align}
    \mathcal L_0=\{\bm a=n_1\bm a_1+n_2\bm a_2\,|\,n_1\in2\mathbb Z\},\qquad \mathcal L_1=\{\bm a=n_1\bm a_1+n_2\bm a_2\,|\,n_1\in2\mathbb Z+1\}.
\end{align}
Acting with $\hat T_{\mu\bm a}$ on the basis $\psi_{\bm k,\tau}$, we have
\begin{align}
    \hat T_{\mu\bm a}\psi_{\bm k,\tau}&=\eta_{\mu\bm a}e^{i\mu\bm a\cdot\bm k}e^{-\frac{i\tau}{2}\bm a_1\wedge\bm a}\psi_{\bm k,\tau},\qquad \bm a\in\mathcal L_0,\\
    \hat T_{\mu\bm a}\psi_{\bm k,\tau}&=\eta_{\mu\bm a-(\tau-\frac{1}{2})\bm a_1}e^{i\mu\bm a\cdot\bm k}e^{-\frac{i}{2}(\tau-\frac{1}{2})\bm a_1\cdot\bm k}e^{-\frac{i}{4}\bm a_1\wedge\bm a}\psi_{\bm k,1-\tau},\qquad \bm a\in\mathcal L_1.
\end{align}
For $\hat h\psi_{\bm k,\tau}=\sum_{\tau'}h_{\tau'\tau}(\bm k)\psi_{\bm k,\tau'}$, the Hamiltonian in the two-component basis can be written as
\begin{align}
    h(\bm k)=\begin{pmatrix}h_{00}(\bm k)&h_{01}(\bm k)\\h_{10}(\bm k)&h_{11}(\bm k)\end{pmatrix},
\end{align}
where
\begin{align}
    h_{00}(\bm k)&=\sum_{\bm a\in\mathcal L_0}\tilde V_{\bm a}\eta_{\mu\bm a}e^{i\mu\bm a\cdot\bm k},\qquad h_{01}(\bm k)=\sum_{\bm a\in\mathcal L_1}\tilde V_{\bm a}\eta_{\mu\bm a-\frac{1}{2}\bm a_1}e^{i\mu\bm a\cdot\bm k}e^{-\frac{i}{4}\bm a_1\cdot\bm k}e^{-\frac{i}{4}\bm a_1\wedge\bm a},\\
    h_{10}(\bm k)&=\sum_{\bm a\in\mathcal L_1}\tilde V_{\bm a}\eta_{\mu\bm a+\frac{1}{2}\bm a_1}e^{i\mu\bm a\cdot\bm k}e^{\frac{i}{4}\bm a_1\cdot\bm k}e^{-\frac{i}{4}\bm a_1\wedge\bm a},\qquad h_{11}(\bm k)=\sum_{\bm a\in\mathcal L_0}\tilde V_{\bm a}\eta_{\mu\bm a}e^{i\mu\bm a\cdot\bm k}e^{-\frac{i}{2}\bm a_1\wedge\bm a}.
\end{align}
The two band dispersions are
\begin{align}
    E_{\pm}(\bm k)=\frac{h_{00}(\bm k)+h_{11}(\bm k)}{2}\pm\sqrt{\left[\frac{h_{00}(\bm k)-h_{11}(\bm k)}{2}\right]^2+|h_{01}(\bm k)|^2}.
\end{align}
Let $c_{\pm}(\bm k)=\left(c_{0,\pm}(\bm k),c_{1,\pm}(\bm k)\right)^T$ be the normalized eigenvectors of $h(\bm k)$. The corresponding single-particle eigenstates are
\begin{align}
    \psi_{\bm k,\pm}(\bar\xi)=\sum_{\tau=0,1}c_{\tau,\pm}(\bm k)\psi_{\bm k,\tau}(\bar\xi).
\end{align}

Since $\psi_{\bm k,\tau}$ is not an eigenstate of $\hat T_{\bm a_2}$, acting with $\hat T_{\bm a_2}$ generates three degenerate states whose momenta differ by $-2\bm\beta_1/3\equiv\bm\beta_1/3$. The three band minima are related by
\begin{align}
    \bm K_\alpha=\bm K_1-\frac{\alpha-1}{3}\bm\beta_1,\qquad \alpha=1,2,3.
\end{align}
Expanding around a minimum, $\bm k=\bm K_\alpha+\bm\kappa$, we have $E_-(\bm K_\alpha+\bm\kappa)=E_{\min}+\frac{|\bm\kappa|^2}{2m^*}+O(\kappa^3)$, where $m^*$ is the effective mass of the lower band. 

\subsection{Translation and rotation symmetries for 2/3 quasihole pairs}

We study how magnetic translations and $C_n$ ($n=3,4,6$) rotations act on the states at the three band minima, $\mathcal H_-=\{\psi_{\bm K_\alpha,-}\}_{\alpha=1}^3$, for $2/3$ quasihole pairs. To do so, it is useful to first consider the six-dimensional space spanned by $\mathcal H=\{\psi_{\bm K_\alpha,\tau}\}_{\alpha=1,2,3;\,\tau=0,1}$. Since the magnetic translation group for $\mathcal B=s=2/3$ has a minimal three-dimensional representation, the representation on $\mathcal H_-$ is irreducible. Therefore, we obtain it as an irreducible component of the six-dimensional representation on $\mathcal H$.

For a symmetry operation $\hat O$, we define its representation matrix $U_O^{(\mathcal H)}$ on $\mathcal H$ and $U_O$ on $\mathcal H_-$ by
\begin{align}
    [\hat O\psi_{\bm K_\alpha,\tau}](\bar\xi)=\sum_{\beta=1}^3\sum_{\tau'=0}^1(U_O^{(\mathcal H)})_{(\beta,\tau'),(\alpha,\tau)}\psi_{\bm K_\beta,\tau'}(\bar\xi),\qquad [\hat O\psi_{\bm K_\alpha,-}](\bar\xi)=\sum_{\beta=1}^3(U_O)_{\beta\alpha}\psi_{\bm K_\beta,-}(\bar\xi).
\end{align}

\subsubsection{Translation symmetry}

Let us first consider the magnetic translation symmetry. The action of the magnetic translation operator on a function $f(\bar\xi)$ is defined by Eq.~\eqref{magnetic-translation-definition}. Its action on the six basis functions $\psi_{\bm K_\alpha,\tau}$ is
\begin{align}
    \hat T_{\bm a_1}\psi_{\bm k,\tau}=-e^{i\bm k\cdot\bm a_1}\psi_{\bm k,\tau},\qquad \hat T_{\bm a_2}\psi_{\bm k,\tau}=e^{\frac{i}{2}\bm k\cdot\bm a_2-\frac{i\pi\tau}{3}}\psi_{\bm k-\frac{2}{3}\bm\beta_1,\tau}.
\end{align}
Restricting to the six states $\{\psi_{\bm K_\alpha,\tau}\}_{\alpha=1,2,3;\,\tau=0,1}$, this magnetic-translation representation consists of two equivalent three-dimensional irreducible representations. We can fix $\tau$ to evaluate the determinant of either irreducible representation.

\subsubsection{Rotation symmetry}

We next consider the rotation symmetry. The action of the $C_n$ rotation operator on a function is defined by
\begin{align}
    [\hat C_n f](\bm\xi)=f(C_n^{-1}\bm\xi),\qquad [\hat C_n f](\xi)=f(\omega_n^{-1}\xi),\qquad [\hat C_n f](\bar\xi)=f(\omega_n\bar\xi),
    \label{rotation-operator-definition}
\end{align}
where $\omega_n=e^{2\pi i/n}$ and $C_n$ denotes a counterclockwise rotation by $2\pi/n$. The valley triplet must be mapped by $C_n$ onto an equivalent representation of the magnetic translation group, which restricts the allowed configurations of the three band minima $\{\bm K_\alpha\}$ up to reduced reciprocal lattice vectors.

\textbf{1. $C_6$ symmetry.} For $C_6$ symmetry, there is only one symmetry-allowed configuration of the three band minima,
\begin{align}
    (\bm K_1,\bm K_2,\bm K_3)=\left(\frac{1}{6},-\frac{1}{6},-\frac{1}{2}\right)\bm\beta_1.
\end{align}
The band minima can be obtained from the first-harmonic Hamiltonian with $V_{\bm a}=V$ for $\bm a=\pm\bm a_1,\pm\bm a_2,\pm(\bm a_1+\bm a_2)$ and zero otherwise. Because there are two bands, the sign of $V$ determines which of the two representations is realized. Throughout, for all three rotational symmetries ($C_6$, $C_4$, and $C_3$), we label the lower-band representations corresponding to $V>0$ and $V<0$ as (1) and (2), respectively.

For the magnetic translations, we have
\begin{align}
    \det U_{T_{\bm a_1}}=\det U_{T_{\bm a_2}}=1.
\end{align}
For the $C_6$ rotation, there are two inequivalent three-dimensional irreducible representations. The lower-band triplet $\{\psi_{\bm K_\alpha,-}\}_{\alpha=1}^3$ carries one of these two representations and cannot switch between them under continuous symmetry-preserving deformations of the Hamiltonian without a band crossing. The two representations nevertheless have the same determinant, $\det U_{C_6}^{(1)}=\det U_{C_6}^{(2)}=\omega_6^5$. Therefore, for the lower-band triplet,
\begin{align}
    \det U_{C_6}=\omega_6^5.
\end{align}

\textbf{2. $C_4$ symmetry.} For $C_4$ symmetry, there are two symmetry-allowed configurations of the three band minima,
\begin{align}
    \mathcal V_1:&\quad (\bm K_1,\bm K_2,\bm K_3)=\left(\frac{1}{6},-\frac{1}{6},-\frac{1}{2}\right)\bm\beta_1,\notag\\
    \mathcal V_2:&\quad (\bm K_1,\bm K_2,\bm K_3)=-\frac{1}{2}\bm\beta_2+\left(0,-\frac{1}{3},\frac{1}{3}\right)\bm\beta_1.
\end{align}
The band minima in $\mathcal V_1$ can be obtained from the first-harmonic Hamiltonian with $V_{\bm a}=V$ for $\bm a=\pm\bm a_1,\pm\bm a_2$ and zero otherwise, and the band minima in $\mathcal V_2$ can be obtained from the second-harmonic Hamiltonian with $V_{\bm a}=V$ for $\bm a=\pm(\bm a_1+\bm a_2),\pm(\bm a_1-\bm a_2)$ and zero otherwise.

For $\mathcal V_1$, the determinants of the magnetic translation operators are
\begin{align}
    \det U_{T_{\bm a_1}}=\det U_{T_{\bm a_2}}=1.
\end{align}
The two inequivalent three-dimensional irreducible representations of the $C_4$ rotation have determinants
\begin{align}
    \det U_{C_4}^{\mathcal V_1,(1)}=\omega_4^3,\qquad \det U_{C_4}^{\mathcal V_1,(2)}=\omega_4.
\end{align}
For $\mathcal V_2$, the determinants of the magnetic translation operators are
\begin{align}
    \det U_{T_{\bm a_1}}=\det U_{T_{\bm a_2}}=-1.
\end{align}
The two inequivalent three-dimensional irreducible representations of the $C_4$ rotation have determinants
\begin{align}
    \det U_{C_4}^{\mathcal V_2,(1)}=1,\qquad \det U_{C_4}^{\mathcal V_2,(2)}=\omega_4^3.
\end{align}

\textbf{3. $C_3$ symmetry.} For $C_3$ symmetry, there are three symmetry-allowed configurations of the three band minima,
\begin{align}
    \mathcal V_1:&\quad (\bm K_1,\bm K_2,\bm K_3)=\left(\frac{1}{6},-\frac{1}{6},-\frac{1}{2}\right)\bm\beta_1,\notag\\
    \mathcal V_2:&\quad (\bm K_1,\bm K_2,\bm K_3)=\frac{1}{3}\bm\beta_2+\left(\frac{5}{18},-\frac{1}{18},-\frac{7}{18}\right)\bm\beta_1,\notag\\
    \mathcal V_3:&\quad (\bm K_1,\bm K_2,\bm K_3)=-\frac{1}{3}\bm\beta_2+\left(\frac{7}{18},\frac{1}{18},-\frac{5}{18}\right)\bm\beta_1.
\end{align}
The three sets of band minima can be obtained from the first-harmonic Hamiltonian with $V_{\bm a}=Ve^{i\phi}$ for $\bm a=\bm a_1,\bm a_2,-(\bm a_1+\bm a_2)$, $V_{\bm a}=Ve^{-i\phi}$ for $\bm a=-\bm a_1,-\bm a_2,\bm a_1+\bm a_2$, and zero otherwise. The configurations $\mathcal V_1$, $\mathcal V_2$, and $\mathcal V_3$ correspond to $\phi=0$, $-\pi/3$, and $\pi/3$, respectively.

For $\mathcal V_1$, the determinants of the magnetic translation operators are
\begin{align}
    \det U_{T_{\bm a_1}}=\det U_{T_{\bm a_2}}=1.
\end{align}
The two inequivalent three-dimensional irreducible representations of the $C_3$ rotation have the same determinant, $\det U_{C_3}^{\mathcal V_1,(1)}=\det U_{C_3}^{\mathcal V_1,(2)}=\omega_3^2$. Therefore, for the lower-band triplet,
\begin{align}
    \det U_{C_3}=\omega_3^2.
\end{align}
For $\mathcal V_2$, the determinants of the magnetic translation operators are
\begin{align}
    \det U_{T_{\bm a_1}}=\det U_{T_{\bm a_2}}=\omega_3.
\end{align}
The two inequivalent three-dimensional irreducible representations of the $C_3$ rotation have the same determinant, $\det U_{C_3}^{\mathcal V_2,(1)}=\det U_{C_3}^{\mathcal V_2,(2)}=1$. Therefore, for the lower-band triplet,
\begin{align}
    \det U_{C_3}=1.
\end{align}
For $\mathcal V_3$, the determinants of the magnetic translation operators are
\begin{align}
    \det U_{T_{\bm a_1}}=\det U_{T_{\bm a_2}}=\omega_3^2.
\end{align}
The two inequivalent three-dimensional irreducible representations of the $C_3$ rotation have the same determinant, $\det U_{C_3}^{\mathcal V_3,(1)}=\det U_{C_3}^{\mathcal V_3,(2)}=1$. Therefore, for the lower-band triplet,
\begin{align}
    \det U_{C_3}=1.
\end{align}

\subsubsection{Symmetry-compatible gauge}

In the previous subsection, we studied the representation of symmetry operations on the three band-minimum states $\mathcal H_-=\{\psi_{\bm K_\alpha,-}\}_{\alpha=1}^3$,
\begin{align}
    [\hat O\psi_{\bm K_\alpha,-}](\bar\xi)=\sum_{\beta=1}^3(U_O)_{\beta\alpha}\psi_{\bm K_\beta,-}(\bar\xi).
\end{align}
We now study the action of symmetry on $\psi_{\bm K_\alpha+\bm\kappa,-}$ in the vicinity of each band minimum. The gauge in each patch around $\bm K_\alpha$ can be chosen independently, so in general the representation depends on $\bm\kappa$. We would like to choose a symmetry-compatible gauge such that the symmetry action is represented by the same matrix $U_O$ throughout each patch.

The symmetry-compatible gauge can be constructed as follows. We first choose a smooth gauge for $\psi_{\bm K_1+\bm\kappa,-}$ in the patch around $\bm K_1$. Let $\hat T_{\bm d_\alpha}$ be a magnetic translation that maps the $\bm K_1$ valley to the $\bm K_\alpha$ valley. If at the band minima $\hat T_{\bm d_\alpha}\psi_{\bm K_1,-}=v_\alpha\psi_{\bm K_\alpha,-}$, we define the gauge in the patch around $\bm K_\alpha$ by
\begin{align}
    \psi_{\bm K_\alpha+\bm\kappa,-}=e^{-i\bm\kappa\cdot\bm d_\alpha}v_\alpha^{-1}\hat T_{\bm d_\alpha}\psi_{\bm K_1+\bm\kappa,-}.
\end{align}
With this gauge choice, the magnetic translations act as
\begin{align}
    \hat T_{\bm a_i}\psi_{\bm K_\alpha+\bm\kappa,-}=e^{i\bm\kappa\cdot\bm a_i}\sum_{\beta=1}^{3}(U_{T_{\bm a_i}})_{\beta\alpha}\psi_{\bm K_\beta+\bm\kappa,-}.
\end{align}
We next consider the rotation symmetry. In the translation-compatible gauge constructed above, the rotation action has the form
\begin{align}
    \hat C_n\psi_{\bm K_\alpha+\bm\kappa,-}=e^{i\phi_n(\bm\kappa)}\sum_{\beta=1}^3(U_{C_n})_{\beta\alpha}\psi_{\bm K_\beta+C_n\bm\kappa,-}.
\end{align}
The phase $\phi_n(\bm\kappa)$ can be removed by the gauge transformation $\psi_{\bm K_\alpha+\bm\kappa,-}\rightarrow e^{i\chi(\bm\kappa)}\psi_{\bm K_\alpha+\bm\kappa,-}$, under which $\phi_n(\bm\kappa)\rightarrow\phi_n(\bm\kappa)+\chi(\bm\kappa)-\chi(C_n\bm\kappa)$. We choose $\chi(\bm\kappa)$ such that $\phi_n(\bm\kappa)=0$. In this symmetry-compatible gauge,
\begin{align}
    \hat C_n\psi_{\bm K_\alpha+\bm\kappa,-}=\sum_{\beta=1}^3(U_{C_n})_{\beta\alpha}\psi_{\bm K_\beta+C_n\bm\kappa,-}.
\end{align}

Defining $\psi_{\bm\eta,\alpha}(\bar\xi)=\int\frac{d^2\bm\kappa}{(2\pi)^2}e^{-i\bm\kappa\cdot\bm\eta}\psi_{\bm K_\alpha+\bm\kappa,-}(\bar\xi)$, the symmetry actions can be written as
\begin{align}
    \hat T_{\bm a_i}\psi_{\bm\eta,\alpha}(\bar\xi)=\sum_{\beta=1}^3(U_{T_{\bm a_i}})_{\beta\alpha}\psi_{\bm\eta-\bm a_i,\beta}(\bar\xi),\qquad \hat C_n\psi_{\bm\eta,\alpha}(\bar\xi)=\sum_{\beta=1}^3(U_{C_n})_{\beta\alpha}\psi_{C_n\bm\eta,\beta}(\bar\xi).
\end{align}

\section{Effective Hamiltonian derivation from geometric quantization}\label{sec:Hamiltonian}

In this section, we derive the effective Hamiltonian for quasihole wavepackets in both the $1/3$-quasihole and $2/3$-quasihole-pair cases using geometric quantization. We first consider the $1/3$-quasihole case and then turn to the $2/3$-quasihole-pair case, where the quasihole-pair dispersion involves multiple bands. Throughout this section, we consider electron filling $\nu=1/q=1/3$ and set $\ell_B=1$. Since the effect of the periodic modulation can be absorbed into the single-particle band structure, we use the Laughlin quasihole states to describe the many-body geometric and statistical properties in the following derivation.

We focus on the dilute limit, $M/N_\Phi\ll1$, where $M$ denotes the number of quasiholes or quasihole pairs and $N_\Phi$ is the number of magnetic flux quanta. The Q-symbol $\mathcal U(\bar\xi,\xi)=\langle\xi|\hat V|\xi\rangle/\langle\xi|\xi\rangle$ generally contains both one-body and many-body contributions in the quasihole coordinates. In the following, we retain only the one-body periodic contribution corresponding to the single-particle potential in Eq.~\eqref{periodic-potential},
\begin{align}
    \mathcal U(\bar\xi,\xi)\simeq\sum_{a=1}^M\mathcal U_1(\bar\xi_a,\xi_a)=\sum_{a=1}^M\sum_{\bm a}V_{\bm a}e^{i\bm a\wedge\bm\xi_a}.
\end{align}
This periodic potential felt by the quasiholes may originate from an external periodic potential experienced by the electrons or from a periodic modulation. The periodic potential leads to a band dispersion for each quasihole or quasihole pair. In the dilute limit, the quasiholes or quasihole pairs occupy states near the bottom of the band, allowing their low-energy dynamics to be described in terms of wavepackets constructed from states near the band minima.

However, in the quasihole case, the low-energy eigenstates cannot be written simply as product states of independent single-quasihole wavepackets because of the statistical gauge fields generated by the other quasiholes. In geometric quantization, the periodic potential is encoded in $\mathcal U$, while the K\"ahler potential determines the statistical gauge field experienced by the quasiholes.

\subsection{1/3-quasihole effective Hamiltonian}

The Laughlin $M$-quasihole state with $N$ electrons at filling $1/q$ is given by
\begin{align}
    \psi_{\xi}(\{\bm z\})=\prod_{a=1}^{M}\prod_{i=1}^{N}(z_i-\xi_a)\prod_{i<j}(z_i-z_j)^q e^{-\frac{1}{4}\sum_i|z_i|^2}.
    \label{Laughlin-quasihole-state}
\end{align}
where $\xi=(\xi_1,\dots,\xi_M)$. The total flux is $N_\Phi=q(N-1)+M$. Its K\"ahler potential is written as
\begin{align}
    e^{\mathcal K(\bar\xi,\xi)}=\langle\xi|\xi\rangle,\qquad \mathcal K(\bar\xi,\xi)=\frac{\bar\xi\xi}{2q}+\delta\mathcal K(\bar\xi,\xi),
\end{align}
where $\bar\xi\xi=\sum_{a=1}^{M}\bar\xi_a\xi_a$. In the dilute limit, when quasiholes are far away from each other, we have
\begin{align}
    \delta\mathcal K(\bar\xi,\xi)\simeq-\frac{1}{q}\sum_{a<b}\ln|\xi_a-\xi_b|^2,\qquad |\xi_a-\xi_b|\gg1.
\end{align}
When two quasiholes approach each other, such that $|\xi_a-\xi_b|\sim O(\ell_B)$, the short-distance behavior of $\delta\mathcal K$ is regularized, and the K\"ahler potential remains a smooth function. In the long-distance limit, the measure factor takes the form
\begin{align}
    \mathcal F(\bar\xi,\xi)\simeq e^{-\delta\mathcal K(\bar\xi,\xi)}\simeq\prod_{a<b}|\xi_a-\xi_b|^{\frac{2}{q}},\qquad |\xi_a-\xi_b|\gg1.
\end{align}
We decompose the K\"ahler and measure factors into smooth and short-distance contributions as
\begin{align}
    e^{\delta\mathcal K(\bar\xi,\xi)}=\mathcal W(\bar\xi,\xi)=\mathcal W_{\rm smooth}(\bar\xi,\xi)+\mathcal W_{\rm short}(\bar\xi,\xi),\qquad \mathcal F(\bar\xi,\xi)=\mathcal F_{\rm smooth}(\bar\xi,\xi)+\mathcal F_{\rm short}(\bar\xi,\xi).
\end{align}
Here, $\mathcal W_{\rm smooth}$ and $\mathcal F_{\rm smooth}$ are obtained from the long-distance forms supplemented by a smooth short-distance regularization, while $\mathcal W_{\rm short}$ and $\mathcal F_{\rm short}$ contain the remaining short-distance contributions. We also define 
\begin{align*}
    e^{W(\bar\xi,\xi)} = \mathcal W_{\rm smooth},
\end{align*}
so that at long distances $e^{-W(\bar\xi,\xi)}\simeq \prod_{a<b}|\xi_a-\xi_b|^{\frac{2}{q}}$

\subsubsection{Envelope effective Hamiltonian}

We first recall the single-quasihole band structure derived in Section~\ref{single-particle-band-structure}. The single-particle Bloch state is denoted by $\psi_{\bm k}(\bar\xi)$. There are three degenerate band minima $\bm K_\alpha$, $\alpha=1,2,3$, and the dispersion near each minimum is $E_{\bm K_\alpha+\bm\kappa}=E_{\min}+\frac{|\bm\kappa|^2}{2m^*}+O(\kappa^3)$. A many-quasihole state constructed from Bloch states near the band minima can then be written as
\begin{align}
    f(\bar\xi)=\sum_{\alpha}\int d^2\bm\kappa\,w_{\alpha}(\bm\kappa)\Psi_{\bm\kappa,\alpha}(\bar\xi),\qquad \Psi_{\bm\kappa,\alpha}(\bar\xi)=\prod_{a=1}^M\psi_{\bm\kappa_a,\alpha_a}(\bar\xi_a).
\end{align}
Here, $\alpha=(\alpha_1,\dots,\alpha_M)$ with $\alpha_a=1,2,3$, $\bm\kappa=(\bm\kappa_1,\dots,\bm\kappa_M)$, $d^2\bm\kappa=\prod_{a=1}^M d^2\bm\kappa_a$, and $\psi_{\bm\kappa,\alpha}(\bar\xi)=\psi_{\bm K_\alpha+\bm\kappa}(\bar\xi)$. The momentum-space envelope $w_{\alpha}(\bm\kappa)$ is concentrated around $\bm\kappa=0$, with $|\bm\kappa_a|\lesssim\Lambda\ll1$ for each $a$. We define the single-particle envelope function by
\begin{align}
    \psi_{\bm\eta,\alpha}(\bar\xi)=\int_{|\bm\kappa|\lesssim\Lambda}\frac{d^2\bm\kappa}{(2\pi)^2}e^{-i\bm\kappa\cdot\bm\eta}\psi_{\bm\kappa,\alpha}(\bar\xi).
\end{align}

Using the Fourier representations of $\mathcal W=e^{\delta\mathcal K}$ and $\mathcal F$ introduced in Eq.~\eqref{F-K-Fourier-transform}, the many-body Hamiltonian obtained from Eq.~\eqref{QuantizationV} acts on an antiholomorphic wavefunction $f(\bar\xi)$ as
\begin{align}
    [\hat H f](\bar\xi)&=\sum_{i=1}^M\sum_{\bm a}V_{\bm a}e^{-\frac{1}{2}a\bar\xi_i}\int d^2\bm p\,d^2\bm l\,F_{\bm p}\mathcal W_{\bm l}e^{\frac{i}{2}l\bar\xi}\int\prod_{j=1}^M\left(\frac{d^2\bm\omega_j}{2\pi q}\right)e^{-\frac{1}{2q}\omega\left(\bar\omega-\bar\xi-q\bar a\,\bm e_i-iq(\bar p+\bar l)\right)}e^{\frac{i}{2}p\bar\omega}f(\bar\omega)\nonumber\\
    &=\sum_{i=1}^M\sum_{\bm a}V_{\bm a}e^{-\frac{1}{2}a\bar\xi_i}\int d^2\bm l\,e^{\frac{i}{2}l\bar\xi}f(\bar\xi+q\bar a\,\bm e_i+iq\bar l)\int d^2\bm p\,F_{\bm p}\mathcal W_{\bm l-\bm p}e^{-\frac{q}{2}p\bar l}e^{\frac{iq}{2}p_i\bar a}.
\end{align}
Here, $\bm e_i$ denotes the unit vector in the $i$-th quasihole coordinate, $(\bm e_i)_j=\delta_{ij}$. Acting with the Hamiltonian on the wavepacket wavefunction, we obtain
\begin{align}
    [\hat H f](\bar\xi)&=\sum_{\alpha}\sum_{i=1}^M\sum_{\bm a}V_{\bm a}\int d^2\bm l\,d^2\bm\kappa\,w_{\alpha}(\bm\kappa)e^{-\frac{1}{2}a\bar\xi_i}e^{\frac{i}{2}l\bar\xi}\Psi_{\bm\kappa,\alpha}(\bar\xi+q\bar a\,\bm e_i+iq\bar l)\int d^2\bm p\,F_{\bm p}\mathcal W_{\bm l-\bm p}e^{-\frac{q}{2}p\bar l}e^{\frac{iq}{2}p_i\bar a}\nonumber\\
    &=\sum_{\alpha}\sum_{i=1}^M\sum_{\bm a}\tilde V_{\bm a}\int d^2\bm\kappa\,d^2\bm l\,w_{\alpha}(\bm\kappa)e^{\frac{q}{4}\sum_{j=1}^M|l_j|^2}e^{\frac{iq}{2}\sum_{j=1}^M(\bm K_{\alpha_j}+\bm\kappa_j)\wedge\bm l_j}e^{-\frac{iq}{2}l_i\bar a}\eta_{q\bm a}e^{iq(\bm K_{\alpha_i}+\bm\kappa_i+\bm l_i)\cdot\bm a}\Psi_{\bm\kappa+\bm l,\alpha}(\bar\xi)\nonumber\\
    &\quad\times\int d^2\bm p\,F_{\bm p}\mathcal W_{\bm l-\bm p}e^{-\frac{q}{2}p\bar l}e^{\frac{iq}{2}p_i\bar a}.
\end{align}

Therefore, defining the envelope effective Hamiltonian by $[\hat H f](\bar\xi)=\sum_{\alpha}\int d^2\bm\kappa\,[\hat H_{\rm eff}w]_{\alpha}(\bm\kappa)\Psi_{\bm\kappa,\alpha}(\bar\xi)$, we obtain
\begin{align}
    [\hat H_{\rm eff}w]_{\alpha}(\bm\kappa)=\sum_{i=1}^M\sum_{\bm a}\tilde V_{\bm a}\eta_{q\bm a}e^{iq(\bm K_{\alpha_i}+\bm\kappa_i)\cdot\bm a}\int d^2\bm l\,w_{\alpha}(\bm\kappa-\bm l)e^{\frac{q}{4}\sum_{j=1}^M|l_j|^2}e^{\frac{iq}{2}\sum_{j=1}^M(\bm K_{\alpha_j}+\bm\kappa_j)\wedge\bm l_j}e^{-\frac{iq}{2}l_i\bar a}\mathcal I_i(\bar a,\bm l),
\end{align}
where
\begin{align}
    \mathcal I_i(\bar a,\bm l)=\int d^2\bm p\,F_{\bm p}\mathcal W_{\bm l-\bm p}e^{-\frac{q}{2}p\bar l}e^{\frac{iq}{2}p_i\bar a}.
\end{align}

To include low-energy processes connecting different band minima, we decompose the total momentum transfer as $\Delta\bm K_{\alpha\beta,j}+\bm l_j$, where $\bm l_j$ is the residual momentum satisfying $|\bm l_j|\lesssim\Lambda$ and $\Delta\bm K_{\alpha\beta,j}=\bm K_{\alpha_j}-\bm K_{\beta_j}$. Projecting onto the low-energy states near the band minima, the effective Hamiltonian can be written as
\begin{align}
    [\hat H_{\mathrm{eff}}w]_{\alpha}(\bm\kappa)=\sum_{\beta}\int d^2\bm l\,H_{\alpha\beta}(\bm\kappa,\bm l)w_{\beta}(\bm\kappa-\bm l),
\end{align}
where
\begin{align}
    H_{\alpha\beta}(\bm\kappa,\bm l)&=\sum_{i=1}^M\sum_{\bm a}\tilde V_{\bm a}\eta_{q\bm a}e^{\frac{q}{4}|\bm l+\Delta\bm K_{\alpha\beta}|^2}e^{\frac{iq}{2}a(\bar l_i+\Delta\bar K_{\alpha\beta,i})}\mathcal I_i(\bar a,\bm l+\Delta\bm K_{\alpha\beta})\nonumber\\
    &\qquad\times e^{iq(\bm K_{\beta_i}+\bm\kappa_i-\bm l_i)\cdot\bm a}e^{\frac{iq}{2}\sum_{j=1}^M(\bm K_{\beta_j}+\bm\kappa_j-\bm l_j)\wedge(\bm l_j+\Delta\bm K_{\alpha\beta,j})}.
    \label{full-effective-Hamiltonian}
\end{align}

\subsubsection{Low-energy expansion of the effective Hamiltonian}

We now expand the effective Hamiltonian to second order, $O(\Lambda^2)$, in the small momenta $\bm\kappa$ and $\bm l$. We note that all dependence on the K\"ahler potential and the integration measure is contained in $\mathcal I_i$. We correspondingly decompose
\begin{align}
    \mathcal I_i(\bar a,\bm l)=\mathcal I_i^{\rm smooth}(\bar a,\bm l)+\mathcal I_i^{\rm short}(\bar a,\bm l),
\end{align}
where $\mathcal I_i^{\rm smooth}$ contains the contribution from the smooth components $\mathcal F_{\rm smooth}$ and $\mathcal W_{\rm smooth}$, while $\mathcal I_i^{\rm short}$ contains all contributions involving at least one short-distance component.

Expanding $\mathcal I_i(\bar a,\bm l)$ in $\bar a$, we obtain
\begin{align}
    \mathcal I_i(\bar a,\bm l)=\sum_{r=0}^{\infty}\frac{1}{r!}\left(\frac{iq}{2}\right)^r\bar a^r M_{i,r}(\bm l),\qquad M_{i,r}(\bm l)=\int d^2\bm p\,p_i^r F_{\bm p}\mathcal W_{\bm l-\bm p}e^{-\frac{q}{2}p\bar l}.
\end{align}
The smooth and short-distance contributions can be expanded in the same way, with the corresponding moments denoted by $M_{i,r}^{\rm smooth}$ and $M_{i,r}^{\rm short}$, respectively.

For momenta within the low-energy window $|\bm l|\lesssim\Lambda$, the zeroth moment is fixed by the reproducing relation,
\begin{align}
    M_{i,0}(\bm l)=\delta^{(2)}(\bm l),
\end{align}
and $M_{i,0}^{\rm smooth}(\bm l)$ has the same scaling. The higher smooth moments are suppressed by powers of the momentum cutoff,
\begin{align}
    \frac{M_{i,r}^{\rm smooth}(\bm l)}{M_{i,0}^{\rm smooth}(\bm l)}=O(\Lambda^r).
\end{align}
The intravalley short-distance contribution $\mathcal I_i^{\rm short}(\bar a,\bm l)$ is of order $O(\Lambda^2)$ relative to the smooth intravalley contribution $\mathcal I_i^{\rm smooth}(\bar a,\bm l)$. In addition, the entire intervalley contribution $\mathcal I_i(\bar a,\bm l+\Delta\bm K_{\alpha\beta})$ for $\alpha\neq\beta$ is also of order $O(\Lambda^2)$ relative to $\mathcal I_i^{\rm smooth}(\bar a,\bm l)$. Therefore, to second order in the low-energy expansion, it is sufficient to retain the smooth intravalley moments up to $r=2$, together with the leading intravalley short-distance and intervalley contributions.

Expanding Eq.~\eqref{full-effective-Hamiltonian} to second order in the low-energy momenta, we obtain
\begin{align}
    [\hat H_{\rm eff}w]_{\alpha}(\bm\kappa)&=\frac{1}{2m^*}\sum_{i=1}^M|\bm\kappa_i|^2w_{\alpha}(\bm\kappa)+\frac{1}{2m^*}\sum_{i=1}^M\int d^2\bm l\,\bar\kappa_i M_{i,1}^{\rm smooth}(\bm l)w_{\alpha}(\bm\kappa-\bm l)\notag\\
    &\qquad+\sum_{\beta}\int d^2\bm l\,[\mathcal U_{\rm short}]_{\alpha\beta}(\bm l)w_{\beta}(\bm\kappa-\bm l)+O(\Lambda^3).
\end{align}
Here, $\mathcal U_{\rm short}$ collects the second-order short-distance contributions generated by the one-body periodic potential $\mathcal U_1$ in combination with the short-distance K\"ahler and measure corrections. It contains the intravalley short-distance correction for $\alpha=\beta$ and the intervalley scattering terms for $\alpha\neq\beta$,
\begin{align}
    [\mathcal U_{\rm short}]_{\alpha\beta}(\bm l)=
    \begin{cases}
    \displaystyle \sum_{i=1}^M\sum_{\bm a}\tilde V_{\bm a}\eta_{q\bm a}e^{iq\bm K_{\alpha_i}\cdot\bm a}\mathcal I_i^{\rm short}(\bar a,\bm l), & \alpha=\beta, \\[3mm]
    \displaystyle \sum_{i=1}^M\sum_{\bm a}\tilde V_{\bm a}\eta_{q\bm a}e^{\frac{q}{4}|\Delta\bm K_{\alpha\beta}|^2}e^{\frac{iq}{2}a\Delta\bar K_{\alpha\beta,i}}e^{iq\bm K_{\beta_i}\cdot\bm a}e^{\frac{iq}{2}\sum_{j=1}^M\bm K_{\beta_j}\wedge\Delta\bm K_{\alpha\beta,j}}\mathcal I_i(\bar a,\bm l+\Delta\bm K_{\alpha\beta}), & \alpha\neq\beta.
    \end{cases}
    \label{Ushort-expression}
\end{align}
The short-range contribution $\mathcal U_{\rm short}$ is in general parametrically independent of the effective mass $m^*$. The full short-range interaction $\mathcal V_{\rm short}$ also contains the contribution generated by the two-body Q-symbol $\mathcal U_2$.

We introduce the Fourier transforms
\begin{align}
    w_{\alpha}(\bm\eta)=\int d^2\bm\kappa\,e^{i\bm\kappa\cdot\bm\eta}w_{\alpha}(\bm\kappa),\qquad M_{i,1}^{\rm smooth}(\bm\eta)=\int d^2\bm l\,e^{i\bm l\cdot\bm\eta}M_{i,1}^{\rm smooth}(\bm l).
\end{align}
In the long-distance limit, where the relevant Fourier momenta are small, we further approximate $M_{i,1}^{\rm smooth}$ by
\begin{align}
    M_{i,1}^{\rm smooth}(\bm\eta)\simeq 2i\,\partial_{\bar\eta_i}W(\bar\eta,\eta).
\end{align}
The envelope effective Hamiltonian therefore takes the form
\begin{align}
    [\hat H_{\rm eff}w]_{\alpha}(\bm\eta)=-\frac{2}{m^*}\sum_{i=1}^M\partial_{\eta_i}\left[\partial_{\bar\eta_i}-\partial_{\bar\eta_i}W(\bar\eta,\eta)\right]w_{\alpha}(\bm\eta)+[\mathcal V_{\rm short}w]_{\alpha}(\bm\eta).
    \label{envelope-effective-Hamiltonian}
\end{align}
Here, $\mathcal V_{\rm short}$ contains $\mathcal U_{\rm short}$, which arises from the one-body periodic potential $\mathcal U_1$ in combination with the short-distance K\"ahler and measure corrections, as well as the short-range contribution generated by the two-body Q-symbol $\mathcal U_2$.

\subsubsection{Hermitian form of the effective Hamiltonian}

The many-quasihole state can be written in the envelope coordinates as
\begin{align}
    f(\bar\xi)=\sum_{\alpha}\int d^2\bm\eta\,w_{\alpha}(\bm\eta)\Psi_{\bm\eta,\alpha}(\bar\xi),\qquad \Psi_{\bm\eta,\alpha}(\bar\xi)=\prod_{i=1}^M\psi_{\bm\eta_i,\alpha_i}(\bar\xi_i).
\end{align}
The envelope functions satisfy
\begin{align}
    \bar\xi_i\,\psi_{\bm\eta_i,\alpha_i}(\bar\xi_i)=\left(\bar\eta_i+q\partial_{\eta_i}-\frac{iq}{2}\bar K_{\alpha_i}\right)\psi_{\bm\eta_i,\alpha_i}(\bar\xi_i).
    \label{xi-eta-relation}
\end{align}
In the long-distance limit, the derivative term and the valley-dependent shift give only subleading microscopic corrections. Therefore, to leading order, we may identify the quasihole coordinate with the envelope coordinate,
\begin{align}
    \bar\xi_i\,\psi_{\bm\eta_i,\alpha_i}(\bar\xi_i)\simeq\bar\eta_i\,\psi_{\bm\eta_i,\alpha_i}(\bar\xi_i).
\end{align}

Then, for the inner product between two many-quasihole states obtained from geometric quantization, $\langle f'|f\rangle=\int d\mu_{\bm\xi}\,\overline{f'(\bar\xi)}f(\bar\xi)$, the long-distance approximation gives
\begin{align}
    \langle f'|f\rangle\simeq\sum_{\alpha}\int d^2\bm\eta\,e^{-W(\bar\eta,\eta)}\overline{w'_{\alpha}(\bm\eta)}w_{\alpha}(\bm\eta).
\end{align}
We therefore define the rescaled envelope amplitude
\begin{align}
    \tilde w_{\alpha}(\bm\eta)=e^{-W(\bar\eta,\eta)/2}w_{\alpha}(\bm\eta).
\end{align}

The corresponding effective Hamiltonian is obtained by the similarity transformation
\begin{align}
    \hat{\tilde H}_{\rm eff}=e^{-W/2}\hat H_{\rm eff}e^{W/2}=\hat{\tilde H}_{\rm ACCS}+\mathcal V_{\rm short}.
\end{align}
Here, the Aharonov--Casher--Chern--Simons Hamiltonian is
\begin{align}
    \hat{\tilde H}_{\rm ACCS}=\frac{1}{2m^*}\sum_{i=1}^M\Pi_i^\dagger\Pi_i,
    \label{hermitian-Aharonov-Casher-Hamiltonian}
\end{align}
with
\begin{align}
    \Pi_i=2\partial_{\bar\eta_i}-\partial_{\bar\eta_i}W(\bar\eta,\eta),\qquad \Pi_i^\dagger=-2\partial_{\eta_i}-\partial_{\eta_i}W(\bar\eta,\eta).
\end{align}
The corresponding statistical magnetic field is
\begin{align}
    [\Pi_i,\Pi_i^\dagger]=-2\,\delta\mathcal B_i,\qquad \delta\mathcal B_i=2\partial_{\eta_i}\partial_{\bar\eta_i}W(\bar\eta,\eta).
\end{align}

\subsection{2/3-quasihole-pair effective Hamiltonian}

In the absence of the periodic potential, a short-range repulsive interaction $V(\bm z)$ between electrons, such as a Yukawa interaction with an appropriate screening length, can induce an effective short-range attraction between two $1/3$ quasiholes in the $\nu=1/q=1/3$ Laughlin state. Two quasiholes tend to form a bound state in the relative angular momentum channel $L=2$, with relative wavefunction $f(\bar\xi_{\rm rel})\propto\bar\xi_{\rm rel}^{L}$, due to this attraction, which we refer to as a $2/3$ quasihole pair. For a dilute system containing many quasiholes, the low-energy states can be viewed as a dilute gas of such quasihole pairs. It is therefore natural to restrict our discussion to the Hilbert subspace spanned by dilute $2/3$ quasihole pairs. The interaction $V(\bm z)$ also induces a short-range repulsive interaction between quasihole pairs, which gives an additional contribution to the short-range interaction $\mathcal V_{\rm short}$ in the effective Hamiltonian.

We now write the Laughlin quasihole-pair state with quasihole-pair center-of-mass coordinates $\xi=(\xi_1,\dots,\xi_M)$. The wavefunction in the bosonic representation is
\begin{align}
    \psi^{(B)}_\xi(\{\bm z\})=\prod_{a=1}^M e_1\!\left((z-\xi_a)^{-2}\right)\prod_{a=1}^M\prod_{i=1}^{N}(z_i-\xi_a)^2\prod_{i<j}(z_i-z_j)^q e^{-\frac14\sum_i|z_i|^2}.
\end{align}
Here, the elementary symmetric polynomial is
\begin{align}
    e_1\!\left((z-\xi_a)^{-2}\right)=\sum_{i=1}^{N}\frac{1}{(z_i-\xi_a)^2}.
\end{align}
The factor $e_1\!\left((z-\xi_a)^{-2}\right)$ arises because the two quasiholes form a bound state with relative angular momentum $L=2$, whose relative wavefunction is $f(\bar\xi_{\rm rel})\propto\bar\xi_{\rm rel}^2$. Although $e_1\!\left((z-\xi_a)^{-2}\right)$ has poles, they are canceled by $\prod_i(z_i-\xi_a)^2$. Thus, the quasihole-pair state itself is free of poles.

For the $2/3$ quasihole-pair problem, we find it more convenient to work in the fermionic representation discussed in Section~\ref{bosonic-and-fermionic-representations}. Therefore, throughout the remainder of this subsection, we use the fermionic representation unless stated otherwise, and $\psi_\xi$ denotes the corresponding fermionic quasihole-pair state,
\begin{align}
    \psi_\xi(\{\bm z\})=\prod_{1\leq a<b\leq M}(\xi_a-\xi_b)\prod_{a=1}^M e_1\!\left((z-\xi_a)^{-2}\right)\prod_{a=1}^M\prod_{i=1}^{N}(z_i-\xi_a)^2\prod_{i<j}(z_i-z_j)^q e^{-\frac14\sum_i|z_i|^2}.
    \label{quasihole-pair-Laughlin-state}
\end{align}

For a $2/3$ quasihole pair, we have $\mu=\frac{q}{2}=\frac{3}{2}$. In the fermionic representation, the K\"ahler potential takes the form
\begin{align}
    \mathcal K_F(\bar\xi,\xi)=\frac{\bar\xi\xi}{2\mu}+\delta\mathcal K_F(\bar\xi,\xi).
\end{align}
Defining
\begin{align}
    \mathcal W_F(\bar\xi,\xi)=e^{\delta\mathcal K_F(\bar\xi,\xi)},
\end{align}
we have at long distances
\begin{align}
    \mathcal W_F(\bar\xi,\xi)\simeq\prod_{a<b}|\xi_a-\xi_b|^{-2\lambda},\qquad |\xi_a-\xi_b|\gg1,\qquad \lambda=\frac{2}{\mu}-1=\frac{4}{q}-1=\frac{1}{3}.
\end{align}
At short distances, Eq.~\eqref{bosonic-and-fermionic-kahler} gives
\begin{align}
    \mathcal W_F(\bar\xi,\xi)\propto|\xi_a-\xi_b|^2,\qquad |\xi_a-\xi_b|\rightarrow0,
\end{align}
when two quasihole pairs approach each other. The measure is written as
\begin{align}
    d\mu_{\bm\xi}=\left(\prod_a\frac{d^2\bm\xi_a}{2\pi\mu}\right)\mathcal F_F(\bar\xi,\xi)e^{-\frac{\bar\xi\xi}{2\mu}},
\end{align}
and its long-distance behavior is
\begin{align}
    \mathcal F_F(\bar\xi,\xi)\simeq\prod_{a<b}|\xi_a-\xi_b|^{2\lambda},\qquad |\xi_a-\xi_b|\gg1.
\end{align}

\subsubsection{Envelope effective Hamiltonian}

We first recall the single-quasihole-pair band structure derived in Section~\ref{single-particle-band-structure}. The single-particle problem has two bands with eigenstates $\psi_{\bm k,\pm}(\bar\xi)$. We retain only the lower band, whose three degenerate minima are denoted by $\bm K_\alpha$, $\alpha=1,2,3$, and whose dispersion near each minimum is $E_-(\bm K_\alpha+\bm\kappa)=E_{\min}+\frac{|\bm\kappa|^2}{2m^*}+O(\kappa^3)$. We define
\begin{align}
    \psi_{\bm\kappa,\alpha}(\bar\xi)=\psi_{\bm K_\alpha+\bm\kappa,-}(\bar\xi).
\end{align}
A many-quasihole-pair state constructed from lower-band states near the band minima can then be written as
\begin{align}
    f(\bar\xi)=\sum_{\alpha}\int d^2\bm\kappa\,w_{\alpha}(\bm\kappa)\Psi_{\bm\kappa,\alpha}(\bar\xi),\qquad \Psi_{\bm\kappa,\alpha}(\bar\xi)=\prod_{i=1}^M\psi_{\bm\kappa_i,\alpha_i}(\bar\xi_i).
\end{align}
Here, $\alpha=(\alpha_1,\dots,\alpha_M)$ with $\alpha_i=1,2,3$, $\bm\kappa=(\bm\kappa_1,\dots,\bm\kappa_M)$, and $d^2\bm\kappa=\prod_{i=1}^M d^2\bm\kappa_i$. The momentum-space envelope $w_{\alpha}(\bm\kappa)$ is concentrated around $\bm\kappa=0$, with $|\bm\kappa_i|\lesssim\Lambda\ll1$ for each $i$.

The effective Hamiltonian is defined by $[\hat H f](\bar\xi)=\sum_{\alpha}\int d^2\bm\kappa\,[\hat H_{\rm eff}w]_{\alpha}(\bm\kappa)\Psi_{\bm\kappa,\alpha}(\bar\xi)$. The derivation of $\hat H_{\rm eff}$ follows the same procedure as in the $1/3$-quasihole case. Virtual transitions to the upper band generate Schrieffer--Wolff corrections to the lower-band Hamiltonian. For the triangular-lattice potential considered here, these corrections are $O(\Lambda^4)$ and can therefore be neglected to $O(\Lambda^2)$. We find that the effective Hamiltonian up to $O(\Lambda^2)$ takes the same form as in the $1/3$-quasihole case, Eq.~\eqref{envelope-effective-Hamiltonian},
\begin{align}
    [\hat H_{\rm eff}w]_{\alpha}(\bm\eta)=-\frac{2}{m^*}\sum_{i=1}^M\partial_{\eta_i}\left[\partial_{\bar\eta_i}-\partial_{\bar\eta_i}W_F(\bar\eta,\eta)\right]w_{\alpha}(\bm\eta)+[\mathcal V_{\rm short}w]_{\alpha}(\bm\eta)+O(\Lambda^3).
\end{align}
The short-range interaction $\mathcal V_{\rm short}$ contains the contribution $\mathcal U_{\rm short}$ generated by the one-body periodic potential in combination with the short-distance K\"ahler and measure corrections, together with the short-range contribution generated by the two-body Q-symbol $\mathcal U_2$. In the present $2/3$ quasihole-pair problem, $\mathcal U_{\rm short}$ differs from Eq.~\eqref{Ushort-expression} because the projection onto the lower band introduces additional band form factors.

Defining the flat-weight envelope wavefunction by $\tilde w_{\alpha}(\bm\eta)=e^{-W_F(\bar\eta,\eta)/2}w_{\alpha}(\bm\eta)$, the corresponding effective Hamiltonian is
\begin{align}
    \hat{\tilde H}_{\rm eff}=e^{-W_F/2}\hat H_{\rm eff}e^{W_F/2}=\hat{\tilde H}_{\rm ACCS}+\mathcal V_{\rm short}+O(\Lambda^3),
\end{align}
where
\begin{align}
    \hat{\tilde H}_{\rm ACCS}=\frac{1}{2m^*}\sum_{i=1}^M(\Pi_i^F)^\dagger\Pi_i^F,\qquad \Pi_i^F=2\partial_{\bar\eta_i}-\partial_{\bar\eta_i}W_F(\bar\eta,\eta).
\end{align}

\section{Ground-state structure of the ACCS Hamiltonian}
\label{ground-state-of-AC-Hamiltonian}

We showed that in the regime where the short-range interaction $\mathcal V_{\rm short}$ is weak, the Aharonov--Casher--Chern--Simons part of the effective Hamiltonian dominates. It is therefore natural to first consider the zero-energy states of the ACCS Hamiltonian. They satisfy
\begin{align}
    \Pi_i\tilde w_\alpha(\bm\eta)=0,\qquad i=1,\dots,M.
\end{align}
The general solution can be written as
\begin{align}
    \tilde w_\alpha(\bm\eta)=e^{W(\bar\eta,\eta)/2}h_\alpha(\eta),
    \label{ground-state-solution}
\end{align}
where $h_\alpha(\eta)$ is holomorphic away from quasihole coincidence points and may, in general, be meromorphic at coincidence points, with the allowed singularities determined by the short-distance behavior of $e^{W/2}$. The discussion in this section applies to both the bosonic and fermionic representations. In each case, we should use the corresponding K\"ahler weight $W$, covariant derivative $\Pi_i$, and flat-weight envelope wavefunction $\tilde w_\alpha$. For notational simplicity, we suppress the representation label below.

The allowed zero-energy functions $h_\alpha(\eta)$ are subject to three constraints:
\begin{enumerate}
    \item Short-distance normalizability constrains the allowed singularities of $h_\alpha(\eta)$. In particular, zeros of $e^{W(\bar\eta,\eta)/2}$ at quasihole coincidence points can compensate poles of $h_\alpha(\eta)$.

    \item The wavefunction should not grow at infinity. When one quasihole coordinate $|\eta_i|$ becomes large while the others are kept fixed, the long-distance behavior of $e^{W(\bar\eta,\eta)/2}$ constrains the maximal power of $\eta_i$ that can appear in $h_\alpha(\eta)$. We do not require the wavefunction to be integrable at long distances, since wavefunctions describing a superconductor are extended, similar to plane-wave states.

    \item $h_\alpha(\eta)$ must satisfy the appropriate exchange symmetry for the chosen representation under the simultaneous exchange of the position and valley labels of any two quasiholes, $(\bm\eta_i,\alpha_i)\leftrightarrow(\bm\eta_j,\alpha_j)$.
\end{enumerate}
These constraints determine the zero-energy subspace of the ACCS Hamiltonian, which is generally degenerate. Different zero-energy states can have different short-distance behavior near quasihole coincidence points. Therefore, a weak short-range interaction $\mathcal V_{\rm short}$ can generally lift this degeneracy.

\subsection{ACCS ground states of 2/3 quasihole pairs}

The ground states of the $2/3$-quasihole-pair problem have an interesting structure. We now work in the fermionic gauge, and write the ground-state solutions in Eq.~\eqref{ground-state-solution} as
\begin{align}
    \tilde w_\alpha(\bm\eta)=\frac{|\Delta(\eta)|}{|\Delta(\eta)|_{\rm reg}^{4/3}}\,h_{F,\alpha}(\eta),
\end{align}
where $\Delta(\eta)=\prod_{i<j}(\eta_i-\eta_j)$ and $|\Delta(\eta)|_{\rm reg}$ denotes a smooth short-distance regularization of $|\Delta(\eta)|$ that approaches $|\Delta(\eta)|$ at long distances and remains finite and nonzero when two quasihole pairs coincide. Equivalently, $e^{W_F(\bar\eta,\eta)/2}=\frac{|\Delta(\eta)|}{|\Delta(\eta)|_{\rm reg}^{4/3}}$. Therefore, short-distance normalizability allows $h_{F,\alpha}(\eta)$ to have at most a simple pole at $\eta_i=\eta_j$, while $\tilde w_\alpha(\bm\eta)$ remains finite there. If $h_{F,\alpha}(\eta)\sim\eta_i^{n_i}$ as $|\eta_i|\to\infty$ with the other coordinates fixed, the condition that the wavefunction does not grow requires $n_i\leq\left\lfloor\frac{M-1}{3}\right\rfloor$. We also require $h_{F,\alpha}(\eta)$ to be antisymmetric under the simultaneous exchange $(\bm\eta_i,\alpha_i)\leftrightarrow(\bm\eta_j,\alpha_j)$.

There are in general many functions $h_{F,\alpha}(\eta)$ satisfying the above constraints. However, if we further require $h_{F,\alpha}(\eta)$ to be holomorphic and take $M=3M_{\rm v}$ with $M_{\rm v}\in\mathbb Z$, there is a unique solution. In this solution, each valley contains $M_{\rm v}$ quasihole pairs, and
\begin{align}
    h_{F,\alpha}(\eta)=\epsilon(\alpha)\prod_{a=1}^3\Delta_a(\eta).
\end{align}
For a given valley configuration $\alpha=(\alpha_1,\ldots,\alpha_M)$, we define $I_a(\alpha)=\{i\,|\,\alpha_i=a\}=\{i_1^a<i_2^a<\cdots<i_{M_{\rm v}}^a\},\quad a=1,2,3$, the permutation $L(\alpha)=(i_1^1,\ldots,i_{M_{\rm v}}^1,i_1^2,\ldots,i_{M_{\rm v}}^2,i_1^3,\ldots,i_{M_{\rm v}}^3)$, and
\begin{align}
    \epsilon(\alpha)=\operatorname{sgn}L(\alpha).
\end{align}
The intravalley Vandermonde factors are
\begin{align}
    \Delta_a(\eta)=\Delta\!\left(\{\eta_i\}_{i\in I_a(\alpha)}\right)=\prod_{\substack{i<j\\ i,j\in I_a(\alpha)}}(\eta_i-\eta_j).
\end{align}
The corresponding ground-state envelope wavefunctions are therefore
\begin{align}
    \tilde w_\alpha(\bm\eta)&=\epsilon(\alpha)\frac{|\Delta(\eta)|}{|\Delta(\eta)|_{\rm reg}^{4/3}}\prod_{a=1}^3\Delta_a(\eta)
    \simeq\epsilon(\alpha)|\Delta(\eta)|^{-1/3}\prod_{a=1}^3\Delta_a(\eta),\notag\\
    w_\alpha(\bm\eta)&=\epsilon(\alpha)\frac{|\Delta(\eta)|^2}{|\Delta(\eta)|_{\rm reg}^{8/3}}\prod_{a=1}^3\Delta_a(\eta)
    \simeq\epsilon(\alpha)|\Delta(\eta)|^{-2/3}\prod_{a=1}^3\Delta_a(\eta).
    \label{ground-state-envelope-wavefunction}
\end{align}
Here, the expressions after $\simeq$ denote the corresponding long-distance behavior.

Similar to the Laughlin state selected by repulsive pseudopotential interactions, this holomorphic solution can be favored by a repulsive short-range interaction. For this $h_{F,\alpha}(\eta)$, the corresponding $\tilde w_\alpha(\bm\eta)$ vanishes whenever any two quasihole pairs coincide. By contrast, any other allowed solution necessarily contains at least one pole at a coincidence point, so that $\tilde w_\alpha(\bm\eta)$ remains finite there. Therefore, within the zero-energy manifold of the ACCS Hamiltonian, repulsive intra-valley and inter-valley components of $\mathcal V_{\rm short}$ naturally select this state.

\subsubsection{Ground-state wavefunction in electron coordinates}

We first reconstruct the many-quasihole-pair state from the ground-state envelope $w_\alpha(\bm\eta)$,
\begin{align}
    f(\bar\xi)=\sum_\alpha\int d^2\bm\kappa\,d^2\bm\eta\,e^{-i\bm\kappa\cdot\bm\eta}w_\alpha(\bm\eta)\Psi_{\bm\kappa,\alpha}(\bar\xi),
\end{align}
where $d^2\bm\eta=\prod_{i=1}^M d^2\bm\eta_i/(2\pi)^2$. After integration by parts, we obtain
\begin{align}
    f(\bar\xi)=\sum_\alpha\left[w_\alpha\!\left(-2i\partial_\kappa,-2i\partial_{\bar\kappa}\right)\Psi_{\bm\kappa,\alpha}(\bar\xi)\right]_{\bm\kappa=0},
\end{align}
where $\bar\eta_i\rightarrow-2i\partial_{\kappa_i}$ and $\eta_i\rightarrow-2i\partial_{\bar\kappa_i}$. We next consider the action of $-2i\partial_{\kappa_i}$ and $-2i\partial_{\bar\kappa_i}$ on the single-particle states $\psi_{\bm K_{\alpha_i}+\bm\kappa_i,-}(\bar\xi_i)$. In the leading long-distance approximation, the momentum derivatives acting on $\Psi_{\bm\kappa,\alpha}(\bar\xi)$ can be replaced by
\begin{align}
    -2i\partial_{\kappa_i}\rightarrow\bar\xi_i,\qquad -2i\partial_{\bar\kappa_i}\rightarrow2\mu\partial_{\bar\xi_i}.
\end{align}
Here we have neglected the $O(1)$ contributions, which are subleading at long distances. The derivative $\partial_{\bar\xi_i}$ should be retained because the single-particle wavefunctions contain the sigma function, whose large-distance behavior satisfies $|\sigma(\xi)|\sim e^{|\xi|^2/(4\mu)}$ and $|\partial_{\bar\xi}\ln\overline{\sigma(\xi)}|\sim|\xi|$. Using the normal ordering in which $\bar\xi_i$ is placed to the left of $\partial_{\bar\xi_i}$, we obtain
\begin{align}
    f(\bar\xi)\simeq\sum_\alpha\epsilon(\alpha):\left[\Delta(\bar\xi)\Delta(2\mu\partial_{\bar\xi})\right]^{-1/3}:\left[\prod_{a=1}^3\det_{\substack{i\in I_a(\alpha)\\r=0,\dots,M_{\rm v}-1}}\left[\partial_{\bar\xi_i}^r\psi_{\bm K_a,-}(\bar\xi_i)\right]\right],
\end{align}
where overall powers of $2\mu$ have been dropped.

The corresponding electron wavefunction is obtained by integrating the many-quasihole-pair state against the fermionic quasihole-pair state in Eq.~\eqref{quasihole-pair-Laughlin-state},
\begin{align}
    \Psi(\{\bm z\})=\int d\mu_{\bm\xi}\,f(\bar\xi)\psi_\xi(\{\bm z\})
    =\psi_{\rm L}(\{\bm z\})\int\prod_i\frac{d^2\bm\xi_i}{2\pi\mu}e^{-\frac{\sum_i|\xi_i|^2}{2\mu}}\mathcal F_F(\bar\xi,\xi)f(\bar\xi)\Delta(\xi)\prod_l E(z;\xi_l).
\end{align}
Here, $\psi_{\rm L}$ is the Laughlin state, and $E(z;\xi_l)$ is the factor associated with a quasihole pair in the relative-angular-momentum channel $L=2$,
\begin{align}
    \psi_{\rm L}(\{\bm z\})=\prod_{i<j}(z_i-z_j)^q e^{-\frac14\sum_i|z_i|^2},\qquad
    E(z;\xi_l)=\sum_v\prod_{u\neq v}(z_u-\xi_l)^2.
\end{align}
In the long-distance limit, after dropping overall normalization factors, we obtain
\begin{align}
    \Psi(\{\bm z\})\simeq\psi_{\rm L}(\{\bm z\})\sum_\alpha\epsilon(\alpha)\int\prod_i\frac{d^2\bm\xi_i}{2\pi\mu}e^{-\frac{\sum_i|\xi_i|^2}{2\mu}}\Psi_{0,\alpha}(\bar\xi)\Delta(\xi)\prod_{a=1}^3\Delta_a(\xi)\prod_l E(z;\xi_l),
\end{align}
where $\Psi_{0,\alpha}(\bar\xi)=\prod_{i=1}^M\psi_{\bm K_{\alpha_i},-}(\bar\xi_i)$. This expression is the Gutzwiller-projected wavefunction obtained from the parton construction discussed in SM.~\ref{matching-the-field-theory-description}.
Using the identity
\begin{align}
    \int\frac{d^2\bm\xi}{2\pi\mu}e^{-\frac{|\xi|^2}{2\mu}}F(\xi)G(\bar\xi)
    =\left[F(2\mu\partial_{\bar\xi})G(\bar\xi)\right]_{\bar\xi=0}
    =\left[G(2\mu\partial_\xi)F(\xi)\right]_{\xi=0},
\end{align}
we can write the electron ground-state wavefunction in the long-distance limit as
\begin{align}
    \Psi(\{\bm z\})&\simeq\psi_{\rm L}(\{\bm z\})\sum_\alpha\epsilon(\alpha)\left[\Delta(\partial_{\bar\xi})\prod_{a=1}^3\Delta_a(\partial_{\bar\xi})\prod_l E(z;2\mu\partial_{\bar\xi_l})\Psi_{0,\alpha}(\bar\xi)\right]_{\bar\xi=0}\notag\\
    &=\psi_{\rm L}(\{\bm z\})\det_{\substack{n=0,\dots,M-1\\a=1,2,3,\;r=0,\dots,M_{\rm v}-1}}
    \left[E(z;2\mu\partial_{\bar\xi})\partial_{\bar\xi}^{n+r}\psi_{\bm K_a,-}(\bar\xi)\big|_{\bar\xi=0}\right].
\end{align}

\section{Bosonic and fermionic representations}
\label{bosonic-and-fermionic-representations}

For the bosonic representation of the Laughlin quasihole state $\psi_\xi(\{\bm z\})$ in Eq.~\eqref{Laughlin-quasihole-state}, which we denote by $\psi_\xi^{(B)}$ in this section, the wavefunction is symmetric under the exchange of any two quasihole coordinates. The K\"ahler contribution $\delta\mathcal K^{(B)}$ and the corresponding statistical magnetic field $\delta\mathcal B^{(B)}$ are smooth when two quasiholes approach each other within a distance of $O(\ell_B)$. We can alternatively introduce a fermionic representation by attaching one unit of statistical flux to each quasihole,
\begin{align}
    \psi_\xi^{(F)}(\{\bm z\})=\Delta(\xi)\psi_\xi^{(B)}(\{\bm z\}),\qquad
    \Delta(\xi)=\prod_{i<j}(\xi_i-\xi_j).
\end{align}
Since $\Delta(\xi)$ changes sign under the exchange of two quasiholes, $\psi_\xi^{(F)}$ is antisymmetric under quasihole exchange.

This change of representation modifies the K\"ahler contribution $\delta\mathcal K(\bar\xi,\xi)$, and hence the exponentiated K\"ahler factor $\mathcal W(\bar\xi,\xi)=e^{\delta\mathcal K(\bar\xi,\xi)}$, as well as the corresponding measure factor $\mathcal F(\bar\xi,\xi)$. The many-quasihole wavefunction $f(\bar\xi)$ also transforms accordingly. We obtain
\begin{align}
    \delta\mathcal K^{(F)}(\bar\xi,\xi)&=\delta\mathcal K^{(B)}(\bar\xi,\xi)+\ln|\Delta(\xi)|^2,\qquad \mathcal W^{(F)}(\bar\xi,\xi)=|\Delta(\xi)|^2\mathcal W^{(B)}(\bar\xi,\xi),\notag\\
    \mathcal F^{(F)}(\bar\xi,\xi)&=|\Delta(\xi)|^{-2}\mathcal F^{(B)}(\bar\xi,\xi),\qquad\qquad f^{(F)}(\bar\xi)=\overline{\Delta(\xi)}\,f^{(B)}(\bar\xi).
    \label{bosonic-and-fermionic-kahler}
\end{align}
In the envelope effective theory, the corresponding K\"ahler weights are related by
\begin{align}
    e^{W_F(\bar\eta,\eta)/2}=|\Delta(\eta)|e^{W_B(\bar\eta,\eta)/2},
\end{align}
and the envelope wavefunctions transform as
\begin{align}
    w_\alpha^{(F)}(\bm\eta)&=\overline{\Delta(\eta)}\,w_\alpha^{(B)}(\bm\eta),\qquad \tilde w_\alpha^{(F)}(\bm\eta)=\frac{\overline{\Delta(\eta)}}{|\Delta(\eta)|}\,\tilde w_\alpha^{(B)}(\bm\eta).
    \label{boson-fermion-transformation-w}
\end{align}
For the zero-mode solutions $\tilde w_\alpha^{(B)}(\bm\eta)=e^{W_B/2}h_\alpha^{(B)}(\eta)$, $\tilde w_\alpha^{(F)}(\bm\eta)=e^{W_F/2}h_\alpha^{(F)}(\eta)$, the holomorphic or meromorphic functions are therefore related by
\begin{align}
    h_\alpha^{(F)}(\eta)=\Delta^{-1}(\eta)h_\alpha^{(B)}(\eta).
    \label{f-gauge-transformation}
\end{align}
This relation also shows that $h_\alpha^{(F)}(\eta)$ can in general be meromorphic, with poles allowed at quasihole coincidence points.

The flux attachment modifies both the long-distance and short-distance behavior of the K\"ahler weight in the envelope effective theory. Suppose that in the bosonic representation,
\begin{align}
    e^{W_B(\bar\eta,\eta)/2}\simeq\prod_{i<j}|\eta_i-\eta_j|^{-\lambda}
\end{align}
at long distances. Attaching one flux changes the long-distance exponent from $\lambda$ to $\lambda-1$, giving $e^{W_F(\bar\eta,\eta)/2}\simeq\prod_{i<j}|\eta_i-\eta_j|^{1-\lambda}$. In Section~\ref{ground-state-of-AC-Hamiltonian}, we show that the short-distance behavior of $e^{W/2}$ is important for determining the degeneracy of the ACCS ground-state subspace. This short-distance behavior cannot be inferred from its long-distance asymptotic form alone. Instead, it is determined by first returning to the bosonic representation, where $e^{W_B/2}$ is smooth and finite when two quasiholes approach each other, and then applying the flux-attachment transformation. Since $e^{W_F(\bar\eta,\eta)/2}=|\Delta(\eta)|e^{W_B(\bar\eta,\eta)/2}$, the fermionic K\"ahler weight $e^{W_F/2}$ vanishes linearly when two quasiholes approach each other.

\section{Order parameter and off-diagonal long-range order}\label{sec:pairs}

For the $2/3$ quasihole-pair problem, we focus on the ground-state solution in Eq.~\eqref{ground-state-envelope-wavefunction} selected by repulsive short-range interactions. In this section, we study the Cooper-pair order parameter and the off-diagonal long-range order (ODLRO) associated with this ground state.

We work in the bosonic representation, $\psi_\xi(\{\bm z\})=\psi_\xi^{(B)}(\{\bm z\})$, throughout this section and omit the superscript $(B)$ when no confusion can arise. The wavefunction for $N$ electrons and $M$ $2/3$ quasihole pairs at electronic filling $\nu=1/q=1/3$ is
\begin{align}
    \psi_{\xi}^{(N,M)}(\{\bm z\})=\prod_{a=1}^M e_1\!\left((z-\xi_a)^{-2}\right)\prod_{a=1}^M\prod_{i=1}^N(z_i-\xi_a)^2\prod_{i<j}^N(z_i-z_j)^q e^{-\frac14\sum_{i=1}^N|z_i|^2}.
    \label{laughlin-quasihole-wavefunction}
\end{align}
Then the ground state is written as
\begin{align}
    |\Psi^{(N,M)}\rangle=\int d\mu_{\bm\xi}\,f(\bar\xi)|\psi_{\xi}^{(N,M)}\rangle,\qquad
    f(\bar\xi)=\sum_{\alpha}\int d^2\bm\eta\,w_\alpha(\bm\eta)\Psi_{\bm\eta,\alpha}(\bar\xi),
\end{align}
where $\Psi_{\bm\eta,\alpha}(\bar\xi)=\prod_{i=1}^M\psi_{\bm\eta_i,\alpha_i}(\bar\xi_i),$ and $\psi_{\bm\eta,\alpha}(\bar\xi)=\int\frac{d^2\bm\kappa}{(2\pi)^2}e^{-i\bm\kappa\cdot\bm\eta}\psi_{\bm\kappa,\alpha}(\bar\xi).$
Here, $d^2\bm\eta=\prod_{i=1}^M d^2\bm\eta_i$. Using the bosonic--fermionic transformation in Eq.~\eqref{boson-fermion-transformation-w}, the envelope wavefunction corresponding to Eq.~\eqref{ground-state-envelope-wavefunction} takes the form
\begin{align}
    w_\alpha(\bm\eta)=\epsilon(\alpha)\frac{\Delta(\eta)}{|\Delta(\eta)|_{\rm reg}^{8/3}}\prod_{\beta=1}^3\Delta_\beta(\eta)
    \simeq\epsilon(\alpha)\frac{\Delta(\eta)}{|\Delta(\eta)|^{8/3}}\prod_{\beta=1}^3\Delta_\beta(\eta),
\end{align}
where the expression after $\simeq$ denotes the long-distance behavior, and $M_{\rm v}=M/3$ is the number of quasihole pairs in each valley.

We consider the Cooper-pair operator $\hat\Delta(\bm R,\bm v)$, together with the corresponding Cooper-pair order parameter,
\begin{align}
    \hat\Delta(\bm R,\bm v)&=\hat c_{\bm R+\bm v}\hat c_{\bm R-\bm v},\qquad \Delta(\bm R,\bm v)=\langle\Psi^{(N-2,M+3)}|\hat\Delta(\bm R,\bm v)|\Psi^{(N,M)}\rangle.
\end{align}
Decomposing the pair operator and the order parameter into relative angular momentum channels, the component in the channel $\ell$ is
\begin{align}
    \hat\Delta_\ell(\bm R)&=g_\ell\int d^2\bm v\,\bar v^\ell\hat\Delta(\bm R,\bm v),\qquad
    \Delta_\ell(\bm R)=\langle\Psi^{(N-2,M+3)}|\hat\Delta_\ell(\bm R)|\Psi^{(N,M)}\rangle=g_\ell\int d^2\bm v\,\bar v^\ell\Delta(\bm R,\bm v),
\end{align}
where the normalization constant $g_\ell=\frac{1}{2^{\ell+1/2}\sqrt{\ell!}}$ is chosen such that $\|\hat\Delta_\ell^\dagger(\bm R)|0\rangle\|=1$. $\ell$ is odd because of fermionic antisymmetry.


\subsection{Symmetry properties of the order parameter}

We now study the symmetry properties of the Cooper-pair order parameter.

\textbf{1. Translation.} To study the translational symmetry of the order parameter, it is convenient to work on a torus. We define the magnetic translation operator for an electron by
\begin{align}
    [\hat T_{\bm d}\psi](\bm r)=e^{-\frac{i}{2}\bm d\wedge\bm r}\psi(\bm r+\bm d).
\end{align}
Then its action on $\psi_{\xi}^{(N,M)}(\{\bm z\})$ is
\begin{align}
    \hat T_{\bm d}\psi_{\xi}^{(N,M)}(\{\bm z\})=e^{i\vartheta_{\bm d}}e^{-\frac{M|d|^2}{4\mu}}e^{\frac{\bar d}{2\mu}\sum_{a=1}^M\xi_a}\psi_{\xi-d}^{(N,M)}(\{\bm z\}),\qquad \mu=\frac{q}{2}=\frac{3}{2}.
\end{align}
Here, $\xi-d$ denotes $\xi_a\rightarrow\xi_a-d$ for all $a$, and $\vartheta_{\bm d}$ is a phase independent of $\xi$, $N$, and $M$. We omit this overall phase below. Acting with $\hat T_{\bm d}$ on $|\Psi^{(N,M)}\rangle$ induces a magnetic translation of the quasihole-pair wavefunction,
\begin{align}
    [\hat T_{\bm d}f](\bar\xi)=e^{-\frac{M|d|^2}{4\mu}}e^{-\frac{d}{2\mu}\sum_{a=1}^M\bar\xi_a}f(\bar\xi+\bar d),
\end{align}
which is precisely the magnetic translation defined in Eq.~\eqref{magnetic-translation-definition}. Therefore,
\begin{align}
    \hat T_{\bm a_i}|\Psi^{(N,M)}\rangle=\left(\det U_{T_{\bm a_i}}\right)^{M_{\rm v}}|\Psi^{(N,M)}\rangle=e^{iM_{\rm v}\theta_i}|\Psi^{(N,M)}\rangle,\qquad \det U_{T_{\bm a_i}}=e^{i\theta_i},\qquad i=1,2.
\end{align}
The electron annihilation operator transforms as $\hat T_{\bm d}\hat c_{\bm r}\hat T_{\bm d}^{-1}=e^{\frac{i}{2}\bm d\wedge\bm r}\hat c_{\bm r-\bm d}$. Then the Cooper-pair operator and its $\ell$-channel component transform as
\begin{align}
    \hat T_{\bm d}\hat\Delta(\bm R,\bm v)\hat T_{\bm d}^{-1}=e^{i\bm d\wedge\bm R}\hat\Delta(\bm R-\bm d,\bm v),\qquad \hat T_{\bm d}\hat\Delta_\ell(\bm R)\hat T_{\bm d}^{-1}=e^{i\bm d\wedge\bm R}\hat\Delta_\ell(\bm R-\bm d).
\end{align}
Therefore, the corresponding order parameters transform under magnetic translations as
\begin{align}
    \Delta(\bm R+\bm a_i,\bm v)=e^{-i\theta_i}e^{i\bm a_i\wedge\bm R}\Delta(\bm R,\bm v),\qquad \Delta_\ell(\bm R+\bm a_i)=e^{-i\theta_i}e^{i\bm a_i\wedge\bm R}\Delta_\ell(\bm R),\qquad i=1,2.
    \label{order-parameter-translation}
\end{align}

\textbf{2. Rotation.} The rotation operator is defined in Eq.~\eqref{rotation-operator-definition} by
\begin{align}
    [\hat C_n f](\bm r)=f(C_n^{-1}\bm r),\qquad [\hat C_n f](z)=f(\omega_n^{-1}z),\qquad [\hat C_n f](\bar z)=f(\omega_n\bar z),
\end{align}
where $\omega_n=e^{2\pi i/n}$. Its action on $\psi_{\xi}^{(N,M)}(\{\bm z\})$ is
\begin{align}
    \hat C_n\psi_{\xi}^{(N,M)}(\{\bm z\})=\omega_n^{-\frac{qN(N-1)}{2}-M(2N-L)}\psi_{\omega_n\xi}^{(N,M)}(\{\bm z\}),
\end{align}
where $\omega_n\xi$ denotes $\xi_a\rightarrow\omega_n\xi_a$ for all $a$. Then for $|\Psi^{(N,M)}\rangle$, the rotation gives
\begin{align}
    \hat C_n|\Psi^{(N,M)}\rangle=\omega_n^{-L_n(N,M)}|\Psi^{(N,M)}\rangle,
\end{align}
where
\begin{align}
    L_n(N,M)=\frac{3N(N-1)}{2}+\frac{M}{3}\left(6N+2M-3-3L-L_U\right)\pmod n,
\end{align}
with $\det U_{C_n}=\omega_n^{L_U}$. The electron annihilation operator transforms as $\hat C_n\hat c_{\bm r}\hat C_n^{-1}=\hat c_{C_n\bm r}$. The Cooper-pair operator and its $\ell$-channel component therefore transform as
\begin{align}
    \hat C_n\hat\Delta(\bm R,\bm v)\hat C_n^{-1}=\hat\Delta(C_n\bm R,C_n\bm v),\qquad \hat C_n\hat\Delta_\ell(\bm R)\hat C_n^{-1}=\omega_n^{\ell}\hat\Delta_\ell(C_n\bm R).
\end{align}
Therefore, the Cooper-pair order parameter transforms as
\begin{align}
    \Delta(C_n\bm R,C_n\bm v)=\omega_n^{L_{\rm tot}}\Delta(\bm R,\bm v),\qquad L_{\rm tot}=3L+L_U
\end{align}
while its $\ell$-channel component transforms as
\begin{align}
    \Delta_\ell(C_n\bm R)=\omega_n^{L_\Delta(\ell)}\Delta_\ell(\bm R),
    \label{order-parameter-rotation}
\end{align}
where
\begin{align}
    L_\Delta(\ell)=L_n(N,M)-L_n(N-2,M+3)-\ell=L_{\rm tot}-\ell\pmod n.
\end{align}
For the $2/3$ quasihole pair considered here, $L=2$, so that $L_\Delta(\ell)=6+L_U-\ell\pmod n$. Using the results of Section~\ref{single-particle-band-structure}, the possible angular momenta of the order parameter are summarized as
\begin{align*}
    C_6:&\qquad L_U=5 \quad\Rightarrow\quad L_{\Delta}(\ell)=11-\ell\pmod 6,\\
    C_4:&\qquad \mathcal V_1:\ L_U=3,1 \quad\Rightarrow\quad L_{\Delta}(\ell)=1-\ell,\,3-\ell\pmod 4,\\
         &\qquad \mathcal V_2:\ L_U=0,3 \quad\Rightarrow\quad L_{\Delta}(\ell)=2-\ell,\,1-\ell\pmod 4,\\
    C_3:&\qquad \mathcal V_1:\ L_U=2 \quad\Rightarrow\quad L_{\Delta}(\ell)=2-\ell\pmod 3,\\
         &\qquad \mathcal V_{2,3}:\ L_U=0 \quad\Rightarrow\quad L_{\Delta}(\ell)=-\ell\pmod 3.
\end{align*}

\textbf{3. Holomorphic structure.} By the lowest-Landau-level structure, the order parameter can be written as
\begin{align}
    \Delta_{\ell}(\bm R)=f_\ell(R)e^{-\frac{|R|^2}{2}},
    \label{order-parameter-holomorphic}
\end{align}
where $f_\ell(R)$ is a holomorphic function.

The translation symmetry in Eq.~\eqref{order-parameter-translation}, together with the holomorphic structure in Eq.~\eqref{order-parameter-holomorphic}, implies that $\Delta_{\ell}(\bm R)$ carries two vortices per unit cell. The locations of the vortices can be further constrained by considering the rotation symmetry in Eq.~\eqref{order-parameter-rotation} around the high-symmetry points. Let $L_P$ denote the angular momentum associated with the $C_m$ rotation about a high-symmetry point $P$. For a high-symmetry point satisfying $C_m\bm P=\bm P+\bm a_P$, where $\bm a_P$ is a lattice vector, we obtain
\begin{align}
    \omega_m^{L_P}=e^{i\theta_{\bm a_P}}e^{-i\bm a_P\wedge\bm P}\omega_m^{L_\Delta(\ell)}.
\end{align}

For $C_6$ symmetry, the six high-symmetry points are grouped into three special Wyckoff positions: $1a$, $2b$, and $3c$, with site symmetries $C_6$, $C_3$, and $C_2$, respectively. They are
\begin{align}
    1a:&\quad 0,\qquad 2b:\quad \frac{2\bm a_1+\bm a_2}{3},\ \frac{\bm a_1+2\bm a_2}{3},\qquad 3c:\quad \frac{\bm a_1}{2},\ \frac{\bm a_2}{2},\ \frac{\bm a_1+\bm a_2}{2}.
\end{align}
The rotation angular momenta about them are
\begin{align}
    L_{1a}=L_\Delta\pmod 6,\qquad L_{2b}=L_\Delta+1\pmod 3,\qquad L_{3c}=L_\Delta\pmod 2.
\end{align}
Therefore, the allowed values of $L_\Delta$ and the corresponding vortex configurations are
\begin{enumerate}
    \item $L_\Delta=0\Longleftrightarrow \ell=5\pmod 6$: one vortex at each site of the $2b$ Wyckoff position.
    \item $L_\Delta=2\Longleftrightarrow \ell=3\pmod 6$: a double vortex at the $1a$ Wyckoff position.
\end{enumerate}

For $C_4$ symmetry, the four high-symmetry points are grouped into three special Wyckoff positions: $1a$, $1b$, and $2c$, with site symmetries $C_4$, $C_4$, and $C_2$, respectively. They are
\begin{align}
    1a:&\quad 0,\qquad 1b:\quad \frac{\bm a_1+\bm a_2}{2},\qquad 2c:\quad \frac{\bm a_1}{2},\ \frac{\bm a_2}{2}.
\end{align}

For $\mathcal V_1$, we have $\theta_1=\theta_2=0$. The rotation angular momenta about the special Wyckoff positions are
\begin{align}
    L_{1a}=L_\Delta\pmod 4,\qquad L_{1b}=L_\Delta+2\pmod 4,\qquad L_{2c}=L_\Delta\pmod 2.
\end{align}
Therefore, the allowed values of $L_\Delta$ and the corresponding vortex configurations are
\begin{enumerate}
    \item $L_\Delta=0\Longleftrightarrow (L_U=3,\,\ell=1\pmod 4)$ or $(L_U=1,\,\ell=3\pmod 4)$: a double vortex at the $1b$ Wyckoff position.
    \item $L_\Delta=2\Longleftrightarrow (L_U=3,\,\ell=3\pmod 4)$ or $(L_U=1,\,\ell=1\pmod 4)$: a double vortex at the $1a$ Wyckoff position.
\end{enumerate}

For $\mathcal V_2$, we have $\theta_1=\theta_2=\pi$. The rotation angular momenta about the special Wyckoff positions are
\begin{align}
    L_{1a}=L_\Delta\pmod 4,\qquad L_{1b}=L_\Delta\pmod 4,\qquad L_{2c}=L_\Delta+1\pmod 2.
\end{align}
Therefore, the allowed values of $L_\Delta$ and the corresponding vortex configurations are
\begin{enumerate}
    \item $L_\Delta=0\Longleftrightarrow (L_U=3,\,\ell=1\pmod 4)$: one vortex at each site of the $2c$ Wyckoff position.
    \item $L_\Delta=1\Longleftrightarrow (L_U=0,\,\ell=1\pmod 4)$: one vortex at the $1a$ Wyckoff position and one vortex at the $1b$ Wyckoff position.
\end{enumerate}

For $C_3$ symmetry, the three high-symmetry points correspond to three special Wyckoff positions: $1a$, $1b$, and $1c$, all with site symmetry $C_3$. They are
\begin{align}
    1a:&\quad 0,\qquad 1b:\quad \frac{2\bm a_1+\bm a_2}{3},\qquad 1c:\quad \frac{\bm a_1+2\bm a_2}{3}.
\end{align}

For $\mathcal V_1$, we have $e^{i\theta_1}=e^{i\theta_2}=1$. The rotation angular momenta about the special Wyckoff positions are
\begin{align}
    L_{1a}=L_\Delta\pmod 3,\qquad L_{1b}=L_\Delta+1\pmod 3,\qquad L_{1c}=L_\Delta+1\pmod 3.
\end{align}
Therefore, the allowed values of $L_\Delta$ and the corresponding vortex configurations are
\begin{enumerate}
    \item $L_\Delta=0\Longleftrightarrow \ell=5\pmod 6$: one vortex at each of the $1b$ and $1c$ Wyckoff positions.
    \item $L_\Delta=2\Longleftrightarrow \ell=3\pmod 6$: a double vortex at the $1a$ Wyckoff position.
\end{enumerate}

For $\mathcal V_2$, we have $e^{i\theta_1}=e^{i\theta_2}=\omega_3$. The rotation angular momenta about the special Wyckoff positions are
\begin{align}
    L_{1a}=L_\Delta\pmod 3,\qquad L_{1b}=L_\Delta\pmod 3,\qquad L_{1c}=L_\Delta+2\pmod 3.
\end{align}
Therefore, the allowed values of $L_\Delta$ and the corresponding vortex configurations are
\begin{enumerate}
    \item $L_\Delta=0\Longleftrightarrow \ell=3\pmod 6$: a double vortex at the $1c$ Wyckoff position.
    \item $L_\Delta=1\Longleftrightarrow \ell=5\pmod 6$: one vortex at each of the $1a$ and $1b$ Wyckoff positions.
\end{enumerate}

For $\mathcal V_3$, we have $e^{i\theta_1}=e^{i\theta_2}=\omega_3^2$. The rotation angular momenta about the special Wyckoff positions are
\begin{align}
    L_{1a}=L_\Delta\pmod 3,\qquad L_{1b}=L_\Delta+2\pmod 3,\qquad L_{1c}=L_\Delta\pmod 3.
\end{align}
Therefore, the allowed values of $L_\Delta$ and the corresponding vortex configurations are
\begin{enumerate}
    \item $L_\Delta=0\Longleftrightarrow \ell=3\pmod 6$: a double vortex at the $1b$ Wyckoff position.
    \item $L_\Delta=1\Longleftrightarrow \ell=5\pmod 6$: one vortex at each of the $1a$ and $1c$ Wyckoff positions.
\end{enumerate}

\begin{figure}[h]
    \centering
    \includegraphics[width=0.75\linewidth]{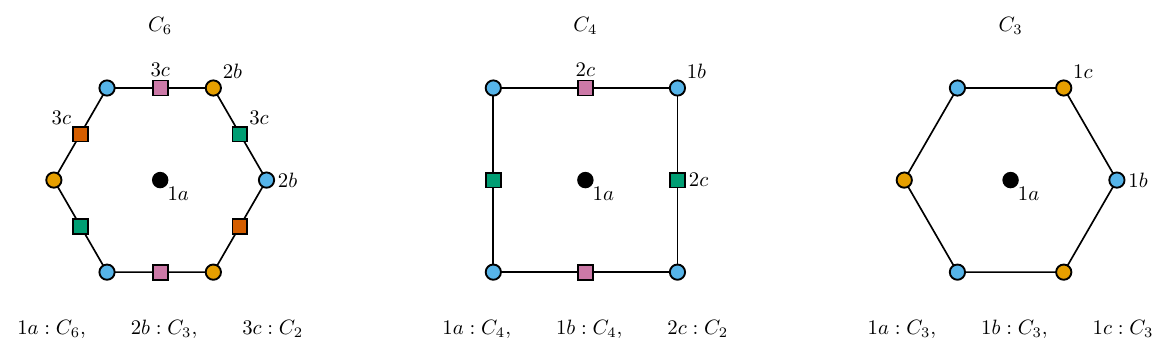}
    \caption{Special Wyckoff positions for $C_6$, $C_4$, and $C_3$ symmetries. Site symmetries are indicated below each panel.}
    \label{fig:wyckoff-position}
\end{figure}

\begin{table}
    \centering
    \resizebox{\textwidth}{!}{
    \begin{tabular}{|c|c|c|c|c|c|c|}
        \hline
        symmetry & band minima $(\bm K_1,\bm K_2,\bm K_3)$ & irrep & $\det U_T$ & $\det U_C$ & allowed $\ell$ & vortex configuration \\
        \hline
        \multirow{2}{*}{$C_6$} & \multirow{2}{*}{$\left(\frac{\bm\beta_1}{6},-\frac{\bm\beta_1}{6},-\frac{\bm\beta_1}{2}\right)$} & $C_6,(1)$ & \multirow{2}{*}{$1$} & \multirow{2}{*}{$\omega_6^5$} & \multirow{2}{*}{$\ell=3,5\pmod 6$} & \multirow{2}{*}{$\ell=5\pmod 6$: two $2b$ vortices; $\ell=3\pmod 6$: double vortex at $1a$} \\
        \cline{3-3}
        & & $C_6,(2)$ & & & & \\
        \hline
        \multirow{4}{*}{$C_4$} & \multirow{2}{*}{$\mathcal V_1:\left(\frac{\bm\beta_1}{6},-\frac{\bm\beta_1}{6},-\frac{\bm\beta_1}{2}\right)$} & $C_4,\mathcal V_1,(1)$ & \multirow{2}{*}{$1$} & $\omega_4^3$ & \multirow{2}{*}{$\ell=1,3\pmod 4$} & $\ell=1\pmod 4$: double vortex at $1b$; $\ell=3\pmod 4$: double vortex at $1a$ \\
        \cline{3-3}\cline{5-5}\cline{7-7}
        & & $C_4,\mathcal V_1,(2)$ & & $\omega_4$ & & $\ell=1\pmod 4$: double vortex at $1a$; $\ell=3\pmod 4$: double vortex at $1b$ \\
        \cline{2-7}
        & \multirow{2}{*}{$\mathcal V_2:\left(-\frac{\bm\beta_2}{2},-\frac{\bm\beta_1}{3}-\frac{\bm\beta_2}{2},\frac{\bm\beta_1}{3}-\frac{\bm\beta_2}{2}\right)$} & $C_4,\mathcal V_2,(1)$ & \multirow{2}{*}{$-1$} & $1$ & \multirow{2}{*}{$\ell=1\pmod 4$} & one vortex at each of $1a$ and $1b$ \\
        \cline{3-3}\cline{5-5}\cline{7-7}
        & & $C_4,\mathcal V_2,(2)$ & & $\omega_4^3$ & & one vortex at each site of $2c$ \\
        \hline
        \multirow{6}{*}{$C_3$} & \multirow{2}{*}{$\mathcal V_1:\left(\frac{\bm\beta_1}{6},-\frac{\bm\beta_1}{6},-\frac{\bm\beta_1}{2}\right)$} & $C_3,\mathcal V_1,(1)$ & \multirow{2}{*}{$1$} & \multirow{2}{*}{$\omega_3^2$} & \multirow{2}{*}{$\ell=3,5\pmod 6$} & \multirow{2}{*}{$\ell=5\pmod 6$: vortices at $1b$ and $1c$; $\ell=3\pmod 6$: double vortex at $1a$} \\
        \cline{3-3}
        & & $C_3,\mathcal V_1,(2)$ & & & & \\
        \cline{2-7}
        & \multirow{2}{*}{$\mathcal V_2:\left(\frac{5\bm\beta_1}{18}+\frac{\bm\beta_2}{3},-\frac{\bm\beta_1}{18}+\frac{\bm\beta_2}{3},-\frac{7\bm\beta_1}{18}+\frac{\bm\beta_2}{3}\right)$} & $C_3,\mathcal V_2,(1)$ & \multirow{2}{*}{$\omega_3$} & \multirow{2}{*}{$1$} & \multirow{2}{*}{$\ell=3,5\pmod 6$} & \multirow{2}{*}{$\ell=3\pmod 6$: double vortex at $1c$; $\ell=5\pmod 6$: vortices at $1a$ and $1b$} \\
        \cline{3-3}
        & & $C_3,\mathcal V_2,(2)$ & & & & \\
        \cline{2-7}
        & \multirow{2}{*}{$\mathcal V_3:\left(\frac{7\bm\beta_1}{18}-\frac{\bm\beta_2}{3},\frac{\bm\beta_1}{18}-\frac{\bm\beta_2}{3},-\frac{5\bm\beta_1}{18}-\frac{\bm\beta_2}{3}\right)$} & $C_3,\mathcal V_3,(1)$ & \multirow{2}{*}{$\omega_3^2$} & \multirow{2}{*}{$1$} & \multirow{2}{*}{$\ell=3,5\pmod 6$} & \multirow{2}{*}{$\ell=3\pmod 6$: double vortex at $1b$; $\ell=5\pmod 6$: vortices at $1a$ and $1c$} \\
        \cline{3-3}
        & & $C_3,\mathcal V_3,(2)$ & & & & \\
        \hline
    \end{tabular}}
\end{table}

\subsection{Off-diagonal long-range order}

To determine whether the ground state $|\Psi^{(N,M)}\rangle$ describes a superconducting state, we study the Cooper-pair correlation function and examine whether it exhibits off-diagonal long-range order. We first consider the action of $\hat\Delta_{\ell}(\bm R)$ on the Laughlin quasihole-pair state in Eq.~\eqref{laughlin-quasihole-wavefunction}. Acting with $\hat c_{\bm R+\bm v}\hat c_{\bm R-\bm v}$ on $\psi_{\xi}^{(N,M)}$ gives
\begin{align}
    &[\hat c_{\bm R+\bm v}\hat c_{\bm R-\bm v}\psi_{\xi}^{(N,M)}](\{\bm z\})=\sqrt{N(N-1)}\,\psi_{\xi}^{(N,M)}(R-v,R+v,z_1,\dots,z_{N-2})\notag\\
    &=\sqrt{N(N-1)}(-2v)^q e^{-\frac{|R|^2+|v|^2}{2}}\prod_{a=1}^M\prod_{i=1}^{N-2}(z_i-\xi_a)^2\prod_{i<j}^{N-2}(z_i-z_j)^q e^{-\frac14\sum_{i=1}^{N-2}|z_i|^2}\prod_{a=1}^M(R-v-\xi_a)^2(R+v-\xi_a)^2\notag\\
    &\quad\times\prod_{i=1}^{N-2}(R-v-z_i)^q(R+v-z_i)^q\prod_{a=1}^M\left[\frac{1}{(R-v-\xi_a)^2}+\frac{1}{(R+v-\xi_a)^2}+\sum_{i=1}^{N-2}\frac{1}{(z_i-\xi_a)^2}\right].
\end{align}

Let $\hat P_{\rm pair}$ denote the projection onto the low-energy subspace in which all quasiholes form $L=2$ quasihole pairs. We now consider the lowest channel $\ell$ of $\hat\Delta_\ell(\bm R)$ that has a nonvanishing projection onto this subspace. For the $q=3$ Laughlin state, the annihilated electron pair contributes the relative-coordinate factor $(-2v)^q=(-2v)^3$, corresponding to relative angular momentum $L_e=q=3$. After projection onto the quasihole-pair subspace, removing two electrons is equivalent to creating three quasihole pairs centered at $R$. Since the relative wavefunction of each quasihole pair is proportional to $\bar\xi_{\rm rel}^{L}$, each pair contributes angular momentum $-L$, and the three quasihole pairs therefore contribute a total internal angular momentum $-3L=-6$. The lowest electron-pair channel compatible with the projected quasihole-pair state is thus
\begin{align}
    \ell_{\rm min}=q+3L=3+6=9.
\end{align}
Therefore, the projected action of $\hat\Delta_{\ell}(\bm R)$ can be written as
\begin{align}
    \hat P_{\rm pair}\hat\Delta_{\ell}(\bm R)|\psi_\xi^{(N,M)}\rangle=e^{-\frac{|R|^2}{2}}D_{\ell}(R,\xi)\prod_{a=1}^M(R-\xi_a)^4|\psi_{\xi,R,R,R}^{(N-2,M+3)}\rangle,
\end{align}
where $D_{\ell}(R,\xi)$ carries angular momentum $9-\ell$. For the lowest channel $\ell=9$, $D_9(R,\xi)$ is a constant, and for $\ell=11$, we have $D_{11}(R,\xi)\propto\sum_{a=1}^M\frac{1}{(R-\xi_a)^2}$.

For the ground state $|\Psi^{(N,M)}\rangle=\int d\mu_{\bm\xi}\,f(\bar\xi)|\psi_{\xi}^{(N,M)}\rangle$, we define the projected Cooper-pair component by
\begin{align}
    |\Gamma_{\ell}(\bm R)\rangle=\hat P_{\rm pair}\hat\Delta_\ell(\bm R)|\Psi^{(N,M)}\rangle=e^{-\frac{|R|^2}{2}}\int d\mu_{\bm\xi}\,f(\bar\xi)D_{\ell}(R,\xi)\prod_{a=1}^M(R-\xi_a)^4|\psi_{\xi,R,R,R}^{(N-2,M+3)}\rangle.
\end{align}
To diagnose ODLRO, we consider the projected pair correlation function
\begin{align}
    C_{\ell}(\bm R',\bm R)=\frac{1}{\mathcal Z_\Psi}\langle\Gamma_{\ell}(\bm R')|\Gamma_{\ell}(\bm R)\rangle,\qquad \mathcal Z_\Psi=\langle\Psi^{(N,M)}|\Psi^{(N,M)}\rangle.
\end{align}
Substituting $|\Gamma_{\ell}(\bm R)\rangle$ into the projected pair correlation function, we obtain
\begin{align}
    C_{\ell}(\bm R',\bm R)=\frac{e^{-\frac{|R|^2+|R'|^2}{2}}}{\mathcal Z_\Psi}\int d\mu_{\bm\xi'}d\mu_{\bm\xi}\,\overline{f(\bar\xi')}f(\bar\xi)\,\overline{D_{\ell}(R',\xi')}D_{\ell}(R,\xi)\prod_{a=1}^M(\bar R'-\bar\xi_a')^4\prod_{b=1}^M(R-\xi_b)^4\,\mathcal G_{M+3}(\bar X_{R'}';X_R),
\end{align}
where $X_R=(\xi_1,\ldots,\xi_M,R_1,R_2,R_3)$ and $X_{R'}'=(\xi_1',\ldots,\xi_M',R_1',R_2',R_3')$, with $R_i=R$ and $R_i'=R'$, and
\begin{align}
    \mathcal G_{M+3}(\bar X_{R'}';X_R)=e^{\mathcal K(\bar X_{R'}';X_R)}=\langle\psi_{\xi',R',R',R'}^{(N-2,M+3)}|\psi_{\xi,R,R,R}^{(N-2,M+3)}\rangle.
\end{align}

We next analyze the overlap $\mathcal G_{M+3}$ using the plasma approximation. For well-separated coordinates, the plasma approximation for the diagonal K\"ahler potential is
\begin{align}
    \mathcal K(\bar X;X)\simeq\frac{1}{3}\sum_i\bar X_iX_i-\frac{4}{3}\sum_{i<j}\ln(\bar X_i-\bar X_j)-\frac{4}{3}\sum_{i<j}\ln(X_i-X_j).
\end{align}
When analytically continuing to the off-diagonal K\"ahler potential $\mathcal K(\bar X';X)$, we have to keep in mind that the exact overlap $e^{\mathcal K(\bar X';X)}=\langle\psi_{X'}|\psi_X\rangle$ is invariant under independent permutations of $X$ and $X'$. In contrast, the off-diagonal plasma approximation
\begin{align}
    \mathcal K^{\rm pl}(\bar X';X)=\frac{1}{3}\sum_i\bar X_i'X_i-\frac{4}{3}\sum_{i<j}\ln(\bar X_i'-\bar X_j')-\frac{4}{3}\sum_{i<j}\ln(X_i-X_j)
\end{align}
is only invariant under simultaneous permutations of $X$ and $X'$. In particular, the first term assumes a specific pairing $X_i\leftrightarrow X_i'$. To understand the meaning of this pairing, we note that the off-diagonal K\"ahler factor always appears together with the Gaussian factor $e^{-\frac{1}{6}\sum_i(|X_i|^2+|X_i'|^2)}$, which combines with the first term to give an object whose magnitude goes as $e^{-\frac{1}{6}\sum_i|X_i-X_i'|^2}$, i.e. it forces $X_i$ and $X_i'$ to be close to each other. This suggests that the proper plasma expression for the off-diagonal factor $e^{\mathcal K(\bar X';X)}$ should involve a sum over all permutations, with the dominant permutations selected in a saddle-point approximation,
\begin{align}
    e^{\mathcal K(\bar X';X)}\simeq\sum_{\sigma\in S_{M+3}}e^{\frac{1}{3}\sum_i\bar X_i'X_{\sigma(i)}}\prod_{i<j}\left[(\bar X_i'-\bar X_j')(X_{\sigma(i)}-X_{\sigma(j)})\right]^{-\frac{4}{3}},
\end{align}
where $S_{M+3}$ denotes the permutation group of the $M+3$ coordinates. Note that this does not affect any of our previous expressions since we always integrate the K\"ahler factor against symmetric functions with a symmetric measure. However, in this case, we should be careful since we are singling out some coordinates and not all coordinates are integrated. The expression above applies when all coordinates in $X$ are well separated.  When some $X_{\sigma(i)}$ and $X_{\sigma(j)}$ are close, the corresponding logarithmic terms on both the $X$ side and the paired $X'$ side should be regularized.

For $X_R=(\xi_1,\ldots,\xi_M,R_1,R_2,R_3)$ and $X_{R'}'=(\xi_1',\ldots,\xi_M',R_1',R_2',R_3')$, we classify the permutation saddles according to the number $k$ of fixed coordinates $R_i'$ paired with the integrated coordinates $\xi_a$. Equivalently, $k$ is also the number of fixed coordinates $R_i$ paired with the integrated coordinates $\xi_a'$. We then decompose
\begin{align}
    \mathcal G_{M+3}=\sum_{k=0}^{3}\mathcal G_{M+3}^{(k)},\qquad C_{\ell}(\bm R',\bm R)=\sum_{k=0}^{3}C_{\ell}^{(k)}(\bm R',\bm R).
\end{align}
The sectors $k=0,1,2,3$ correspond to zero, one, two, or three cross-pairings between the fixed coordinates $R_i,R_i'$ and the integrated quasihole coordinates $\xi_a,\xi_a'$, respectively.

\subsubsection{The fully exchanged sector}

We focus on the fully exchanged sector with $k=3$. We separate the last three integrated quasihole-pair coordinates by writing
\begin{align*}
    \xi=(\eta_1,\ldots,\eta_{M-3},\zeta_1,\zeta_2,\zeta_3),\qquad \xi'=(\eta_1',\ldots,\eta_{M-3}',\zeta_1',\zeta_2',\zeta_3').
\end{align*}
We choose a reference pairing in which $R_i$ is paired with $\zeta_i'$ and $\zeta_i$ is paired with $R_i'$ for $i=1,2,3$, while the remaining coordinates are paired as $\eta_a\leftrightarrow\eta_a'$ for $a=1,\ldots,M-3$. Since the integration measure and the envelope function $f$ are symmetric under permutations of the quasihole-pair coordinates, all other $k=3$ pairings are related to this reference pairing by relabeling the integration variables and therefore give identical contributions. The total $k=3$ contribution is thus obtained by multiplying the reference contribution by the combinatorial factor
\begin{align*}
    \mathcal N_3=\binom{M}{3}^2(3!)^2(M-3)!=\frac{(M!)^2}{(M-3)!}.
\end{align*}
For the reference pairing, the overlap is given by
\begin{align}
    \mathcal G_{M+3,{\rm ref}}^{(3)}(\bar X_{R'}';X_R)&\simeq C^{(3)} e^{\frac{1}{3}\sum_{a=1}^{M-3}\bar\eta_a'\eta_a+\frac{1}{3}\sum_{i=1}^{3}\left(\bar\zeta_i'R+\bar R'\zeta_i\right)}\prod_{a<b}^{M-3}\left[(\bar\eta_a'-\bar\eta_b')(\eta_a-\eta_b)\right]^{-\frac{4}{3}}\prod_{a=1}^{M-3}\prod_{i=1}^{3}\left[(\bar\eta_a'-\bar\zeta_i')(\eta_a-\zeta_i)\right]^{-\frac{4}{3}}\notag\\
    &\qquad\times\prod_{a=1}^{M-3}\left[(\bar\eta_a'-\bar R')(\eta_a-R)\right]^{-4}\prod_{i=1}^{3}\left[(\bar\zeta_i'-\bar R')(\zeta_i-R)\right]^{-4}.
\end{align}
where $C^{(3)}$ is an $O(1)$ constant from the short-distance regularization of the coincident quasihole-pair coordinates, which we will ignore. 

Therefore, the fully exchanged contribution is
\begin{align}
    C_{\ell}^{(3)}(\bm R',\bm R)&=\frac{\mathcal N_3}{\mathcal Z_\Psi}e^{-\frac{|R|^2+|R'|^2}{2}}\int d\mu_{\bm\eta,\bm\zeta}\,d\mu_{\bm\eta',\bm\zeta'}\,\overline{f(\bar\eta',\bar\zeta')}f(\bar\eta,\bar\zeta)\,\overline{D_{\ell}(R',\eta',\zeta')}D_{\ell}(R,\eta,\zeta)e^{\frac{1}{3}\sum_{a=1}^{M-3}\bar\eta_a'\eta_a+\frac{1}{3}\sum_{i=1}^{3}\left(\bar\zeta_i'R+\bar R'\zeta_i\right)}\notag\\
    &\qquad\times\prod_{a<b}^{M-3}\left[(\bar\eta_a'-\bar\eta_b')(\eta_a-\eta_b)\right]^{-\frac{4}{3}}\prod_{a=1}^{M-3}\prod_{i=1}^{3}\left[(\bar\eta_a'-\bar\zeta_i')(\eta_a-\zeta_i)\right]^{-\frac{4}{3}}.
\end{align}
In the dilute limit, the measure for the well-separated quasihole-pair coordinates is approximated by
\begin{align}
    d\mu_{\bm\eta,\bm\zeta}\simeq\prod_{a=1}^{M-3}\frac{d^2\bm\eta_a}{3\pi}e^{-\frac{|\eta_a|^2}{3}}\prod_{i=1}^{3}\frac{d^2\bm\zeta_i}{3\pi}e^{-\frac{|\zeta_i|^2}{3}}\prod_{a<b}^{M-3}|\eta_a-\eta_b|^{\frac{8}{3}}\prod_{a=1}^{M-3}\prod_{i=1}^{3}|\eta_a-\zeta_i|^{\frac{8}{3}},
\end{align}
where the factors involving $\zeta_i-\zeta_j$ are in the short-distance regime and have been regularized to an $O(1)$ constant, which we ignore. The ground-state coefficient $f(\bar\xi)$ can be written as
\begin{align}
    f(\bar\eta,\bar\zeta)=\sum_{\alpha,\beta}\int d^2\bm\gamma\,d^2\bm\lambda\,w_{\alpha,\beta}(\bm\gamma,\bm\lambda)\prod_{a=1}^{M-3}\psi_{\bm\gamma_a,\alpha_a}(\bar\eta_a)\prod_{i=1}^{3}\psi_{\bm\lambda_i,\beta_i}(\bar\zeta_i),
\end{align}
where $d^2\bm\gamma=\prod_{a=1}^{M-3}d^2\bm\gamma_a$ and $d^2\bm\lambda=\prod_{i=1}^{3}d^2\bm\lambda_i$.
Substituting the envelope expansion into $C_{\ell}^{(3)}(\bm R',\bm R)$, we obtain
\begin{align}
    &C_{\ell}^{(3)}(\bm R',\bm R)=\frac{\mathcal N_3}{\mathcal Z_\Psi}e^{-\frac{|R|^2+|R'|^2}{2}}\sum_{\alpha,\beta,\alpha',\beta'}\int d^2\bm\gamma\,d^2\bm\lambda\,d^2\bm\gamma'\,d^2\bm\lambda'\,\overline{w_{\alpha',\beta'}(\bm\gamma',\bm\lambda')}w_{\alpha,\beta}(\bm\gamma,\bm\lambda)\notag\\
    &\quad\times\int\prod_{a=1}^{M-3}\frac{d^2\bm\eta_a\,d^2\bm\eta_a'}{(3\pi)^2}e^{-\frac{|\eta_a|^2+|\eta_a'|^2}{3}}\prod_{i=1}^{3}\frac{d^2\bm\zeta_i\,d^2\bm\zeta_i'}{(3\pi)^2}e^{-\frac{|\zeta_i|^2+|\zeta_i'|^2}{3}}\,\overline{D_{\ell}(R',\eta',\zeta')}D_{\ell}(R,\eta,\zeta)e^{\frac{1}{3}\sum_{a=1}^{M-3}\bar\eta_a'\eta_a+\frac{1}{3}\sum_{i=1}^{3}\left(\bar\zeta_i'R+\bar R'\zeta_i\right)}\notag\\
    &\quad\times\prod_{a<b}^{M-3}(\bar\eta_a-\bar\eta_b)^{\frac{4}{3}}(\eta_a'-\eta_b')^{\frac{4}{3}}\prod_{a=1}^{M-3}\prod_{i=1}^{3}(\bar\eta_a-\bar\zeta_i)^{\frac{4}{3}}(\eta_a'-\zeta_i')^{\frac{4}{3}}\prod_{a=1}^{M-3}\overline{\psi_{\bm\gamma_a',\alpha_a'}(\bar\eta_a')}\psi_{\bm\gamma_a,\alpha_a}(\bar\eta_a)\prod_{i=1}^{3}\overline{\psi_{\bm\lambda_i',\beta_i'}(\bar\zeta_i')}\psi_{\bm\lambda_i,\beta_i}(\bar\zeta_i).
\end{align}
In the long-distance limit, we use the approximation $\bar\eta_a\psi_{\bm\gamma_a,\alpha_a}(\bar\eta_a)\simeq\bar\gamma_a\psi_{\bm\gamma_a,\alpha_a}(\bar\eta_a)$ and $\bar\zeta_i\psi_{\bm\lambda_i,\beta_i}(\bar\zeta_i)\simeq\bar\lambda_i\psi_{\bm\lambda_i,\beta_i}(\bar\zeta_i)$, together with the Bargmann reproducing identity $\int\frac{d^2\bm\xi}{3\pi}e^{-\frac{|\xi|^2}{3}+\frac{\bar z\xi}{3}}F(\bar\xi)=F(\bar z)$. After integrating out $\eta,\zeta$ and $\eta',\zeta'$, the fully exchanged contribution becomes
\begin{align}
    C_{\ell}^{(3)}(\bm R',\bm R)&\simeq\frac{\mathcal N_3}{\mathcal Z_\Psi}e^{-\frac{|R|^2+|R'|^2}{2}}\sum_{\alpha,\beta,\beta'}\int d^2\bm\gamma\,d^2\bm\lambda\,d^2\bm\lambda'\,\overline{w_{\alpha,\beta'}(\bm\gamma,\bm\lambda')}w_{\alpha,\beta}(\bm\gamma,\bm\lambda)\,\overline{D_{\ell}(R',\gamma,\lambda')}D_{\ell}(R,\gamma,\lambda)\notag\\
    &\quad\times\prod_{a<b}^{M-3}|\gamma_a-\gamma_b|^{\frac{8}{3}}\prod_{a=1}^{M-3}\prod_{i=1}^{3}(\bar\gamma_a-\bar\lambda_i)^{\frac{4}{3}}(\gamma_a-\lambda_i')^{\frac{4}{3}}\prod_{i=1}^{3}\overline{\psi_{\bm\lambda_i',\beta_i'}(\bar R)}\psi_{\bm\lambda_i,\beta_i}(\bar R').
\end{align}
We define the modified envelope wavefunction for the $k$-exchanged sector by
\begin{align}
    \omega_{\alpha,\beta}^{(k)}(\bm\gamma,\bm\lambda)=\prod_{a<b}^{M-k}(\bar\gamma_a-\bar\gamma_b)^{\frac{4}{3}}\prod_{a=1}^{M-k}\prod_{i=1}^{k}(\bar\gamma_a-\bar\lambda_i)^{\frac{4}{3}}w_{\alpha,\beta}(\bm\gamma,\bm\lambda),
\end{align}
where $\alpha=(\alpha_1,\ldots,\alpha_{M-k})$ and $\beta=(\beta_1,\ldots,\beta_k)$. We define the corresponding dressed anyon 3RDM by
\begin{align}
    \rho_{\ell;\beta'\beta}^{(3)}(\bm R',\bm R;\bm\lambda',\bm\lambda)=\frac{1}{\mathcal Z_\Psi}\frac{M!}{(M-3)!}\sum_{\alpha}\int\prod_{a=1}^{M-3}d^2\bm\gamma_a\,\overline{D_{\ell}(R',\gamma,\lambda)\omega_{\alpha,\beta}^{(3)}(\bm\gamma,\bm\lambda)}D_{\ell}(R,\gamma,\lambda')\omega_{\alpha,\beta'}^{(3)}(\bm\gamma,\bm\lambda'),
\end{align}
and the convolution kernel
\begin{align}
    \mathcal L_{\beta\beta'}^{(3)}(\bm R',\bm R;\bm\lambda',\bm\lambda)=e^{-\frac{|R|^2+|R'|^2}{2}}\prod_{i=1}^{3}\overline{\psi_{\bm\lambda_i,\beta_i}(\bar R)}\psi_{\bm\lambda_i',\beta_i'}(\bar R').
\end{align}
The fully exchanged contribution can then be written as
\begin{align}
    C_{\ell}^{(3)}(\bm R',\bm R)\simeq M!\sum_{\beta,\beta'}\int\prod_{i=1}^{3}d^2\bm\lambda_i\,d^2\bm\lambda_i'\,\rho_{\ell;\beta'\beta}^{(3)}(\bm R',\bm R;\bm\lambda',\bm\lambda)\mathcal L_{\beta\beta'}^{(3)}(\bm R',\bm R;\bm\lambda',\bm\lambda).
\end{align}
Since $\psi_{\bm\lambda,\beta}(\bar R)$ is constructed from a Fourier transform over a narrow momentum window $|\bm\kappa|\lesssim\Lambda$ around the valley minimum $\bm K_\beta$, it is localized around $\bm\lambda\simeq\bm R$ with a spatial width of order $\Lambda^{-1}$. The convolution kernel is therefore appreciable only when $\bm\lambda_i\simeq\bm R$ and $\bm\lambda_i'\simeq\bm R'$. Hence, in the limit $|\bm R-\bm R'|\to\infty$, the convolution probes the long-distance regime of the dressed anyon 3RDM in which the two triplets $\bm\lambda$ and $\bm\lambda'$ are widely separated.

\textbf{Plasma analogy.} We now use the plasma analogy~\cite{Girvin1990, Seo2026} to analyze the dressed anyon 3RDM $\rho_{\ell;\beta'\beta}^{(3)}(\bm R',\bm R;\bm\lambda',\bm\lambda)$. In the dilute long-distance regime, the modified ground-state envelope wavefunction $\omega_{\alpha}^{(0)}$ takes the form
\begin{align}
    \omega_{\alpha}^{(0)}(\bm\gamma)\simeq\epsilon(\alpha)\frac{\prod_{\nu=1}^{3}\Delta_\nu(\gamma)}{\Delta(\gamma)^{1/3}}=\epsilon(\alpha)\prod_{a<b}(\gamma_a-\gamma_b)^{\mathsf B_{\alpha_a\alpha_b}},
\end{align}
where the coupling matrix is
\begin{align}
    \mathsf B=\begin{pmatrix}
    \frac{2}{3} & -\frac{1}{3} & -\frac{1}{3} \\
    -\frac{1}{3} & \frac{2}{3} & -\frac{1}{3} \\
    -\frac{1}{3} & -\frac{1}{3} & \frac{2}{3}
    \end{pmatrix}.
\end{align}
The normalization $\mathcal Z_\Psi$ can then be written as
\begin{align}
    \mathcal Z_\Psi\simeq \sum_{\alpha}\int\prod_{a=1}^{M}d^2\bm\gamma_a\,|\omega^{(0)}_{\alpha}(\bm\gamma)|^2=\mathcal N_\Psi\int\prod_{\nu=1}^{3}\prod_{i=1}^{M_{\rm v}}d^2\bm\gamma_i^{(\nu)}\,e^{-H_{\rm pl}(\bm\gamma)},
\end{align}
where $\mathcal N_\Psi=M!/(M_{\rm v}!)^3$. The plasma Hamiltonian is
\begin{align}
    H_{\rm pl}(\bm\gamma)&=-\frac{4}{3}\sum_{\nu=1}^{3}\sum_{i<j}^{M_{\rm v}}\ln|\gamma_i^{(\nu)}-\gamma_j^{(\nu)}|+\frac{2}{3}\sum_{\nu<\mu}\sum_{i=1}^{M_{\rm v}}\sum_{j=1}^{M_{\rm v}}\ln|\gamma_i^{(\nu)}-\gamma_j^{(\mu)}|=\int d^2\bm\gamma\,d^2\bm\gamma'\,\rho^T(\bm\gamma)\mathsf B \rho(\bm\gamma')\,v(\bm\gamma-\bm\gamma'),
\end{align}
where $\rho_\nu(\bm\gamma)=\sum_{i=1}^{M_{\rm v}}\delta^{(2)}(\bm\gamma-\bm\gamma_i^{(\nu)})$, $\rho(\bm\gamma)=(\rho_1(\bm\gamma),\rho_2(\bm\gamma),\rho_3(\bm\gamma))^T$, and $v(\bm\gamma-\bm\gamma')=-\ln|\gamma-\gamma'|$. To diagonalize $\mathsf B$, we define one charge channel $\rho_c$ and two relative channels $\rho_{r_1}$ and $\rho_{r_2}$,
\begin{align*}
    \rho_c=\rho_1+\rho_2+\rho_3,\qquad \rho_{r_1}=\rho_1-\rho_2,\qquad \rho_{r_2}=\frac{\rho_1+\rho_2-2\rho_3}{\sqrt{3}}.
\end{align*}
The charge channel $\rho_c$ corresponds to the null eigenvector $(1,1,1)^T$ of $\mathsf B$ and therefore does not contribute to the plasma Hamiltonian.

We now turn to the numerator of the dressed anyon 3RDM. For the valley-balanced sector $\beta,\beta'\in S_3$, with one particle in each valley, we obtain
\begin{align}
    &\mathcal Z_{\ell;\beta'\beta}^{(3)}(\bm R',\bm R;\bm\lambda',\bm\lambda)=\frac{M!}{(M-3)!}\sum_{\alpha}\int\prod_{a=1}^{M-3}d^2\bm\gamma_a\,\overline{D_{\ell}(R',\gamma,\lambda)\omega_{\alpha,\beta}^{(3)}(\bm\gamma,\bm\lambda)}D_{\ell}(R,\gamma,\lambda')\omega_{\alpha,\beta'}^{(3)}(\bm\gamma,\bm\lambda')\notag\\
    &\simeq \operatorname{sgn}(\beta')\operatorname{sgn}(\beta)\frac{M!}{[(M_{\rm v}-1)!]^3}\,\overline{\mathcal J(\bm\lambda)}\mathcal J(\bm\lambda')\int\prod_{\nu=1}^{3}\prod_{i=1}^{M_{\rm v}-1}d^2\bm\gamma_i^{(\nu)}\,e^{-H_{\rm pl}(\bm\gamma)-\overline{\Delta H_{\beta}(\bm\gamma;\bm\lambda)}-\Delta H_{\beta'}(\bm\gamma;\bm\lambda')}\,\overline{D_{\ell}(R',\gamma,\lambda)}D_{\ell}(R,\gamma,\lambda'),
\end{align}
where $\Delta H_{\beta}(\bm\gamma;\bm\lambda)$ is the complex interaction between the impurity triplet and the background plasma,
\begin{align}
    \Delta H_{\beta}(\bm\gamma;\bm\lambda)=-\sum_{\nu=1}^{3}\sum_{a=1}^{M_{\rm v}-1}\sum_{i=1}^{3}\mathsf B_{\nu\beta_i}\ln(\gamma_a^{(\nu)}-\lambda_i).
\end{align}
The factor $\mathcal J(\bm\lambda)$ is the internal factor of the impurity triplet,
\begin{align}
    \mathcal J(\bm\lambda)=\frac{\Delta(\lambda)}{|\Delta(\lambda)|_{\rm reg}^{8/3}},
\end{align}
where $|\Delta(\lambda)|_{\rm reg}^{8/3}$ denotes the short-distance regularization from the ground-state envelope wavefunction in Eq.~\eqref{ground-state-envelope-wavefunction}.

In the plasma analogy, $e^{-\overline{\Delta H_{\beta}(\bm\gamma;\bm\lambda)}}D_{\ell}(R,\gamma,\lambda')$ corresponds to inserting a dressed triplet of impurities at $\bm\lambda$. Since the convolution kernel localizes $\bm\lambda$ near $\bm R$ and $\bm\lambda'$ near $\bm R'$, the terms in $D_{\ell}(R,\gamma,\lambda')$ involving $R-\lambda_i'$ are suppressed in the limit $|\bm R-\bm R'|\to\infty$. We therefore approximate
\begin{align}
    D_{\ell}(R',\gamma,\lambda)\simeq D_{\ell}(R',\gamma),\qquad D_{\ell}(R,\gamma,\lambda')\simeq D_{\ell}(R,\gamma),
\end{align}
and define the dressed triplet insertion
\begin{align}
    \mathcal O_{\ell,\beta}(\bm R,\bm\lambda)=e^{-\overline{\Delta H_{\beta}(\bm\gamma;\bm\lambda)}}D_{\ell}(R,\gamma).
\end{align}
The numerator corresponds to inserting two dressed triplets, at $\bm\lambda$ and $\bm\lambda'$, respectively. In the sector considered here, each triplet contains one particle from each valley, corresponding to the null vector $(1,1,1)^T$ of $\mathsf B$. Therefore, each triplet is neutral with respect to both the interacting plasma channels and the statistical gauge field, allowing long-range correlations to survive. Since the plasma is in a screening phase, the connected correlation between two well-separated neutral dressed triplets vanishes at long distances, and the full correlation factorizes into the product of their one-point functions.

We denote the $M$-particle plasma partition function by
\begin{align}
    Z_{\rm pl}^{(M)}=\int\prod_{\nu=1}^{3}\prod_{i=1}^{M_{\rm v}}d^2\bm\gamma_i^{(\nu)}\,e^{-H_{\rm pl}(\bm\gamma)}.
\end{align}
Then the dressed anyon 3RDM can be written as
\begin{align}
    \rho_{\ell;\beta'\beta}^{(3)}(\bm R',\bm R;\bm\lambda',\bm\lambda)\simeq\operatorname{sgn}(\beta')\operatorname{sgn}(\beta)A_M\,\overline{\mathcal J(\bm\lambda)}\mathcal J(\bm\lambda')\left\langle\overline{\mathcal O_{\ell,\beta'}(\bm R',\bm\lambda')}\mathcal O_{\ell,\beta}(\bm R,\bm\lambda)\right\rangle_{{\rm pl},M-3},
\end{align}
where $A_M=M_{\rm v}^3\frac{Z_{\rm pl}^{(M-3)}}{Z_{\rm pl}^{(M)}}$ is an $O(1)$ quantity determined by the particle density in each valley, and $\langle O\rangle_{{\rm pl},M-3}=\frac{1}{Z_{\rm pl}^{(M-3)}}\int\prod_{\nu=1}^{3}\prod_{i=1}^{M_{\rm v}-1}d^2\bm\gamma_i^{(\nu)}\,e^{-H_{\rm pl}(\bm\gamma)}O$.

The long-distance 3RDM takes the factorized form
\begin{align}
    \rho_{\ell;\beta'\beta}^{(3)}(\bm R',\bm R;\bm\lambda',\bm\lambda)\xrightarrow{|\bm R-\bm R'|\to\infty}\overline{\Phi_{\ell,\beta'}(\bm R',\bm\lambda')}\Phi_{\ell,\beta}(\bm R,\bm\lambda),
\end{align}
where
\begin{align}
    \Phi_{\ell,\beta}(\bm R,\bm\lambda)=\operatorname{sgn}(\beta)\sqrt{A_M}\,\overline{\mathcal J(\bm\lambda)}\left\langle\mathcal O_{\ell,\beta}(\bm R,\bm\lambda)\right\rangle_{{\rm pl},M-3}.
\end{align}
Using the localization $\bm\lambda_i\simeq\bm R$, we write $\bm\lambda_i=\bm R+\bm u_i$. We then obtain the corresponding projected Cooper-pair order parameter
\begin{align}
    \tilde{\Delta}_{\ell}(\bm R)&=\sqrt{M!}\,e^{-\frac{|R|^2}{2}}\sum_{\beta}\int\prod_{i=1}^{3}d^2\bm u_i\,\Phi_{\ell,\beta}(\bm R,\bm R+\bm u)\prod_{i=1}^{3}\overline{\psi_{\bm R+\bm u_i,\beta_i}(\bar R)}\notag\\
    &=\sqrt{M!A_M}\,e^{-\frac{|R|^2}{2}}\sum_{\beta}\operatorname{sgn}(\beta)\int\prod_{i=1}^{3}d^2\bm u_i\,\overline{\mathcal J(\bm R+\bm u)}\left\langle\mathcal O_{\ell,\beta}(\bm R,\bm R+\bm u)\right\rangle_{{\rm pl},M-3}\prod_{i=1}^{3}\overline{\psi_{\bm R+\bm u_i,\beta_i}(\bar R)}.
    \label{tilde-Delta-definition}
\end{align}
Therefore,
\begin{align}
    C_{\ell}^{(3)}(\bm R',\bm R)\xrightarrow{|\bm R-\bm R'|\to\infty}\overline{\tilde{\Delta}_{\ell}(\bm R')}\tilde{\Delta}_{\ell}(\bm R).
\end{align}

\subsubsection{The unexchanged and partially exchanged sectors}

Having studied the fully exchanged sector with $k=3$, we now consider the remaining sectors with $k=0,1,2$. We define the dressed anyon $k$RDM by
\begin{align}
    \rho_{\ell;\beta'\beta}^{(k)}(\bm R',\bm R;\bm\lambda',\bm\lambda)=\frac{1}{\mathcal Z_\Psi}\frac{M!}{(M-k)!}\sum_{\alpha}\int\prod_{a=1}^{M-k}d^2\bm\gamma_a\,\overline{D_{\ell}(R',\gamma,\lambda)\omega_{\alpha,\beta}^{(k)}(\bm\gamma,\bm\lambda)}D_{\ell}(R,\gamma,\lambda')\omega_{\alpha,\beta'}^{(k)}(\bm\gamma,\bm\lambda'),
\end{align}
the convolution kernel by
\begin{align}
    \mathcal L_{\beta\beta'}^{(k)}(\bm R',\bm R;\bm\lambda',\bm\lambda)=e^{-\frac{k}{6}(|R|^2+|R'|^2)}\prod_{i=1}^{k}\overline{\psi_{\bm\lambda_i,\beta_i}(\bar R)}\psi_{\bm\lambda_i',\beta_i'}(\bar R'),
\end{align}
and the combinatorial factor $\mathcal C_k=\binom{3}{k}^2(3-k)!M!$.

For $k=0,1,2,3$, the corresponding contribution can be written as
\begin{align}
    C_{\ell}^{(k)}(\bm R',\bm R)\simeq\mathcal C_k\,e^{-\frac{3-k}{6}(|R|^2+|R'|^2)+\frac{3-k}{3}\bar R'R}\sum_{\beta,\beta'}\int\prod_{i=1}^{k}d^2\bm\lambda_i\,d^2\bm\lambda_i'\,\rho_{\ell;\beta'\beta}^{(k)}(\bm R',\bm R;\bm\lambda',\bm\lambda)\mathcal L_{\beta\beta'}^{(k)}(\bm R',\bm R;\bm\lambda',\bm\lambda).
\end{align}
For $k<3$, the prefactor $\left|e^{-\frac{3-k}{6}(|R|^2+|R'|^2)+\frac{3-k}{3}\bar R'R}\right|=e^{-\frac{3-k}{6}|R-R'|^2}$ is Gaussian-suppressed at large separation. Moreover, for $k=1,2$, the plasma analogy shows that the corresponding anyon $k$RDM cannot develop off-diagonal long-range order because the impurity cluster is not neutral with respect to either the interacting plasma channels or the statistical gauge field. Therefore, $C_{\ell}^{(k)}(\bm R',\bm R)$ decays at long distances for $k=0,1,2$.

\subsubsection{Order parameter from the ODLRO}

The analysis above shows that only the fully exchanged sector of the pair correlation function $C_{\ell}(\bm R',\bm R)$ contributes to the long-distance pair correlation. At long distances, we obtain
\begin{align}
    C_{\ell}(\bm R',\bm R)\xrightarrow{|\bm R-\bm R'|\to\infty}\overline{\widetilde{\Delta}_{\ell}(\bm R')}\widetilde{\Delta}_{\ell}(\bm R),
\end{align}
where $\widetilde{\Delta}_{\ell}(\bm R)$ is defined in Eq.~\eqref{tilde-Delta-definition}.

We expect $\widetilde{\Delta}_{\ell}(\bm R)$ to satisfy the same symmetry and holomorphic structure as $\Delta_{\ell}(\bm R)$. A direct calculation shows that $\widetilde{\Delta}_{\ell}(\bm R)$ indeed obeys the same translation and rotation symmetries as $\Delta_{\ell}(\bm R)$ in Eqs.~\eqref{order-parameter-translation} and \eqref{order-parameter-rotation}, and has the same holomorphic structure as in Eq.~\eqref{order-parameter-holomorphic}.

\subsection{Matching the field theory description} \label{matching-the-field-theory-description}

Here we discuss the field theory description to the anyon SC and establish the correspondence between the wavefunction and the field theory operators.

We start from the standard description to Laughlin state
\begin{align}
    {\cal L} = - \frac{3}{4\pi}\alpha\mathrm{d}\alpha + \frac{1}{2\pi} \alpha \mathrm{d} (A+\frac{3}{2}\omega) + {\cal L} [\phi; -\alpha]
\end{align}
Here $\phi$ is a bosonic field sourcing the doped quasiholes with charge $-1/3$ (electron charge is $1$ in our units), $A$ is the electromagnetic probe field, and $\omega$ is the $SO(2)$ spin connection of the underlying manifold, or its crystalline analogue, which tracks orbital angular-momentum response. The mixed term $\alpha \mathrm{d}\omega$ is the Wen-Zee coupling: its coefficient fixes the orbital spin of the anyons~\cite{WenZeeShift,PhysRevB.90.115139}. The physical electric charge density (per crystalline unit cell) is 
\begin{align}
    \rho_e = \frac{\langle\nabla\times \vec{ \alpha}\rangle}{2\pi} = \frac{1}{3} - \delta ,
\end{align}
and the doping from the $1/3$ filling is induced by anyons
\begin{align}
    \rho_\phi = 3\delta . 
\end{align}

In this work we consider the parameter regime where the $-1/3$ quasiholes tend to form $-2/3$ anyon bound states with angular momentum $l=2$. This means that the low-energy description should involve a composite field $\Phi \sim \phi^2$:
\begin{align}
    {\cal L} = - \frac{3}{4\pi}\alpha\mathrm{d}\alpha + \frac{1}{2\pi} \alpha \mathrm{d} (A+\frac{3}{2}\omega) + {\cal L} [\Phi; -2\alpha + l\omega] , 
\end{align}
where the density  $\rho_\Phi = 3\delta/2$. The effective flux it sees is $-\frac{2}{3} + 2\delta$, which creates three low-energy valleys $v=0,1,2$ within each magnetic Brillouin zone. 

The rotation symmetry action on the valley degrees of freedom is worth commenting before moving forward. For $-2/3 = 1/3$ (mod $1$) flux per unit cell, the magnetic translations satisfy $T_1T_2=\omega T_2T_1$, with $\omega=e^{2\pi i/3}$, and the rotation $C_k$ symmetry further enriches the algebra by acting projectively. The complete projective space group algebra can admit multiple irreducible representations with the three-dimensional valley multiplet as a basis.  Upon diagonalizing the unitary matrix representing $C_k$, which we call $U_{C_k}$, its eigenvalues $e^{2\pi i\ell_va/k}$ define the angular-momentum channels $\ell_va$ of the valley multiplet. In particular we can always choose a basis in which $C_k$ act diagonally for the following discussion. 


Then, the valley-dependent angular momenta can be effectively captured by
\begin{align}
    {\cal L} = - \frac{3}{4\pi}\alpha\mathrm{d}\alpha + \frac{1}{2\pi} \alpha \mathrm{d} (A+\frac{3}{2}\omega) + \sum_v {\cal L} [\Phi_v; -2\alpha + (l+\ell_v)\omega] .
\end{align}

Our wavefunction corresponds to a parton decomposition to the $\Phi$ field:
\begin{align}
    \Phi_v = f_v\cdot g
\end{align}
where $f$ and $g$ are both fermions with equal density $\rho_f=\rho_g = \rho_\Phi$, and the field theory reduces to 
\begin{align}\label{eq:fg}
    {\cal L}[\Phi; -2\alpha + l\omega] \rightarrow {\cal L}[f_v; - a -\alpha + \ell_v \omega ] +{\cal L}[g; a -\alpha + l\omega].
\end{align}
This assignment means that all the valley dependences of $\Phi_v$ are inherited by $f_v$, and can be achieved by assigning internal mean-field flux 
\begin{align}
    \frac{\langle \nabla\times \vec{a}\rangle}{2\pi} = \frac{1}{3} +\frac{\delta }{2} , 
\end{align}
which indicates that $f$ particles see $- \frac{2}{3} + \frac{\delta}{2}$ flux density whereas $g$ particles see $3\delta /2$ flux density. This means we can put $g$ in a $\nu=1$ IQH state with no valley structure:
\begin{align}
    {\cal L}[g; a -\alpha + l\omega] \rightarrow& -\frac{1}{4\pi} \beta\mathrm{d}\beta + \frac{1}{2\pi}\beta\mathrm{d}(a -\alpha + l\omega + \frac{1}{2}\omega)\nonumber\\
    \rightarrow &\frac{1}{4\pi} (a -\alpha) \mathrm{d} (a-\alpha) + \frac{l+ \frac{1}{2}}{2\pi} (a -\alpha ) \mathrm{d} \omega  +\dots .
\end{align}
On the other hand, since $f$ sees a flux density commensurate with the lattice, $-2/3$, there is an emergent valley degree of freedom in each reduced magnetic Brillouin zone. Evenly distributing the $f$ particles into those valleys, we find that the natural state to consider is to put $f$ in the lowest Landau level produced by the small incommensurate part of the statistical gauge field, $\delta/2$. More rigorously, this is a lattice Hofstadter state. The corresponding field theory description is:
\begin{align}
     {\cal L}[f_v; - a -\alpha +\ell_v\omega]  \rightarrow & \sum_v -\frac{1}{4\pi} \beta^{(v)} \mathrm{d} \beta^{(v)}  + \frac{1}{2\pi} \beta^{(v)}  \mathrm{d} [-a -\alpha +(\frac{1}{2}+\ell_v)\omega]\nonumber\\
     \rightarrow  &\frac{3}{4\pi} (a +\alpha)\mathrm{d} (a +\alpha) - \frac{\frac{3}{2}+\sum_v \ell_v}{2\pi}  (a +\alpha) \mathrm{d} \omega +\dots
\end{align}

Assembling terms together, we reach an effective field theory description:
\begin{align}
    {\cal L} = \frac{1}{4\pi} (2a +\alpha) \mathrm{d} (2a+\alpha) + \frac{l-1-\sum_v \ell_v}{2\pi} a\mathrm{d} \omega  + \frac{1}{2\pi} \alpha \mathrm{d} [A-(l+ \frac{1}{2}+\sum_v \ell_v)\omega] +\dots . 
\end{align}
Shifting $\alpha\rightarrow \alpha - 2a$, we find 
\begin{align}
    {\cal L} = \frac{1}{4\pi} \alpha \mathrm{d} \alpha + \frac{1}{2\pi} a\mathrm{d} [-2A + (3l+\sum_v \ell_v)\omega ]   + \frac{1}{2\pi} \alpha \mathrm{d} [A-(l+ \frac{1}{2}+\sum_v \ell_v)\omega] +\dots . 
\end{align}
Since $a$ doesn't have self Chern-Simons field, its dynamics is governed by Maxwell theory and effectively describes a superconductor. From the mutual coupling coefficient, we read that the $2\pi$ monopole creation operator of $a$, ${\cal M}_a$ is the Cooper-pair annihilation operator, which carries charge $-2$ and angular momentum $3l+\sum_v \ell_v$. This means that the SC order is in the angular momentum $L=-3l-\sum_v \ell_v$ channel. (We note that in this section the definition of angular momentum and electric charge differs from the other sections by a minus sign.)


With this understanding, we are now ready to interpret how adding a Cooper pair modifies the wavefunction. We note that the ground-state wavefunction was written exactly according to the parton interpretation, where the $-2/3$ anyon wavefunction could be viewed as a product of an overall Vandermonde multiplying a Hofstadter wavefunction. Tracking back from the final field theory, we note that at the level of Eq.~\ref{eq:fg}, adding a cooper pair locally is equivalent to adding $2\pi$ flux to $a$ and $-4\pi$ to $\alpha$, which further means that $f$ locally sees one extra flux, whereas $g$ locally sees three extra fluxes. These extra fluxes creates vortices in the corresponding wavefunctions, exactly agree with form of the wavefunction with a cooper pair. 

\section{Scale-invariant fixed points of a dilute three-valley \texorpdfstring{$2e/3$}{2e/3} anyon gas}
\label{sec:sm_fixed_points}
In this section, we consider the dilute gas of three-valley $2e/3$ anyons directly as an infrared theory. This description is appropriate when the typical interparticle separation is much larger than the pair size $l_{\rm pair}$, so that the short-distance structure of quasihole scattering can be integrated out and incorporated into short-ranged terms in the effective theory. The anyon statistics is encoded by the usual Chern--Simons statistical interaction. Throughout this section, we work in the bosonic statistical gauge, in which each particle carries $2\pi p$, with $p=4/3$, statistical flux. The long-distance
Hamiltonian, therefore, has the form
\begin{equation}
H_{\rm IR}
=
\frac{1}{2m^*}\sum_{a} {\boldsymbol \Pi}_{a}^2+H_{\rm ct},
\qquad
{\boldsymbol \Pi}_{a}
=
-i{\boldsymbol \nabla}_{a}
-p\sum_{b\neq a}
\frac{\hat{z}\times \mathbf r_{ab}}{r_{ab}^2},\qquad p=\frac43.
\label{eq:HIR_AB}
\end{equation}
In complex notation, $\mathbf r_a\leftrightarrow\eta_a$, the combination $\Pi_{a,+}\equiv\Pi_{a,x}+i\Pi_{a,y}$ equals $-i\,\Pi^{(\rm IR)}_a$ introduced in the main text. Below, $\Psi$ denotes the flat-weight wave function $\tilde w$ of the main text.
Here $H_{\rm ct}$ denotes short-ranged operators associated with particle collisions, including the Pauli contact transferred from the factorized kinetic term and the residual $V_{\rm short}$ of the main text. These terms can depend on the valley channel and, in a microscopic description, are determined by the corresponding few-body problem. In the strict point-particle limit, their effect is encoded in the boundary conditions imposed on the Aharonov--Bohm Hamiltonian (the first term in Eq.~\eqref{eq:HIR_AB}) at particle coincidence. We will be interested in the scale-invariant choices of this UV boundary data, which define the zero-density fixed points of the dilute anyon gas. This problem has appeared in the literature before \cite{Bergman1994,Giovanni1995,Ghosh1996,Kim1997,Nishida2008,Turner2020,boisvert2025revisiting}, and here we review the aspects relevant to the present three-valley case.

We first analyze the two-particle problem and classify the scale-invariant boundary conditions in the two independent collision channels. This recovers the four fixed points $(\sigma_2,\sigma_1)$ of the main text, and we give explicit regularized Hamiltonians that realize each of them (Sec.~\ref{sec:sm_two_body}). We then turn to the higher-body sectors, which are not fixed by the two-body data (Sec.~\ref{sec:sm_higher_body}). After setting up the hyperspherical form of the $M$-body problem, we solve the angular three-body problem at the fixed point $(+,-)$ and show that it contains exactly one channel in which an additional three-body boundary condition can be imposed (Sec.~\ref{sec:three_body_contact}). This is the statement used in the main text to identify the marginal valley-singlet coupling $g_3$.

\subsection{Two-particle boundary conditions}
\label{sec:sm_two_body}
 
We begin with the two-particle problem. In the bosonic statistical gauge,
exchange symmetry requires a symmetric valley wave function to be
accompanied by even relative angular momentum $\ell$, while an
antisymmetric valley wave function must be accompanied by odd $\ell$.
Away from particle coincidence, the relative Hamiltonian is simply the
Aharonov--Bohm Hamiltonian, 
\begin{equation}
H_{\rm rel}
=
\frac{1}{m^*}
\left[
-\partial_r^2-\frac{1}{r}\partial_r
+\frac{(\ell-p)^2}{r^2}
\right].
\label{eq:AB_relative}
\end{equation}
A nontrivial short-distance boundary condition is possible whenever
$\nu_\ell\equiv|\ell-p|<1$, since in this case both independent solutions are
locally normalizable at the origin. For $p=4/3$, there is one such
partial wave in each valley channel,
\begin{equation}
\text{symmetric}:\quad \ell=2,\qquad \nu_{2}=\left|2-\frac{4}{3}\right|=\frac{2}{3},
\qquad
\text{antisymmetric}:\quad \ell=1,\qquad \nu_{1}=\left|1-\frac{4}{3}\right|=\frac{1}{3}.
\label{eq:sa_channels}
\end{equation}
All other partial waves have a unique
short-distance boundary condition. At zero energy, the two solutions in the two nontrivial channels are
therefore
\begin{equation}
R_{\ell =2}(r)
=
A_{2} r^{-2/3}+B_{2} r^{2/3},
\qquad
R_{\ell =1}(r)
=
A_{1} r^{-1/3}+B_{1} r^{1/3}.
\label{eq:two_body_solutions}
\end{equation}
For a generic microscopic core, the ratios $A_2/B_2$ and $A_1/B_1$
are determined by matching to the short-distance problem. A finite non-zero value of either ratio introduces a length scale, the detuning length $r_\ell$, and breaks classical scale invariance. Scale invariance is recovered at the two
endpoints, where only one of the two power laws is present. We label these
boundary conditions by $R_\ell(r)\sim r^{\sigma_\ell\nu_\ell}$ with $\sigma_\ell=\pm1$ for $\ell=1,2$. Since the
two collision channels are independent, the four scale-invariant
two-particle fixed points $(\sigma_2,\sigma_1)=(+,+),\;(+,-),\;(-,+),\;(-,-)$, where $\sigma_\ell=+$ denotes the regular and $\sigma_\ell=-$ the singular branch, as in the main text. Here and in the main text the short-range couplings are taken to be valley independent, so that each exchange channel carries a single ratio $A_\ell/B_\ell$. For a lower valley symmetry, the six valley-symmetric and three valley-antisymmetric pair states would split into several irreducible components with independent boundary conditions, and the number of fixed points would be larger. The valley-blind $(+,-)$ Hamiltonian remains a possible reference point, but lower valley symmetry can require more independent pair tunings to reach it.

The $(+,-)$ fixed point has a particularly simple realization as the Aharonov--Casher-Chern-Simons Hamiltonian (the contact term is just an attractive Pauli term with $g=2$)
\begin{equation}
H_{\rm ACCS}
=\frac{1}{2m^*}\sum_{a}\Big[ {\boldsymbol \Pi}_{a}^2 - \pi p\, g\sum_{b\neq a} \delta(\mathbf r_{a}-\mathbf r_{b})\Big],\quad g=2,\label{eq:H_AC}
\end{equation}
which for $g=2$ is $\frac{1}{2m^*}\sum_a(\Pi^{(\rm IR)}_a)^\dagger\Pi^{(\rm IR)}_a$. It should be understood as the zero-radius limit
of a regularized statistical flux tube problem, with both the vector potential
and the associated magnetic field smeared over a core of radius
$\epsilon\sim l_{\rm pair}$. Specifically, we may write
\begin{equation}
a_{\theta,\epsilon}(r)=\frac{p}{r}f_\epsilon(r),
\qquad
b_\epsilon(r)=\frac{p}{r}f_\epsilon'(r).
\end{equation}
where $f_\epsilon(0)=0$ and $f_\epsilon(r)\rightarrow1$ for
$r\gg\epsilon$. The radial zero-energy equation becomes
\begin{equation}
\left[
-\partial_r^2-\frac{1}{r}\partial_r
+\frac{\left[\ell-p f_\epsilon(r)\right]^2}{r^2}
-\frac{p}{r}f_\epsilon'(r)
\right]R_{\ell,\epsilon}(r)=0.\label{eq:radial}
\end{equation}
For $\ell\geq0$, the regular zero-energy solution is
\begin{equation}
R_{\ell,\epsilon}(r)
\propto
r^\ell
\exp\left[
-p\int_0^r\frac{dr'}{r'}f_\epsilon(r')
\right].
\end{equation}
It is regular inside the core, while for $r\gg\epsilon$ it approaches
$R_\ell(r)\sim r^{\ell-p}$. Hence, in the two nontrivial channels,
\begin{equation}
R_{\ell=2}(r)\sim r^{2/3},
\qquad
R_{\ell=1}(r)\sim r^{-1/3},
\end{equation}
showing that the $g=2$ Aharonov--Casher Hamiltonian realizes the
$\mathrm{AC}_{(+,-)}$ fixed point in a particularly simple way. The value $g=2$ is special in that the Hamiltonian factorizes, $H_{\rm ACCS}=\frac{1}{2m^*}\sum_a\Pi_{a,+}^\dagger\Pi_{a,+}$, so that its zero-energy modes are determined by the first-order equations $\Pi_{a,+}\psi=0$. The fixed point itself does not require this. More generally, one could regularize the contact term in Eq.~\eqref{eq:H_AC} differently compared to the vector potential, which amounts to replacing the last term in Eq.~\eqref{eq:radial} by $m^*V_\epsilon(r)$ for an independent short-range interaction, e.g. $V_\epsilon(r)
=
-\frac{1}{m^*\epsilon^2}
v\left(\frac{r}{\epsilon}\right)$, 
whose profile $v(x)\geq 0$ need not be related to the flux profile
$f_\epsilon(r)$. For any short-ranged profile, we can vary the magnitude until we cross a bound state formation threshold in the
$\ell=1$ channel. At this critical point one has the asymptotic form $R_{\ell=1,\epsilon}(r)\sim r^{-1/3},
\quad r\gg\epsilon$, while the $\ell=2$ channel is generically
nonresonant and therefore approaches the regular boundary condition
$R_2(r)\sim r^{2/3}$. We also note that the $\ell=0$ valley-symmetric channel may, in principle, bind before the $\ell=1$ resonance is reached. Such a state is not associated with an additional point-particle boundary
condition, since the corresponding exterior behavior
$r^{-4/3}$ is not locally normalizable at the origin, and becomes a
short-distance core state as $\epsilon\rightarrow0$. A particularly
clean generic realization of the $(+,-)$ fixed point is therefore
obtained by retaining only the
valley-antisymmetric component of $V_\epsilon(r)$.

For a monotonic flux profile, any fixed $g<2$ in
Eq.~\eqref{eq:H_AC} generically gives regular boundary conditions for both $\ell=1$ and $\ell =2$. The reason is positivity: for monotonic $f_\epsilon$ one has $b_\epsilon\geq0$, so that $H_g=H_{\rm AC}+\frac{2-g}{4m^*}\sum_{a\neq b}b_\epsilon(r_{ab})\geq H_{\rm AC}\geq0$ for $g\leq2$, no channel can bind, and the $\ell=1$ channel sits on the nonresonant side of its threshold.
This is most easily illustrated for $g=-2$, which is the operator $\frac{1}{2m^*}\sum_a\Pi_{a,+}\Pi^\dagger_{a,+}$ used for the regular fixed point of the main text, in which case the last term in the zero-energy radial equation \eqref{eq:radial} just changes sign, and, for the two channels $\ell=1,2$, the solution normalizable at the origin is
\begin{equation}
R_{\ell,\epsilon}(r)
\propto
\widetilde R_{\ell,\epsilon}(r)
\int_0^r
\frac{dr'}
{r'\,\widetilde R_{\ell,\epsilon}^2(r')} ,\quad \widetilde R_{\ell,\epsilon}(r)
=
r^{-\ell}
\exp\left[
p\int_0^r\frac{dr'}{r'}f_\epsilon(r')
\right]\;.
\label{eq:gminus2_solution}
\end{equation}
For $r\ll\epsilon$ this behaves as $R_{\ell,\epsilon}\sim r^\ell$.
Outside the core, the solution
generically contains both power laws. However, for
$g=-2$, the core is nonresonant, and the coefficient of the singular
branch vanishes relative to the regular one as $\epsilon\rightarrow0$.
Consequently, Eq.~\eqref{eq:gminus2_solution} gives $R_{\ell=2}(r)\sim r^{2/3},
\quad
R_{\ell=1}(r)\sim r^{1/3}$. Thus, $g=-2$ provides a simple, explicit realization of the fully
regular fixed point $(+,+)$. More generally, any fixed $g<2$ in this family is detuned from the Aharonov--Casher
resonance and flows to the same regular boundary condition as
$\epsilon\rightarrow0$.

The remaining two fixed points can be understood in the same way.
To realize the $(-,+)$ fixed point, it is sufficient to introduce an attractive interaction in the symmetric valley channel,
\begin{equation}
V_{s,\epsilon}(r)
=
-\frac{1}{m^*\epsilon^2}
v_s\left(\frac{r}{\epsilon}\right)P_s ,
\qquad v_s(x)\geq0 ,
\end{equation}
where $P_s$ is the projector onto the valley-symmetric sector (in the pseudopotential language of the main text, this is an attractive $u_2$ tuned to its binding threshold). As the magnitude of $v_s$ is increased, a bound state eventually forms in the
$\ell=2$ channel. At the critical point, the exterior wave
function approaches $R_{\ell=2}(r)\sim r^{-2/3}$.
The antisymmetric channel, generically being away from a resonance, instead flows
to its regular boundary condition, $R_{\ell=1}(r)\sim r^{1/3}$. Thus, tuning a single short-range coupling to the $\ell=2$ binding
threshold realizes the $(-,+)$ fixed point. The projector $P_s$ is useful here because, for a valley-independent
central attraction, the $\ell=1$ channel has a lower effective
centrifugal barrier than the $\ell=2$ channel and therefore reaches a
binding threshold first. Indeed, 
$H_{\rm rel,\ell=2}-H_{\rm rel,\ell=1}=[3-2p f_\epsilon]/(m^*r^2)\geq1/(3m^*r^2)>0$ for $0\leq f_\epsilon\leq1$. Thus, projecting the attraction onto the symmetric
sector avoids the unwanted $\ell=1$ bound state. No analogous ordering with the $\ell=0$ threshold follows, since $H_{\rm rel,\ell=2}-H_{\rm rel,\ell=0}=4[1-p f_\epsilon]/(m^*r^2)$ changes sign across the core. Its location must be checked for the chosen potential. Finally, the $(-,-)$ fixed point is obtained by independently tuning
both collision channels to their respective zero-energy binding
thresholds, giving
$R_{\ell=2}\sim r^{-2/3}$ and
$R_{\ell=1}\sim r^{-1/3}$.
 
Although no special relation
between the short-range interaction profile and the regularization of the
statistical vector potential is required, one can still construct a particularly simple, exactly solvable representative of the $\ell=2$ resonance. In the symmetric
$\ell=2$ channel, choose
\begin{equation}
V_{s,\epsilon}(r)
=
\frac{1}{m^*}\Big[(p-4)\frac{f_\epsilon'(r)}{r}
+
8(p-2)
\frac{f_\epsilon(r)\left[1-f_\epsilon(r)\right]}{r^2}\Big].
\label{eq:factorized_Vs}
\end{equation}
With this choice, the radial $\ell=2$ operator factorizes as
\begin{equation}
-\partial_r^2-\frac{1}{r}\partial_r
+\frac{\left[2-p f_\epsilon(r)\right]^2}{r^2}
+m^*V_{s,\epsilon}(r)
=
q_{s,\epsilon}^\dagger q_{s,\epsilon},\quad \quad q_{s,\epsilon}
=
\partial_r
-
\frac{2+(p-4)f_\epsilon(r)}{r}\;,
\end{equation}
and the adjoint operator is taken with respect to the radial measure $r\,dr$.
The corresponding zero mode is
\begin{equation}
R_{2,\epsilon}(r)
\propto
r^2
\exp\left[
(p-4)
\int_0^r\frac{dr'}{r'}f_\epsilon(r')
\right].
\end{equation}
It behaves as $R_{2,\epsilon}\sim r^2$ inside the core and as $R_{2,\epsilon}(r)\sim r^{p-2}=r^{-2/3}$ for
$r\gg\epsilon$.

We therefore obtain a simple physical interpretation of the four
scale-invariant boundary conditions. A nonresonant collision channel flows to the regular branch, while each singular branch corresponds to tuning the associated two-body channel precisely to a zero-energy binding threshold. Thus, the $(+,+)$ point requires no tuning, the two crossed fixed points each require one resonant channel, and the
$(-,-)$ point requires both channels to be tuned. The factorizable
Hamiltonians discussed above provide particularly simple analytic
representatives of these more general threshold fixed points.

\subsection{Higher-body boundary conditions}
\label{sec:sm_higher_body}
 
Fixing the two-body boundary conditions determines the two-body spectrum of the
point-particle theory. It does not, by itself, determine the
$M$-body sectors because a short-range interaction that acts only when three or more particles come together is a distinct operator. In order to fully specify the
many-body problem, we must determine whether a higher-body collision channel admits two locally normalizable radial behaviors. Such a channel supplies an additional microscopic boundary parameter. This is a stronger question than classifying weak perturbations around the chosen AC point of the main text. Specifically, the general Hamiltonian is
\begin{equation}
\begin{aligned}
H={}&\frac{1}{2m^*}\sum_{a}\boldsymbol\Pi_{a}^2
+\sum_{a<b}V^{(2)}_\epsilon(\boldsymbol r_{ab})
+\sum_{a<b<c}V^{(3)}_{\epsilon}(\boldsymbol r_{a},\boldsymbol r_{b},\boldsymbol r_{c})
\\&+\sum_{a<b<c<d}V^{(4)}_{\epsilon}(\boldsymbol r_{a},\boldsymbol r_{b},\boldsymbol r_{c},\boldsymbol r_{d})+\dots ,
\label{eq:H_general}
\end{aligned}
\end{equation}
where $\boldsymbol\Pi_a=\mathbf p_a-\mathbf a_a$ is the operator of Eq.~\eqref{eq:HIR_AB},
$V^{(2)}_\epsilon$ is tuned to produce the boundary conditions
$(\sigma_{2},\sigma_{1})$, and $V^{(M)}_{\epsilon}$ is a regularized potential that
is nonzero only when $M$ particles approach each other, and it may carry
a projector onto a representation of the valley group. At the Aharonov--Casher
point $(+,-)$, the first two terms together form $H_{\rm ACCS}$, and we need to analyze 
what the $V^{(M)}$ with $M\ge3$ do. The scaling analysis of the main text identifies one marginal three-body singlet and makes all $M\geq4$ contacts irrelevant about this AC point. Here we determine the full negative three-body angular spectrum and derive the flow of the singlet coupling.

For our analysis, it will be convenient to introduce the Jacobi coordinates
\begin{equation}
\boldsymbol\xi_k=\sqrt{\frac{k}{k+1}}\,\Big(\mathbf r_{k+1}-\frac{1}{k}\sum_{a=1}^{k}\mathbf r_{a}\Big),
\qquad k=1,\dots,M-1 ,
\label{eq:jacobi}
\end{equation}
so that $\boldsymbol\xi_1=(\mathbf r_2-\mathbf r_1)/\sqrt2$ is the rescaled separation
of the first pair, $\boldsymbol\xi_2=\sqrt{2/3}\,[\mathbf r_3-\tfrac12(\mathbf r_1+\mathbf r_2)]$
the rescaled separation of the third particle from the center of mass of that pair,
and so on. The prefactors are chosen so that, together with
$\mathbf X=\sqrt{M}\,\mathbf r_{\rm cm}$, $\mathbf r_{\rm cm}=\frac1{M}\sum_{a}\mathbf r_{a}$,
the linear map $(\mathbf r_1,\dots,\mathbf r_{M})\mapsto(\mathbf X,\boldsymbol\xi_1,\dots,\boldsymbol\xi_{M-1})$
is orthogonal. In terms of these coordinates, we also have $\sum_{k=1}^{M-1}|\boldsymbol\xi_k|^2=\sum_{a}|\mathbf r_{a}|^2-M|\mathbf r_{\rm cm}|^2
=\sum_{a}|\mathbf r_{a}-\mathbf r_{\rm cm}|^2=\frac1{M}\sum_{a<b}|\mathbf r_a-\mathbf r_b|^2 $, which is the square of the hyperradius $R$ defined in the main text. Moreover, the kinetic energy splits without
cross terms, $\frac{1}{2m^*}\sum_{a}\mathbf p_{a}^2=\frac{\mathbf P^2}{2Mm^*}+\frac{1}{2m^*}\sum_{k=1}^{M-1}\mathbf p_{\xi_k}^2 $,
with $\mathbf P$ the total momentum. We write $\mathbf R=(\boldsymbol\xi_1,\dots,\boldsymbol\xi_{M-1})$, $R=|\mathbf R|$ and $\Omega=\mathbf R/R\in S^{D-1}$, where $D=2(M-1)$. The hyperradius $R$ measures the size of the cluster, while $\Omega$ determines its shape and orientation. The statistical vector potential has no
center-of-mass component, $\sum_{a}\mathbf a_{a}=p\sum_{a\ne b}\hat{\mathbf z}\times\mathbf r_{ab}/r_{ab}^2=0$, and thus the center of mass 
decouples. In addition, $
\mathbf R\cdot\mathbf a=\sum_{a}\mathbf r_{a}\cdot\mathbf a_{a}
=0 $ 
because $\hat{\mathbf z}\times\mathbf r_{ab}$ is perpendicular to $\mathbf r_{ab}$. Away from coincidences the statistical vector potential obeys
\begin{equation}
\sum_{a}\mathbf a_{a}\cdot d\mathbf r_{a}=p\,d\arg\Delta(\eta),\qquad \Delta(\eta)=\prod_{a<b}(\eta_{a}-\eta_{b}),
\label{eq:hc_statistical_connection}
\end{equation}
with $\eta_a$ the complex coordinate of $\mathbf r_a$, as in the main text.
In combination with scale invariance $\mathbf a(\lambda\mathbf R)=\lambda^{-1}\mathbf a(\mathbf R)$, this leads to a factorization $\mathbf a(\mathbf R)\cdot d\mathbf R=\mathbf a(\Omega)\cdot d\Omega$, where $\mathbf a(\Omega)$
is the potential evaluated on the unit sphere, a tangent vector field on $S^{D-1}$. The wave function $\Psi(\mathbf R)$ carries the valley labels of the $M$ particles and, in the
bosonic gauge used throughout, is symmetric under simultaneous permutation of
positions and labels.

The relative part of the general $M$-body kinetic term
acquires the following form
\begin{equation}
\sum_{a}\boldsymbol\Pi_{a}^2\Big|_{\rm rel}
=-\partial_R^2-\frac{D-1}{R}\,\partial_R+\frac{1}{R^2}\,K ,
\qquad
K=-\big(\nabla_\Omega-i\,\mathbf a(\Omega)\big)^2.
\label{eq:separation}
\end{equation}
$K$ is the covariant Laplacian on $S^{D-1}$ with the statistical gauge field. Just as the two-body problem is the
Aharonov--Bohm problem of a single flux tube in the plane, $K$ describes a
particle on the sphere threaded by a flux tube of strength $p$ through
each of the codimension-two submanifolds where a pair of particles coincides.
$\mathbf a$ is smooth away from these submanifolds and behaves as
$p/\rho_{ab}$ near them, with $\rho_{ab}$ the geodesic distance on $S^{D-1}$ to the
coincidence. Every term in Eq.~\eqref{eq:separation} scales as $R^{-2}$: the
kinetic operator is homogeneous of degree $-2$ under $\mathbf R\to\lambda\mathbf R$
and contains no length.

As in the case of a two-body problem, the local collision behavior follows from the zero-energy equation
\begin{equation}
\sum_{a}\boldsymbol\Pi_{a}^2\,\Psi=0 ,
\label{eq:zero_energy}
\end{equation}
with the two-body boundary condition of the fixed point imposed at every pair
coincidence $\mathbf r_a=\mathbf r_b$. For $M\geq4$, the angular domain also requires the chosen boundary conditions at proper-subcluster collisions, including the no-logarithm three-body condition at the AC point. Since this equation contains no length scale, we can search for a solution in the form
\begin{equation}
\Psi(\mathbf R)=R^{s}\,\Phi(\Omega).
\label{eq:homogeneous}
\end{equation}
Substituting into Eq.~\eqref{eq:separation} gives
$-s(s-1)R^{s-2}\Phi-(D-1)sR^{s-2}\Phi+R^{s-2}K\Phi=0$, i.e.
\begin{equation}
K\Phi=\Lambda\,\Phi,\qquad \Lambda=s\,(s+D-2).
\label{eq:hyperradial}
\end{equation}
For a given $\Lambda$, the two roots of Eq.~\eqref{eq:hyperradial} are
\begin{equation}
s_\pm=-(M-2)\pm\sqrt{(M-2)^2+\Lambda},\qquad s_++s_-=-(2M-4).
\label{eq:roots}
\end{equation}
For the Aharonov--Casher zero modes, the exponent $s$ is $s_M$ given in the main text.
Near $R=0$ the volume element is $R^{D-1}dR$, so $R^{s}\Phi$ is square
integrable there if and only if $2s+D>0$, i.e.\ $s>-(M-1)$. There are four cases
to consider. If only $s_+$ satisfies this, the $M$-body wave function is forced to
behave as $R^{s_+}$, and the $M$-body boundary condition is unique. If both roots
are normalizable, any combination $c_+R^{s_+}+c_-R^{s_-}$ is admissible and the
ratio $c_-/c_+$, which carries units of (length)$^{s_+-s_-}$, is a new parameter
of the $M$-body problem, the analog of a two-body scattering length. Self-adjointness requires a fixed real, energy-independent ratio in each isolated channel. Its
scale-invariant values are $c_-=0$ (regular, $s_+$) and $c_+=0$ (resonant,
$s_-$). With $\nu=\sqrt{(M-2)^2+\Lambda}$, the relative weight of the minus branch at radius $L$ is $(c_-/c_+)L^{-2\nu}$. For $0<\nu<1$ it decreases at the plus endpoint even though both branches are normalizable. The extra boundary parameter can therefore be irrelevant about that endpoint and still allow a separately tuned resonance. If $\Lambda=-(M-2)^2$ the roots coincide at $s=2-M$, and the second
solution is $R^{2-M}\ln(R/R_0)$, where $R_0$ is an arbitrary reference length. If $\Lambda<-(M-2)^2$ the roots are complex, the
solutions oscillate as $R^{2-M}\cos(\sqrt{-\Lambda-(M-2)^2}\ln(R/R_0)+\delta)$, and every self-adjoint point-particle radial extension is unbounded below. A short-distance completion is then required.

\subsubsection{Angular three-body problem}
\label{sec:three_body_contact}
 
For three particles, $D=4$ and $\Omega\in S^3$. Writing $\boldsymbol\xi_{1,2}$ as complex numbers $\xi_{1,2}$,
\begin{equation}
\begin{gathered}R^2=|\xi_1|^2+|\xi_2|^2
=\frac13\sum_{a<b}|\eta_a-\eta_b|^2,\\
\xi_1=R\cos\frac\theta2\,e^{i(\chi+\phi/2)},\qquad
\xi_2=R\sin\frac\theta2\,e^{i(\chi-\phi/2)},\qquad
0\leq\theta\leq\pi,\end{gathered}
\label{eq:hc_coordinates}
\end{equation}
where we parametrize $S^3$ by the angles $(\theta,\phi,\chi)$
\cite{Ganchev2015} with periodicity
$(\chi,\phi)\sim(\chi+2\pi,\phi)\sim(\chi+\pi,\phi+2\pi)$. Here $\chi$ is the overall rotation angle conjugate to the total angular momentum, and $(\theta,\phi)$ parametrizes the shape of the triangle, while $\Omega=(\theta,\phi,\chi)$ denotes the full angular point. All three
particles coincide only at $R=0$.
Direct differentiation of the Jacobi coordinates gives 
\begin{equation}
ds^2=dR^2+R^2ds_\Omega^2,\qquad
ds_\Omega^2=\frac14(d\theta^2+\sin^2\theta\,d\phi^2)
 +\left(d\chi+\frac12\cos\theta\,d\phi\right)^2.
\label{eq:hc_metric_redline}
\end{equation}
This metric fixes both the angular kinetic operator and its measure,
$d\Omega=(\sin\theta/4)d\theta\,d\phi\,d\chi$.
 In these coordinates the Vandermonde factor of Eq.~\eqref{eq:hc_statistical_connection} reads
\begin{equation}
\Delta(\eta)=\frac{R^3}{\sqrt2}\,e^{3i\chi}P(\theta,\phi),\qquad
P(\theta,\phi)=
\cos^3\frac\theta2\,e^{3i\phi/2}
-3\cos\frac\theta2\sin^2\frac\theta2\,e^{-i\phi/2},
\label{eq:hc_vandermonde}
\end{equation}
and the three pair points are the zeros of $P$: the south pole $\theta=\pi$, where $\eta_1=\eta_2$ and $P\propto(\pi-\theta)e^{-i\phi/2}$, and the two points $(\theta,\phi)=(\pi/3,0)$ and $(\pi/3,\pi)$, where $\eta_2=\eta_3$ and $\eta_1=\eta_3$, respectively. By Eq.~\eqref{eq:hc_statistical_connection}, the angular one-form components of the statistical potential are $a_\chi=3p=4$ and $a_\mu=p\operatorname{Im}[(\partial_\mu P)/P]$ for $\mu=\theta,\phi$. Explicitly,
\begin{equation}
a_\theta=
\frac{4\sin\theta\sin2\phi}
{5-8\cos\theta+5\cos^2\theta-3\sin^2\theta\cos2\phi},\quad 
a_\phi=
\frac{-2+8\cos\theta-2\cos^2\theta-2\sin^2\theta\cos2\phi}
{5-8\cos\theta+5\cos^2\theta-3\sin^2\theta\cos2\phi}.
\label{eq:hc_connection_components}
\end{equation}

We now determine the spectrum $\{\Lambda\}$ of the angular operator $K$ of Eq.~\eqref{eq:separation} for three anyons at the fixed point $\mathrm{AC}_{(+,-)}$. Rotational and scale invariance allow each separated channel to be written as
\begin{equation}
\Psi(R,\theta,\phi,\chi)
=F(R)e^{iJ\chi}\Phi_J(\theta,\phi),\qquad
\mathcal K_J\Phi_J=\Lambda_J\Phi_J,
\label{eq:hc_separated_channel}
\end{equation}
where integer $J$ is the total orbital angular momentum. The second identification in Eq.~\eqref{eq:hc_coordinates} makes $\Phi_J$ periodic in $\phi$ for even $J$ and antiperiodic for odd $J$. The covariant Laplacian of Eq.~\eqref{eq:hc_metric_redline}, with $D_\chi=i(J-4)$, gives
\begin{equation}
\mathcal K_J=-4\left[
\frac1{\sin\theta}D_\theta(\sin\theta D_\theta)
+\frac{\left(D_\phi-\frac{i(J-4)}2\cos\theta\right)^2}
{\sin^2\theta}\right]+(J-4)^2,
\label{eq:hc_angular_operator}
\end{equation}
where $D_\theta=\partial_\theta-ia_\theta$,
$D_\phi=\partial_\phi-ia_\phi$, and $D_\chi=\partial_\chi-4i$ are covariant derivatives. This eigenproblem is supplemented by the $(+,-)$ conditions at the three
pair points.

A useful first-order form of this angular eigenproblem is 
\begin{equation}
\begin{aligned}
\mathcal Q_J&=-D_\theta+\frac{i}{\sin\theta}D_\phi
+\frac{J-4}{2}\frac{\cos\theta}{\sin\theta},\qquad
\mathcal Q_J^\dagger=D_\theta+\frac{i}{\sin\theta}D_\phi
+\frac{J-2}{2}\frac{\cos\theta}{\sin\theta},\\
\mathcal K_J&=4\mathcal Q_J^\dagger\mathcal Q_J+(J-4)(J-2),
\qquad
\mathcal K_{J+2}\mathcal Q_J=\mathcal Q_J\mathcal K_J.
\end{aligned}
\label{eq:hc_angular_factorization}
\end{equation}
Here the dagger denotes the formal adjoint in
$\sin\theta\,d\theta\,d\phi$, including a sum over valley indices. Both relations on the last line hold away from the pair points, where $\partial_\theta a_\phi=\partial_\phi a_\theta$.

Below we demonstrate two facts about $\Lambda_J$: (1) its minimal value is $-1$ for $J=3$ and the corresponding eigenfunction is unique, and (2) there are no states with $-1<\Lambda_J<0$ for any $J$. Together they show that the three-body sector at $\mathrm{AC}_{(+,-)}$ contains exactly one channel with an additional boundary parameter.

First, let us explicitly construct a state at $J=3$ and $\Lambda_{J=3}=-1$. We will search for this state in the form
\begin{equation}
\Phi_3(\theta,\phi)=|P(\theta,\phi)|^{-4/3}Y(\theta,\phi), 
\label{eq:hc_real_weight}
\end{equation}
where $Y(\theta,\phi)$ is to be determined and $P$ is defined in Eq.~\eqref{eq:hc_vandermonde}. Up to a power of $R$, the singled-out prefactor is $|\Delta(\eta)|^{-4/3}$, i.e. the weight $e^{W/2}$ of the zero modes restricted to the sphere. This is natural from a two-body perspective, since near the south-pole collision
$P\propto(\pi-\theta)e^{-i\phi/2}$, and so for the full wave-function to behave as 
$(\pi-\theta)^{-1/3}$ we should have a simple zero in $Y$. The same 
holds at the other two pair points. This also simplifies the first-order equation:
\begin{equation}
\mathcal Q_3\bigl(|P|^{-4/3}Y\bigr)
=-|P|^{-4/3}\left[
\partial_\theta-\frac{i}{\sin\theta}\partial_\phi
-\frac{3\cos\theta}{2\sin\theta}\right]Y.
\label{eq:hc_weight_cancellation}
\end{equation}
This identity follows from $a_\theta=p\,\partial_\theta\arg P$, $a_\phi=p\,\partial_\phi\arg P$ (cf. Eq.~\eqref{eq:hc_statistical_connection}).  The pair conditions make $Y$ bounded and vanishing at the collision points. Thus the remaining equation
$\mathcal Q_3\Phi_3=0$ can be further simplified using $Y\propto e^{ik\phi}$, and then it becomes
\begin{equation}
\partial_\theta Y
=\frac{\frac32\cos\theta-k}{\sin\theta}\,Y,
\qquad
Y\propto
\cos^{3/2+k}\frac\theta2\,
\sin^{3/2-k}\frac\theta2\,e^{ik\phi}.
\label{eq:hc_angular_kernel_ode}
\end{equation}
Since $\Phi_3$ is antiperiodic in $\phi$, $k$ is half-integer. Boundedness
at both poles requires $3/2+k\geq0$ and $3/2-k\geq0$, so
$k=3/2,1/2,-1/2,-3/2$. The complete regular numerator is consequently
\begin{equation}
Y={}b_0\cos^3\frac\theta2\,e^{3i\phi/2}
+b_1\cos^2\frac\theta2\sin\frac\theta2\,e^{i\phi/2}+b_2\cos\frac\theta2\sin^2\frac\theta2\,e^{-i\phi/2}
+b_3\sin^3\frac\theta2\,e^{-3i\phi/2}.
\label{eq:hc_four_angular_harmonics}
\end{equation}
The coefficients may carry valley
indices. The zero at $\theta=\pi$
sets $b_3=0$. The two remaining
zero conditions at $(\theta,\phi)=(\pi/3,0)$ and $(\pi/3,\pi)$ are
\begin{equation}
3\sqrt3\,b_0+3b_1+\sqrt3\,b_2=0,\qquad
-3\sqrt3\,b_0+3b_1-\sqrt3\,b_2=0,\qquad \rightarrow \qquad
b_1=0,\qquad b_2=-3b_0.
\label{eq:hc_pair_zero_constraints}
\end{equation}
Thus $Y=b_0P$ and every solution of $\mathcal Q_3\Phi_3=0$ is proportional to
$P/|P|^{4/3}$. Restoring $e^{3i\chi}$ makes its numerator
proportional to $\Delta(\eta)$, Eq.~\eqref{eq:hc_vandermonde}, which is
antisymmetric under particle exchange. Bose symmetry therefore forces
the constant valley tensor to be fully antisymmetric. For three
valleys there is only one such tensor:
\begin{equation}
(\Phi_3)_{\gamma_1\gamma_2\gamma_3}
=\mathcal N\varepsilon_{\gamma_1\gamma_2\gamma_3} \frac{ P}{|P|^{4/3}},
\qquad \mathcal Q_3\Phi_3=0,\qquad \Lambda_3=-1.
\label{eq:hc_critical_state}
\end{equation}
Here 
$\mathcal N$ is a normalization factor. Near any pair point, $\Phi_3$ has the required $\rho^{-1/3}$ branch with corrections two powers higher and no independent $\rho^{1/3}$ branch. Its angular norm is finite because the transverse integral is $\int_0 d\rho\,\rho^{1/3}$. This establishes existence and uniqueness
of the critical angular mode without choosing its radial boundary
condition.

Next, we discuss other possible eigenstates. After multiplying $\mathcal K_J\Phi_J=\Lambda_J\Phi_J$ by $\Phi^*_J$ and integrating by parts, we find
\begin{equation}
\Lambda_J\|\Phi_J\|^2
=4\|\mathcal Q_J\Phi_J\|^2
+(J-4)(J-2)\|\Phi_J\|^2,
\qquad \rightarrow \qquad\Lambda_J\geq(J-3)^2-1,
\label{eq:hc_angular_lower_bound}
\end{equation}
where we used that $\|\mathcal Q_J\Phi_J\|^2 /\|\Phi_J\|^2\geq 0 $. Because $J$ is an integer, all sectors except $J=3$ are nonnegative.
There is nothing below $-1$, and a state at $-1$ must satisfy
$\mathcal Q_3\Phi_3=0$.
However, there is still a possibility of a state with $-1<\Lambda_3<0$. In order to exclude this, we use the
intertwining relation in Eq.~\eqref{eq:hc_angular_factorization}: if
$\Phi_3$ is an eigenfunction of $\mathcal K_3$ with $u\equiv\mathcal Q_3\Phi_3\neq0$,
then $u$ satisfies the $\mathcal K_5$ eigenvalue equation away from particle collision points with the same
eigenvalue. If the identity of Eq.~\eqref{eq:hc_angular_lower_bound} can be applied to $u$ in the sector $J=5$, we conclude that $\Lambda_3\geq (5-3)^2-1=3$, which is what we need. The state $u$ is normalizable, since
$4\|u\|^2=(\Lambda_3+1)\|\Phi_3\|^2$ by Eq.~\eqref{eq:hc_angular_lower_bound}. 
 
What remains to be checked is that the integration by parts behind Eq.~\eqref{eq:hc_angular_lower_bound} produces no surface terms at the collision points, both for $\Phi_J$ and for $u$, and that $\|\mathcal Q_5u\|$ is finite. To this end, we expand near $\theta=\pi$, and define $\rho=\pi-\theta$ as the small deviation from the collision point. The vector potential components are expanded as
$a_\theta=O(\rho)$ and $a_\phi=-2/3+O(\rho^2)$.
The pair angular momentum of $\rho^{q} e^{ik\phi}$ in sector $J$
is $\ell=k+J/2$, as follows by rotating the first Jacobi coordinate
while holding the second fixed. Therefore
\begin{equation}
\mathcal Q_J(\rho^{q} e^{ik\phi})
=\left(q-\ell+p\right)\rho^{q-1}e^{ik\phi}
+O(\rho^{q+1}).
\label{eq:hc_pair_image}
\end{equation}
Corrections to each leading branch start two powers higher in the
local eigenvalue equation and, because $a_\theta$ and $a_\phi$ depend on $\phi$ through $\sin2\phi$ and $\cos2\phi$, Eq.~\eqref{eq:hc_connection_components}, they also carry $k\to k\pm2$. Thus the allowed $(\ell,q)=(1,-1/3)$,
$(2,2/3)$ and $(0,4/3)$ branches give images of orders
$\rho^{2/3}$, $\rho^{5/3}$ and $\rho^{1/3}$, respectively.
All other partial waves give higher powers. Consequently
$\Phi_J=O(\rho^{-1/3})$ and $\mathcal Q_J\Phi_J=O(\rho^{1/3})$.
The boundary measure is proportional to $\rho$, so the surface term in Eq.~\eqref{eq:hc_angular_lower_bound}, $\rho\,\Phi_J^*\,\mathcal Q_J\Phi_J$, is $O(\rho)$ and vanishes. Particle permutations give the same estimate at the other two pair points. For $u=\mathcal Q_3\Phi_3=O(\rho^{1/3})$, applying Eq.~\eqref{eq:hc_pair_image} once more gives $\mathcal Q_5u=O(\rho^{-2/3})$. The corresponding surface term is $\rho\cdot O(\rho^{1/3})\cdot O(\rho^{-2/3})=O(\rho^{2/3})$ and vanishes as well, and $\|\mathcal Q_5u\|^2$ is finite because its transverse integral $\int_0 d\rho\,\rho\,\rho^{-4/3}$ converges. This completes the argument above. Since $\mathcal Q_3$ changes the exchange parity, $u$ need not be a bosonic state, and the second integration is performed on labeled configuration space. The asymptotic bounds just derived make it valid without requiring $u$ to obey the original bosonic pair domain.
 
To summarize, the angular spectrum of three anyons at $\mathrm{AC}_{(+,-)}$ is contained in $\{-1\}\cup[0,\infty)$. The only channel which requires additional microscopic matching is $\Lambda=-1$, with possible radial solutions $R^{-1}$ and $R^{-1}\ln(R/R_0)$. Writing $F(R)=[A+B\ln(R/l_{\rm pair})]/R$, the normalization-independent logarithmic derivative gives
\begin{equation}
g_3(L)\equiv1+\frac{LF'(L)}{F(L)}
=\frac{B}{A+B\ln(L/l_{\rm pair})},\qquad
\frac{dg_3}{d\ln L}=-g_3^2.
\label{eq:hc_triplet_flow_redline}
\end{equation}
The derivative holds $A/B$ fixed. The AC choice follows from its positive radial form, $q_{\rm AC,3}=(2m^*)^{-1}\int dR\,R|\partial_R(RF)|^2$. A logarithmic term contributes $|B|^2(2m^*)^{-1}\int dR/R$, so the AC domain selects $B=0$, or $g_3=0$. Other self-adjoint radial conditions describe the three-body contact. Positive $g_3$ decreases with $L$, while a weak negative bare value produces the length $R_{\rm trimer}=l_{\rm pair}e^{-1/g_3(l_{\rm pair})}$. Restoring $R$ and $\chi$ in Eq.~\eqref{eq:hc_critical_state} at $g_3=0$, the wave function is $\Psi\propto\varepsilon_{\gamma_1\gamma_2\gamma_3}\Delta(\eta)/|\Delta(\eta)|^{4/3}$, the three-body Aharonov--Casher zero mode in the fully valley-antisymmetric channel. This establishes the statement used in the main text that, with the pair conditions of $(+,-)$ imposed, the three-body sector has exactly one valley-singlet channel with two locally normalizable radial behaviors. The irrelevance of $M\geq4$ contacts about this AC domain follows from the separate finite-form argument in the main text. Excluding additional resonant endpoints in those sectors would require their angular spectra and is not established by the three-body calculation.




\section{Effective superfluid action from the Aharonov--Casher-Chern-Simons wavefunction}
\label{sec:seft}
 
 In this section, we derive the coefficients of the superfluid theory of the main text from the AC wavefunction. Its density correlations fix the structure factor and the gradient term of the action. Its pair and triplet collision amplitudes fix the energy generated by $\delta u_1$ and $g_3$, including the values of $C_2$ and $C_3$. We also discuss relative-valley excitations and their energy gap.
 
Let $\eta_a$ and $\alpha_a\in\{1,2,3\}$ be the coordinate and valley of anyon $a$, with $M_v=M/3$ anyons in each valley and $\eta_{ab}=\eta_a-\eta_b$. The component of the AC zero mode with valley labels $\alpha_1,\dots,\alpha_M$, and its probability weight are
\begin{equation}
\tilde w=\epsilon(\alpha)\Delta(\eta)\prod_{\beta=1}^{3}\Delta_\beta(\eta)
\prod_{a<b}|\eta_{ab}|^{-4/3},\qquad
W_M=|\tilde w|^2=\prod_{a<b}|\eta_{ab}|^{2\mathsf B_{\alpha_a\alpha_b}},\qquad
\mathsf B_{\alpha\beta}=\delta_{\alpha\beta}-\tfrac13.
\label{eq:seft_ac_wavefunction}
\end{equation}
Here $\Delta(\eta)=\prod_{a<b}\eta_{ab}$ and $\Delta_\beta(\eta)$ is the same product restricted to the anyons of valley $\beta$. The weight is the plasma weight $e^{-H_{\rm pl}}$ of the main text, with $\mathbf q_1=(\sqrt{2/3},0)$, $\mathbf q_2=(-1/\sqrt6,1/\sqrt2)$ and $\mathbf q_3=-\mathbf q_1-\mathbf q_2$, so that $\mathbf q_\alpha\cdot\mathbf q_\beta=\mathsf B_{\alpha\beta}$.
As before, the partition function is $Z^{(M)}_{\rm pl}=\int_A\prod_a d^2\eta_a\,W_M$ for the normalization in the area $A$. In addition, $\langle\cdot\rangle_{\rm pl}$ denotes averages with the probability weight $W_M$ in the bulk limit $M,A\to\infty$ at fixed $\rho=M/A$. We will be interested in calculating density observables in this three-component classical plasma.
 
\subsection{Equal-time structure factor \texorpdfstring{$S_{\rm ACCS}(k)$}{S AC(k)}}
\label{sec:seft_structure}
 
First, we calculate the equal-time density correlations. To this end, define $\rho_\alpha(\mathbf x)=\sum_{a:\alpha_a=\alpha}\delta^{(2)}(\mathbf x-\boldsymbol\eta_a)$, $\rho=\sum_\alpha \rho_\alpha$ and
\begin{equation}
\begin{gathered}
\mathcal C_{\alpha\beta}(k)=\int d^2r\,e^{-i\mathbf k\cdot\mathbf r}
\Big[\langle \rho_\alpha(\mathbf r)\rho_\beta(\mathbf 0)\rangle_{\rm pl}-\frac{\rho^2}{9}\Big],\qquad
S_{\rm ACCS}(k)=\frac1\rho\sum_{\alpha\beta}\mathcal C_{\alpha\beta}(k).
\end{gathered}
\label{eq:seft_covariance_definition}
\end{equation}
The diagonal correlator includes the same-particle term
$(\rho/3)\delta^{(2)}(\mathbf r)$. We take the thermodynamic limit
before $k\to0$ in the screening phase of the plasma. 
 
As we show below, using a field-theory representation, the exact small-$k$ expansion is
\begin{equation}
\begin{gathered}
\mathcal C_{\alpha\beta}(k)
=\frac\rho6+\frac{k^2}{32\pi}(8\delta_{\alpha\beta}-3)+O(k^4/\rho),\\ S_{\rm ACCS}(k)=\frac32-\frac{3k^2}{32\pi\rho}+O(k^4/\rho^2).
\end{gathered}
\label{eq:seft_small_k_summary}
\end{equation}

To obtain the field-theory representation, we define a grand partition function $\Xi$ by summing over valley-balanced particle numbers
\begin{equation}
\Xi=\sum_{M=0,3,6,\ldots}\frac{\zeta^M }{[(M/3)!]^3}Z^{(M)}_{\rm pl},\qquad
\rho=\frac1A\frac{\partial\ln\Xi}{\partial\ln\zeta}.
\label{eq:seft_activity}
\end{equation}
Here $\rho$ is the prescribed density, and its value determines the auxiliary fugacity $\zeta$. We note that it is distinct from the physical anyon chemical potential.

Now we rewrite Eq.~\eqref{eq:seft_activity} in terms of a certain field integral (our construction is a direct generalization of the two-component plasma mapping developed in \cite{TwoComponentPlasma2000,Tellez_2005}). First, for two Gaussian fields $\boldsymbol\phi=(\phi_1,\phi_2)$ with action $(8\pi)^{-1}\int(\nabla\boldsymbol\phi)^2$ the bare propagator is $\langle\phi_i(\mathbf r)\phi_j(\mathbf 0)\rangle_0=-2\delta_{ij}\ln r$, and thus the following identities for vertex operators hold
\begin{equation}
V_\alpha=:e^{i\mathbf q_\alpha\cdot\boldsymbol\phi}:,\qquad
\Big\langle\prod_a V_{\alpha_a}(\boldsymbol\eta_a)\Big\rangle_0
=\prod_{a<b}|\eta_{ab}|^{2\mathbf q_{\alpha_a}\cdot\mathbf q_{\alpha_b}}=W_M,
\label{eq:seft_wick}
\end{equation}
 Here normal ordering removes contractions at the same point, the Gaussian average $\langle ...\rangle_0$ is normalized to one, and we only allow for neutral configurations $\sum_a\mathbf q_{\alpha_a}=0$. The observation in Eq.~\eqref{eq:seft_wick}, combined with the fact that expanding the exponential of $\zeta\int\sum_\alpha V_\alpha$ reproduces the particle integrals and factorials in $\Xi$, allows us to rewrite the expectation values in the AC state as the interacting field integral over $\phi$ with action 
\begin{equation}
S_{\rm T}[h_\alpha]=\int d^2x\left[
\frac{(\nabla\boldsymbol\phi)^2}{8\pi}
-\zeta\sum_{\alpha=1}^3e^{h_\alpha(\mathbf x)}V_\alpha(\mathbf x)\right],\qquad \Xi[h_\alpha] = \left\langle \exp \left\{\zeta\sum_{\alpha=1}^3\int d^2x e^{h_\alpha(\mathbf x)}V_\alpha(\mathbf x) \right\} \right\rangle_0
\label{eq:seft_toda_action}
\end{equation}
We denote the corresponding generating functional by $\Xi[h_\alpha]$, with $\Xi[0]$ equal to the expression given in Eq.~\eqref{eq:seft_activity}. The dimensionless functions $h_\alpha$ are density sources. In the particle integral they multiply the weight by $\exp[\sum_a h_{\alpha_a}(\boldsymbol\eta_a)]$. The vectors $\sqrt3\mathbf q_\alpha$ are the simple and affine roots of $A_2$, and so Eq.~\eqref{eq:seft_toda_action} is imaginary $A_2$ affine Toda theory \cite{Fateev2002}. This theory is integrable, and we will be able to make use of some of its known properties. At zero source, the first two functional derivatives give
\begin{equation}
\begin{aligned}
\left.\frac{\delta\ln\Xi}{\delta h_\alpha(\mathbf x)}\right|_{h=0}
&=\langle \rho_\alpha\rangle_{\rm pl}
=\zeta\langle V_\alpha\rangle_{\rm T}=\frac\rho3,\\
\left.\frac{\delta^2\ln\Xi}{\delta h_\alpha(\mathbf r)\delta h_\beta(\mathbf0)}\right|_{h=0}
&=\langle \rho_\alpha(\mathbf r)\rho_\beta(\mathbf0)\rangle_{{\rm pl},c}=\frac\rho3\delta_{\alpha\beta}\delta^{(2)}(\mathbf r)
+\zeta^2\langle V_\alpha(\mathbf r)V_\beta(\mathbf0)\rangle_{{\rm T},c}.
\end{aligned}
\label{eq:seft_toda_density_sources}
\end{equation}
Here $\langle...\rangle_{\rm T}= \int D\phi ... e^{-S_T[h=0]}/\int D\phi e^{-S_T[h=0]}$ is the interacting Toda average, and $c$ denotes a connected correlator. We choose $\zeta$ so that the zero-source mean density equals the
prescribed $\rho$ (the precise relation is given in Eq.~\eqref{eq:seft_density_fugacity}). In the homogeneous screening plasma phase, the resulting local
correlators coincide with those of the fixed-density AC state
in the thermodynamic limit. The following identities will also be useful
\begin{equation}
    \left\langle e^{\sum\limits_{a}h_{\gamma_a}(\eta_a)} \right\rangle_{\rm pl} = \frac{\Xi[h]}{\Xi[0]},\qquad     \left\langle \rho_\nu({\bf r})e^{\sum\limits_{a} h_{\gamma_a}(\eta_a)} \right\rangle_{\rm pl} = \frac{1}{\Xi[0]} \frac{\delta \Xi[h]}{\delta h_\nu({\bf r})}\;.
    \label{eq:useful_identities_Toda}
\end{equation}

Differentiating the normalized Toda average gives
$\partial_\zeta\langle V_\alpha\rangle_{\rm T}
=\sum_\beta\int d^2r\,\langle V_\alpha(\mathbf r)V_\beta(\mathbf0)\rangle_{{\rm T},c}
=\langle V_\alpha\rangle_{\rm T}/(2\zeta)$, which follows from the scaling dimension $2/3$ for the vertex operator. Summing over valleys, and then adding the same-particle term in
Eq.~\eqref{eq:seft_toda_density_sources}, gives
\begin{equation}
\zeta^2\sum_{\alpha\beta}\int d^2r\,
\langle V_\alpha(\mathbf r)V_\beta(\mathbf0)\rangle_{{\rm T},c}
=\frac\rho2,\qquad
\int d^2r\,\langle \rho(\mathbf r)\rho(\mathbf0)\rangle_{{\rm pl},c}
=\frac{3\rho }{2},\qquad S_{\rm ACCS}(k\rightarrow 0)=\frac32.
\label{eq:seft_S_zero}
\end{equation}

For the second spatial moment, we find
\begin{equation}
\zeta^2\sum_{\alpha\beta}\int d^2r\,r^2
\langle V_\alpha(\mathbf r)V_\beta(\mathbf0)\rangle_{{\rm T},c}
=\frac{2}{3\pi(2-\Delta_V)^2}\Big|_{\Delta_V=2/3}=\frac{3}{8\pi}.
\label{eq:seft_second_moment}
\end{equation}
This is the specialization of the general moment result
\cite{Jancovici2000} to two Gaussian fields and
perturbing vertices of dimension $\Delta_V=2/3$. The identity uses decay of the connected correlations in the
massive screened phase. The same-particle term has zero second moment.
The angular average of $e^{-i\mathbf k\cdot\mathbf r}$ is
$1-k^2r^2/4+O(k^4r^4)$, so
Eqs.~\eqref{eq:seft_S_zero} and \eqref{eq:seft_second_moment}
immediately give $S_{\rm ACCS}(k)$ in
Eq.~\eqref{eq:seft_small_k_summary}.

The large-momentum behavior follows from the short-distance
intervalley correlator. Vertex fusion gives
$\langle\rho_\alpha(\mathbf r)\rho_\beta(\mathbf0)\rangle_{\rm pl}
\simeq\zeta^2G_-r^{-2/3}$ for $\alpha\ne\beta$, where
$G_-=\langle:e^{-i\mathbf q_\gamma\cdot\boldsymbol\phi}:
\rangle_{\rm T}$ is independent of $\gamma$ by valley symmetry. There are six ordered intervalley pairs, while the same-valley correlators are less singular. Using
$\int d^2r\,e^{-i\mathbf k\cdot\mathbf r}r^{-2/3}
=2^{4/3}\pi\Gamma(2/3)/[\Gamma(1/3)k^{4/3}]$ therefore gives
\begin{equation}
 S_{\rm ACCS}(k)
 =1+\frac{12\pi\,2^{1/3}\zeta^2G_-}{\rho}
       \frac{\Gamma(2/3)}{\Gamma(1/3)}
       \frac1{k^{4/3}}+\cdots,
 \qquad
 \sqrt{\rho}\ll k\ll l_{\rm pair}^{-1}.
 \label{eq:seft_large_k}
\end{equation}
The constant comes from the same-particle contribution.
The $k^{-4/3}$ tail directly reflects the singular
$r^{-2/3}$ probability of an intervalley pair collision. The explicit form of $\zeta$ and $G_-$ will be calculated in Sec.~\ref{sec:seft_contact_evaluation}.

\subsection{Corrections to the energy density}
\label{sec:seft_eos}
 
Next, we discuss corrections to the energy per unit area, denoted as $e(\rho)$ in the main text. For the AC wave function this quantity is exactly zero, but a pair detuning $\delta u_1$ or a three-body coupling $g_3$ generates a finite value. In order to calculate their respective energy shifts, we first recall that the pair and triplet boundary conditions of the main text are $R_1(r)=A_1r^{-1/3}+B_1r^{1/3}$ and $F(R)=[A+B\ln(R/l_{\rm pair})]/R$, with running couplings $\delta u_1(L)=L^{2/3}B_1/A_1$ and $g_3(L)=1+LF'(L)/F(L)$. The former is the linearized deviation of the logarithmic slope $u_1(L)$ from its fixed-point value, $u_1(L)=-\tfrac13+\tfrac23\,\delta u_1(L)+O(\delta u_1^2)$. Here $R^2=(r^2+r'^2+|\mathbf r-\mathbf r'|^2)/3$ for a triplet at $\mathbf r,\mathbf r',\mathbf 0$. Both couplings vanish at the AC point.
 
To convert the boundary conditions into energies, we represent each detuning by a thin shell potential at a radius $\epsilon$ with $l_{\rm pair}\ll\epsilon\ll L_\rho$, inside which the AC collision profiles hold. A pair shell $v_2\delta(r-\epsilon)$ gives the jump $R_1'(\epsilon^+)-R_1'(\epsilon^-)=m^*v_2R_1(\epsilon)$. The logarithmic slope $rR_1'/R_1$ is $-1/3$ inside the shell and $[-1+\delta u_1(\epsilon)]/[3(1+\delta u_1(\epsilon))]$ outside, so that its jump is $\tfrac23\delta u_1(\epsilon)=m^*\epsilon v_2$ to first order. For the triplet the radial kinetic operator is $-(\partial_R^2+3R^{-1}\partial_R)/2m^*$, a shell $v_3\delta(R-\epsilon)$ gives $\epsilon[F'(\epsilon^+)-F'(\epsilon^-)]/F(\epsilon)=2m^*\epsilon v_3$, and the jump of $RF'/F$, from $-1$ inside to $-1+g_3(\epsilon)$ outside, is $g_3(\epsilon)$. To first order,
\begin{equation}
\delta V_2=\frac{2\delta u_1(\epsilon)}{3m^*\epsilon}\,\delta(r-\epsilon),\qquad
\delta V_3=\frac{g_3(\epsilon)}{2m^*\epsilon}\,\delta(R-\epsilon),
\label{eq:seft_shell_matching}
\end{equation}
where the pair shell acts on each of the three intervalley combinations and the triplet shell on each one-of-each-valley triple. We define the contact amplitudes, as in the main text,
\begin{equation}
\begin{gathered}
C_2=\frac{4\pi}{\rho^{5/3}}\lim_{r\to0}r^{2/3}
\langle \rho_1(\mathbf r)\rho_2(\mathbf 0)\rangle_{\rm pl},\\
C_3=\lim_{\epsilon\to0}\frac1{\rho^2\epsilon^2}
\int_{R<\epsilon}d^2r\,d^2r'\,
\langle \rho_1(\mathbf r)\rho_2(\mathbf r')\rho_3(\mathbf 0)\rangle_{\rm pl}.
\end{gathered}
\label{eq:seft_contact_definitions}
\end{equation}
These coefficients are evaluated exactly in Sec.~\ref{sec:seft_contact_evaluation} through the Toda representation of the plasma. The limits are taken at fixed $\rho$ in the zero-range AC state. In a microscopic calculation they mean the scale window $l_{\rm pair}\ll r,\epsilon\ll L_\rho$. 
The powers of $r$ and $\epsilon$ are those of the squared AC collision profiles, $r^{-2/3}$ for a pair and degree $-2$ in the four relative dimensions of a triplet. The powers of $\rho$ then follow by dimensional analysis. First-order perturbation theory in $\delta V_2$ and $\delta V_3$ then gives
\begin{equation}
\begin{gathered}
\delta e_2=\frac{4\pi}{m^*}\,\delta u_1(\epsilon)
\langle \rho_1(\boldsymbol\epsilon)\rho_2(\mathbf 0)\rangle_{\rm pl},\qquad
\delta e_3=\frac{g_3(\epsilon)}{2m^*\epsilon}\frac{d}{d\epsilon}
\int_{R<\epsilon}d^2r\,d^2r'\,
\langle \rho_1(\mathbf r)\rho_2(\mathbf r')\rho_3(\mathbf 0)\rangle_{\rm pl},
\end{gathered}
\label{eq:seft_shell_energies}
\end{equation}
which with Eq.~\eqref{eq:seft_contact_definitions} become $\delta e_2=C_2\,\rho^{5/3}\,\delta u_1(\epsilon)\epsilon^{-2/3}/m^*$ and $\delta e_3=C_3\,\rho^2 g_3(\epsilon)/m^*$.

With the bare values $\delta u_1=\delta u_1(l_{\rm pair})$ and $g_3^{(0)}=g_3(l_{\rm pair})$ of the main text, fixed boundary data give $d\delta u_1/d\ln L=2\delta u_1/3$ and $dg_3/d\ln L=-g_3^2$, and thus
\begin{equation}
\delta u_1(L_\rho)=\delta u_1\,(l_{\rm pair}^2 \rho)^{-1/3},\qquad
g_3(L_\rho)=\frac{g_3^{(0)}}{1+\tfrac12 g_3^{(0)}\ln[1/(\rho l_{\rm pair}^2)]}.
\label{eq:seft_running_contacts}
\end{equation}
The triplet flow is taken at pair resonance and mixed corrections are of higher order. The pair term is independent of the shell radius, since $\delta u_1(\epsilon)\epsilon^{-2/3}=B_1/A_1=\delta u_1(L_\rho)\rho^{1/3}$. The triplet term depends on $\epsilon$ only through $g_3(\epsilon)$. This dependence is removed by the higher orders in $g_3$, and the resummed result is the first-order expression with the coupling evaluated at the interparticle distance $L_\rho$. Thus
\begin{equation}
e(\rho)=\frac{\rho^2}{m^*}
\Big[C_2\,\delta u_1(L_\rho)+C_3\,g_3(L_\rho)
+\cdots\Big].
\label{eq:seft_eos_matched}
\end{equation}
 For an attractive triplet coupling this expansion breaks down when $|g_3(L_\rho)|$ becomes of order one, before the formal pole at $R_{\rm trimer}=l_{\rm pair}e^{-1/g_3^{(0)}}$. Density derivatives at fixed $l_{\rm pair}$, $\delta u_1$ and $g_3^{(0)}$ use $d\delta u_1(L_\rho)/d\ln \rho=-\delta u_1(L_\rho)/3$ and $dg_3(L_\rho)/d\ln\rho=g_3(L_\rho)^2/2$, which give
\begin{equation}
\begin{gathered}
\mu=e'(\rho)=\frac{\rho}{m^*}
\Big[\frac53C_2\,\delta u_1(L_\rho)+2C_3\,g_3(L_\rho)+\cdots\Big],\\
e''(\rho)=\frac1{m^*}
\Big[\frac{10}{9}C_2\,\delta u_1(L_\rho)+2C_3\,g_3(L_\rho)+\cdots\Big].
\end{gathered}
\label{eq:seft_eos_derivatives}
\end{equation}
The omitted terms are assumed subleading on the branch considered. The $O(g_3^2)$ term generated by differentiating the running coupling is of the same order as the omitted second-order energy and is dropped consistently.
 
\subsection{Density--phase action}
\label{sec:seft_dynamics}
 
Having determined the static correlations, we use the same Toda density source to construct the smooth charge dynamics. The physical time evolution is supplied by the factorized ACCS Hamiltonian of the main text. We work at wavelengths large compared with $L_\rho$, and focus only on the total density and the charge-$2e$ phase $\varphi$ modes. Relative-valley excitations will be discussed in Sec.~\ref{sec:seft_relative_relaxation}.

Apply a slowly varying real source $h(\mathbf x,t)$, common to all
valleys, and the phase $\varphi$ to the normalized AC state:
\begin{equation}
\tilde w[h,\varphi]=
\frac{\exp[\tfrac12\sum_a h(\boldsymbol\eta_a)
+\tfrac i3\sum_a\varphi(\boldsymbol\eta_a)]}
{\langle\exp[\sum_a h(\boldsymbol\eta_a)]\rangle_{\rm pl}^{1/2}}
\,\tilde w.
\label{eq:seft_deformed_state}
\end{equation}
Each anyon carries charge $2e/3$ and hence phase $\varphi/3$. The probability weight contains the same Toda source insertion $\exp[\sum_a h(\boldsymbol\eta_a)]$, now normalized at fixed $M$. Smooth common sources preserve the AC collision conditions. The homogeneous local response to $h$ is $d\rho/dh=\rho S_{\rm ACCS}(0^+)=3\rho/2$, and using
Eqs.~\eqref{eq:seft_toda_density_sources},\eqref{eq:seft_S_zero}, we find $\rho(\bm{r})=\rho e^{3h(\bm{r})/2}$ for a source varying slowly compared with the local screening length. Consequently $\nabla h(\bm{r})=2\nabla \rho(\bm{r})/(3\rho(\bm{r}))$ to leading order in gradients.

The real source drops out of the Berry term: its time derivative cancels the time derivative of the normalization in
Eq.~\eqref{eq:seft_deformed_state}. Thus
$i\langle\tilde w|\partial_t\tilde w\rangle
=-\int d^2r\,\rho(\bm{r})\,\partial_t\varphi (\bm{r})/3$.
For the energy, $\Pi_a^{(\rm IR)}\tilde w=0$ and the product rule give $\Pi_a^{(\rm IR)}\tilde w[h,\varphi]
=[\partial_{\bar\eta_a}h+(2i/3)\partial_{\bar\eta_a}\varphi]
\tilde w[h,\varphi]$. Substitution in $H_{\rm ACCS}=(2m^*)^{-1}\sum_a(\Pi_a^{(\rm IR)})^\dagger
\Pi_a^{(\rm IR)}$ then yields
\begin{equation}
\begin{aligned}
E[h,\varphi]&=\frac1{2m^*}\int d^2r\,\rho(\bm{r})
\Big[\frac{(\nabla h)^2}{4}+\frac{(\nabla\varphi)^2}{9}
-\frac13 (\nabla h\times\nabla\varphi)_z\Big]\\
&=\int d^2r\Big[\frac{(\nabla \rho)^2}{18m^*\rho (\bm{r})}
+\frac{\rho(\bm{r})(\nabla\varphi)^2}{18m^*}
-\frac{(\nabla \rho\times\nabla\varphi)_z}{9m^*}\Big]
+\cdots.
\end{aligned}
\label{eq:seft_energy_from_state}
\end{equation}
The second line here uses $\rho(\bm{r})=\rho e^{3h(\bm{r})/2}$. Adding the contact energy of Eq.~\eqref{eq:seft_eos_matched} gives
the leading smooth Lagrangian
\begin{equation}
\mathcal L=\rho(\bm{r})X-e(\rho(\bm{r}))-\frac{(\nabla \rho)^2}{18m^*\rho(\bm{r})}+\cdots,\qquad
X=\mu-\frac{\partial_t\varphi}{3}-\frac{(\nabla\varphi)^2}{18m^*}.
\label{eq:seft_nonlinear_action}
\end{equation}
We have omitted the term $(\nabla\rho\times\nabla\varphi)_z/(9m^*)$ because it is a spatial total derivative for smooth single-valued $\varphi$ (we ignore vortex configurations for now).
For $e''>0$ and at wavelengths where the density gradient is negligible, $e'(\rho)=X$ determines $\rho(X)$ and Eq.~\eqref{eq:seft_nonlinear_action} becomes 
\begin{equation}
\mathcal L_\varphi=P(X),\qquad
P(X)=X\rho(X)-e(\rho(X)),\qquad
P'(\mu)=\rho,\qquad P''(\mu)=\frac1{e''(\rho)},
\label{eq:seft_pressure_action}
\end{equation}
 which is analogous to the superfluid EFT construction of Ref.~\cite{SON2006197}. For repulsive pair detuning and $g_3^{(0)}=0$, with $r_1=l_{\rm pair}(\delta u_1)^{-3/2}$, the energy is $e(\rho)=C_2\rho^{5/3}/(m^*r_1^{2/3})$ and
\begin{equation}
P(X)=\frac25\Big(\frac{3}{5C_2}\Big)^{3/2}(m^*)^{3/2}\,r_1X^{5/2},\qquad X>0.
\label{eq:seft_pair_pressure}
\end{equation}
The exponent is that of the three-dimensional unitary Fermi gas. Here it follows from the anyonic pair exponent $2/3$ and the prefactor retains $r_1$. At pair resonance with weak repulsive $g_3^{(0)}$ the leading relations are $X=2C_3\rho(X)g_3(L_{\rho(X)})/m^*$ and $P(X)=m^*X^2/[4C_3g_3(L_{\rho(X)})]$, the form of the dilute two-dimensional Bose gas with the running three-body coupling in place of the contact. At the AC point $e(\rho)=0$ and the density gradient of Eq.~\eqref{eq:seft_nonlinear_action} must be kept.
We also define $\rho(r)=\rho+\delta \rho(r)$, where $\delta\rho(r)$ is the fluctuating density. At $\mu=e'(\rho)$ the linear term vanishes. The quadratic action in Euclidean time is
\begin{equation}
S=\int d\tau\,d^2x\left[
\frac{i\delta\rho(r)}{3}\partial_\tau\varphi
+\frac{e''(\rho)}2\delta\rho^2(r)
+\frac{(\nabla \delta\rho)^2}{18m^*\rho}
+\frac{\rho(\nabla\varphi)^2}{18m^*}\right].
\label{eq:seft_quadratic_redline}
\end{equation}
Evaluating the correlators $G_{\varphi\varphi}^{}$ and $G_{\rho\rho}^{}$ gives the results stated in the main text. This action also reproduces $S_{\rm ACCS}(k\rightarrow 0)=3/2$.

For the electromagnetic response we use the substitution $\Pi_a^{(\rm IR)}\mapsto\Pi_a^{(\rm IR)}-i(2e/3)(A_x+iA_y)$. In Eq.~\eqref{eq:seft_energy_from_state} this replaces $\nabla\varphi$ by $\nabla\varphi-2e\mathbf A$. The cross term, which was a boundary term at zero field, now contributes $[2e/(9m^*)]\int (\nabla \delta\rho\times\mathbf A)_z=-[2e/(9m^*)]\int_r \delta\rho(r) B(r)$, with $B=\partial_xA_y-\partial_yA_x$. For a static transverse field the quadratic energy is therefore
\begin{equation}
E=\int d^2x\left[
\frac{e''(\rho)}2 \delta\rho^2(r)
+\frac{(\nabla \delta\rho)^2}{18m^*\rho}
+\frac{2e^2\rho}{9m^*}|\mathbf A_T|^2
-\frac{2e}{9m^*}\delta \rho(r) B\right].
\label{eq:seft_orbital_energy_redline}
\end{equation}
Minimizing over $\delta\rho(r)$ and using $|B_{\mathbf k}|^2=k^2|\mathbf A_{T,\mathbf k}|^2$ gives the coefficient of $|\mathbf A_T|^2/2$,
\begin{equation}
K_T(k)=\frac{4e^2\rho}{9m^*}
-\frac{(2e/9m^*)^2k^2}{e''(\rho)+k^2/(9m^*\rho)}
=\frac{4e^2\rho}{9m^*}\frac{k_*^2}{k^2+k_*^2},\qquad  k_*=3\sqrt{m^*\rho e''(\rho)}\;.
\label{eq:seft_magnetic_redline}
\end{equation}
For $e''>0$, $K_T(0)=4e^2J_s$. At AC, density rearrangement cancels the static cost at nonzero $k$, while a uniform twist still has stiffness $J_s=\rho/(9m^*)$.

\subsection{Relative-valley excitations}
\label{sec:seft_relative_relaxation}

So far the effective description has involved only the total density
and its conjugate phase. Another type of excitation, specific to the
anyon problem, is the relative-valley density mode, which changes the
valley composition while preserving the total anyon number. The total
density--phase description is therefore justified only at scales below
the gap of this mode. The standard Chern--Simons mean-field theory
gives a gap of order $\rho/m^*$, which by scale invariance is the
only energy scale at $\delta u_1=g_3=0$. However, special care is
required near the AC point, because a relative-valley excitation can
lower its energy by depleting the total density around it. At the AC
point such a depletion has no bulk energy cost. A finite $g_3>0$ restores a finite
compressibility and cuts off the resulting divergence at the healing momentum
$k_*\propto\sqrt{g_3\rho}$. As we show below using RG, this effect reduces the mean-field gap by a factor of order
$1/\ln[1/g_3(L_\rho)]\ll1$. We also corroborate this conclusion using a variational argument in Sec.~\ref{sec:seft_relative_variational}. Nevertheless, the gap remains
parametrically larger than the sound crossover energy
$ck_*\propto g_3 \rho/m^*$, so the analysis of the sound regime
above is unaffected.

We consider the pair resonance $\delta u_1=0$, with $g_3=g_3(L_\rho)>0$, and $L_\rho=\rho^{-1/2}$. 
The relative valley mode can be described by a complex field $\psi_v$,
and the leading action for the density $\rho$, phase $\varphi$, and
$\psi_v$ becomes
\begin{equation}
\begin{aligned}
 S=\int d\tau\,d^2x\bigg[&
 \frac{i\rho(r)}{3}\partial_\tau\varphi
 +\frac{\rho(r)}{18m^*}(\nabla\varphi)^2
 +\frac{(\nabla \rho)^2}{18m^*\rho(r)}+e(\rho(r))-\mu \rho(r)\\
 &+\psi_v^\dagger\left(\partial_\tau-\frac{\nabla^2}{2m_v}
                       +\frac{\gamma_v \rho(r)}{m^*}\right)\psi_v
 \bigg],\qquad
 e(\rho)=\frac{C_3g_3}{m^*}\rho^2.
\end{aligned}
\label{eq:seft_relative_local_action}
\end{equation}
As before, the chemical potential is $\mu=e'(\rho)$.
At the matching momentum $\Lambda\sim 1/L_\rho$, $\gamma_v$ and
$m_v$ are finite positive parameters describing the relative excitation
before relaxation of the long-wavelength density. Their Chern--Simons
mean-field values are $2\pi/3$ and $m^*$. Scale invariance fixes the local
gap to $\gamma_v \rho(r)/m^*$, which gives the density coupling in
Eq.~\eqref{eq:seft_relative_local_action}. 

We set $\rho(r)=\rho+\delta\rho(r)$ and $\Delta_0=\gamma_v\rho/m^*$.
The quadratic action and the cubic vertices needed below are
\begin{equation}
\begin{aligned}
 S_2=\int d\tau\,d^2x\bigg[&
 \frac{i\delta\rho}{3}\partial_\tau\varphi
 +\frac{\rho}{18m^*}(\nabla\varphi)^2
 +\frac{(\nabla \delta\rho)^2}{18m^*\rho}
 +\frac{e''(\rho)}2(\delta\rho)^2+\psi_v^\dagger\left(\partial_\tau-\frac{\nabla^2}{2m_v}
                         +\Delta_0\right)\psi_v\bigg],\\
 S_3=\int d\tau\,d^2x\bigg[&
 \frac{\delta\rho(\nabla\varphi)^2}{18m^*}
 -\frac{\delta\rho(\nabla \delta\rho)^2}{18m^*\rho^2}
 +\frac{e'''(\rho)}6\delta\rho^3
 +\frac{\gamma_v}{m^*}\delta\rho|\psi_v|^2\bigg].
\end{aligned}
\label{eq:seft_relative_expanded_action}
\end{equation}
Eliminating the phase at quadratic order gives
\begin{equation}
 G_{\rho\rho}(\Omega,q)
 =\frac{\rho q^2/m^*}{\Omega^2+\omega_q^2},\qquad
 \omega_q=\frac{q\sqrt{q^2+k_*^2}}{3m^*},\qquad
 k_*^2=9m^*\rho e''(\rho)\approx18C_3g_3\rho.
\label{eq:seft_relative_density_propagator}
\end{equation}
The same elimination determines the cubic density vertex.
Use $Q=(\Omega,\mathbf q)$,
$\int_Q\equiv\int d\Omega\,d^2q/(2\pi)^3$, and the Fourier convention
$\delta\rho(\tau,\mathbf x)=
\int_Qe^{i\mathbf q\cdot\mathbf x-i\Omega\tau}\delta \rho(Q)$.
The linear phase equation and its momentum-space solution are
$-\frac{i}{3}\partial_\tau \delta\rho
-\frac{\rho}{9m^*}\nabla^2\varphi=0$ and
$\varphi^{(1)}(Q)=\frac{3m^*\Omega}{\rho q^2}\delta\rho(Q)$.
Substitute this solution into the charge terms of $S_3$.
The quadratic correction to $\varphi$ makes no contribution through
$S_2$, since its first variation already vanishes. The resulting
three-density interaction is
\begin{equation}
 S_{3\rho}={}\int_Q\int_{Q'}
 \delta\rho(Q) \delta\rho(Q') \delta\rho(-Q-Q')\left[
 -\frac{m^*\Omega\Omega'\,\mathbf q\cdot\mathbf q'}
              {2\rho^2q^2q'^2}
 +\frac{\mathbf q\cdot\mathbf q'}{18m^*\rho^2}
 +\frac{e'''(\rho)}6\right].
\label{eq:seft_relative_cubic_density_action}
\end{equation}
Here $Q'=(\Omega',\mathbf q')$. The first term is the dynamical
contribution from eliminating the phase. The third density factor
ensures momentum and frequency conservation.

The one-loop self-energy has two $\delta\rho|\psi_v|^2$ vertices joined by
one density line and one relative line. The bare Green's function is
$G_v(\Omega,\mathbf p)=[i\Omega-\Delta_0-p^2/(2m_v)]^{-1}$,
with the self-energy contributing to its inverse as
$G_v^{-1}-\Sigma_v$. 
The corresponding diagram is
\begin{equation}
 \Sigma_v(\Omega,\mathbf p)
 =\frac{\gamma_v^2}{m^{*2}}
 \int\frac{d^2q}{(2\pi)^2}\frac{d\Omega'}{2\pi}
 \frac{\rho q^2/m^*}{\Omega'^2+\omega_q^2}
 \frac{1}{i\Omega-i\Omega'-\Delta_0
                       -(\mathbf p-\mathbf q)^2/(2m_v)}.
\label{eq:seft_relative_frequency_integral}
\end{equation}

We integrate thin momentum shells $ke^{-ds}<q<k$, with
$s=\ln(\Lambda/k)$ and $k_*\ll k\ll\Lambda$. In this range,
$\omega_q\simeq q^2/(3m^*)$.
From here $\gamma_v,m_v$ and $\Delta_v$ denote running coefficients.
After each shell, we retain the local density coupling and the
quadratic recoil term. The gap estimate below is obtained within
this local one-loop approximation.
At zero external momentum the shell self-energy and its on-pole
frequency derivative are
\begin{equation}
\begin{aligned}
 d\Delta_v\equiv d\Sigma_v(-i\Delta_v,0)
 &=-\frac{3\rho\gamma_v^2}{2m^{*2}}
 \int_{\rm shell}\frac{d^2q}{(2\pi)^2A_q},\\
 -i\partial_\Omega d\Sigma_v(-i\Delta_v,0)
 &=-\frac{3\rho\gamma_v^2}{2m^{*2}}
 \int_{\rm shell}\frac{d^2q}{(2\pi)^2A_q^2},
\end{aligned}
\label{eq:seft_relative_shell_energy}
\end{equation}
where we defined $A_q=q^2/(3m^*)+q^2/(2m_v)$, and the correction to the gap, $d\Delta_v$, is determined from the real part of the on-shell self-energy. The frequency derivative determines the residue $Z_v=(1+i\partial_\Omega d\Sigma_v)^{-1}$. 

There are two one-loop corrections to the coupling
$\rho|\psi_v|^2$.
First, the external density can attach to the internal density line
through Eq.~\eqref{eq:seft_relative_cubic_density_action}.
The external density leg is static, with its frequency set to zero
before its momentum. Omitting the subleading $e'''(\rho)$
term, we find
\begin{equation}
\begin{aligned}
 d\Gamma_{\rho |\psi|^2}^{(3\rho)}(\Omega)
 ={}&\frac{\gamma_v^2}{m^{*2}}
 \int_{\rm shell}\frac{d^2q}{(2\pi)^2}
 \int\frac{d\Omega'}{2\pi}
 \left[\frac{m^*\Omega'^2}{\rho^2q^2}
                   +\frac{q^2}{9m^* \rho^2}\right]\\
 &\times
 \frac{(\rho q^2/m^*)^2}{(\Omega'^2+\omega_q^2)^2}
 \frac1{i\Omega-i\Omega'-\Delta_v-q^2/(2m_v)},\\
 d\Gamma_{\rho |\psi|^2}^{(3\rho)}(-i\Delta_v)
 ={}&-\frac{3\gamma_v^2}{2m^{*2}}
 \int_{\rm shell}\frac{d^2q}{(2\pi)^2A_q}.
\end{aligned}
\label{eq:seft_relative_density_line_diagram}
\end{equation}
Second, the usual triangle attaches the external density to the
internal relative line. It contains three density--relative vertices,
one density propagator and two relative propagators:
\begin{equation}
\begin{aligned}
 d\Gamma_{\rho |\psi|^2}^{(\triangle)}(\Omega)
 &=\frac{\gamma_v^3}{m^{*3}}
 \int_{\rm shell}\frac{d^2q}{(2\pi)^2}
 \int\frac{d\Omega'}{2\pi}\,
 \frac{\rho q^2/m^*}{\Omega'^2+\omega_q^2}
 \frac{1}{
 \left[i\Omega-i\Omega'-\Delta_v-q^2/(2m_v)\right]^2},\\
 d\Gamma_{\rho |\psi|^2}^{(\triangle)}(-i\Delta_v)
 &=\frac{i\gamma_v}{m^*}\partial_\Omega d\Sigma_v.
\end{aligned}
\label{eq:seft_relative_triangle_diagram}
\end{equation}

Now we are in a position to obtain the full $\rho |\psi_v|^2$ vertex renormalization, which combines $d \Gamma_{\rho |\psi|^2}^{(3\rho)}$, $d \Gamma_{\rho |\psi|^2}^{(\triangle)}$, and the residue $Z_v$.
Each external relative leg contributes $\sqrt{Z_v}$, so we obtain
\begin{equation}
 \frac{\gamma_v+d\gamma_v}{m^*}
 =Z_v\left[\frac{\gamma_v}{m^*}
        +d\Gamma_{\rho |\psi|^2}^{(\triangle)}+d\Gamma_{\rho |\psi|^2}^{(3\rho)}\right]=\frac{\gamma_v}{m^*}+d\Gamma_{\rho |\psi|^2}^{(3\rho)}+O((ds)^2).
\label{eq:seft_relative_vertex_cancellation}
\end{equation}
The triangle diagram cancels the tree vertex times the residue correction,
including its entire $A_q^{-2}$ integral (consequence of a Ward identity).
The remaining shell integral is
\begin{equation}
 \int_{\rm shell}\frac{d^2q}{(2\pi)^2A_q}
 =\frac{3m^*}{2\pi\left(1+3m^*/(2m_v)\right)}\,ds,
 \qquad
 \frac{d\gamma_v}{ds}
 =-\frac{9\gamma_v^2}{4\pi\left(1+3m^*/(2m_v)\right)}.
\label{eq:seft_relative_density_vertex_flow}
\end{equation}
Crucially, the three-density vertex is thus the only source of running of $\gamma_v$.
Comparison with Eq.~\eqref{eq:seft_relative_shell_energy} then gives
$d\Delta_v=(\rho/m^*)d\gamma_v$. Starting from
$\Delta_v(0)=\rho\gamma_v(0)/m^*$, this implies
$\Delta_v(s)=\rho\gamma_v(s)/m^*$.

The momentum dependence of the same self-energy also renormalizes
the mass. Expanding to order $p^2$ and normalizing the frequency
term gives
\begin{equation}
 \frac{dm_v}{ds}
 =\frac{81m^*\rho\gamma_v^2}
 {4\pi k^2\left(1+3m^*/(2m_v)\right)^3},
 \qquad k=\Lambda e^{-s}.
\label{eq:seft_relative_recoil_flow}
\end{equation}
It describes the extra inertia of the density deformation that
accompanies the relative excitation. The mass is increasing, while
the vertex flow implies that $\gamma_v$ decreases no faster than $1/s$.
The mass equation therefore forces $m_v$ to grow at least as
$e^{2s}/s^2$. Recoil becomes negligible at large $s$, and integrating
the coupling flow gives
\begin{equation}
 \frac1{\gamma_v(s)}=\frac{9s}{4\pi}+O(1).
\label{eq:seft_relative_integrated_flow}
\end{equation}
The mass renormalization contribution changes only the constant. From Eq.~\eqref{eq:seft_relative_shell_energy}, we find $ d\Delta_v/ds= \rho(d\gamma_v/ds)/m^*$, and thus
\begin{equation}
 \Delta_v
 \approx\frac{\rho\gamma_v(s_*)}{m^*}
 =\frac{8\pi\rho}{9m^*\,[\ln(1/g_3)+O(1)]},
 \qquad
 \frac{\Delta_v}{ck_*}
 \sim\frac{1}{g_3\ln(1/g_3)}\longrightarrow\infty,
\label{eq:seft_relative_quantum_gap}
\end{equation}
where $s_*=\ln(\Lambda/k_*)$. Below $k_*$, the remaining energy
correction is $O[\rho\gamma_v(s_*)^2/m^*]$, smaller than the
retained gap by a factor $\gamma_v(s_*)$. The second relation in Eq.~\eqref{eq:seft_relative_quantum_gap} shows that the renormalized gap is still
parametrically separated from the sound crossover energy
$ck_*\simeq6C_3g_3\rho/m^*$. When the density logarithm dominates
the bare inverse coupling, accounting for logarithmic running of
$g_3$ gives $\Delta_v\propto\rho/
\{m^*\ln\ln[1/(\rho l_{\rm pair}^2)]\}$. Two additional comments about Eq.~\eqref{eq:seft_relative_quantum_gap} are in order. First, the RG flow only depends on $g_3$ via the IR cutoff at the healing length $1/k_*$. This is because the suppression of the gap is a property of the exact $\delta u_1=g_3=0$ AC point, while the explicit $g_3$-dependent corrections are weaker than the effect that we accounted for. Second, the inclusion of the three-density vertex was essential because without it $\gamma_v$ would not flow, and from Eq.~\eqref{eq:seft_relative_shell_energy} the gap would cross zero at a finite RG scale, indicating a breakdown of a perturbative one-loop RG.

\subsubsection{Variational argument}
\label{sec:seft_relative_variational}

The reduction of the relative-valley energy found above has a simple
physical interpretation. The relative-valley excitation can lower its energy by
occupying a region of reduced total density. At the AC point,
$e(\rho(r))=0$, so this depletion costs only gradient energy. We calculate
that cost directly from the AC wavefunction and then include the
restoring effect of $g_3>0$. 

As before,  the normalized valley-singlet AC state is denoted $\tilde w$, and
$\eta_a$ are the anyon coordinates. The first-order equations
$\Pi_a\tilde w=0$ give the energy of any trial state
$\mathcal{F}\tilde w$ as
\begin{equation}
 \frac{\langle \mathcal{F}\tilde w|H_{\rm ACCS}|\mathcal{F}\tilde w\rangle}
      {\langle \mathcal{F}\tilde w|\mathcal{F}\tilde w\rangle}
 =\frac{2}{m^*}
 \frac{\sum_a\langle|\partial_{\bar\eta_a}\mathcal{F}|^2\rangle_{\rm pl}}
      {\langle|\mathcal{F}|^2\rangle_{\rm pl}}.
 \label{eq:seft_variational_ac_form}
\end{equation}
Here $\mathcal{F}$ is a fixed function of all anyon coordinates and valley indices, which we now specify. First, we fix two length scales $R_{\rm core}\ll R_{\rm cloud}$ and
smooth, real, radially symmetric functions $f(r)$ and $h(r)$.
The function $f$ is supported inside $r<R_{\rm core}/2$.
The function $h$ is a negative constant for $r<R_{\rm core}$
and vanishes for $r>R_{\rm cloud}$. Our unnormalized variational state is then given by
\begin{equation}
\begin{aligned}
 \tilde w[f,h]&=\sigma_v[f]D[h]\tilde w,\qquad
 D[h]=\exp\!\left[\frac12\sum_a h(|\eta_a|)\right],\\
 \sigma_v[f]&=\int d^2r\,f(r)
                  \frac{\rho_1(\mathbf r)-\rho_2(\mathbf r)}{\sqrt2}=\frac1{\sqrt2}\sum_{a=1}^M
   \bigl(\delta_{\alpha_a,1}-\delta_{\alpha_a,2}\bigr)
   f(|\eta_a|).
\end{aligned}\label{eq:seft_variational_state}
\end{equation}
The scalar factor $D[h]$ depletes the total density in the core,
while $\sigma_v[f]$ creates a relative-valley excitation inside it. The factor $1/2$ in the support radius for $f$ simply leaves a buffer
between the excitation and the region where the density varies, so that $\nabla f$ and $\nabla h$ have non-overlapping supports.

To calculate the energy, we note that
\begin{equation}
 \partial_{\bar\eta_a}\bigl(\sigma_v[f]D[h]\bigr)
 =D[h]\left[
 \frac{\delta_{\alpha_a,1}-\delta_{\alpha_a,2}}{\sqrt2}\,
 \partial_{\bar\eta_a}f(|\eta_a|)
 +\frac{\sigma_v[f]}2\,
 \partial_{\bar\eta_a}h(|\eta_a|)
 \right].
\end{equation}
After squaring this expression to get Eq.~\eqref{eq:seft_variational_ac_form}, the cross term vanishes
because $\nabla f$ and $\nabla h$ have disjoint supports.
For a real function, $|\partial_{\bar\eta}f|^2=|\nabla f|^2/4$. Thus
\begin{equation}
 E[f,h]={}
 \frac{\displaystyle
 \left\langle D[h]^2
       \sum_a\bigl[f'(|\eta_a|)\bigr]^2\right\rangle_{\rm pl}}
 {6m^*\left\langle\sigma_v[f]^2D[h]^2\right\rangle_{\rm pl}}+
 \frac{\displaystyle
 \left\langle\sigma_v[f]^2D[h]^2
       \sum_a\bigl[h'(|\eta_a|)\bigr]^2\right\rangle_{\rm pl}}
 {8m^*\langle\sigma_v[f]^2D[h]^2\rangle_{\rm pl}}.
\label{eq:seft_variational_exact_energy}
\end{equation}
The two terms arise from differentiating the relative-valley
insertion and the density deformation, respectively. We evaluate the resulting averages by
assuming local equilibrium in the screened plasma.
The homogeneous density response to this weight is
$d\ln \rho(r)/dh(r)=S_{\rm ACCS}(0^+)=3/2$, so the leading local density is
$\rho e^{3h(r)/2}$. The surrounding fluid fixes the density
$\rho$ outside the cloud. We denote the lower core density by
$\tilde{\rho}=\rho e^{3h(0)/2}$ and choose $f$ to vary on the scale
$R_{\rm core}$, with $R_{\rm core}/L_{\tilde{\rho}}\sim R_{\rm core}\sqrt{\tilde{\rho}}\gg1$.
The first numerator then counts particles in an approximately
uniform core, which gives density $\tilde{\rho}$. Thus, we find
\begin{equation}
 \begin{aligned}
 \frac{\left\langle D[h]^2
       \sum_a[f'(|\eta_a|)]^2\right\rangle_{\rm pl}}
      {\langle D[h]^2\rangle_{\rm pl}}
 &\simeq \tilde{\rho}\int d^2r\,[f'(r)]^2,\\
 \frac{\left\langle\sigma_v[f]^2D[h]^2\right\rangle_{\rm pl}}
      {\langle D[h]^2\rangle_{\rm pl}}
 &\simeq\frac{1}{2}\int d^2r d^2r'\,f(r)f(r')
                  \left[\mathcal{C}_{11}(r-r')+\mathcal{C}_{22}(r-r')-2\mathcal{C}_{12}(r-r')\right]\\
                  &\simeq
 \int\frac{d^2k}{(2\pi)^2}\,|f_{\mathbf k}|^2\frac{k^2}{4\pi}
 =\frac1{4\pi}\int d^2r\,[f'(r)]^2.
 \end{aligned}
 \label{eq:seft_variational_core_averages}
\end{equation}
The omitted $k^4$ term produces a relative correction of order
$1/(\tilde{\rho} R_{\rm core}^2)$.

For the second numerator in Eq.~\eqref{eq:seft_variational_exact_energy}, $\sigma_v[f]$ depends only on particles
inside the core, whereas $h'$ is supported in the surrounding
annulus. Screening allows their averages to factorize when the
buffer is much wider than the local screening length. This gives
\begin{equation}
 \begin{aligned}
 &\frac{\left\langle\sigma_v[f]^2D[h]^2
       \sum_a[h'(|\eta_a|)]^2\right\rangle_{\rm pl}}
      {\left\langle\sigma_v[f]^2D[h]^2\right\rangle_{\rm pl}}\simeq
 \frac{\left\langle D[h]^2
       \sum_a[h'(|\eta_a|)]^2\right\rangle_{\rm pl}}
      {\langle D[h]^2\rangle_{\rm pl}}
 \simeq
\rho\int d^2r\,e^{3h(r)/2}[h'(r)]^2.
 \end{aligned}
 \label{eq:seft_variational_annulus_average}
\end{equation}
The first approximation neglects connected correlations between
the core and the annulus. The second replaces the particle sum
by its local-equilibrium average. Both require the plasma
reweighted by $D[h]^2$ to remain screened, with the profiles
varying slowly compared with the local screening length.
Substituting into Eq.~\eqref{eq:seft_variational_exact_energy},
the dependence on the shape and normalization of $f$ cancels
to leading order, leaving
\begin{equation}
 E[f,h]\simeq
 \frac{2\pi \tilde{\rho}}{3m^*}
 +\frac{\pi \rho}{4m^*}
 \int_{R_{\rm core}}^{R_{\rm cloud}}
 dr\,r\,e^{3h(r)/2}[h'(r)]^2.
 \label{eq:seft_variational_profile_energy}
\end{equation}

We first minimize the annulus energy at fixed $\tilde{\rho}$, with the boundary conditions
$h(R_{\rm core})=(2/3)\ln(\tilde{\rho}/\rho)$ (fixes the inner density to $\tilde{\rho}$) and
$h(R_{\rm cloud})=0$. Variation with respect to $h$ gives
$h''+h'/r+(3/4)(h')^2=0$, or equivalently
$\frac{d}{dr}[r\frac{d}{dr}e^{3h/4}]=0$.
Thus 
\begin{equation}
 e^{3h(r)/4}
 =\sqrt{\tilde{\rho}/\rho}
 +\left(1-\sqrt{\tilde{\rho}/\rho}\right)
 \frac{\ln(r/R_{\rm core})}
      {\ln(R_{\rm cloud}/R_{\rm core})},\qquad R_{\rm core}\leq r\leq R_{\rm cloud}.
 \label{eq:seft_variational_log_profile}
\end{equation}
Substituting this into
Eq.~\eqref{eq:seft_variational_profile_energy} gives
\begin{equation}
 E[f,h]\simeq
 \frac{2\pi\tilde{\rho}}{3m^*}
 +\frac{4\pi(\sqrt{\rho}-\sqrt{\tilde{\rho}})^2}
 {9m^*\ln(R_{\rm cloud}/R_{\rm core})}.
 \label{eq:seft_variational_core_cloud_energy}
\end{equation}
Lowering $\tilde{\rho}$
reduces the first term but increases the second. Minimizing
their sum with respect to $\tilde{\rho}$ gives
\begin{equation}
 \frac{\tilde{\rho}}{\rho}
 =\left[1+\frac32\ln\frac{R_{\rm cloud}}{R_{\rm core}}\right]^{-2},
 \qquad
 E\simeq
 \frac{4\pi \rho}
 {9m^*\left[\ln(R_{\rm cloud}/R_{\rm core})+2/3\right]}.
 \label{eq:seft_variational_ac_upper}
\end{equation}
At the AC point, the cloud can grow without a bulk density cost:
for each fixed ratio $R_{\rm cloud}/R_{\rm core}$, we first take
$R_{\rm core}\sqrt{\tilde{\rho}}\gg1$ and the sample much larger than
$R_{\rm cloud}$, so that the preceding plasma approximations
apply. Increasing the radius ratio then lowers both the optimal
core density and the trial energy. This suggests the valley-adjoint AC sector has a gap that approaches zero in the thermodynamic limit.

For $g_3>0$, the energy density acquires the following dependence
$e(\rho(r))=C_3g_3 \rho^2(r)/m^*+\cdots$, where $\rho(r)=\rho e^{3h(r)/2}$, and $g_3$ is evaluated at
$L_\rho$. The resulting bulk energy of the density depletion is
\begin{equation}
 E_{\rm contact}[h]
 =\int d^2r\,
 \left[e(\rho(r))-e(\rho)-e'(\rho)(\rho(r)-\rho)\right]
 \simeq\frac{C_3g_3 \rho^2}{m^*}\int d^2r\,[e^{3h(r)/2}-1]^2.
 \label{eq:seft_variational_contact_cost}
\end{equation}
The subtraction in Eq.~\eqref{eq:seft_variational_contact_cost} removes the energy of the uniform ground state with the same total number of particles.
 Minimizing the sum of
Eqs.~\eqref{eq:seft_variational_profile_energy}
and~\eqref{eq:seft_variational_contact_cost} gives a density recovery
length $k_*^{-1}=(18C_3g_3 \rho)^{-1/2}$, which sets $R_{\rm cloud}$.
We choose $R_{\rm core}=L_\rho[\ln(1/g_3)]^2$ so that the
locally uniform plasma calculation remains valid inside the
depleted core. Minimizing over the core density then gives
$\tilde{\rho}\sim \rho/[\ln(1/g_3)]^2$, and hence
$R_{\rm core}\sqrt{\tilde{\rho}}\sim\ln(1/g_3)\gg1$.
The contact cost inside the core is
$O(g_3 \rho[\ln(1/g_3)]^4/m^*)$, which is negligible compared
with the leading energy.
This choice of core radius enters the gradient energy through
$\ln(R_{\rm cloud}/R_{\rm core})
=\tfrac12\ln(1/g_3)-2\ln\ln(1/g_3)+O(1)$.
Its precise logarithmic growth therefore affects only subleading
terms. The resulting variational energy is
\begin{equation}
 E\simeq
 \frac{8\pi\rho}{9m^*\ln(1/g_3)}
 \left[
 1+O\!\left(
 \frac{\ln\ln(1/g_3)}{\ln(1/g_3)}
 \right)
 \right].
 \label{eq:seft_variational_finite_g}
\end{equation}
The repulsive contact therefore limits the size of the depletion
cloud and cuts off the logarithmic reduction of the relative-valley
energy. This result is consistent with our RG analysis in Eq.~\eqref{eq:seft_relative_quantum_gap}.

\subsection{Contact coefficients $C_2$ and $C_3$ from Toda theory}
\label{sec:seft_contact_evaluation}

The coefficients $C_2$ and $C_3$ defined in Eq.~\eqref{eq:seft_contact_definitions} can be calculated exactly. We evaluate them with the Toda theory representation of the plasma introduced in Sec.~\ref{sec:seft_structure}. One-point functions are $\langle \rho_\alpha\rangle=\zeta\langle V_\alpha\rangle_{\rm T}$. Two particles of different valleys fuse to the charge $\mathbf q_1+\mathbf q_2=-\mathbf q_3$ and one of each valley fuses to the identity. With $G_\pm=\langle:e^{\pm i\mathbf q_3\cdot\boldsymbol\phi}:\rangle_{\rm T}$,
\begin{equation}
\begin{gathered}
\rho=3\zeta G_+,\qquad
\langle \rho_1(\mathbf r)\rho_2(\mathbf 0)\rangle_{\rm pl}\approx \zeta^2G_-\,r^{-2/3},\\
\langle \rho_1(\mathbf r)\rho_2(\mathbf r')\rho_3(\mathbf 0)\rangle_{\rm pl}
\approx\frac{\zeta^3}{(|\mathbf r|\,|\mathbf r'|\,|\mathbf r-\mathbf r'|)^{2/3}},
\end{gathered}
\label{eq:seft_fusion_limits}
\end{equation}
where we used the standard fusion rules for normal ordered vertex operators
\begin{equation}
 V_\nu(\mathbf r)V_\alpha(\mathbf r')
 =
 |\mathbf r-\mathbf r'|^{2\mathsf B_{\nu\alpha}}
 :e^{ i\mathbf q_\nu\cdot\boldsymbol\phi(\mathbf r)
 +i\mathbf q_\alpha\cdot\boldsymbol\phi(\mathbf r')}:.
 \label{eq:pairing_vertex_product}
\end{equation}

The remaining unknown correlators $G_{\pm}$ are given by the reflection relation of the Toda theory and its minimal meromorphic solution~\cite{Fateev2002}. We summarize the argument of Ref.~\cite{Fateev2002} in the present notation. Let $G(\mathbf p)$ be the normalized average of an insertion whose weight with a valley-$\alpha$ particle is $r^{2p_{\alpha}}$, with $\sum_{\alpha}p_{\alpha}=0$ and $G(\mathbf 0)=1$. An insertion of vertex charge $\mathbf b$ has $p_{\alpha}=\mathbf b\cdot\mathbf q_{\alpha}$ and $\mathbf b^2=\sum_{\alpha}p_{\alpha}^2$. With $\sigma=p_1+2/3$ and $\gamma=\Gamma(2/3)/\Gamma(1/3)$ the reflection relation reads
\begin{equation}
\begin{gathered}
\mathbf p'=(-p_1-\tfrac43,\ p_2+p_1+\tfrac23,\ p_3+p_1+\tfrac23),\\
\frac{G(\mathbf p)}{G(\mathbf p')}
=(9\pi\zeta\gamma)^{-3\sigma}
\frac{\Gamma(1+3\sigma)\Gamma(1-\sigma)}{\Gamma(1-3\sigma)\Gamma(1+\sigma)},
\label{eq:seft_scalar_reflection}
\end{gathered}
\end{equation}
and the other reflections follow by permuting valleys. Its normalization can be checked at $p_1=-1$. There the insertion fused with one valley-1 particle has exactly the charges $\mathbf p'$, so that $G(\mathbf p)\simeq G(\mathbf p')\,\zeta\int d^2r\,r^{2p_1}$ near this point, and both sides of Eq.~\eqref{eq:seft_scalar_reflection} have a simple pole with residue $\pi\zeta$, from the collapsing integral on the left and from $\Gamma(1+3\sigma)$ on the right.
 
To solve it, use $\ln\Gamma(1+x)-\ln\Gamma(1-x)=\int_0^\infty\frac{ds}{s}[2xe^{-s}-2\sinh(xs)/(e^s-1)]$ with $s=2t/3$ and $s=2t$ in the two Gamma ratios, and $\int_0^\infty(e^{-at}-e^{-bt})\,dt/t=\ln(b/a)$. This gives
\begin{equation}
\ln\frac{G(\mathbf p)}{G(\mathbf p')}
=-3\sigma\ln(\pi\zeta\gamma)
+\int_0^\infty\frac{dt}{t}
\Big[4\sigma e^{-2t}
-\frac{\sinh(2t/3)\sinh(2\sigma t)}{\sinh t\,\sinh(t/3)}\Big],
\label{eq:seft_integral_reflection}
\end{equation}
first for $|\operatorname{Re}\sigma|<1/3$ and then by continuation. We then define the kernel
\begin{equation}
F(\mathbf p,t)=
\frac{\sinh(2t/3)\big[\sum_{\alpha}\cosh((2p_{\alpha}-2/3)t)-3\cosh(2t/3)\big]}
{2\sinh t\,\sinh(t/3)\sinh(2t)}.
\label{eq:seft_constructed_kernel}
\end{equation}
The reflection exchanges two of the three cosines and changes the third by $-2\sinh(2\sigma t)\sinh(2t)$, while $\sum_{\alpha}(p_{\alpha}')^2-\sum_{\alpha}p_{\alpha}^2=4\sigma$. Therefore
\begin{equation}
\ln G(\mathbf p)=\frac34\sum_{\alpha}p_{\alpha}^2\,\ln(\pi\zeta\gamma)
-\int_0^\infty\frac{dt}{t}\Big[\sum_{\alpha}p_{\alpha}^2\,e^{-2t}-F(\mathbf p,t)\Big]
\label{eq:seft_constructed_vev}
\end{equation}
solves Eq.~\eqref{eq:seft_integral_reflection} and its permutations. The integrand is regular at $t\to0$, since $F=\sum_{\alpha}p_{\alpha}^2+O(t^2)$, and converges for $-1<\operatorname{Re}p_{\alpha}<5/3$. The minimal-solution prescription selects this solution among its products with reflection-invariant functions.
 
For $\mathbf p_\pm=\pm(-\tfrac13,-\tfrac13,\tfrac23)$, with $\sum_{\alpha}p_{\alpha}^2=2/3$, the kernel reduces to
\begin{equation}
\begin{gathered}
G_\pm=(\pi\zeta\gamma)^{1/2}
\exp\Big[-\int_0^\infty\frac{dt}{t}
\Big(\frac23e^{-2t}-f_\pm(t)\Big)\Big],\\
f_+(t)=\frac{2\sinh(2t/3)}{\sinh(2t)},\qquad
f_-(t)=\frac{\sinh(2t/3)\big[\sinh(2t/3)\sinh(4t/3)-2\sinh^2(t/3)\big]}
{\sinh t\,\sinh(t/3)\sinh(2t)}.
\end{gathered}
\label{eq:seft_toda_vev}
\end{equation}
The plus integral equals $\frac23\ln2+2\ln\gamma$, by the substitution $s=4t$, the integral representation of $\ln\gamma$ and Frullani's identity. Hence
\begin{equation}
G_+=2^{-2/3}\sqrt{\pi\zeta}\,\gamma^{-3/2},\qquad
\zeta=\frac{2^{4/9}\gamma}{3^{2/3}\pi^{1/3}}\,\rho^{2/3}.
\label{eq:seft_density_fugacity}
\end{equation}
This fixes the auxiliary activity at the prescribed density. For the pair contact, the ratio $G_-/G_+$ with $x=e^{-2t/3}$ gives
\begin{equation}
\begin{gathered}
\mathcal J\equiv\ln\frac{G_-}{G_+}
=-\int_0^1\frac{(1-x^2)^2\,dx}{\ln x\,(1+x+x^2)(1+x^2+x^4)}=0.2802122813\ldots,\\
C_2=\frac{4\cdot2^{4/9}\pi^{2/3}}{3^{5/3}}\,\gamma\,e^{\mathcal J}
=1.2516146782\ldots,
\end{gathered}
\label{eq:seft_C2_exact}
\end{equation}
obtained by inserting $\zeta^2G_-$ into the definition of $C_2$ and using $\rho=3\zeta G_+$ and Eq.~\eqref{eq:seft_density_fugacity}.
 
For $C_3$ only a geometric integral remains, since by Eq.~\eqref{eq:seft_fusion_limits} the triplet correlator is $\zeta^3$ times a fixed function. Write the triplet coordinates as $z=r_x+ir_y$ and $z'=r'_x+ir'_y$ and set $\xi_1=z/\sqrt2$ and $\xi_2=(2z'-z)/\sqrt6$, the Jacobi coordinates of Sec.~\ref{sec:sm_higher_body} for particles ordered as $0,z,z'$, so that $d^2r\,d^2r'=3\,d^2\xi_1d^2\xi_2$ and $R^2=|\xi_1|^2+|\xi_2|^2$. The unitary rotation $u=(\xi_1+i\xi_2)/\sqrt2$, $v=(\xi_1-i\xi_2)/\sqrt2$ gives $zz'(z-z')=u^3+v^3$. With $u=R\cos\vartheta\,e^{i\chi_1}$ and $v=R\sin\vartheta\,e^{i\chi_2}$, where $\vartheta$ is the angle in this rotated Jacobi frame, the measure is $R^3\sin\vartheta\cos\vartheta\,dR\,d\vartheta\,d\chi_1d\chi_2$, the radial integral gives $\epsilon^2/2$, and the angular average $\frac1{2\pi}\int_0^{2\pi}|1+te^{i\chi}|^{-2/3}d\chi={}_2F_1(\tfrac13,\tfrac13;1;t^2)$ follows from the two binomial series. Using $u\leftrightarrow v$ to restrict $\vartheta\le\pi/4$ and setting $x=\tan^2\vartheta$,
\begin{equation}
\begin{aligned}
\frac1{\epsilon^2}\int_{R<\epsilon}
\frac{d^2r\,d^2r'}{(|\mathbf r|\,|\mathbf r'|\,|\mathbf r-\mathbf r'|)^{2/3}}
&=6\pi^2\int_0^1\frac{{}_2F_1(\tfrac13,\tfrac13;1;x^3)}{1+x}\,dx,\\
C_3&=\frac{2^{7/3}\pi}{3}\,\gamma^3
\int_0^1\frac{{}_2F_1(\tfrac13,\tfrac13;1;x^3)}{1+x}\,dx
=0.4887669732\ldots.
\end{aligned}
\label{eq:seft_C3_exact}
\end{equation}
The second line uses the triplet coefficient $\zeta^3$ and Eq.~\eqref{eq:seft_density_fugacity}.

Using these results, we can also calculate the expectation values
of the dressed trimer insertions associated with the
$\ell=9$ and $\ell=11$ electronic Cooper pairs.
To this end, we set the trimer center to $R=0$ and denote the position
of the inserted anyon in valley $\alpha$ as  $u_\alpha$, with $\sum_\alpha  u_\alpha=0 $.
The spatial extent of the trimer, $d_{\rm trimer}$, is fixed by the electronic projection, and we assume
with $\rho d_{\rm trimer}^2\ll1$.
The corresponding plasma operators are
\begin{equation}
 \mathcal O_9(\mathbf u)
 =\prod_a \mathcal W_{\gamma_a}(\eta_a,\mathbf u),\qquad
 \mathcal O_{11}(\mathbf u)=\mathcal O_9(\mathbf u)\sum_a\frac1{\eta_a^2}
 =\mathcal O_9(\mathbf u)
   \int\frac{d^2\eta}{\eta^2}\,
        \rho(\boldsymbol\eta),
 \label{eq:pairing_plasma_operators}
\end{equation}
where $\gamma_a$ is the valley of background particle $a$.
We also defined
\begin{equation}
 \mathcal W_\nu(\bm{\eta},\mathbf u)
 =\prod_{\alpha=1}^3
   (\bar\eta-\bar u_\alpha)^{\mathsf B_{\nu\alpha}}
 =\frac{\bar\eta-\bar u_\nu}
 {\bigl[\prod_{\alpha=1}^3
       (\bar\eta-\bar u_\alpha)\bigr]^{1/3}},
 \label{eq:pairing_spectator_factor}
\end{equation}
and $\mathcal W_\nu\to1$ at
infinity. Multiplying the weight of each valley-$\alpha$ particle by
$\mathcal W_\alpha$ changes its local activity from $\zeta$ to
$\zeta \mathcal W_\alpha$ (cf. Eq.~\eqref{eq:seft_toda_density_sources},\eqref{eq:useful_identities_Toda}). The Toda representation therefore gives
\begin{equation}
 \begin{aligned}
 \langle\mathcal O_9(\mathbf u)\rangle_{\rm pl}
 &=
 \left\langle
 \exp\left[
 \zeta\sum_\alpha\int d^2x\,
       (\mathcal W_\alpha(x,\mathbf u)-1)V_\alpha(x)
 \right]\right\rangle_{\rm T},
 \\
 \langle \rho_\nu(\bm{\eta})\mathcal O_9(\mathbf u)\rangle_{\rm pl}
 &=
 \zeta \mathcal W_\nu(\bm{\eta},\mathbf u)
 \left\langle V_\nu(\eta)
 \exp\left[
 \zeta\sum_\alpha\int d^2x\,
       (\mathcal W_\alpha(x,\mathbf u)-1)V_\alpha(x)
 \right]\right\rangle_{\rm T}.
 \end{aligned}
 \label{eq:pairing_toda_insertions}
\end{equation}
We then expand the exponential while retaining the
interacting expectation values of its vertex operators. Using $\zeta\langle V_\nu\rangle_{\rm T}=\rho/3$ gives
\begin{equation}
 \begin{aligned}
 \langle\mathcal O_9(\mathbf u)\rangle_{\rm pl}
 &=1+\frac{\rho}{3}
       \sum_\nu\int d^2\eta\,
          [\mathcal W_\nu(\eta,\mathbf u)-1]+\cdots,
 \\
 \langle \rho_\nu(\bm{\eta})\mathcal O_9(\mathbf u)\rangle_{\rm pl}
 &=\frac{\rho}{3}\mathcal W_\nu(\bm{\eta},\mathbf u)+\cdots.
 \end{aligned}
\label{eq:pairing_first_density_terms}
\end{equation}
The first local intervalley corrections follow from
$\zeta^2\langle V_\alpha(x)V_\beta(y)\rangle_{\rm T}
\simeq C_2\rho^{5/3}/(4\pi|x-y|^{2/3})$,
valid for $\alpha\ne\beta$ and
$|x-y|\sqrt{\rho}\ll1$.
Integrating over a region of size $d_{\rm trimer}$ gives corrections
of order $\rho^{5/3}d_{\rm trimer}^{10/3}$ to
$\langle\mathcal O_9\rangle_{\rm pl}$ and
$\rho^{5/3}d_{\rm trimer}^{4/3}$ to
$\langle\mathcal O_{11}\rangle_{\rm pl}$. 
Therefore, at leading order in density, we find
\begin{equation}
\begin{aligned}
    \langle\mathcal O_9(\mathbf u)\rangle_{\rm pl}
&=1+\rho I_9+\cdots,\qquad I_9=\frac13\int d^2\eta\,
\left[\sum_\nu \mathcal W_\nu(\eta,\mathbf u)-3\right]\\
 \langle\mathcal O_{11}(\mathbf u)\rangle_{\rm pl}
 &=\rho I_{11}+\cdots,
 \qquad I_{11}=\frac13\int \frac{d^2\eta}{\eta^2}\,
\sum_\nu \mathcal W_\nu(\eta,\mathbf u).
\end{aligned}
\label{eq:pairing_insertion_integrals}
\end{equation}
Here the integrals are assumed to be regulated in a rotationally-symmetric way on the scale of $d_{\rm trimer}$. It long-distance behavior is convergent since
\begin{equation}
 \sum_\nu \mathcal W_\nu(\eta,\mathbf u)
 =
 \frac{3\bar\eta-\sum_\alpha\bar u_\alpha}
 {\bigl[\prod_\alpha
       (\bar\eta-\bar u_\alpha)\bigr]^{1/3}}
 =
 3+O\left(\frac{d_{\rm trimer}^2}{\bar\eta^2}\right),
 \label{eq:pairing_far_field}
\end{equation}
The constant term has zero angular momentum, so it integrates to zero against
$\eta^{-2}$, while the remaining integrand decays as $d^2_{\rm trimer}/|\eta|^4$.

\end{document}